%% file: main.tex
\def\wpesversion{1}

\ifnum\wpesversion=0
\documentclass{article}
\fi

\ifnum\wpesversion=1
\documentclass[sigconf]{acmart}
\fi

\usepackage{graphicx} 
\usepackage{url}
\usepackage{hyperref}
\usepackage{amscd}
\usepackage{amsfonts}
\usepackage{amsmath}
\usepackage{enumitem}
\usepackage{framed}
\usepackage{xcolor}
\usepackage{xspace}
\usepackage{totcount}
\usepackage{siunitx}
\usepackage{booktabs}
\usepackage{tabularx}

\ifnum\wpesversion=1
\input{WPES0Abstract}
\fi

\title{Outfox: a Postquantum Packet Format for Layered Mixnets}

\ifnum\wpesversion=1
\author{Alfredo Rial}
\orcid{0000-0003-1107-4841}
\affiliation{\institution{Nym Technologies}
\country{}
}
\email{alfredo@nymtech.net}

\author{Ania M. Piotrowska}
\orcid{}
\affiliation{\institution{Nym Technologies}
\country{}
}
\email{ania@nymtech.net}

\author{Harry Halpin}
\orcid{}
\affiliation{\institution{Nym Technologies}
\country{}
}
\email{harry@nymtech.net}
\fi

\input{macros}

\begin{document}

\maketitle

\ifnum\wpesversion=0
\input{0Abstract}
\fi


\ifnum\wpesversion=0
\newpage
\tableofcontents
\newpage
\fi

\ifnum\wpesversion=0
\input{1introduction}

\input{2Background}

\input{3SecurityFramework}

\input{4SecDefs}

\input{5Preliminaries}

\input{6OutfoxPacketFormat}

\input{7SecAnalysis}

\input{8Efficiency}

\input{9Conclusion}

\input{main.bbl}

\fi

\ifnum\wpesversion=1
\input{WPES1introduction}
\input{WPES1introduction2OurContribution}
\input{WPES1introduction3Outline}
\input{WPES2systemmodel}

\input{WPES2systemmodel1Parties}
\input{WPES2systemmodel2Phases}
\input{WPES2systemmodel3SecurityProperties}

\input{WPES3construction}

\input{WPES3construction1buildingblocks}
\input{WPES3construction2Outfox}

\input{WPES3construction3Protocol}

\input{WPES4securityanalysis}
\input{WPES5efficiencyanalysis}
\input{WPES5efficiencyanalysis1Outfox}
\input{WPES5efficiencyanalysis2Sphinxcomparison}

\input{WPES6Conclusion}


\input{main.bbl}
\appendix

\input{3SecurityFramework}
\input{4SecDefs}
\input{5Preliminaries}
\input{6OutfoxPacketFormat}
\input{7SecAnalysis}
\fi

\end{document}

%% file: WPES0Abstract.tex
\begin{abstract}
Anonymous communication relies on encrypted packet formats that resist traffic analysis and ensure unlinkability. Sphinx, the current standard for mixnets, provides strong anonymity but relies on classical public-key cryptography, making it vulnerable to quantum attacks. In this paper, we present Outfox, a simplified variant of Sphinx tailored for mixnets with fixed-length routes and designed for post-quantum security. Outfox eliminates a full key exchange and introduces a compact, per-hop header structure, reducing both computational and communication costs. We formally define Outfox and prove its security in the Universal Composability (UC) framework. Our evaluation shows that Outfox retains strong anonymity guarantees while offering improved efficiency and adaptability to quantum-resistant cryptographic primitives.
\end{abstract}

%% file: macros.tex
\newcommand{\Adv}{\textrm{Adv}}
\newcommand{\A}{\mathcal{A}}
\newcommand{\Prob}{\textrm{Pr}}
\newcommand{\myvar}[1]{\ensuremath{\mathit{#1}}}
\newcommand{\alg}[1]{\ensuremath{\mathsf{#1}}}

\newcommand{\securityparameter}{\myvar{k}}

\newcommand{\Gam}{\textbf{Game}\ }

\newcommand{\ssid}{\myvar{qid}}
\newcommand{\tid}{\myvar{lpid}}
\newcommand{\ppid}{\myvar{gpid}}
\newcommand{\rpid}{\myvar{rpid}}

\newcommand\getsdollar{\mathrel{{\leftarrow}\vcenter{\hbox{\tiny\rmfamily\upshape\$}}}}

\newtheorem{definition}{Definition}
\newtheorem{theorem}{Theorem}

\newcommand{\Environment}{\ensuremath{\mathcal{Z}}\xspace}
\newcommand{\Adversary}{\ensuremath{\mathcal{A}}\xspace}
\newcommand{\Functionality}{\ensuremath{\mathcal{F}}\xspace}
\newcommand{\FunctionalityG}{\ensuremath{\mathcal{G}}\xspace}
\newcommand{\Simulator}{\ensuremath{\mathcal{S}}\xspace}
\newcommand{\Party}{\ensuremath{\mathcal{P}}\xspace}

\newcommand{\qid}{\myvar{qid}}
\newcommand{\sid}{\myvar{sid}}
\newcommand{\pid}{\myvar{pid}}

\newcommand{\IdealEnsemble}{\ensuremath{\mathrm{IDEAL}}\xspace}
\newcommand{\RealEnsemble}{\ensuremath{\mathrm{REAL}}\xspace}

\newcommand{\LRM}{\mathrm{LMC}}

\newcommand{\flrmNode}{\ensuremath{\mathcal{N}}\xspace}
\newcommand{\flrmNodeArequest}{\ensuremath{\mathcal{N}_1}\xspace}
\newcommand{\flrmNodeBrequest}{\ensuremath{\mathcal{N}_2}\xspace}
\newcommand{\flrmNodeCrequest}{\ensuremath{\mathcal{N}_3}\xspace}
\newcommand{\flrmNodeAresponse}{\ensuremath{\mathcal{\hat{N}}_1}\xspace}
\newcommand{\flrmNodeBresponse}{\ensuremath{\mathcal{\hat{N}}_2}\xspace}
\newcommand{\flrmNodeCresponse}{\ensuremath{\mathcal{\hat{N}}_3}\xspace}
\newcommand{\flrmGateway}{\ensuremath{\mathcal{W}}\xspace}
\newcommand{\flrmGatewayentry}{\ensuremath{\mathcal{W}_e}\xspace}
\newcommand{\flrmGatewayexit}{\ensuremath{\mathcal{W}_x}\xspace}
\newcommand{\flrmGatewayentryreply}{\ensuremath{\mathcal{\hat{W}}_e}\xspace}
\newcommand{\flrmGatewayexitreply}{\ensuremath{\mathcal{\hat{W}}_x}\xspace}
\newcommand{\flrmSender}{\ensuremath{\mathcal{T}}\xspace}
\newcommand{\flrmSenderreply}{\ensuremath{\mathcal{\hat{T}}}\xspace}
\newcommand{\flrmReceiver}{\ensuremath{\mathcal{R}}\xspace}
\newcommand{\flrmParty}{\ensuremath{\mathcal{P}}\xspace}
\newcommand{\flrmUser}{\ensuremath{\mathcal{U}}\xspace}

\newcommand{\flrmposition}{\myvar{pos}}
\newcommand{\flrmindex}{\myvar{ind}}
\newcommand{\flrmindexhonest}{\myvar{ind_h}}
\newcommand{\flrmMessageSpace}{\mathbb{M}}
\newcommand{\flrmSetA}{\mathbb{A}}
\newcommand{\flrmSetB}{\mathbb{B}}
\newcommand{\flrmSetC}{\mathbb{C}}
\newcommand{\flrmSetW}{\mathbb{W}}
\newcommand{\flrmSetR}{\mathbb{R}}
\newcommand{\flrmSetS}{\mathbb{S}}
\newcommand{\flrmSetU}{\mathbb{U}}
\newcommand{\flrmNodesmin}{\myvar{N_{min}}}
\newcommand{\flrmmessagerequest}{\myvar{m}}
\newcommand{\flrmmessageresponse}{\myvar{\hat{m}}}
\newcommand{\flrmGatewaysmin}{\myvar{W_{min}}}
\newcommand{\flrmdestroymessage}{\myvar{d}}

\newcommand{\flrmfirstmessage}{\myvar{c}}
\newcommand{\flrmstring}{\myvar{str}}

\newcommand{\flrmverify}{\mathsf{nextvf}}

\newcommand{\flrmroutelist}{\myvar{LR}}

\newcommand{\flrmpriorvf}{\mathsf{priorvf}}
\newcommand{\flrmCompLeakList}{\mathsf{CompLeakList}}
\newcommand{\flrmCompIndex}{\mathsf{CompIndex}}

\newcommand{\flrmlistleakage}{\myvar{LK}}

\newcommand{\flrmsetupini}{\mathsf{lmc.setup.ini}}
\newcommand{\flrmsetupsim}{\mathsf{lmc.setup.sim}}
\newcommand{\flrmsetupend}{\mathsf{lmc.setup.end}}
\newcommand{\flrmsetuprep}{\mathsf{lmc.setup.rep}}

\newcommand{\flrmregisterini}{\mathsf{lmc.register.ini}}
\newcommand{\flrmregistersim}{\mathsf{lmc.register.sim}}
\newcommand{\flrmregisterend}{\mathsf{lmc.register.end}}
\newcommand{\flrmregisterrep}{\mathsf{lmc.register.rep}}
\newcommand{\flrmsend}{\mathsf{lmc.send}}
\newcommand{\flrmsendini}{\mathsf{lmc.send.ini}}
\newcommand{\flrmsendsim}{\mathsf{lmc.send.sim}}
\newcommand{\flrmsendend}{\mathsf{lmc.send.end}}
\newcommand{\flrmsendrep}{\mathsf{lmc.send.rep}}

\newcommand{\flrmforwardini}{\mathsf{lmc.forward.ini}}
\newcommand{\flrmforwardsim}{\mathsf{lmc.forward.sim}}
\newcommand{\flrmforwardend}{\mathsf{lmc.forward.end}}
\newcommand{\flrmforwardrep}{\mathsf{lmc.forward.rep}}
\newcommand{\flrmreply}{\mathsf{lmc.send}}
\newcommand{\flrmreplyini}{\mathsf{lmc.send.ini}}
\newcommand{\flrmreplysim}{\mathsf{lmc.send.sim}}
\newcommand{\flrmreplyend}{\mathsf{lmc.send.end}}
\newcommand{\flrmreplyrep}{\mathsf{lmc.send.rep}}

\newcommand{\KEM}{\mathrm{KEM}}

\newcommand{\KEMpublickeyspace}{\mathcal{PK}}
\newcommand{\KEMsecretkeyspace}{\mathcal{SK}}
\newcommand{\KEMciphertextspace}{\mathcal{C}}
\newcommand{\KEMsharedkeyspace}{\mathcal{K}}
\newcommand{\KEMsecretkey}{\myvar{sk}}
\newcommand{\KEMpublickey}{\myvar{pk}}
\newcommand{\KEMciphertext}{\myvar{c}}
\newcommand{\KEMsharedkey}{\myvar{shk}}

\newcommand{\KEMKeyGen}{\alg{KEM.KeyGen}}
\newcommand{\KEMEnc}{\alg{KEM.Enc}}
\newcommand{\KEMDec}{\alg{KEM.Dec}}
\newcommand{\KEMOracleDec}{\alg{Dec}}

\newcommand{\FregT}{\ensuremath{\mathcal{T}}\xspace}

\newcommand{\Freg}{\mathrm{REG}}
\newcommand{\fregvalue}{\myvar{v}}
\newcommand{\fregmessagespace}{\mathcal{M}}

\newcommand{\fregregister}{\mathsf{reg.register}}
\newcommand{\fregregisterini}{\mathsf{reg.register.ini}}
\newcommand{\fregregisterend}{\mathsf{reg.register.end}}
\newcommand{\fregregistersim}{\mathsf{reg.register.sim}}
\newcommand{\fregregisterrep}{\mathsf{reg.register.rep}}
\newcommand{\fregretrieve}{\mathsf{reg.retrieve}}
\newcommand{\fregretrieveini}{\mathsf{reg.retrieve.ini}}
\newcommand{\fregretrieveend}{\mathsf{reg.retrieve.end}}
\newcommand{\fregretrievesim}{\mathsf{reg.retrieve.sim}}
\newcommand{\fregretrieverep}{\mathsf{reg.retrieve.rep}}

\newcommand{\FpregT}{\ensuremath{\mathcal{T}}\xspace}

\newcommand{\Fpreg}{\mathrm{PREG}}
\newcommand{\fpregvalue}{\myvar{v}}
\newcommand{\fpregmessagespace}{\mathcal{M}}

\newcommand{\fpregregister}{\mathsf{preg.register}}
\newcommand{\fpregregisterini}{\mathsf{preg.register.ini}}
\newcommand{\fpregregisterend}{\mathsf{preg.register.end}}
\newcommand{\fpregregistersim}{\mathsf{preg.register.sim}}
\newcommand{\fpregregisterrep}{\mathsf{preg.register.rep}}
\newcommand{\fpregretrieve}{\mathsf{preg.retrieve}}
\newcommand{\fpregretrieveini}{\mathsf{preg.retrieve.ini}}
\newcommand{\fpregretrieveend}{\mathsf{preg.retrieve.end}}
\newcommand{\fpregretrievesim}{\mathsf{preg.retrieve.sim}}
\newcommand{\fpregretrieverep}{\mathsf{preg.retrieve.rep}}

\newcommand{\routinginformation}{\myvar{I}}

\newcommand{\route}{\myvar{rt}}
\newcommand{\routeresponse}{\myvar{rt'}}
\newcommand{\receiverinfo}{\myvar{Info_{\flrmReceiver}}}
\newcommand{\senderinfo}{\myvar{Info_{\flrmSender}}}
\newcommand{\surb}{\myvar{surb}}
\newcommand{\surbidentifier}{\myvar{idsurb}}
\newcommand{\surbsecrets}{\myvar{sksurb}}
\newcommand{\payloadsecrets}{\myvar{skep}}

\newcommand{\packet}{\myvar{packet}}
\newcommand{\node}[1]{\ensuremath{N_{#1}}}
\newcommand{\routing}[1]{\ensuremath{R_{#1}}\xspace}

\newcommand{\lrmSurbCreate}{\alg{LMC.SurbCreate}}
\newcommand{\lrmSurbUse}{\alg{LMC.SurbUse}}
\newcommand{\lrmSurbCheck}{\alg{LMC.SurbCheck}}
\newcommand{\lrmSurbRecover}{\alg{LMC.SurbRecover}}
\newcommand{\lrmPacketCreate}{\alg{LMC.PacketCreate}}
\newcommand{\lrmPacketProcess}{\alg{LMC.PacketProcess}}

\newcommand{\Sender}{\ensuremath{\mathcal{T}}\xspace}
\newcommand{\Receiver}{\ensuremath{\mathcal{R}}\xspace}

\newcommand{\SMT}{\mathrm{SMT}}
\newcommand{\SMTmessage}{\myvar{m}}
\newcommand{\SMTfleakage}{\myvar{l}}
\newcommand{\fsmtmessagespace}{\mathcal{M}}

\newcommand{\fsmtsend}{\mathsf{smt.send}}
\newcommand{\fsmtsendini}{\mathsf{smt.send.ini}}
\newcommand{\fsmtsendsim}{\mathsf{smt.send.sim}}
\newcommand{\fsmtsendrep}{\mathsf{smt.send.rep}}
\newcommand{\fsmtsendend}{\mathsf{smt.send.end}}

\newcommand{\KDFname}{\ensuremath{\mathsf{KDF}}\xspace}
\newcommand{\KDF}{\ensuremath{\mathsf{KDF}}\xspace}

\newcommand{\AEAD}{\ensuremath{\mathsf{AEAD}}\xspace}
\newcommand{\AEADE}{\ensuremath{\mathsf{AEAD.Enc}}\xspace}
\newcommand{\AEADD}{\ensuremath{\mathsf{AEAD.Dec}}\xspace}
\newcommand{\AEADSetMessage}{\ensuremath{\mathsf{Message}}\xspace}
\newcommand{\AEADSetNonce}{\ensuremath{\mathsf{Nonce}}\xspace}
\newcommand{\AEADSetHeader}{\ensuremath{\mathsf{Header}}\xspace}
\newcommand{\AEADKeyspace}{\ensuremath{\mathcal{K}}\xspace}
\newcommand{\AEADmessage}{\myvar{M}}
\newcommand{\AEADheader}{\myvar{H}}
\newcommand{\AEADkey}{\myvar{K}}
\newcommand{\AEADnonce}{\myvar{N}}
\newcommand{\AEADciphertext}{\myvar{C}}

\newcommand{\blockcipher}{\ensuremath{\mathsf{SE}}\xspace}
\newcommand{\blockcipherenc}{\ensuremath{\mathsf{SE.Enc}}\xspace}
\newcommand{\blockcipherdec}{\ensuremath{\mathsf{SE.Dec}}\xspace}
\newcommand{\blockcipherkeyspace}{\ensuremath{\mathcal{K}}}
\newcommand{\blockciphermessagespace}{\ensuremath{\mathcal{M}}}
\newcommand{\payload}[1]{\myvar{ep_{#1}}\xspace}
\newcommand{\payloa}{\myvar{ep}}


%% file: 0Abstract.tex
\begin{abstract}
We present Outfox, a post-quantum-enabled packet format designed for efficient and secure communication in mixnets. Outfox employs layered encryption and is built on a key encapsulation mechanism (KEM), ensuring quantum resistance when instantiated with a quantum-safe KEM. Optimized for mixnets with fixed path lengths and single-layer mix nodes, Outfox eliminates unnecessary padding and reduces computational overhead by halving the number of public key operations per node compared to the Sphinx packet format. To evaluate its security, we define an ideal functionality for a layered replyable mixnet that enforces reply-request indistinguishability and demonstrate how Outfox realizes this functionality. Outfox provides provable security, enhanced performance for modern mixnets, and seamless adaptability to post-quantum environments, making it a compelling alternative to Sphinx.
\end{abstract}

%% file: 1introduction.tex
\section{Introduction}
\label{sec:introduction}

Encrypted packet formats are essential for securing the transport of message-based data through one or more servers, yet there is no provably secure packet format that can be used against adversaries with quantum computers. Encrypted packet formats can feature the key property of \emph{bitwise unlinkability}, so that a message that is input to a server cannot be correlated to the output of the message by the pattern of bits produced by encrypting the original message. The bitwise unlinkable message format Sphinx~\cite{DBLP:conf/sp/DanezisG09} has seen a number of uses, ranging from payments transfers in Bitcoin using the Lightning network to anonymous communication networks such as mixnets~\cite{DBLP:journals/cacm/Chaum81}.

However, there is no packet format that is unlinkable pattern format resistant to an adversary with a quantum computer. An adversary with a quantum computer could thus possibly decrypt existing Sphinx messages as well as discovering the path a message travels within a network, and so link messages with senders and receivers. We present \emph{Outfox}, an evolution of the Sphinx packet format that is both more efficient and capable of being used with postquantum cryptography due to its usage of key encapsulation mechanisms (KEMs).  We concentrate on the usage of Outfox in anonymous communication networks, in particular mixnets, with unlinkable replies to messages, although like Sphinx, the encrypted packet format may be useful in other settings.

%% file: 2Background.tex
\section{Background}\label{sec:background}
A mix network or mixnet~\cite{DBLP:journals/cacm/Chaum81} is a communication network that enables anonymous communication. We describe a mixnet with three types of entities: users, mix nodes and gateways. 

Users $\flrmUser$ use the mixnet to communicate anonymously. Every user can act both as a sender $\flrmSender$ and as a receiver $\flrmReceiver$. To send an anonymous message $\flrmmessagerequest$ to $\flrmReceiver$, $\flrmSender$ encrypts the message by computing a request packet $\packet$. The sender may or may not enable the receiver to send a reply. To enable a reply, $\flrmSender$ computes a single-use reply block $\surb$ and encrypts it along with $\flrmmessagerequest$ during the computation of $\packet$. Once  $\flrmReceiver$ obtains $\flrmmessagerequest$ and $\surb$, $\flrmReceiver$ uses $\surb$ to encrypt a reply message $\flrmmessageresponse$ and sends a reply packet to $\flrmSender$.

In a mixnet, every request packet sent from $\flrmSender$ to $\flrmReceiver$ is routed through a different path chosen by $\flrmSender$. We consider a mixnet with three layers of mix nodes, where each mix node belongs to a single layer. In our mixnet, every path from $\flrmSender$ to $\flrmReceiver$ consists of an entry gateway $\flrmGatewayentry$, a first-layer node $\flrmNodeArequest$, a second-layer node $\flrmNodeBrequest$, a third-layer node $\flrmNodeCrequest$, and an exit gateway $\flrmGatewayexit$. When $\flrmReceiver$ sends a reply packet, the reply packet uses a path chosen by $\flrmSender$ during the computation of $\surb$ that is, in general, different from the path of the associated request packet. The path of a reply packet consists of an exit gateway $\flrmGatewayexitreply$, a first-layer node $\flrmNodeAresponse$, a second-layer node $\flrmNodeBresponse$, a third-layer node $\flrmNodeCresponse$, and an entry gateway $\flrmGatewayentryreply$. We remark that every gateway acts both as an entry and as an exit gateway. We also remark that every node and gateway communicates both request and reply packets.

The aim of our mixnet is to provide anonymous communications against an adversary that controls the communication network and that may corrupt a subset of users, a subset of nodes and a subset of gateways. To guarantee anonymity, our mixnet uses strategies against both traffic analysis and cryptanalysis. Strategies against traffic analysis comprise mixing and cover traffic. Every mix node performs mixing by applying a different delay to every packet it sends, so that the order of the packets received by the mix node is different from the order in which they are sent by the mix node. Regarding cover traffic, every user sends packets that do not convey any information in order to hide from the adversary when actual communication is happening and thereby disguise communication patterns. 

In this paper, we focus on strategies against cryptanalysis. Our mixnet uses secure communication channels, which provide confidentiality and authentication, for every link between two parties in the path of a packet. However, using a secure communication channel is not enough to guarantee anonymity. For example, consider a path $(\flrmGatewayentry, \allowbreak \flrmNodeArequest, \allowbreak \flrmNodeBrequest, \allowbreak \flrmNodeCrequest, \allowbreak \flrmGatewayexit)$ between $\flrmSender$ and $\flrmReceiver$ in which only the node $\flrmNodeBrequest$ is honest. If $\flrmNodeBrequest$ forwards to $\flrmNodeCrequest$ the packet received from $\flrmNodeArequest$ unaltered, it is clear that the adversary finds out the identity of the receiver $\flrmReceiver$ with whom the sender $\flrmSender$ wishes to communicate. 

To provide anonymity, it is necessary to guarantee that the adversary cannot link the packet sent to $\flrmNodeCrequest$ with the packet received from $\flrmNodeArequest$. To provide such unlinkability, our mixnet uses layered encryption. The basic idea of layered encryption is the following. A sender $\flrmSender$ computes a packet on input the message $\flrmmessagerequest$ and the public keys of the receiver and of each of the parties in the path of the packet. First, $\flrmSender$ encrypts the message $\flrmmessagerequest$ on input the public key of the receiver and obtains a ciphertext.\footnote{Here we describe the use of layered encryption in the client model. Later we refer to the service model.} Second, $\flrmSender$ encrypts that ciphertext on input the public key of the exit gateway $\flrmGatewayexit$. That process is repeated until reaching the first party in the path. When $\flrmSender$ sends the packet, each party in the path uses its secret key to remove one layer of encryption. By removing one layer of encryption, it is guaranteed that the packet forwarded by the party is unlinkable to the packet received by the party. We remark that, in our mixnet, entry gateways $\flrmGatewayentry$ do not process request packets, while exit gateways $\flrmGatewayexitreply$ do not process reply packets, and thus their public keys are not used when computing a request packet or the single-use reply block for the reply. In other words, the first layer of encryption uses the public key of the first-layer mix node.

A desirable property in a layered encryption scheme is request-reply indistinguishability, i.e.\ nodes and gateways should not be able to distinguish request packets from reply packets. To achieve request-reply indistinguishability, it is necessary to guarantee that the processing (i.e., the removal of a layer of encryption) is the same for request and reply packets. We remark that, in our mixnet, a gateway should not be able to distinguish a request packet forwarded to a receiver from a reply packet forwarded to a sender (in both cases, the gateway removes a layer of encryption). Similarly, a gateway should not be able to distinguish a request packet received from a sender from a reply packet received from a receiver (in both cases, the gateway does not remove any layer of encryption and simply forwards the packet to a first-layer node).

\subsection{The Sphinx Packet Format}
\label{sec:previouswork}

The Sphinx packet format~\cite{DBLP:conf/sp/DanezisG09} is a packet format based on the concept of layered encryption. It has been used in the design of several onion routing protocols and mixnets. The Sphinx packet format could be used in the design of our mixnet, but there are some shortcomings:
\begin{itemize}

    \item Sphinx is based on a non-interactive key agreement protocol whose security is based on a variant of the Diffie-Hellman assumption~\cite{DBLP:journals/popets/SchererWS24a}. Therefore, it is not secure against adversaries equipped with a sufficiently powerful quantum computer.

    \item Sphinx hides the length of the path, i.e.\ the number of layers of encryption that were used to compute a packet. In our mixnet, all the paths have the same length, so this is not necessary.

    \item Sphinx hides the position of a layer. I.e., when a mix node processes a packet, it cannot find out how many layers are left to decrypt and thus it cannot find out its position in the path. In our mixnet, there are three layers and every mix node is assigned to one of them. Therefore, hiding that information is not necessary.

    \item Sphinx is designed to minimize communication cost. The processing of a layer requires two public key operations.

\end{itemize}

A Sphinx packet consists of a header and a payload. Sphinx protects the integrity of the header, i.e., if a corrupt party modifies the header of a packet, the next honest party that processes the packet will detect it. However, the integrity of the payload is not protected, i.e., if a corrupt party modifies the payload of a packet, this will only be detected by the party that removes the last layer of encryption, and not by honest parties that process the packet along its path. 

Regarding its use of layered encryption, a mixnet can operate in two models referred to as the client model and the service model. In the client model, the last layer of encryption for request packets is processed by the receiver, while the last layer of encryption for reply packets is processed by the sender. In the service model, the last layer of encryption for request packets is processed by an exit gateway $\flrmGatewayexit$, while the last layer of encryption for reply packets is processed by an entry gateway $\flrmGatewayentryreply$.

In~\cite{DBLP:conf/sp/KuhnBS20}, a tagging attack against sender anonymity is described. This attack takes advantage of the fact that Sphinx does not protect the integrity of the payload. In the client model, the attack would work as follows. When a sender sends a packet to a corrupt gateway $\flrmGatewayentry$, the gateway would modify the encrypted payload. When the corrupt receiver decrypts the payload, it finds out that the message is destroyed. If the receiver and the gateway collude, they could find out the identity of the sender. In the service model, the exit gateway would be the one that finds out that the message is destroyed. Therefore, in this model the adversary could find out that the sender communicates with one of the receivers associated with that gateway. 

In~\cite{DBLP:journals/iacr/KloossRSSW24}, a solution is proposed against that attack. It requires duplicating the size of the payload. More efficient solutions seem difficult without sacrificing request-reply indistinguishability. 

We also remark that similar attacks that do not rely on the packet format can be launched. In a typical communication, the sender needs to send more than one packet to the receiver. If the gateway drops one of the packets sent by the sender, the receiver would notice it. If the receiver and the gateway collude, then the receiver could also find out the identity of the sender.

\subsection{Our Contribution: The Outfox Packet Format}
\label{sec:contribution}

We propose Outfox, a packet format based on layered encryption that is suitable for layered mixnets in which all paths have the same length and every mix node is associated to a single layer. Outfox can be regarded as a variant of Sphinx that is optimized for its use in such layered mixnets. Moreover, Outfox aim is to minimize the computation cost of packet processing by mix nodes. The main differences of Outfox in comparison to Sphinx are the following:
\begin{itemize}

    \item Outfox uses a key encapsulation mechanism (KEM) instead of a non-interactive key agreement protocol. Outfox can be instantiated with different KEM schemes and, in particular, with quantum-safe KEM schemes. Moreover, the number of public key operations is reduced from 2 to 1.

    \item Outfox removes padding used in Sphinx to hide path lengths or positions of mix nodes in the path.

\end{itemize}

We analyze the security of Outfox in the ideal-world/real-world paradigm~\cite{DBLP:conf/focs/Canetti01}. For that purpose, we describe an ideal functionality $\Functionality_{\LRM}$ for a layered mixnet with replies. $\Functionality_{\LRM}$ guarantees request-reply indistinguishability. We remark that $\Functionality_{\LRM}$ does not perform mixing, but is nevertheless suitable to analyze the security of a packet format.

We propose a construction $\mathrm{\Pi}_{\LRM}$ that realizes $\Functionality_{\LRM}$. $\mathrm{\Pi}_{\LRM}$ uses the algorithms associated with the Outfox packet format as building blocks. Moreover, $\mathrm{\Pi}_{\LRM}$ operates in the hybrid model and uses as building blocks the ideal functionality $\Functionality_{\SMT}$ for a secure communication channel, and the functionalities $\Functionality_{\Freg}$ and $\Functionality_{\Fpreg}$ for registration. $\Functionality_{\SMT}$ is used to establish a secure communication channel for every link between two parties in the path of a packet, whereas $\Functionality_{\Freg}$ and $\Functionality_{\Fpreg}$ are needed to ensure that all honest parties use the same public key for a given party.

As Sphinx, Outfox does not protect the integrity of the payload and thus is vulnerable to the attack described above. To take that into account, $\Functionality_{\LRM}$ allows corrupt parties to destroy the payload of a packet. We intend to analyze the security of Outfox both in the client model and in the service model.  $\Functionality_{\LRM}$ operates in the client model, but a modified functionality for the service model will be added in a future update of this paper. 

Previous work has used the ideal-world/real-world paradigm to define the security of mixnets or onion routing protocols~\cite{DBLP:journals/iacr/KloossRSSW24,DBLP:journals/popets/SchererWS24a}. However, to analyze the security of a packet format, those works use game-based definitions that are equivalent to the ideal functionality. We deviate from that approach and provide a security analysis that shows indistinguishablity between the real-world protocol $\mathrm{\Pi}_{\LRM}$ and the ideal-world protocol defined by $\Functionality_{\LRM}$. An advantage of this approach is that $\mathrm{\Pi}_{\LRM}$ describes all the building blocks and operations needed to realize $\Functionality_{\LRM}$. For example, $\mathrm{\Pi}_{\LRM}$ uses $\Functionality_{\SMT}$, while in previous work the use of secure channels is assumed but does not form part of the description of the construction.

\subsection{Organization of the paper}
\label{sec:organization}

Section~\ref{sec:background} provides an overview of the background and prior work on mixnets and the Sphinx packet format, emphasizing their relevance and limitations.
The ideal-world/real-world paradigm, forming the basis of our security framework, is summarized in Section~\ref{sec:ucsecurity}.
We define the ideal functionality $\Functionality_{\LRM}$, which models a layered mixnet with unlinkable replies, in Section~\ref{sec:functionality}.
Key cryptographic components, including the key encapsulation mechanism (KEM) and other building blocks used in the Outfox packet format, are introduced in Section~\ref{sec:preliminaries}.
The Outfox packet format itself, along with the construction $\mathrm{\Pi}{\LRM}$ that realizes $\Functionality{\LRM}$, is detailed in Section~\ref{sec:construction}.
In Section~\ref{sec:securityanalysis}, we formally analyze the security of $\mathrm{\Pi}_{\LRM}$.
The efficiency of Outfox is evaluated in Section~\ref{sec:implementationefficiency}.
Finally, Section~\ref{sec:conclusion} concludes the paper with a summary of contributions and suggestions for future research directions.


%% file: 3SecurityFramework.tex
\section{Ideal-World/Real-World Paradigm}
\label{sec:ucsecurity}

The security of a protocol $\varphi$ is analyzed by comparing the view of an environment $\Environment$ in a real execution of $\varphi$ against the view of $\Environment$ in the ideal protocol defined by the ideal functionality $\Functionality_{\varphi}$~\cite{DBLP:conf/focs/Canetti01}. $\Environment$ chooses the inputs of the parties and collects their outputs. In the real world, $\Environment$ can communicate freely with an adversary $\Adversary$ who controls both the network and any corrupt parties.
In the ideal world, $\Environment$ interacts with dummy parties, who simply relay inputs and outputs between $\Environment$ and $\Functionality_{\varphi}$, and a simulator $\Simulator$.
We say that a protocol $\varphi$ securely realizes $\Functionality_{\varphi}$ if $\Environment$ cannot distinguish the real world from the ideal world, i.e., $\Environment$ cannot distinguish whether it is interacting with $\Adversary$ and parties running protocol $\varphi$ or with $\Simulator$ and dummy parties relaying to $\Functionality_{\varphi}$.

A protocol $\varphi^{\FunctionalityG}$ securely realizes $\Functionality_{\varphi}$ in the $\FunctionalityG$-hybrid model when $\varphi$ is allowed to invoke the ideal functionality $\FunctionalityG$. Therefore, for any protocol $\psi$ that securely realizes $\FunctionalityG$, the composed protocol $\varphi^{\psi}$, which is obtained by replacing each invocation of an instance of $\FunctionalityG$ with an invocation of an instance of $\psi$, securely realizes $\Functionality_{\varphi}$.

In the ideal functionalities described in this paper, we consider static corruptions. Every functionality is parameterized by a set of corrupt parties $\flrmSetS$, which includes the symbol $\Adversary$. When describing ideal functionalities, we use the conventions introduced in~\cite{DBLP:conf/crypto/CamenischDR16}.

\begin{description}

\item[Interface Naming Convention.]
An ideal functionality can be invoked by using one or more interfaces. The name of a message in an interface consists of three fields separated by dots, e.g., $\flrmsetupini$ in the functionality $\Functionality_{\LRM}$ in~\S\ref{sec:functionality}.
The first field indicates the name of the functionality and is the same in all interfaces of the functionality.
This field is useful for distinguishing between invocations of different functionalities in a hybrid protocol that uses two or more different functionalities.
The second field indicates the kind of action performed by the functionality and is the same in all messages that the functionality exchanges within the same interface.
The third field distinguishes between the messages that belong to the same interface, and can take the following different values.
 A message $\flrmsetupini$ is the incoming message received by the functionality, i.e., the message through which the interface is invoked.
 A message $\flrmsetupend$ is the outgoing message sent by the functionality, i.e., the message that ends the execution of the interface.
 The message $\flrmsetupsim$ is used by the functionality to send a message to $\Simulator$, and the message $\flrmsetuprep$ is used to receive a message from $\Simulator$.  

\item[Network communication vs local communication.] The identity of an interactive Turing machine instance (ITI) consists of a party identifier $\pid$ and a session identifier $\sid$. A set of parties in an execution of a system of interactive Turing machines is a protocol instance if they have the same session identifier $\sid$. 
    ITIs can pass direct inputs to and outputs from ``local'' ITIs that have the same $\pid$.
    An ideal functionality $\Functionality$ has $\pid=\bot$ and is considered local to all parties. An instance of $\Functionality$ with the session identifier $\sid$ only accepts inputs from and passes outputs to machines with the same session identifier $\sid$. Some functionalities require the session identifier to have some structure. Those functionalities check whether the session identifier possesses the required structure in the first message that invokes the functionality. For the subsequent messages, the functionality implicitly checks that the session identifier equals the session identifier used in the first message.
    Communication between ITIs with different party identifiers must take place over the network.
    The network is controlled by $\Adversary$, meaning that he can arbitrarily delay, modify, drop, or insert messages.

\item[Query identifiers.] It is possible that an interface of a functionality can be invoked more than once. That is the case of, e.g., the interface $\flrmsend$. When the functionality sends a message $\flrmsendsim$ to $\Simulator$, a query identifier $\qid$ is included in the message. The query identifier must also be included in the response $\flrmsendrep$ sent by $\Simulator$. The query identifier is used to identify the message $\flrmsendsim$ to which $\Simulator$ replies with a message $\flrmsendrep$. We note that, typically, $\Simulator$ in the security proof may not be able to provide an immediate answer to the functionality after receiving a message $\flrmsendsim$. The reason is that $\Simulator$ typically needs to interact with the copy of $\Adversary$ it runs in order to produce the message $\flrmsendrep$, but $\Adversary$ may not provide the desired answer or may provide a delayed answer. In such cases, when the functionality sends more than one message $\flrmsendsim$ to $\Simulator$, $\Simulator$ may provide delayed replies, and the order of those replies may not follow the order of the messages received.

\item[Aborts.]  When an ideal functionality $\Functionality$ aborts after being activated with a message sent by a party, we mean that $\Functionality$ halts the execution of its program and sends a special abortion message to the party that invoked the functionality. When an ideal functionality $\Functionality$ aborts after being activated with a message sent by $\Simulator$, we mean that $\Functionality$ halts the execution of its program and sends a special abortion message to the party that receives the outgoing message from $\Functionality$ after $\Functionality$ is activated by $\Simulator$.

\end{description}





%% file: 4SecDefs.tex
\section{Ideal Functionality for a Layered Mixnet in the Client Model}
\label{sec:functionality}

We describe our ideal functionality $\Functionality_{\LRM}$ for a layered mixnet in the client model. In~\S\ref{sec:functionalityParties}, we describe the parties that interact with $\Functionality_{\LRM}$. 
In~\S\ref{sec:functionalityInterfaces}, we describe the interfaces of $\Functionality_{\LRM}$. In~\S\ref{sec:functionalityParameters}, we describe the parameters used by $\Functionality_{\LRM}$. In~\S\ref{sec:functionalityDefinition}, we give the formal definition of $\Functionality_{\LRM}$. 

\input{4SecDefs1Parties}

\input{4SecDefs2Interfaces}

\input{4SecDefs3Parameters}

\input{4SecDefs4Definition}

%% file: 4SecDefs1Parties.tex
\subsection{Parties}
\label{sec:functionalityParties}

Our ideal functionality $\Functionality_{\LRM}$  interacts with users, gateways and nodes.
\begin{description}

    \item[Users.] A user $\flrmUser$ is a party that uses the mixnet to send messages to other users and to receive messages from other users. Therefore, a user acts both as a sender $\flrmSender$ and a receiver $\flrmReceiver$. A packet produced by a sender is referred to as a \emph{request}, whereas a packet produced by a receiver is referred to as a \emph{reply}. 

    \begin{description}

        \item[Senders.] A sender $\flrmSender$ performs two tasks. On the one hand, a sender $\flrmSender$ computes a request to communicate a message $\flrmmessagerequest$ to a receiver $\flrmReceiver$ and sends the request to an entry gateway $\flrmGatewayentry$. On the other hand, a sender $\flrmSender$ receives a reply from an entry gateway $\flrmGatewayentryreply$ and processes it in order to retrieve a message $\flrmmessageresponse$.  When a sender $\flrmSender$ computes a request, the sender $\flrmSender$ decides whether the receiver $\flrmReceiver$ can reply to it or not. If the sender $\flrmSender$ is corrupt, it can choose that the reply be sent to another user $\flrmSenderreply$.

        \item[Receivers.] A receiver $\flrmReceiver$ performs two tasks. On the one hand, a receiver $\flrmReceiver$ receives a request from an exit gateway $\flrmGatewayexit$ and processes it to retrieve a message $\flrmmessagerequest$. On the other hand, a receiver $\flrmReceiver$ computes a reply to communicate a message $\flrmmessageresponse$ to a sender $\flrmSender$ and sends the reply to an exit gateway $\flrmGatewayexitreply$. We remark that a receiver $\flrmReceiver$ can compute a reply only if the sender $\flrmSender$ enabled that possibility when computing the request associated with that reply.

    \end{description}

     \item[Gateways.] A gateway $\flrmGateway$ is a party that provides senders and receivers with access to the mixnet. Every gateway acts as both an entry gateway and an exit gateway. 

    \begin{description}

        \item[Entry gateway.] An entry gateway performs two tasks. On the one hand, an entry gateway $\flrmGatewayentry$ receives a request from a sender $\flrmSender$ and relays it to a first-layer node $\flrmNodeArequest$. On the other hand, an entry gateway $\flrmGatewayentryreply$ receives a reply from a third-layer node $\flrmNodeCresponse$, processes it, and sends it to a sender $\flrmSender$. We remark that entry gateways do not process requests, i.e.\ they simply relay them from a sender to a first-layer node.

        \item[Exit gateway.] An exit gateway performs two tasks. On the one hand, an exit gateway $\flrmGatewayexit$ receives a request from a third-layer node $\flrmNodeCrequest$, processes it, and sends it to a receiver $\flrmReceiver$. On the other hand, an exit gateway $\flrmGatewayexitreply$ receives a reply from a receiver $\flrmReceiver$ and relays it to a first-layer node $\flrmNodeAresponse$. We remark that exit gateways do not process replies, i.e.\ they simply relay them from a receiver to a first-layer node.    
    
    \end{description}

    A gateway $\flrmGateway$ should not be able to distinguish between requests and replies. When a gateway $\flrmGateway$ receives a packet from a user $\flrmUser$, the gateway $\flrmGateway$ cannot distinguish whether the user acts as a sender $\flrmSender$ that sends a request (and thus the gateway acts as an entry gateway $\flrmGatewayentry$), or whether the user acts as a receiver $\flrmReceiver$ that sends a reply (and thus the gateway acts as an exit gateway $\flrmGatewayexitreply$). We remark that, in both cases, the gateway simply relays the packet to a  first-layer node. 
    
    Similarly, when a gateway $\flrmGateway$ receives a packet from a third-layer node, the gateway  $\flrmGateway$ cannot distinguish whether it is a request from a third-layer node $\flrmNodeCrequest$ (and thus the gateway acts as an exit gateway $\flrmGatewayexit$) or a reply from a third-layer node $\flrmNodeCresponse$ (and thus the gateway acts as an entry gateway $\flrmGatewayentryreply$).  We remark that, in both cases, the gateway processes the packet and sends it to a user.\footnote{The remarks regarding request-reply indistinguishability for gateways are valid when users send both requests and replies. If a user is e.g.\ a web host, then the gateway knows that any packet sent to that user is a request and any packet received from that user is a reply.}
    

    \item[Nodes.] A node $\flrmNode$ is a party that receives, processes and sends request packets and reply packets. Our mixnet consists of three layers, and each node is assigned to one of the layers. We denote nodes assigned to the first, second and third layers by  $\flrmNodeArequest$, $\flrmNodeBrequest$, and $\flrmNodeCrequest$, when they are processing a request, and by $\flrmNodeAresponse$, $\flrmNodeBresponse$, and $\flrmNodeCresponse$, when they are processing a reply.
    \begin{description}

        \item[First-layer nodes.] First-layer nodes perform two tasks. On the one hand, a first-layer node $\flrmNodeArequest$ receives a request from an entry gateway $\flrmGatewayentry$, processes it and sends it to a second-layer node $\flrmNodeBrequest$. On the other hand, a first-layer node $\flrmNodeAresponse$ receives a reply from an exit gateway $\flrmGatewayexitreply$, processes it and sends it to a second-layer node $\flrmNodeBresponse$.

        \item[Second-layer nodes.] These nodes perform two tasks. On the one hand, a second-layer node $\flrmNodeBrequest$ receives a request from a first-layer node $\flrmNodeArequest$, processes it and sends it to a third-layer node $\flrmNodeCrequest$. On the other hand, a second-layer node $\flrmNodeBresponse$ receives a reply from a first-layer node $\flrmNodeAresponse$, processes it and sends it to a third-layer node $\flrmNodeCresponse$.

        \item[Third-layer nodes.] Third-layer nodes perform two tasks. On the one hand, a third-layer node $\flrmNodeCrequest$ receives a request from a second-layer node $\flrmNodeBrequest$, processes it and sends it to an exit gateway $\flrmGatewayexit$. On the other hand, a third-layer node $\flrmNodeCresponse$ receives a reply from a second-layer node $\flrmNodeBresponse$, processes it and sends it to an entry gateway $\flrmGatewayentryreply$. 

    \end{description}
     Every node processes both requests and replies and  should not be able to distinguish between requests and replies.

 \end{description}
In Table~\ref{tab:parties}, we describe the notation for all the parties that communicate with $\Functionality_{\LRM}$.

\begin{table}
    \caption{Parties that communicate with $\Functionality_{\LRM}$}
    \label{tab:parties}
    \centering
    \begin{tabular}{|l|l|} \hline
        \multicolumn{2}{|c|}{\textbf{Parties}} \\ \hline
        $\flrmUser$ & User \\ \hline 
        $\flrmSender$ & Sender \\ \hline
        $\flrmReceiver$ & Receiver \\ \hline
        $\flrmSenderreply$ & Receiver of a reply ($\flrmSender = \flrmSenderreply$ if $\flrmSender$ is honest) \\ \hline
        $\flrmGateway$ & Gateway \\ \hline
        $\flrmGatewayentry$ & Entry Gateway processing a request \\ \hline
        $\flrmGatewayentryreply$ & Entry Gateway processing a reply \\ \hline
        $\flrmGatewayexit$ & Exit Gateway processing a request \\ \hline
        $\flrmGatewayexitreply$ & Exit Gateway processing a reply \\ \hline
        $\flrmNode$ & Node \\ \hline
        $\flrmNodeArequest$ & First-layer node processing a request \\ \hline
        $\flrmNodeAresponse$ & First-layer node processing a reply \\ \hline
        $\flrmNodeBrequest$ & Second-layer node processing a request \\ \hline
        $\flrmNodeBresponse$ & Second-layer node processing a reply \\ \hline
        $\flrmNodeCrequest$ & Third-layer node processing a request \\ \hline
        $\flrmNodeCresponse$ & Third-layer node processing a reply \\ \hline
        $\flrmParty$ & A party \\ \hline
    \end{tabular}
\end{table}

%% file: 4SecDefs2Interfaces.tex
\subsection{Interfaces}
\label{sec:functionalityInterfaces}

$\Functionality_{\LRM}$ consists of the following interfaces: setup, registration, request, reply and forward. In this section, we give a brief description of each of the interfaces. The formal definition of $\Functionality_{\LRM}$ is given in~\S\ref{sec:functionalityDefinition}.

In the request and reply interfaces, the messages that the functionality $\Functionality_{\LRM}$ sends and receives use the same name $\flrmreply$. The reason is that $\Functionality_{\LRM}$ needs to provide request-reply indistinguishability. If the messages had a different name, a gateway could distinguish between a request and a reply simply by using the message name. $\Functionality_{\LRM}$ is able to distinguish whether a $\flrmreplyini$ message refers to the request or to the reply interface because the list of inputs sent along with the message is different in each case.
\begin{description}
    
    \item[Setup.] The  $\flrmsetupini$ message  is sent by any node or gateway. $\Functionality_{\LRM}$ records the fact that the node or gateway has run the setup interface.

    \item[Registration.] The $\flrmregisterini$ message is sent by any user. $\Functionality_{\LRM}$ records the fact that the user has run the registration interface.

    \item[Request.] The $\flrmsendini$ message is sent by any sender $\flrmSender$ on input a global packet identifier $\ppid$, an entry gateway $\flrmGatewayentry$, an exit gateway $\flrmGatewayexit$, a receiver identifier $\flrmReceiver$, a request message $\flrmmessagerequest$, a first-layer node $\flrmNodeArequest$, a second-layer node $\flrmNodeBrequest$, and a third-layer node $\flrmNodeCrequest$. When the sender $\flrmSender$ wishes to enable a reply, the input also includes an entry gateway $\flrmGatewayentryreply$, an exit gateway $\flrmGatewayexitreply$, a first-layer node $\flrmNodeAresponse$, a second-layer node $\flrmNodeBresponse$, and a third-layer node $\flrmNodeCresponse$, and a sender $\flrmSenderreply$. ($\flrmSenderreply = \flrmSender$ if $\flrmSender$ is honest.) $\Functionality_{\LRM}$ creates a local packet identifier $\tid$ and stores $\tid$ along with the message $\flrmmessagerequest$, the global identifier $\ppid$, the routing information $\flrmroutelist$, and the index $\flrmindex$ and the party identifier $\flrmposition$ of the next party that should interact with $\Functionality_{\LRM}$ regarding this packet. Then $\Functionality_{\LRM}$ sends $\tid$, the sender identifier $\flrmSender$ and the first-layer node identifier $\flrmNodeArequest$ to the entry gateway $\flrmGatewayentry$. If $\flrmGatewayentry$ and subsequent parties in the route are corrupt, $\Functionality_{\LRM}$ also sends a list $\flrmlistleakage$ that contains the identifiers of those parties and, if it exists, the next honest party in the route. $\Functionality_{\LRM}$ also leaks the global packet identifier $\ppid$ and the message $\flrmmessagerequest$ when both the sender $\flrmSender$ and the gateway $\flrmGatewayentry$ are corrupt. The message $\flrmmessagerequest$ is also leaked if $\flrmSender$ is honest but $\flrmGatewayentry$ and subsequent parties until and including the receiver $\flrmReceiver$ are corrupt.

    \item[Reply.] The $\flrmreplyini$ message is sent by the receiver $\flrmReceiver$ on input a local packet identifier $\tid$ and the reply message $\flrmmessageresponse$. If $\flrmReceiver$ is corrupt, $\Functionality_{\LRM}$ can also receive a reply packet identifier $\rpid$, the identifier $\flrmGatewayexitreply'$ of an exit gateway, and the identifier $\flrmNodeAresponse'$ of a first-layer node. The reason is that, in our construction, a corrupt receiver may choose a gateway $\flrmGatewayexitreply'$ different from the gateway $\flrmGatewayexitreply$ chosen by the sender when generating a request packet, and may also instruct that gateway to forward the packet to a different first-layer node $\flrmNodeAresponse'$. $\Functionality_{\LRM}$ checks whether it stores a packet identifier $\tid$ associated with routing information that enables the receiver to send a reply (i.e.\ the tuple $\langle \flrmGatewayexitreply,  \allowbreak \flrmNodeAresponse, \allowbreak \flrmNodeBresponse, \allowbreak \flrmNodeCresponse, \allowbreak \flrmGatewayentryreply, \allowbreak \flrmSenderreply \rangle$ is not empty), and whether $\flrmReceiver$ is the receiver associated with $\tid$.  After that, $\Functionality_{\LRM}$ creates a new $\tid$  and sends $\tid$, the receiver identifier $\flrmReceiver$ and the first-layer node identifier $\flrmNodeAresponse$ to the exit gateway $\flrmGatewayexitreply$. If $\flrmGatewayexitreply$ and subsequent parties in the route are corrupt, but the party that computed the associated request packet is honest, $\Functionality_{\LRM}$ also sends a list $\flrmlistleakage$ that contains the identifiers of those parties until and including the next honest party in the route. If $\flrmGatewayexitreply$ and the party that computed the associated request packet are both corrupt, then $\Functionality_{\LRM}$  leaks the global packet identifier $\ppid$ and the message $\flrmmessageresponse$ to $\flrmGatewayexitreply$. 
    
    \item[Forward.] The $\flrmforwardini$ message is sent by a node or a gateway $\flrmParty$ along with a local packet identifier $\tid$. When $\flrmParty$ is corrupt, $\Functionality_{\LRM}$ also receives a bit $\flrmdestroymessage$ and a tuple $\{\ppid, \allowbreak \flrmGatewayexit, \allowbreak \flrmReceiver, \allowbreak \flrmmessagerequest, \allowbreak \flrmNodeArequest, \allowbreak \flrmNodeBrequest, \allowbreak \flrmNodeCrequest, \allowbreak \langle   \flrmGatewayentryreply, \allowbreak \flrmGatewayexitreply, \allowbreak \flrmNodeAresponse, \allowbreak \flrmNodeBresponse, \allowbreak \flrmNodeCresponse, \allowbreak \flrmSenderreply \rangle\}$. The bit $\flrmdestroymessage$ allows a corrupt node or gateway $\flrmParty$ to destroy the message sent in a packet. The tuple allows a corrupt node or gateway $\flrmParty$ to input a new packet with a global packet identifier $\ppid$. 
    
    $\Functionality_{\LRM}$ either checks that it stores routing information for a packet with local packet identifier $\tid$ or uses the new routing information provided by a corrupt node or gateway $\flrmParty$ in order to determine the next party that should receive the packet. Then $\Functionality_{\LRM}$ proceeds as follows:
    \begin{itemize}

        \item If the party that receives the packet is a sender $\flrmSenderreply$, $\Functionality_{\LRM}$ sends $\flrmSenderreply$ the reply message associated with $\tid$, the party identifier of the entry gateway $\flrmGatewayentryreply$ that forwards the reply packet to $\flrmSenderreply$, and the global  packet identifier $\ppid$, which allows $\flrmSenderreply$ to match the reply packet with its corresponding request packet.

        \item If the party that receives the packet is a receiver $\flrmReceiver$, $\Functionality_{\LRM}$ sends the request message associated with $\tid$, and the identifier of the exit gateway $\flrmGatewayexit$ that forwards the request packet to $\flrmReceiver$. If the packet route associated with $\tid$ enables a reply, $\Functionality_{\LRM}$ also sends a new local packet identifier $\tid'$ to allow $\flrmReceiver$ to input a reply packet through the reply interface, along with the identifiers $\flrmGatewayexitreply$ and $\flrmNodeAresponse$ of the exit gateway and the first-layer node in the route of the reply packet. ($\flrmReceiver$ learns $\flrmNodeAresponse$ because exit gateways do not process reply packets, so $\flrmReceiver$ needs to send $\flrmNodeAresponse$ to $\flrmGatewayexitreply$.) Additionally, when $\flrmReceiver$ is corrupt, $\flrmReceiver$ receives a leakage list $\flrmlistleakage$ that contains the identifiers of the next corrupt parties (if any) in the route of the reply packet. A corrupt $\flrmReceiver$ receives the global packet identifier $\ppid$ when the party that generated the request packet is also corrupt. 

        \item  If the party that receives the packet is a gateway or a node, $\Functionality_{\LRM}$ sends a new local packet identifier $\tid'$ along with the identifier of the party that forwards the packet and the identifier of the next party in the route of the packet. $\tid'$ allows the gateway or node that receives the packet to subsequently forward it to the next party in the route. If the gateway or node that receives the packet is corrupt, $\Functionality_{\LRM}$ also sends a leakage list $\flrmlistleakage$ to disclose the next corrupt parties (if any) in the route. A corrupt gateway or node that receives the packet learns the global packet identifier $\ppid$ and the message associated to the packet when the party that generated the request packet is also corrupt. A corrupt first-layer node $\flrmNodeAresponse$ learns a reply packet identifier $\rpid$ if a corrupt receiver input it to  $\Functionality_{\LRM}$ through the reply interface.
    
    \end{itemize}

\end{description}

%% file: 4SecDefs3Parameters.tex
\subsection{Parameters, Variables and Algorithms}
\label{sec:functionalityParameters}

$\Functionality_{\LRM}$ uses the parameters, variables and algorithms depicted in Table~\ref{tab:parameters}.  In~\ref{sec:packetroute}, we describe how $\Functionality_{\LRM}$ stores the route of a packet. In~\ref{sec:packetidentifiers}, we describe the packet identifiers used by $\Functionality_{\LRM}$. In~\ref{sec:leakagelist}, we explain how $\Functionality_{\LRM}$ sets the list of parties in the route that need to be leaked to a corrupt party.

\begin{table}
    \caption{Parameters used by $\Functionality_{\LRM}$}
    \label{tab:parameters}
    \centering
    \begin{tabular}{|l|l|} \hline
        \multicolumn{2}{|c|}{\textbf{Parameters}} \\ \hline
        $\flrmNodesmin$ & Minimum number of nodes \\ \hline
        $\flrmGatewaysmin$ & Minimum number of gateways \\ \hline
        $\flrmMessageSpace$ & Message space \\ \hline
        $\flrmSetS$ & Set of corrupt parties \\ \hline
        $\flrmSetU$ & Set of users \\ \hline
        \multicolumn{2}{|c|}{\textbf{Variables}} \\ \hline
        $\flrmroutelist$ & Packet route \\ \hline
        $\flrmindex$ & Packet index \\ \hline
        $\ppid$ & Global packet identifier \\ \hline
        $\tid$ & Local packet identifier \\ \hline
        $\rpid$ & Reply packet identifier \\ \hline
        $\flrmlistleakage$ & Leakage list \\ \hline
        $\flrmSetA$ & Set of first-layer nodes \\ \hline
        $\flrmSetB$ & Set of second-layer nodes \\ \hline
        $\flrmSetC$ & Set of third-layer nodes \\ \hline
        $\flrmSetW$ & Set of gateways \\ \hline
        $\flrmSetR$ & Receiver set of a sender  \\ \hline
        $\sid$ & Session identifier \\ \hline
        $\qid$ & Query identifier \\ \hline
        \multicolumn{2}{|c|}{\textbf{Algorithms}} \\ \hline
        $\flrmCompLeakList$ & Computation of leakage list \\ \hline
        $\flrmCompIndex$ & Computation of index \\ \hline
       
    \end{tabular}
\end{table}

\subsubsection{Packet Route and Packet Index}
\label{sec:packetroute}

The route  of a packet is given by a list of party identifiers $\flrmroutelist \allowbreak = \allowbreak [\flrmSender, \allowbreak \flrmGatewayentry, \allowbreak \flrmNodeArequest, \allowbreak \flrmNodeBrequest, \allowbreak \flrmNodeCrequest, \allowbreak \flrmGatewayexit, \allowbreak \flrmReceiver, \langle \flrmGatewayexitreply, \allowbreak \flrmNodeAresponse, \allowbreak \flrmNodeBresponse, \allowbreak \flrmNodeCresponse, \allowbreak  \flrmGatewayentryreply, \allowbreak \flrmSenderreply \rangle]$. The party identifiers in the tuple $\langle \flrmGatewayexitreply, \allowbreak \flrmNodeAresponse, \allowbreak \flrmNodeBresponse, \allowbreak \flrmNodeCresponse, \allowbreak  \flrmGatewayentryreply, \allowbreak \flrmSenderreply \rangle$ are set to $\bot$ if the packet does not enable a reply.

$\Functionality_{\LRM}$ uses a variable $\flrmindex$ to store the index of a packet along the route. The variable $\flrmindex$ takes a value in $[1, \allowbreak 13]$ according to the index in $\flrmroutelist$ of the party that holds the packet. I.e., $\flrmindex \allowbreak = \allowbreak 1$ if the packet is held by the sender $\flrmSender$, $\flrmindex \allowbreak = \allowbreak 2$ if the packet is held by the entry gateway $\flrmGatewayentry$, and so on.

\subsubsection{Packet Identifiers}
\label{sec:packetidentifiers}
We explain how each of the packet identifiers in Table~\ref{tab:parameters} is used by $\Functionality_{\LRM}$:
\begin{description}
    
    \item[Global packet identifier $\ppid$.]  The global packet identifier $\ppid$ is received as input by $\Functionality_{\LRM}$ from an honest sender $\flrmSender$ or from any corrupt party that sends a packet to an honest party. In the case of an honest sender, $\Functionality_{\LRM}$ reveals  $\ppid$ to $\flrmSender$ when the sender receives a reply packet associated with the request packet identified by $\ppid$, so that $\flrmSender$ can link the reply packet to the request packet. We remark that, when $\flrmSender$ is honest,  $\ppid$ is not revealed to any party in the route of the packet, even if the party in the route is corrupt.
    
    When the packet is input to $\Functionality_{\LRM}$ by a corrupt sender or another corrupt party, $\Functionality_{\LRM}$ reveals $\ppid$ when the packet is forwarded to any corrupt party. Therefore, the simulator $\Simulator$ learns that the packet received by the corrupt party is linked to the packet previously input by another corrupt party. In our construction $\mathrm{\Pi}_{\LRM}$, when a packet is computed by a corrupt party, the adversary can trace the packet whenever it is forwarded to any corrupt party in the route.

    In our security analysis, when a packet computed by a corrupt party is forwarded by corrupt parties to honest parties more than once, the simulator $\Simulator$ inputs to $\Functionality_{\LRM}$ different global packet identifiers $\ppid$ each time the packet is forwarded to an honest party. The reason is that $\Simulator$ cannot know whether the two packets are associated or not. 
    Consider the following example. The route of a request packet consists of $[\flrmSender, \allowbreak \flrmGatewayentry, \allowbreak \flrmNodeArequest, \allowbreak \flrmNodeBrequest, \allowbreak \flrmNodeCrequest, \allowbreak \flrmGatewayexit, \allowbreak \flrmReceiver]$. The sender $\flrmSender$ is corrupt, the parties $(\flrmGatewayentry, \allowbreak \flrmNodeArequest)$ are honest, the parties $(\flrmNodeBrequest, \allowbreak \flrmNodeCrequest, \allowbreak \flrmGatewayexit)$ are corrupt, and $\flrmReceiver$ is honest. When the corrupt sender $\flrmSender$ sends the packet to the honest gateway $\flrmGatewayentry$, $\Simulator$ picks up a global packet identifier $\ppid$ and inputs $\ppid$ to $\Functionality_{\LRM}$. When the honest first-layer node $\flrmNodeArequest$ forwards the packet to the corrupt second-layer node $\flrmNodeBrequest$, $\Simulator$ receives $\ppid$ from $\Functionality_{\LRM}$, which allows $\Simulator$ to simulate the packet towards the adversary. Afterwards, when the corrupt gateway $\flrmGatewayexit$ forwards the packet to the honest receiver $\flrmReceiver$, $\Simulator$ is not able to link this packet to the one previously received from the corrupt sender $\flrmSender$ or to the one previously sent to the corrupt second-layer node $\flrmNodeBrequest$. Therefore, $\Simulator$ picks up another global packet identifier $\ppid'$ and inputs it to $\Functionality_{\LRM}$. Although $\Functionality_{\LRM}$ regards it as a different packet, our security analysis shows that the outputs of honest parties (in this example, the honest receiver $\flrmReceiver$) in the ideal protocol are indistinguishable from their outputs in the real protocol. 
    

    \item[Local packet identifier $\tid$.] When a party receives a packet, the functionality $\Functionality_{\LRM}$ picks up a local packet identifier $\tid$ and sends it to that party. Afterwards, when the packet is forwarded, $\Functionality_{\LRM}$ receives $\tid$ as input to identify the packet that should be forwarded, and creates a new identifier $\tid$ for the new party that receives the packet. $\Functionality_{\LRM}$ creates a different identifier $\tid$ for each of the parties that receives the packet, and thus $\tid$ does not reveal any information that allows parties to link a packet to a packet previously received by another party.

    \item[Reply packet identifier $\rpid$.] The reply packet identifier $\rpid$ is received as input by $\Functionality_{\LRM}$ when a corrupt receiver sends a reply packet to an honest exit gateway $\flrmGatewayexitreply$, and that reply packet is computed by using a single-use reply block that was encrypted in a request packet computed by an honest sender. $\Functionality_{\LRM}$ reveals $\rpid$ to the simulator $\Simulator$ when that reply packet is forwarded by $\flrmGatewayexitreply$ to a corrupt first-layer node $\flrmNodeAresponse$.

    The reason why $\rpid$ is needed is the following. In our construction, since $\flrmGatewayexitreply$ does not process reply packets, the adversary can link a reply packet received by a corrupt first-layer node $\flrmNodeAresponse$ to the reply packet computed by a corrupt receiver. However, if the request packet associated with the reply packet was computed by an honest sender, the adversary cannot link it to the request packet computed by the sender if at least one honest party processes the packet in the route after the sender. Therefore, the simulator $\Simulator$ should not learn the global packet identifier $\ppid$, and thus the reply packet identifier  $\rpid$ is used for the leakage described above.
    
\end{description}

\subsubsection{Leakage List}
\label{sec:leakagelist}

When a packet computed by an honest sender $\flrmSender$ is forwarded to a corrupt party, $\Functionality_{\LRM}$ leaks to the simulator $\Simulator$ the identifiers of all the subsequent parties in the route that are also corrupt, until and including the next honest party in the route. If the receiver $\flrmReceiver$ is among those parties, the request message $\flrmmessagerequest$ is also leaked. There is an exception to the rule of leaking all the subsequent corrupt parties until and including the next honest party, which is that the exit gateway $\flrmGatewayexitreply$ should be disregarded. The reason is that exit gateways do not process reply packets, and thus, if $\flrmNodeAresponse$ and subsequent parties in the route are corrupt, their identities should also be leaked by $\Functionality_{\LRM}$.

\paragraph{Algorithm $\flrmCompLeakList$.} To compute the list of parties that needs to be leaked to the corrupt party, $\Functionality_{\LRM}$ runs the algorithm $\flrmCompLeakList$. The algorithm  $\flrmCompLeakList$ receives as input a route $\flrmroutelist$, the index $\flrmindex$ of the corrupt party $\flrmroutelist[\flrmindex]$ in the route that receives the packet, the sets of nodes, gateways and users $(\flrmSetA, \allowbreak \flrmSetB, \allowbreak \flrmSetC, \allowbreak \flrmSetW, \allowbreak \flrmSetU)$, and the set of corrupt parties $\flrmSetS$. The algorithm  $\flrmCompLeakList$ outputs a leakage list $\flrmlistleakage$ that contains the identifiers of all the parties in the route that should be leaked to the corrupt party. It also outputs the index $\flrmindexhonest$ of the last honest party in $\flrmlistleakage$. $\flrmCompLeakList$ is described in Figure~\ref{fig:algcompleaklist}.

For example, consider a request packet with route $\flrmroutelist \allowbreak = \allowbreak [\flrmSender, \allowbreak \flrmGatewayentry, \allowbreak \flrmNodeArequest, \allowbreak \flrmNodeBrequest, \allowbreak \flrmNodeCrequest, \allowbreak \flrmGatewayexit, \allowbreak \flrmReceiver, \langle \bot, \allowbreak \bot, \allowbreak \bot, \allowbreak \bot, \allowbreak  \bot, \allowbreak \bot \rangle]$, where the sender $\flrmSender$ is honest and the entry gateway  $\flrmGatewayentry$ is corrupt. Let $\flrmNodeCrequest$ be the next honest party. In that case, the list leaked to the corrupt entry gateway $\flrmGatewayentry$ is $\flrmlistleakage \allowbreak = \allowbreak [\flrmNodeArequest, \allowbreak \flrmNodeBrequest, \allowbreak \flrmNodeCrequest]$.

Consider now a request packet with route $\flrmroutelist \allowbreak = \allowbreak [\flrmSender, \allowbreak \flrmGatewayentry, \allowbreak \flrmNodeArequest, \allowbreak \flrmNodeBrequest, \allowbreak \flrmNodeCrequest, \allowbreak \flrmGatewayexit, \allowbreak \flrmReceiver, \langle \flrmGatewayexitreply, \allowbreak \flrmNodeAresponse, \allowbreak \flrmNodeBresponse, \allowbreak \flrmNodeCresponse, \allowbreak  \flrmGatewayentryreply, \allowbreak \flrmSenderreply \rangle]$, where the sender $\flrmSender$ is honest and the entry gateway  $\flrmGatewayentry$ is corrupt. Let $\flrmNodeBresponse$ be the next honest party. Then the list leaked to the corrupt entry gateway $\flrmGatewayentry$ is $\flrmlistleakage \allowbreak = \allowbreak [\flrmNodeArequest, \allowbreak \flrmNodeBrequest, \allowbreak \flrmNodeCrequest, \allowbreak \flrmGatewayexit, \allowbreak \flrmReceiver, \langle \flrmGatewayexitreply, \allowbreak \flrmNodeAresponse, \allowbreak \flrmNodeBresponse \rangle]$. We remark that, if $\flrmGatewayexitreply$ is honest, $\flrmlistleakage$ would not change, because $\flrmGatewayexitreply$ is disregarded when finding the next honest party.

The reason why $\flrmlistleakage$  needs to be leaked to a corrupt entry gateway $\flrmGatewayentry$ is that all adversarial parties are assumed to collude with each other. Therefore, if a corrupt $\flrmGatewayentry$ receives a packet, $\flrmGatewayentry$ learns the next parties to which the party will be forwarded whenever those parties are corrupt. Note that, if all the parties in the route of the request packet are corrupt, including the receiver $\flrmReceiver$, then $\flrmGatewayentry$ also learns the message $\flrmmessagerequest$. Moreover, if all the aforementioned parties are corrupt, and a reply is enabled, $\flrmGatewayentry$ also learns the parties included in the path for the reply until and including the next honest party after the exit gateway $\flrmGatewayexitreply$.

\begin{figure}
    \begin{framed}
    \small
Algorithm $(\flrmlistleakage, \flrmindexhonest) \gets \flrmCompLeakList(\flrmroutelist, \flrmindex, \flrmSetA, \flrmSetB, \flrmSetC, \flrmSetW, \flrmSetU, \flrmSetS)$
\begin{itemize}

    \item Abort if $\flrmroutelist$ is not a list of size 13.
    
    \item Abort if $\flrmroutelist[1] \allowbreak \notin \allowbreak \flrmSetU$, or  if $\flrmroutelist[2] \allowbreak \notin \allowbreak \flrmSetW$, or if $\flrmroutelist[3] \allowbreak \notin \allowbreak \flrmSetA$, or if $\flrmroutelist[4] \allowbreak \notin \allowbreak \flrmSetB$, or if $\flrmroutelist[5] \allowbreak \notin \allowbreak \flrmSetC$, or if $\flrmroutelist[6] \allowbreak \notin \allowbreak \flrmSetW$, or if $\flrmroutelist[7] \allowbreak \notin \allowbreak \flrmSetU$.

    \item Abort if $\flrmroutelist[1] \allowbreak \in \allowbreak \flrmSetS$. (This algorithm should only be run when the sender is honest.)

    \item If $\flrmroutelist[1] \allowbreak \neq \allowbreak \flrmroutelist[13]$, abort if $(\flrmroutelist[8] \allowbreak = \allowbreak \flrmroutelist[9] \allowbreak = \allowbreak \flrmroutelist[10] \allowbreak = \allowbreak \flrmroutelist[11] \allowbreak = \allowbreak \flrmroutelist[12] \allowbreak = \allowbreak \flrmroutelist[13] \allowbreak = \allowbreak \bot)$ does not hold. (If the destination of the reply does not equal the sender, then it must be the case that replies are not enabled and thus the parties in the route of the reply should be set to $\bot$.)

    \item If $\flrmroutelist[1] \allowbreak = \allowbreak \flrmroutelist[13]$, abort if $\flrmroutelist[8] \allowbreak \notin \allowbreak \flrmSetW$, or if $\flrmroutelist[9] \allowbreak \notin \allowbreak \flrmSetA$, or if $\flrmroutelist[10] \allowbreak \notin \allowbreak \flrmSetB$, or if $\flrmroutelist[11] \allowbreak \notin \allowbreak \flrmSetC$, or if $\flrmroutelist[12] \allowbreak \notin \allowbreak \flrmSetW$. 
    
    \item Abort if $\flrmindex \allowbreak \notin \allowbreak [2,13]$.

    \item Abort if $\flrmroutelist[\flrmindex] \allowbreak \notin \allowbreak \flrmSetS$.

    \item Set $\flrmlistleakage \allowbreak \gets \allowbreak \flrmroutelist$.

    \item Set $\flrmlistleakage[8]$ to a random party in $\flrmSetS$. (The index 8 is the index of the exit gateway $\flrmGatewayexitreply$, which should be disregarded when finding the next honest party.)

    \item For $i = 1$ to $\flrmindex$, set $\flrmlistleakage[i] \allowbreak \gets \allowbreak \bot$.

    \item Set $\flrmindexhonest \allowbreak \gets \allowbreak 0$ and $i \allowbreak \gets \allowbreak \flrmindex + 1$.

    \item While $\flrmindexhonest = 0$ do
    \begin{itemize}
    
        \item If $\flrmlistleakage[i] \allowbreak \in \allowbreak \flrmSetS$ do
        \begin{itemize}
        
            \item Set $i \allowbreak \gets \allowbreak i+1$.
            
        \end{itemize}
        \item Else if $\flrmlistleakage[i] \allowbreak = \bot$ do
        \begin{itemize}
        
            \item Set $\flrmindexhonest \allowbreak \gets \allowbreak \bot$. (This happens when there are not honest parties in $\flrmlistleakage$, which can only happen when the packet does not enable replies.)
            
        \end{itemize}
        \item Else do
        \begin{itemize}

            \item Set $\flrmindexhonest \allowbreak \gets \allowbreak i$.
            
        \end{itemize}
        
    \end{itemize}
    \item If $\flrmindexhonest \allowbreak \neq \allowbreak \bot$ do
    \begin{itemize}

        \item For $i = \flrmindexhonest + 1$ to $13$, set $\flrmlistleakage[i] \allowbreak \gets \allowbreak \bot$.  
    
        \item If $\flrmindex \allowbreak \leq \allowbreak 7$ and $\flrmindexhonest \geq 9$, set $\flrmlistleakage[8] \allowbreak \gets \allowbreak \flrmroutelist[8]$.
        
    \end{itemize}

    \item Erase all the components in  $\flrmlistleakage$ that are equal to $\bot$. (This is done so that $\flrmlistleakage$ does not leak whether the packet is a request packet or a reply packet.)

    \item Output $\flrmlistleakage$ and $\flrmindexhonest$.

\end{itemize}
    \end{framed}
    \Description{Algorithm $\flrmCompLeakList$}
    \caption{Algorithm $\flrmCompLeakList$}
    \label{fig:algcompleaklist}
\end{figure}

\paragraph{Algorithm $\flrmCompIndex$.} When a packet computed by an honest sender $\flrmSender$ is forwarded to a corrupt party, $\Functionality_{\LRM}$ uses the algorithm $\flrmCompIndex$ to compute the index $\flrmindex'$ of the next party that should interact with  $\Functionality_{\LRM}$ regarding that packet. For example, if an honest sender $\flrmSender$ computes a packet and sends it to a corrupt entry gateway $\flrmGatewayentry$, and subsequent parties on the route are corrupt until reaching an honest receiver $\flrmReceiver$, the algorithm $\flrmCompIndex$ outputs $\flrmindex' \allowbreak = \allowbreak 6$, which is the index of the corrupt exit gateway $\flrmGatewayexit$ that should forward the packet to the honest receiver $\flrmReceiver$.

When the subsequent parties in the route are corrupt including the receiver $\flrmReceiver$, and the packet enables a reply, the algorithm $\flrmCompIndex$ outputs $\flrmindex' \allowbreak = \allowbreak 7$, regardless of whether the parties after $\flrmReceiver$ are corrupt or honest. The reason is that $\Functionality_{\LRM}$ needs to receive the reply message $\flrmmessageresponse$ from the receiver $\flrmReceiver$ through the reply interface. 

The algorithm $\flrmCompIndex$ receives as input the index $\flrmindex$ of the corrupt party that receives the packet and the index $\flrmindexhonest$ of the last honest party in the leakage list $\flrmlistleakage$ computed by algorithm $\flrmCompLeakList$.  The algorithm $\flrmCompIndex$ outputs the index $\flrmindex'$ of the next party that should interact with $\Functionality_{\LRM}$ regarding that packet. We describe the algorithm $\flrmCompIndex$ in Figure~\ref{fig:algcompindex}.

\begin{figure}
    \begin{framed}
    \small
Algorithm $(\flrmindex') \gets \flrmCompIndex(\flrmindex, \flrmindexhonest)$
\begin{itemize}

    \item If $\flrmindexhonest = \bot$, set $\flrmindex' = \bot$. (If all the remaining parties in the route are corrupt, the functionality will not process this packet again.)

    \item Abort if $\flrmindex \allowbreak \notin \allowbreak [2,13]$, or if $\flrmindexhonest \allowbreak \notin \allowbreak [3,13]$, or if $\flrmindexhonest \allowbreak \leq \allowbreak \flrmindex$.

    \item If $\flrmindexhonest \leq 7$, output $\flrmindex' \gets \flrmindexhonest - 1$.

    \item If $\flrmindexhonest \geq 9$, do the following: (We remark that $\flrmindexhonest$ never equals 8.)
    \begin{itemize}

        \item If $\flrmindex \leq 7$, output $\flrmindex' \gets 7$.

        \item If $\flrmindex \geq 8$, output $\flrmindex' \gets \flrmindexhonest - 1$.

    \end{itemize}
    
    \end{itemize}
    \end{framed}
    \Description{Algorithm $\flrmCompIndex$}
    \caption{Algorithm $\flrmCompIndex$}
    \label{fig:algcompindex}
\end{figure}






\subsubsection{Algorithms to Check Party Set Membership}

$\Functionality_{\LRM}$ performs checks to guarantee that parties belong to their correct sets. 
When an honest party forwards a packet, $\Functionality_{\LRM}$ checks whether the party that receives the packet belongs to the correct set and aborts if that is not the case. For instance, if the packet is forwarded by an honest second-layer node $\flrmNodeBrequest$, $\Functionality_{\LRM}$ checks that the packet is received by a party in $\flrmSetC$. Similarly, when an honest party receives a packet, $\Functionality_{\LRM}$ checks whether the party that forwards the packet belongs to the correct set and aborts if that is not the case. For instance, if the packet is received by an honest first-layer node $\flrmNodeArequest$,  $\Functionality_{\LRM}$ checks that the packet was forwarded by a party in $\flrmSetW$.

\paragraph{Algorithm $\flrmverify$.} In the forward interface, $\Functionality_{\LRM}$ uses the algorithm $\flrmverify$ in Figure~\ref{fig:algverify} to check that the intended recipient of a packet belongs to the correct set. In the request and reply interfaces, $\Functionality_{\LRM}$ performs similar checks but does not use the algorithm $\flrmverify$.

The algorithm $\flrmverify$ receives as input the packet route $\flrmroutelist$, the index $\flrmindex$ of the party that forwards the packet, and the sets of parties $\flrmSetA$, $\flrmSetB$, $\flrmSetC$, $\flrmSetW$ and $\flrmSetU$. The algorithm checks if the party with index $\flrmindex + 1$ in the packet route belongs to the correct set. If that is the case,  $\flrmverify$ outputs $1$, else outputs $0$.

\begin{figure}
    \begin{framed}
    \small
Algorithm $b \gets \flrmverify(\flrmroutelist, \flrmindex, \flrmSetA, \flrmSetB, \flrmSetC, \flrmSetW, \flrmSetU)$
\begin{itemize}

    \item Abort if $\flrmroutelist$ is not a list of size 13.

    \item Abort if $\flrmindex \allowbreak \notin \allowbreak [2,6] \cup [8,12]$.

    \item Parse $\flrmroutelist$ as $[\flrmSender, \allowbreak \flrmGatewayentry, \allowbreak \flrmNodeArequest, \allowbreak \flrmNodeBrequest, \allowbreak \flrmNodeCrequest, \allowbreak \flrmGatewayexit, \allowbreak \flrmReceiver, \langle \flrmGatewayexitreply, \allowbreak \flrmNodeAresponse, \allowbreak \flrmNodeBresponse, \allowbreak \flrmNodeCresponse, \allowbreak  \flrmGatewayentryreply, \allowbreak \flrmSenderreply \rangle]$.

    \item If $\flrmindex = 2$, output $1$ if $\flrmNodeArequest \in \flrmSetA$, else output $0$.
    
    \item If $\flrmindex = 3$, output $1$ if $\flrmNodeBrequest \in \flrmSetB$, else output $0$.
    
    \item If $\flrmindex = 4$, output $1$ if $\flrmNodeCrequest \in \flrmSetC$, else output $0$.
    
    \item If $\flrmindex = 5$, output $1$ if $\flrmGatewayexit \in \flrmSetW$, else output $0$.
    
    \item If $\flrmindex = 6$, output $1$ if $\flrmReceiver \in \flrmSetU$, else output $0$.
    
    \item If $\flrmindex = 8$, output $1$ if $\flrmNodeAresponse \in \flrmSetA$, else output $0$.
    
    \item If $\flrmindex = 9$, output $1$ if $\flrmNodeBresponse \in \flrmSetB$, else output $0$.
    
    \item If $\flrmindex = 10$, output $1$ if $\flrmNodeCresponse \in \flrmSetC$, else output $0$.
    
    \item If $\flrmindex = 11$, output $1$ if $\flrmGatewayexitreply \in \flrmSetW$, else output $0$.
    
    \item If $\flrmindex = 12$, output $1$ if $\flrmSenderreply \in \flrmSetU$, else output $0$.

\end{itemize}
    \end{framed}
    \Description{Algorithm $\flrmverify$}
    \caption{Algorithm $\flrmverify$}
    \label{fig:algverify}
\end{figure}

\paragraph{Algorithm $\flrmpriorvf$.} In the forward interface, $\Functionality_{\LRM}$ uses the algorithm $\flrmpriorvf$ in Figure~\ref{fig:algpriorvf} to check that the party that forwards a packet belongs to the correct set. In the request and reply interfaces, $\Functionality_{\LRM}$ performs similar checks but does not use the algorithm $\flrmpriorvf$.

The algorithm $\flrmpriorvf$ receives as input the identifier $\flrmposition$ of the party that forwards the packet, the index $\flrmindex$ of the party that receives the packet, and the sets of parties $\flrmSetA$, $\flrmSetB$, $\flrmSetC$ and $\flrmSetW$. The algorithm checks if $\flrmposition$ belongs to the correct set. If that is the case,  $\flrmpriorvf$ outputs $1$, else outputs $0$.

\begin{figure}
    \begin{framed}
    \small
Algorithm $b \gets \flrmpriorvf(\flrmposition, \flrmindex, \flrmSetA, \flrmSetB, \flrmSetC, \flrmSetW)$
\begin{itemize}

    \item Abort if $\flrmindex \allowbreak \notin \allowbreak [3,7] \cup [9,13]$.

    \item If $\flrmindex = 3$, output $1$ if $\flrmposition \in \flrmSetW$, else output $0$.
    
    \item If $\flrmindex = 4$, output $1$ if $\flrmposition \in \flrmSetA$, else output $0$.
    
    \item If $\flrmindex = 5$, output $1$ if $\flrmposition \in \flrmSetB$, else output $0$.
    
    \item If $\flrmindex = 6$, output $1$ if $\flrmposition \in \flrmSetC$, else output $0$.
    
    \item If $\flrmindex = 7$, output $1$ if $\flrmposition \in \flrmSetW$, else output $0$.
    
    \item If $\flrmindex = 9$, output $1$ if $\flrmposition \in \flrmSetW$, else output $0$.
    
    \item If $\flrmindex = 10$, output $1$ if $\flrmposition \in \flrmSetA$, else output $0$.
    
    \item If $\flrmindex = 11$, output $1$ if $\flrmposition \in \flrmSetB$, else output $0$.
    
    \item If $\flrmindex = 12$, output $1$ if $\flrmposition \in \flrmSetC$, else output $0$.
    
    \item If $\flrmindex = 13$, output $1$ if $\flrmposition \in \flrmSetW$, else output $0$.

\end{itemize}
    \end{framed}
    \Description{Algorithm $\flrmpriorvf$}
    \caption{Algorithm $\flrmpriorvf$}
    \label{fig:algpriorvf}
\end{figure}

%% file: 4SecDefs4Definition.tex
\subsection{Formal Definition of the Functionality}
\label{sec:functionalityDefinition}

We define the security properties of a layered mixnet in the client model in the ideal-world/real-world paradigm, which is summarized in~\S\ref{sec:ucsecurity}. To this end, we define an ideal functionality $\Functionality_{\LRM}$ with 5 interfaces.  In~\S\ref{sec:functionalitySetup}, we define the setup interface. In~\S\ref{sec:functionalityRegistration}, we define the registration interface. In~\S\ref{sec:functionalityRequest}, we define the request interface. In~\S\ref{sec:functionalityReply}, we define the reply interface. In~\S\ref{sec:functionalityForward}, we define the forward interface.

\input{4SecDefs4Definition1Setup}
\input{4SecDefs4Definition2Registration}
\input{4SecDefs4Definition3Request}
\input{4SecDefs4Definition4Reply}
\input{4SecDefs4Definition5Forward}

%% file: 4SecDefs4Definition1Setup.tex
\subsubsection{Setup Interface}
\label{sec:functionalitySetup}

\paragraph{Definition of the setup interface.} $\Functionality_{\LRM}$ interacts with the parties described in~\S\ref{sec:functionalityParties} and uses the parameters, variables and algorithms defined in~\S\ref{sec:functionalityParameters}.  
        \begin{enumerate}

             \item[1.] On input $(\flrmsetupini, \allowbreak \sid)$ from a node or gateway $\flrmParty$:

             \begin{itemize}

                \item Abort if $\sid \neq (\flrmSetA, \allowbreak \flrmSetB, \allowbreak \flrmSetC, \allowbreak \flrmSetW, \allowbreak \sid')$, or if $\flrmParty \notin \flrmSetA \cup \allowbreak \flrmSetB \cup \allowbreak \flrmSetC \cup \flrmSetW$, or if $|\flrmSetA| \allowbreak < \allowbreak \flrmNodesmin$, or if $|\flrmSetB| \allowbreak < \allowbreak \flrmNodesmin$, or if $|\flrmSetC| \allowbreak < \allowbreak \flrmNodesmin$, or if $|\flrmSetW| \allowbreak < \allowbreak \flrmGatewaysmin$, or if $\flrmSetA \allowbreak \cap \allowbreak \flrmSetB \neq \emptyset$, or if $\flrmSetA \allowbreak \cap \allowbreak \flrmSetC \neq \emptyset$, or if $\flrmSetA \allowbreak \cap \allowbreak \flrmSetW \neq \emptyset$, or if $\flrmSetB \allowbreak \cap \allowbreak \flrmSetC \neq \emptyset$, or if $\flrmSetB \allowbreak \cap \allowbreak \flrmSetW \neq \emptyset$, or if $\flrmSetC \allowbreak \cap \allowbreak \flrmSetW \neq \emptyset$, or if $(\sid, \flrmParty', 0)$ such that $\flrmParty' \allowbreak = \allowbreak \flrmParty$ is already stored.

                \item Store $(\sid, \flrmParty, 0)$.
                
                \item Send $(\flrmsetupsim, \allowbreak \sid, \allowbreak \flrmParty)$ to $\Simulator$.

             \end{itemize}

             \item[S.] On input $(\flrmsetuprep, \allowbreak \sid, \allowbreak \flrmParty)$ from $\Simulator$:
        
            \begin{itemize}
        
                \item Abort if $(\sid, \flrmParty', 0)$ such that $\flrmParty' \allowbreak = \allowbreak \flrmParty$ is not stored, or if $(\sid, \flrmParty', 1)$ such that $\flrmParty' \allowbreak = \allowbreak \flrmParty$ is already stored.
        
                \item Store $(\sid, \flrmParty, 1)$.
                
                \item Send $(\flrmsetupend, \allowbreak \sid)$ to $\flrmParty$.
        
            \end{itemize} 
            
        \end{enumerate}

\paragraph{Explanation and discussion of the setup interface.} After receiving the $\flrmsetupini$ message, $\Functionality_{\LRM}$ aborts if the session identifier $\sid$ does not contain one set of nodes for each layer and a set of gateways. Including the sets of nodes and gateways in the session identifier allows any sets of nodes and gateways to create their own instance of the functionality.
    
$\Functionality_{\LRM}$ also aborts if the party $\flrmParty$ that invokes the interface is not in the sets of nodes and gateways included in the session identifier $\sid$. This ensures that only nodes and gateways in those sets can invoke this interface.  

Moreover, $\Functionality_{\LRM}$ aborts if the size of any of the sets of nodes is lower than a parameter $\flrmNodesmin$ or if the size of the set of gateways is lower than a parameter $\flrmGatewaysmin$. This restriction is needed because, if the sets are too small (e.g.\ $|\flrmSetA| = \allowbreak |\flrmSetB| = \allowbreak |\flrmSetC| = \allowbreak |\flrmSetW| = \allowbreak 1$), then $\Functionality_{\LRM}$ does not provide the intended privacy properties. $\Functionality_{\LRM}$ also aborts if the sets $\flrmSetA$, $\flrmSetB$, $\flrmSetC$ and $\flrmSetW$ are not disjoint, i.e., each node or gateway can only belong to one of the sets.

In addition, $\Functionality_{\LRM}$ aborts if the node or gateway $\flrmParty$ has already run the setup interface. This means that the functionality describes an execution of the protocol in which nodes and gateways can only run the setup interface once. In a practical system, nodes and gateways need to renew their keys periodically, and thus run the setup interface more than once. To analyze the security of such a system, we would consider that, each time nodes and gateways run the setup interface, a new instance of the protocol $\mathrm{\Pi}_{\LRM}$ is run. Because of the UC theorem, if a protocol $\mathrm{\Pi}_{\LRM}$ securely realizes $\Functionality_{\LRM}$, then the protocol that consists of the sequential composition of multiple instances of $\mathrm{\Pi}_{\LRM}$ is also secure. 

If $\Functionality_{\LRM}$ does not abort, $\Functionality_{\LRM}$ records the fact that the party $\flrmParty$ has invoked the setup interface. After that, $\Functionality_{\LRM}$ sends a message $\flrmsetupsim$ to the simulator. This message lets the simulator know that the party $\flrmParty$ is running the setup interface. If $\Functionality_{\LRM}$ did not disclose that information to the simulator, then any protocol that realizes $\Functionality_{\LRM}$ would need to hide from the real-world adversary the fact that $\flrmParty$ is running the setup interface, which would be an unnecessary privacy property.

After sending  the message $\flrmsetupsim$ to the simulator, $\Functionality_{\LRM}$ receives a $\flrmsetuprep$ message from the simulator.  This is needed because the real-world adversary controls the network and thus has the ability to drop messages. Therefore, for any protocol  $\mathrm{\Pi}_{\LRM}$ in which the setup interface requires communication between parties, the real-world adversary can impede the finalization of the execution of the setup interface, and thus the simulator must be able to do the same in the ideal world. We remark the following:
    \begin{itemize}

        \item $\Functionality_{\LRM}$ resumes the execution of the setup interface when the simulator sends the $\flrmsetuprep$ message, but $\Functionality_{\LRM}$ does not wait for the simulator to send the message. This is needed because the real-world adversary may not allow the finalization of the execution of the setup interface immediately, i.e.\ the adversary can delay messages for an indefinite amount of time, and in that case the simulator is not able to immediately reply to $\Functionality_{\LRM}$ with a $\flrmsetuprep$ message after receiving a $\flrmsetupsim$ message. If $\Functionality_{\LRM}$ waited for the simulator, and the simulator is not able to provide the reply, then the execution of the ideal-world protocol would be blocked, whereas in the real-world the execution of the protocol continues even if the real-world adversary delays the finalization of the setup interface for one of the nodes or gateways. That would enable the environment to distinguish between the real-world and the ideal-world.

        \item $\Functionality_{\LRM}$ does not need to create a query identifier $\qid$ to match the message $\flrmsetupsim$ to the reply $\flrmsetuprep$ from the simulator. The reason is that $\Functionality_{\LRM}$ ensures that the setup interface is only run once by every node or gateway $\flrmParty$. That allows $\Functionality_{\LRM}$ to use the identifier $\flrmParty$ to match the message $\flrmsetupsim$ with its corresponding reply $\flrmsetuprep$.

    \end{itemize}

Once $\Functionality_{\LRM}$ receives a message $\flrmsetuprep$ with an identifier $\flrmParty$, $\Functionality_{\LRM}$ aborts if the party $\flrmParty$ did not invoke the setup interface, or if the execution of the setup interface for $\flrmParty$ had already finished.
    
If $\Functionality_{\LRM}$ does not abort, $\Functionality_{\LRM}$ records the fact that the setup interface is executed for $\flrmParty$. Finally, $\Functionality_{\LRM}$ sends a $\flrmsetupend$ message to the party $\flrmParty$ to inform the party that the setup interface has been executed.

%% file: 4SecDefs4Definition2Registration.tex
\subsubsection{Registration Interface}
\label{sec:functionalityRegistration}

\paragraph{Definition of the registration interface.} $\Functionality_{\LRM}$ interacts with the parties described in~\S\ref{sec:functionalityParties} and uses the parameters, variables and algorithms defined in~\S\ref{sec:functionalityParameters}.

    \begin{enumerate}
    
        \item[2.] On input $(\flrmregisterini, \allowbreak \sid)$ from a user $\flrmUser$:
        
            \begin{itemize}

                \item Abort if $\flrmUser \notin \flrmSetU$.
        
                \item Abort if a tuple $(\sid, \allowbreak \flrmUser', \allowbreak 0)$ such that $\flrmUser' = \flrmUser$ is already stored.

                \item If $\flrmUser$ is honest, parse $\sid$ as $(\flrmSetA, \allowbreak \flrmSetB, \allowbreak \flrmSetC, \allowbreak \flrmSetW, \allowbreak \sid')$ and abort if, for any $\flrmParty \in \{\flrmSetA, \allowbreak \flrmSetB, \allowbreak \flrmSetC, \allowbreak \flrmSetW\}$, the tuple $(\sid, \flrmParty, 1)$ is not stored.

                \item If $\flrmUser$ is corrupt, abort if $\sid \neq (\flrmSetA, \allowbreak \flrmSetB, \allowbreak \flrmSetC, \allowbreak \flrmSetW, \allowbreak \sid')$,  or if $|\flrmSetA| \allowbreak < \allowbreak \flrmNodesmin$, or if $|\flrmSetB| \allowbreak < \allowbreak \flrmNodesmin$, or if $|\flrmSetC| \allowbreak < \allowbreak \flrmNodesmin$, or if $|\flrmSetW| \allowbreak < \allowbreak \flrmGatewaysmin$.
        
                \item Store $(\sid, \allowbreak \flrmUser, \allowbreak 0)$.
                
                \item Send $(\flrmregistersim, \allowbreak \sid, \allowbreak \flrmUser)$ to $\Simulator$.
        
            \end{itemize}

            \item[S.] On input $(\flrmregisterrep, \allowbreak \sid, \allowbreak \flrmUser)$ from  $\Simulator$:
        
            \begin{itemize}
        
                \item Abort if $(\sid, \allowbreak \flrmUser', \allowbreak 0)$ such that $\flrmUser' = \flrmUser$ is not stored, or if $(\sid, \allowbreak \flrmUser', \allowbreak 1)$ such that $\flrmUser' = \flrmUser$ is stored.
        
                \item Store $(\sid, \allowbreak \flrmUser, \allowbreak 1)$. 
                
                \item Send $(\flrmregisterend, \allowbreak \sid)$ to $\flrmUser$.
        
            \end{itemize}
            
    \end{enumerate}

\paragraph{Explanation and discussion of the registration interface.} After receiving the $\flrmregisterini$ message from a user $\flrmUser$, $\Functionality_{\LRM}$ aborts if $\flrmUser$ does not belong to the set of users $\flrmSetU$ that parameterizes $\Functionality_{\LRM}$, or if $\Functionality_{\LRM}$ had already invoked the registration interface. This means that users can invoke the registration interface only once. In a practical system, users may need to renew their keys periodically. We consider that, each time users renew their keys, a new execution of the protocol begins, as described in~\S\ref{sec:functionalitySetup} for nodes and gateways when they run the setup interface.

   If $\flrmUser$ is honest, $\Functionality_{\LRM}$ also aborts if there is any node or gateway that has not run the setup interface yet. Therefore, $\Functionality_{\LRM}$ enforces that honest users register after all the nodes and gateways run the setup interface. The reason why $\Functionality_{\LRM}$ enforces that is the protection of user privacy. In $\mathrm{\Pi}_{\LRM}$, users need to retrieve the public keys of nodes and gateways. When they do that, the real-world adversary learns the fact that a user is retrieving a certain public key. If users retrieve all of the public keys during registration, they do not leak to the adversary any information about the nodes or gateways that they intend to use. However, if some of the public keys are not ready during registration and the user retrieves them afterwards (e.g. when the user needs to send a request), then the adversary could infer that the user is using a certain node or gateway. To avoid that, all the nodes and gateways should have run the setup interface before the users run the registration interface.

    If $\flrmUser$ is corrupt, $\Functionality_{\LRM}$ performs the same checks on the session identifier $\sid$ that were conducted in the setup interface. The reason is that a corrupt user might invoke the registration interface before any node or gateway invoke the setup interface.


   If $\Functionality_{\LRM}$ does not abort, $\Functionality_{\LRM}$ records the fact that $\flrmUser$ has invoked the registration interface. Then the functionality $\Functionality_{\LRM}$ sends a message $\flrmregistersim$ to the simulator and leaks the identifier $\flrmUser$ to the simulator.  If $\Functionality_{\LRM}$ did not disclose that information to the simulator, then any protocol that realizes $\Functionality_{\LRM}$ would need to hide from the real-world adversary the fact that $\flrmUser$ is running the registration interface, which could be an unnecessary privacy property.

   After sending  the message $\flrmregistersim$ to the simulator, $\Functionality_{\LRM}$ receives a $\flrmregisterrep$ message from the simulator. As described above for the setup interface,  this is needed because the real-world adversary controls the network and thus has the ability to drop messages. Therefore, for any protocol  $\mathrm{\Pi}_{\LRM}$ in which the registration interface requires communication between parties, the real-world adversary can impede the finalization of the execution of the setup interface. The same remarks made in~\S\ref{sec:functionalitySetup} for the setup interface regarding both the fact that $\Functionality_{\LRM}$ does not wait for the simulator to send the $\flrmregisterrep$ message, and the fact that a query identifier $\qid$ is not needed, apply to the registration interface.

   When  $\Functionality_{\LRM}$ receives a $\flrmregisterrep$ message, $\Functionality_{\LRM}$ aborts if $\flrmUser$ did not invoke the registration interface or if the execution of the registration interface for $\flrmUser$ has already finished.
    
   If $\Functionality_{\LRM}$ does not abort, the functionality $\Functionality_{\LRM}$ records the fact that the registration interface is executed for $\flrmUser$. Finally, $\Functionality_{\LRM}$ sends a $\flrmregisterend$ message to $\flrmUser$ to inform the user that the registration interface has been executed.

%% file: 4SecDefs4Definition3Request.tex
\subsubsection{Request Interface}
\label{sec:functionalityRequest}

\paragraph{Definition of the request interface.} $\Functionality_{\LRM}$ interacts with the parties described in~\S\ref{sec:functionalityParties} and uses the parameters, variables and algorithms defined in~\S\ref{sec:functionalityParameters}.

   \begin{enumerate}
    \item[3.] On input the  message $(\flrmsendini, \allowbreak \sid, \allowbreak \ppid, \allowbreak \flrmGatewayentry, \allowbreak \flrmGatewayexit, \allowbreak \flrmReceiver, \allowbreak \flrmmessagerequest, \allowbreak \flrmNodeArequest, \allowbreak \flrmNodeBrequest, \allowbreak \flrmNodeCrequest, \allowbreak \langle   \flrmGatewayentryreply, \allowbreak \flrmGatewayexitreply, \allowbreak \flrmNodeAresponse, \allowbreak \flrmNodeBresponse, \allowbreak \flrmNodeCresponse, \allowbreak \flrmSenderreply \rangle)$ from a sender $\flrmSender$:

            \begin{itemize}

                \item If $\flrmSender \allowbreak \notin \allowbreak \flrmSetS$, abort if $\flrmmessagerequest \allowbreak \notin \allowbreak \flrmMessageSpace$, or if $(\sid, \allowbreak \flrmUser, \allowbreak 1)$ such that $\flrmUser = \flrmSender$ is not stored, or if  $(\sid, \allowbreak \flrmUser, \allowbreak 1)$ such that $\flrmUser = \flrmReceiver$ is not stored. 

                \item If $\flrmSender \allowbreak \notin \allowbreak \flrmSetS$, do the following:

                \begin{itemize}

                    \item Abort if $\flrmGatewayentry \notin \flrmSetW$, or if $\flrmGatewayexit \notin \flrmSetW$, or if $\flrmNodeArequest \allowbreak \notin \flrmSetA$, or if $\flrmNodeBrequest \allowbreak \notin \flrmSetB$, or if $\flrmNodeCrequest \allowbreak \notin \flrmSetC$.
                    

                    \item If  $\langle   \flrmGatewayentryreply, \allowbreak \flrmGatewayexitreply, \allowbreak \flrmNodeAresponse, \allowbreak \flrmNodeBresponse, \allowbreak \flrmNodeCresponse, \allowbreak \flrmSenderreply \rangle \neq \bot$, abort if $\flrmGatewayentryreply \notin \flrmSetW$, or if $\flrmGatewayexitreply \notin \flrmSetW$, or if $\flrmNodeAresponse \allowbreak \notin \flrmSetA$, or if $\flrmNodeBresponse \allowbreak \notin \flrmSetB$, or if $\flrmNodeCresponse \allowbreak \notin \flrmSetC$, or if $\flrmSenderreply \allowbreak \neq \allowbreak \flrmSender$.


                    \item If $\langle   \flrmGatewayentryreply, \allowbreak \flrmGatewayexitreply, \allowbreak \flrmNodeAresponse, \allowbreak \flrmNodeBresponse, \allowbreak \flrmNodeCresponse, \allowbreak \flrmSenderreply \rangle \allowbreak \neq \allowbreak \bot$, abort if a tuple of the form $(\tid, \allowbreak \ppid', \allowbreak \flrmposition,  \allowbreak \flrmindex, \allowbreak \flrmroutelist, \allowbreak \flrmmessagerequest)$ with $\flrmroutelist \allowbreak = \allowbreak [\flrmSender', \allowbreak \flrmGatewayentry, \allowbreak \flrmNodeArequest, \allowbreak \flrmNodeBrequest, \allowbreak \flrmNodeCrequest, \allowbreak \flrmGatewayexit, \allowbreak \flrmReceiver, \langle \flrmGatewayexitreply, \allowbreak \flrmNodeAresponse, \allowbreak \flrmNodeBresponse, \allowbreak \flrmNodeCresponse, \allowbreak  \flrmGatewayentryreply, \allowbreak \flrmSenderreply \rangle]$ such that $\ppid' \allowbreak = \allowbreak \ppid$ and $\flrmSender' \allowbreak = \allowbreak \flrmSender$ is stored, or if a tuple of the form $(\qid, \allowbreak \ppid', \allowbreak \flrmposition,  \allowbreak \flrmindex, \allowbreak \flrmroutelist, \allowbreak \flrmmessagerequest)$ with $\flrmroutelist \allowbreak = \allowbreak [\flrmSender', \allowbreak \flrmGatewayentry, \allowbreak \flrmNodeArequest, \allowbreak \flrmNodeBrequest, \allowbreak \flrmNodeCrequest, \allowbreak \flrmGatewayexit, \allowbreak \flrmReceiver, \langle \flrmGatewayexitreply, \allowbreak \flrmNodeAresponse, \allowbreak \flrmNodeBresponse, \allowbreak \flrmNodeCresponse, \allowbreak  \flrmGatewayentryreply, \allowbreak \flrmSenderreply \rangle]$ such that $\ppid' \allowbreak = \allowbreak \ppid$ and $\flrmSender' \allowbreak = \allowbreak \flrmSender$ is stored.       
                    
                \end{itemize}

                \item If $\flrmSender \allowbreak \in \allowbreak \flrmSetS$, abort if there is a tuple of the form $(\tid, \allowbreak \ppid', \allowbreak \flrmposition,  \allowbreak \flrmindex, \allowbreak \flrmroutelist, \allowbreak \flrmmessagerequest)$ with packet route $\flrmroutelist \allowbreak = \allowbreak [\flrmSender', \allowbreak \flrmGatewayentry, \allowbreak \flrmNodeArequest, \allowbreak \flrmNodeBrequest, \allowbreak \flrmNodeCrequest, \allowbreak \flrmGatewayexit, \allowbreak \flrmReceiver, \langle \flrmGatewayexitreply, \allowbreak \flrmNodeAresponse, \allowbreak \flrmNodeBresponse, \allowbreak \flrmNodeCresponse, \allowbreak  \flrmGatewayentryreply, \allowbreak \flrmSenderreply \rangle]$ such that $\ppid' \allowbreak = \allowbreak \ppid$ and $\flrmSender' \allowbreak \in \allowbreak \flrmSetS$, or if there is a tuple   $(\qid, \allowbreak \ppid', \allowbreak \flrmposition,  \allowbreak \flrmindex, \allowbreak \flrmroutelist, \allowbreak \flrmmessagerequest)$ with $\flrmroutelist \allowbreak = \allowbreak [\flrmSender', \allowbreak \flrmGatewayentry, \allowbreak \flrmNodeArequest, \allowbreak \flrmNodeBrequest, \allowbreak \flrmNodeCrequest, \allowbreak \flrmGatewayexit, \allowbreak \flrmReceiver, \langle \flrmGatewayexitreply, \allowbreak \flrmNodeAresponse, \allowbreak \flrmNodeBresponse, \allowbreak \flrmNodeCresponse, \allowbreak  \flrmGatewayentryreply, \allowbreak \flrmSenderreply \rangle]$ such that $\ppid' \allowbreak = \allowbreak \ppid$ and $\flrmSender' \allowbreak \in \allowbreak \flrmSetS$.       

                \item If $\flrmSender \allowbreak \notin \allowbreak \flrmSetS$ and $(\sid, \allowbreak \flrmSender', \allowbreak \flrmSetR)$ such that $\flrmSender' \allowbreak = \allowbreak \flrmSender$ is stored, do the following:
                \begin{itemize}

                    \item If $\flrmReceiver \allowbreak \in \allowbreak \flrmSetR$, set $\flrmfirstmessage \gets 0$. 
                    
                    \item Else, set $\flrmfirstmessage \allowbreak \gets \allowbreak 1$, set $\flrmSetR' \allowbreak \gets \allowbreak \flrmSetR \allowbreak \cup \allowbreak \{\flrmReceiver\}$ and replace $\flrmSetR$ by $\flrmSetR'$ in the tuple $(\sid, \allowbreak \flrmSender, \allowbreak \flrmSetR)$.
                    
                \end{itemize}

                \item If $\flrmSender \allowbreak \notin \allowbreak \flrmSetS$ and $(\sid, \allowbreak \flrmSender', \allowbreak \flrmSetR)$ such that $\flrmSender' \allowbreak = \allowbreak \flrmSender$ is not stored, do the following:
                \begin{itemize}

                    \item Set $\flrmSetR \allowbreak \gets \allowbreak \{\flrmReceiver\}$ and store $(\sid, \allowbreak \flrmSender, \allowbreak \flrmSetR)$.

                    \item Set $\flrmfirstmessage \gets 1$.
                
                \end{itemize}

                \item Create a fresh $\qid$, set $\flrmposition \allowbreak \gets \allowbreak \flrmSender$, set $\flrmindex \allowbreak \gets \allowbreak 1$, and set $\flrmroutelist \allowbreak \gets \allowbreak [\flrmSender, \allowbreak \flrmGatewayentry, \allowbreak \flrmNodeArequest, \allowbreak \flrmNodeBrequest, \allowbreak \flrmNodeCrequest, \allowbreak \flrmGatewayexit, \allowbreak \flrmReceiver, \langle \flrmGatewayexitreply, \allowbreak \flrmNodeAresponse, \allowbreak \flrmNodeBresponse, \allowbreak \flrmNodeCresponse, \allowbreak  \flrmGatewayentryreply, \allowbreak \flrmSenderreply \rangle]$ with the party identifiers received as input.

                \item If the following conditions are satisfied
                \begin{itemize}
                    
                    \item $\flrmSender \allowbreak \in \allowbreak \flrmSetS$.

                    \item There is a tuple $(\tid, \allowbreak \ppid', \allowbreak \flrmposition',  \allowbreak \flrmindex', \allowbreak \flrmroutelist', \allowbreak \flrmmessagerequest')$ such that $\tid \allowbreak = \allowbreak \ppid$ and $\ppid \allowbreak \neq \allowbreak \bot$, where $\ppid$ is the global packet identifier received as input.


                    \item Given $\flrmroutelist' \allowbreak = \allowbreak [\flrmSender', \allowbreak \flrmGatewayentry', \allowbreak \flrmNodeArequest', \allowbreak \flrmNodeBrequest', \allowbreak \flrmNodeCrequest', \allowbreak \flrmGatewayexit', \allowbreak \flrmReceiver', \langle \flrmGatewayexitreply', \allowbreak \flrmNodeAresponse', \allowbreak \flrmNodeBresponse', \allowbreak \flrmNodeCresponse', \allowbreak  \flrmGatewayentryreply', \allowbreak \flrmSenderreply' \rangle]$ in that tuple,  $\flrmSender' \allowbreak \notin \allowbreak \flrmSetS$ and $\flrmposition' \allowbreak \in \allowbreak \flrmSetS$. 
                    
                \end{itemize}
                do the following:
                \begin{itemize}

                    \item If either (1) $\flrmindex' \allowbreak \leq \allowbreak 7$,  or (2) $\flrmindex' \allowbreak > \allowbreak 7$, and $\langle \flrmGatewayexitreply, \allowbreak \flrmNodeAresponse, \allowbreak \flrmNodeBresponse, \allowbreak \flrmNodeCresponse, \allowbreak  \flrmGatewayentryreply, \allowbreak \flrmSenderreply \rangle \allowbreak \neq \allowbreak \bot$ in $\flrmroutelist$, and $\flrmroutelist'[7] \allowbreak \in \allowbreak \flrmSetS$, do the following:
                    \begin{itemize}

                        \item For $i \allowbreak \in \allowbreak [2, \allowbreak \flrmindex']$, check that $\flrmroutelist[i] \allowbreak \notin \allowbreak \flrmSetS$ and that each party $\flrmroutelist[i]$ belongs to the correct set. Abort if that is not the case.

                        \item For $i \allowbreak \in \allowbreak [\flrmindex' + 1, \allowbreak 13]$, set $\flrmroutelist[i] \allowbreak \gets \allowbreak \flrmroutelist'[i]$.

                        \item Set $\flrmroutelist[1] \allowbreak \gets \allowbreak \flrmroutelist'[1]$.

                        \item Set $\ppid \allowbreak \gets \allowbreak \ppid'$.
                        
                        \item If $\flrmindex' \allowbreak \leq \allowbreak 6$ and $\flrmmessagerequest \allowbreak \neq \allowbreak \top$, set $\flrmmessagerequest \allowbreak \gets \allowbreak \flrmmessagerequest'$. If $\flrmindex' \allowbreak \leq \allowbreak 6$ and $\flrmmessagerequest \allowbreak = \allowbreak \top$,  set $\flrmmessagerequest \allowbreak \gets \allowbreak \bot$.

                    \end{itemize}
                        
                    \item If (3) $\flrmindex' \allowbreak > \allowbreak 7$ and $\langle \flrmGatewayexitreply, \allowbreak \flrmNodeAresponse, \allowbreak \flrmNodeBresponse, \allowbreak \flrmNodeCresponse, \allowbreak  \flrmGatewayentryreply, \allowbreak \flrmSenderreply \rangle \allowbreak = \allowbreak \bot$ in $\flrmroutelist$, do the following:
                    \begin{itemize}

                        \item For $i \allowbreak \in \allowbreak [2, \allowbreak \flrmindex'-6]$, check that $\flrmroutelist[i] \allowbreak \notin \allowbreak \flrmSetS$ and that each party $\flrmroutelist[i]$ belongs to the correct set. Abort if that is not the case.

                        \item For $i \allowbreak \in \allowbreak [8, \allowbreak \flrmindex']$, set $\flrmroutelist[i] \allowbreak \gets \allowbreak \flrmroutelist[i-6]$.

                        \item For $i \allowbreak \notin \allowbreak [8, \allowbreak \flrmindex']$, set $\flrmroutelist[i] \allowbreak \gets \allowbreak \flrmroutelist'[i]$.

                        \item Set $\ppid \allowbreak \gets \allowbreak \ppid'$.
                        
                        \item If $\flrmmessagerequest \allowbreak \neq \allowbreak \top$, set $\flrmmessagerequest \allowbreak \gets \allowbreak \flrmmessagerequest'$, else set $\flrmmessagerequest \allowbreak \gets \allowbreak \bot$.

                        \item Set $\flrmindex \allowbreak \gets \allowbreak 7$.

                        \item Mark $\qid$ as a query identifier for the reply interface.

                    \end{itemize}
                    
                \end{itemize}

                \item Store the tuple $(\qid, \allowbreak \ppid, \allowbreak \flrmposition, \allowbreak \flrmindex, \flrmroutelist, \allowbreak \flrmmessagerequest)$.

                \item Send $(\flrmsendsim, \allowbreak \sid, \allowbreak \qid, \allowbreak \flrmSender, \allowbreak \flrmGatewayentry, \allowbreak \flrmfirstmessage)$ to $\Simulator$.
                
            \end{itemize}

            \item[S.] On input $(\flrmsendrep, \allowbreak \sid, \allowbreak \qid)$ from $\Simulator$:

            \begin{itemize}

                \item Abort if a tuple $I \allowbreak = \allowbreak (\qid', \allowbreak \ppid, \allowbreak \flrmposition, \allowbreak \flrmindex, \flrmroutelist, \allowbreak \flrmmessagerequest)$ such that $\qid' \allowbreak = \allowbreak \qid$ is not stored.

                \item Parse $\flrmroutelist$ as $[\flrmSender, \allowbreak \flrmGatewayentry, \allowbreak \flrmNodeArequest, \allowbreak \flrmNodeBrequest, \allowbreak \flrmNodeCrequest, \allowbreak \flrmGatewayexit, \allowbreak \flrmReceiver, \langle \flrmGatewayexitreply, \allowbreak \flrmNodeAresponse, \allowbreak \flrmNodeBresponse, \allowbreak \flrmNodeCresponse, \allowbreak  \flrmGatewayentryreply, \allowbreak \flrmSenderreply \rangle]$.

                \item If $\flrmGatewayentry \allowbreak \notin \allowbreak \flrmSetS$, abort if $\flrmposition \allowbreak \notin \flrmSetU$ or if $\flrmGatewayentry \allowbreak \notin \allowbreak \flrmSetW$.

                \item If $\flrmGatewayentry \allowbreak \notin \allowbreak \flrmSetS$, abort if a tuple $(\sid, \allowbreak \flrmParty, \allowbreak 1)$ such that $\flrmParty = \flrmGatewayentry$ is not stored.

                \item If $\flrmSender \allowbreak \notin \allowbreak \flrmSetS$ and $\flrmGatewayentry \allowbreak \in \allowbreak \flrmSetS$, do the following:
                \begin{itemize}
                
                    \item Set $(\flrmlistleakage, \flrmindexhonest) \gets \flrmCompLeakList(\flrmroutelist, 2, \flrmSetA, \flrmSetB, \flrmSetC, \flrmSetW, \flrmSetU, \allowbreak \flrmSetS)$, where $\flrmSetA$, $\flrmSetB$, $\flrmSetC$ and $\flrmSetW$ are taken from $\sid$.

                    \item Set $\flrmindex' \gets \flrmCompIndex(2, \flrmindexhonest)$.

                    \item If $\flrmindex' \allowbreak = \allowbreak \bot$, set $\flrmposition' \allowbreak \gets \allowbreak \bot$ and $\tid \allowbreak \gets \allowbreak \bot$, else set $\flrmposition' \allowbreak \gets \allowbreak \flrmroutelist[\flrmindex']$ and create a fresh random $\tid$.

                    \item If $\flrmindex' \allowbreak = \allowbreak \bot$ or $\flrmindex' \allowbreak = \allowbreak 7$, set $\flrmmessagerequest' \allowbreak \gets \allowbreak \flrmmessagerequest$, else set $\flrmmessagerequest' \gets \bot$.

                    \item Set $\ppid' \gets \bot$.
                    
                \end{itemize}

                \item If $\flrmGatewayentry \allowbreak \notin \allowbreak \flrmSetS$, set $\flrmlistleakage \allowbreak \gets \allowbreak \bot$,  set $\flrmindex' \allowbreak \gets \allowbreak 2$, set $\flrmposition' \allowbreak \gets \allowbreak \flrmGatewayentry$, set $\flrmmessagerequest' \allowbreak \gets \allowbreak \bot$, set $\ppid' \allowbreak \gets \allowbreak \bot$, and create a fresh random $\tid$.

                \item If $\flrmSender \in \flrmSetS$ and $\flrmGatewayentry \in \flrmSetS$,  set $\flrmlistleakage \allowbreak \gets \allowbreak \bot$, set $\flrmposition' \allowbreak \gets \allowbreak \bot$, set $\flrmindex' \allowbreak \gets \allowbreak \bot$, set $\flrmmessagerequest' \allowbreak \gets \allowbreak \flrmmessagerequest$, set $\ppid' \allowbreak \gets \allowbreak \ppid$, and set $\tid \allowbreak \gets \allowbreak \bot$.

                \item Store the tuple $J \allowbreak = \allowbreak (\tid, \allowbreak \ppid, \allowbreak \flrmposition', \allowbreak \flrmindex', \flrmroutelist, \allowbreak \flrmmessagerequest)$, where $\tid$, $\flrmposition'$ and $\flrmindex'$ are set as described above and the remaining values are taken from the tuple $I$.


                \item Delete the tuple $I$.

                \item Send $(\flrmsendend, \allowbreak \sid, \allowbreak \tid, \allowbreak  \ppid', \allowbreak \flrmmessagerequest', \allowbreak \flrmSender, \allowbreak \flrmNodeArequest, \allowbreak \flrmlistleakage)$ to $\flrmGatewayentry$.

            \end{itemize}

     \end{enumerate}

\paragraph{Explanation and discussion of the request interface.} A sender $\flrmSender$ sends the $\flrmsendini$ message along with a global packet identifier $\ppid$, a message $\flrmmessagerequest$, and routing information $(\flrmGatewayentry,  \allowbreak \flrmNodeArequest, \allowbreak \flrmNodeBrequest, \allowbreak \flrmNodeCrequest, \allowbreak \flrmGatewayexit, \allowbreak \flrmReceiver)$. Optionally, the input also includes routing information $\langle   \flrmGatewayentryreply, \allowbreak \flrmGatewayexitreply, \allowbreak \flrmNodeAresponse, \allowbreak \flrmNodeBresponse, \allowbreak \flrmNodeCresponse, \allowbreak \flrmSenderreply \rangle$ when the receiver is enabled to send a reply packet. 

   
   After receiving the $\flrmsendini$ message, $\Functionality_{\LRM}$ aborts if $\flrmSender$ is honest and the message $\flrmmessagerequest$ does not belong to the message space $\flrmMessageSpace$. $\Functionality_{\LRM}$ also aborts when $\flrmSender$ is honest and $\flrmSender$ or $\flrmReceiver$ did not run the registration interface. 

   Moreover, $\Functionality_{\LRM}$ aborts when $\flrmSender$ is honest and any of the nodes or gateways included in the routing information does not belong to the right set or has not run the setup interface. $\Functionality_{\LRM}$ also aborts if $\flrmSender$ is honest and $\flrmSender \allowbreak \neq \allowbreak \flrmSenderreply$. Therefore, $\Functionality_{\LRM}$ guarantees that a reply packet associated to a request packet from the honest sender $\flrmSender$ is directed to the same honest sender $\flrmSender$. However, if the sender is corrupt, $\Functionality_{\LRM}$ does not guarantee that.  

   Furthermore, $\Functionality_{\LRM}$ aborts when $\flrmSender$ is honest, the packet enables a reply, and  the global packet identifier $\ppid$ is already stored by $\Functionality_{\LRM}$ in relation to another packet that was input by $\flrmSender$ to  $\Functionality_{\LRM}$. If a reply  is not enabled, $\Functionality_{\LRM}$ does not abort for this reason because $\ppid$ will not be used by $\Functionality_{\LRM}$.

   $\Functionality_{\LRM}$ also aborts when $\flrmSender$ is corrupt and the global packet identifier $\ppid$ is already stored by $\Functionality_{\LRM}$ in relation to another packet that was input by any corrupt party to $\Functionality_{\LRM}$. We remark that the party identifier $\flrmSender$ is set to the adversary symbol $\Adversary$ when any corrupt node or gateway inputs a packet to the functionality through the forward interface.

   If $\Functionality_{\LRM}$ does not abort, $\Functionality_{\LRM}$ checks, when the sender $\flrmSender$ is honest, if $\flrmSender$ has already sent a packet to the receiver $\flrmReceiver$. To this end, $\Functionality_{\LRM}$ checks if the party identifier $\flrmReceiver$ is included into a set $\flrmSetR$ that contains all the receivers to whom the sender $\flrmSender$ has sent a packet. If $\flrmSender$ already sent a packet to $\flrmReceiver$, then a bit $\flrmfirstmessage$ is set to $0$, else $\flrmfirstmessage$ is set to $1$ and $\flrmReceiver$ is included in $\flrmSetR$. The reason why $\Functionality_{\LRM}$ does that is to leak the bit $\flrmfirstmessage$ to the simulator. In our construction $\mathrm{\Pi}_{\LRM}$, the sender  $\flrmSender$ needs to retrieve the public key of the receiver $\flrmReceiver$ when $\flrmSender$ sends a packet to $\flrmReceiver$ for the first time, and the real-world adversary learns that a key is being retrieved. Importantly, the adversary  does not learn the identifier $\flrmReceiver$ of the receiver.
   
   $\Functionality_{\LRM}$ sets the route $\flrmroutelist$ with the information received from the sender. It sets the index $\flrmindex$ to $1$, which means that the packet is held by the sender, and the identifier $\flrmposition$ of the party that holds the packet to $\flrmSender$. It is not necessary to include $\flrmposition$ in this tuple because it is equal to $\flrmroutelist[\flrmindex]$. However, in other interfaces that use tuples with the same structure, it can happen that the party that holds the packet is different from $\flrmroutelist[\flrmindex]$ when $\flrmroutelist[\flrmindex]$ is corrupt.
   
   $\Functionality_{\LRM}$ creates a query identifier $\qid$ and stores $\qid$ along with the global packet identifier $\ppid$, the message $\flrmmessagerequest$, the index $\flrmindex$, the party identifier $\flrmposition$, and the route $\flrmroutelist$. To set $\ppid$, $\flrmmessagerequest$, and $\flrmroutelist$, $\Functionality_{\LRM}$ uses the values received as input, except in the following situation: the sender is corrupt, and there is a packet received by a corrupt party $\flrmposition'$ with index $\flrmindex'$ such that the local packet identifier $\tid$ received by $\flrmposition'$  is equal to the global packet identifier $\ppid$ received as input by $\Functionality_{\LRM}$. Additionally, in the packet route $\flrmroutelist'$ for that packet, the sender is honest. In our construction, a corrupt sender can use a packet computed by an honest party as part of the computation of a packet. When that happens, the simulator in our security analysis inputs a global packet identifier that is equal to the local packet identifier obtained by the corrupt party $\flrmposition'$ that received the packet computed by the honest party. Then $\Functionality_{\LRM}$ evaluates the following conditions:
    \begin{itemize}

        \item Conditions (1) are (2) correspond to the case in which the adversary constructs a request packet as follows:
        \begin{itemize}

            \item If $\flrmindex' = 2$, a corrupt sender sends a request packet received by a corrupt gateway. The packet was computed by an honest sender.

            \item If $\flrmindex' \in [3,6]$, a corrupt sender sends a request packet computed by adding layers of encryption to a request packet computed by an honest sender. 

            \item If $\flrmindex' = 7$, a corrupt sender computes a request packet on input a single-use replied block received by a corrupt receiver, such that the single-use reply block was computed by an honest sender.

            \item If $\flrmindex' > 7$, the corrupt sender sends part of the route for a reply, and the receiver of the packet computed by the honest sender is corrupt, a corrupt sender computes a request packet on input a single-use reply block computed as follows: the corrupt sender removes the layers of encryption associated to corrupt parties until $\flrmindex'$ and replaces them by layers of encryption associated to honest parties. The condition that the receiver of the packet computed by the honest sender is corrupt is needed to ensure that the adversary received the single-use reply block.

        \end{itemize}
        For conditions (1) and (2), $\Functionality_{\LRM}$ proceeds as follows:
        \begin{itemize}
            
            \item $\Functionality_{\LRM}$ checks that the layers of encryption added by the adversary, i.e.\ the party identifiers received as input by the functionality, are associated to honest parties from the correct sets. If that is not the case, $\Functionality_{\LRM}$ aborts. We remark that, if the adversary added layers of encryption associated to corrupt parties, the simulator in our security analysis inputs the packet to the functionality as a new packet computed by the adversary.

            \item $\Functionality_{\LRM}$ sets a packet route $\flrmroutelist$ that uses the party identifiers received as input until $\flrmindex'$, and the party identifiers from the packet route $\flrmroutelist'$ of the packet or single-use reply block computed by an honest sender after $\flrmindex'$. The sender of the packet route $\flrmroutelist$ is set to the honest sender in the packet route $\flrmroutelist'$, so that $\Functionality_{\LRM}$ processes the packet as a packet computed by an honest sender.

            \item When the corrupt sender sends a request packet computed by an honest sender, possibly adding layers of encryption to it (i.e.\ when $\flrmindex' \allowbreak \leq \allowbreak 6$), the message $\flrmmessagerequest$ is set to the message $\flrmmessagerequest'$ chosen by the honest sender, unless $\flrmmessagerequest$ is equal to the special symbol $\top$, which indicates that the adversary destroyed the payload. When the corrupt sender computes a new request packet on input a single-use reply block computed by an honest sender, then  the message $\flrmmessagerequest$ is set to message received as input by $\Functionality_{\LRM}$.
            
        \end{itemize}
        Condition (3) corresponds to the case in which the adversary sends a reply packet computed by adding layers of encryption to a reply packet computed by an honest receiver, or by using a single-use reply block computed by an honest sender (in this case, the adversary has removed layers of encryption associated to corrupt parties until $\flrmindex'$ and has replaced them by layers of encryption associated to honest parties). The adversary possibly adds layers of encryption to the reply packet. Because the simulator in our security analysis cannot distinguish reply packets from request packets, it uses the request interface and sends the party identifiers for the route of the reply packet as a request packet.

        $\Functionality_{\LRM}$ sets the route $\flrmroutelist$ by using those party identifiers in the route of the reply packet. It also sets the index $\flrmindex \allowbreak \gets \allowbreak 7$, so that the packet is processed as a reply packet, and marks the query identifier $\qid$ as a query identifier for the reply interface. The message $\flrmmessagerequest$ is set to the message $\flrmmessagerequest'$ chosen by the honest receiver, unless the adversary destroyed the payload, which is indicated by the special symbol $\top$ received as input by $\Functionality_{\LRM}$ in $\flrmmessagerequest$.
        
        We remark that our functionality does not consider the cases in which the adversary, to compute a single-use reply block, uses a packet header computed by an honest sender (i.e., a packet header that is not part of a single-use reply block from an honest sender received by the adversary). 

    \end{itemize}

   $\Functionality_{\LRM}$ sends the $\flrmsendsim$ message along with the bit $\flrmfirstmessage$ and the identifiers $\flrmSender$ and $\flrmGatewayentry$. This means that any construction $\mathrm{\Pi}_{\LRM}$ for $\Functionality_{\LRM}$ can leak $\flrmfirstmessage$, $\flrmSender$ and $\flrmGatewayentry$ to the adversary, but the message $\flrmmessagerequest$ and the routing information (except $\flrmSender$ and $\flrmGatewayentry$) must remain hidden. Importantly, if $\flrmSender$ needs to retrieve the public key of the receiver $\flrmReceiver$, such retrieval needs to be done in a way that does not leak the identifier $\flrmReceiver$ to the adversary.

   When $\Functionality_{\LRM}$ receives the message $\flrmsendrep$ along with $\qid$ from the simulator, $\Functionality_{\LRM}$ aborts if there is not a tuple with that $\qid$ stored. If $\Functionality_{\LRM}$ does not abort, $\Functionality_{\LRM}$ proceeds as follows:
   \begin{itemize}

        \item When the sender $\flrmSender$ is honest and the gateway $\flrmGatewayentry$ is corrupt, $\Functionality_{\LRM}$ runs the algorithm $\flrmCompLeakList$  in~\S\ref{sec:leakagelist} to compute the leakage list $\flrmlistleakage$ of party identifiers in the route that should be leaked to $\flrmGatewayentry$. $\Functionality_{\LRM}$ also runs $\flrmCompIndex$ in~\S\ref{sec:leakagelist} to compute the index $\flrmindex'$ of the next party that should interact with $\Functionality_{\LRM}$ regarding this packet.

        If $\flrmindex' \allowbreak = \allowbreak \bot$, which happens when all the parties in the route after $\flrmGatewayentry$ are corrupt, then $\Functionality_{\LRM}$ sets $\tid$ and $\flrmposition'$ to $\bot$ because it will not process this packet again. We remark that this can only happen to request packets that do not enable replies. When replies are enabled, the honest sender is the destination of the reply packet, and thus there is at least one honest party after $\flrmGatewayentry$.  If $\flrmindex' \allowbreak \neq \allowbreak \bot$, $\Functionality_{\LRM}$ creates a local packet identifier $\tid$ and sets to $\flrmroutelist[\flrmindex']$ the party $\flrmposition'$ that should interact with $\Functionality_{\LRM}$ regarding this packet. For example, if the next honest party after $\flrmGatewayentry$ is the third-layer node $\flrmNodeCrequest$, then $\flrmindex' \allowbreak = \allowbreak 4$ and $\flrmroutelist[\flrmindex']$ is the second-layer node  $\flrmNodeBrequest$.

        If $\flrmindex' \allowbreak = \allowbreak \bot$ or $\flrmindex' \allowbreak = \allowbreak 7$, the request message $\flrmmessagerequest$ is leaked because the receiver is also corrupt. We recall that the algorithm $\flrmCompIndex$ outputs $7$ whenever the first honest party after $\flrmGatewayentry$ is in the route of the reply packet.

        The global packet identifier $\ppid$ is not leaked by $\Functionality_{\LRM}$ because the sender is honest. We recall that leaking $\ppid$ would enable the adversary to trace the packet along its route.

        \item When the gateway $\flrmGatewayentry$ is honest, $\flrmlistleakage$ is set to $\bot$, since in this case the route is not leaked beyond $\flrmNodeArequest$. The index $\flrmindex'$ is set to $2$ and the party $\flrmposition'$ is set to $\flrmGatewayentry$ since this is the party that will forward the packet, and $\Functionality_{\LRM}$ also creates a local identifier $\tid$. Neither the global packet identifier $\ppid$ nor the request message $\flrmmessagerequest$ are leaked.

        \item When both $\flrmSender$ and $\flrmGatewayentry$ are corrupt, the global packet identifier $\ppid$ and the message $\flrmmessagerequest$ are leaked. $\ppid$ already leaks all the information about the packet to the adversary since the packet was input by a corrupt sender, so computing a leakage list is not necessary. Moreover, $\flrmindex'$, $\flrmposition'$ and $\tid'$ are set to $\bot$. The reason is that, in general, when a packet computed by a corrupt party is received by another corrupt party, the routing information that $\Functionality_{\LRM}$ receives may be invalid after that corrupt party. (For instance, if $\flrmNodeArequest$ is corrupt, the simulator $\Simulator$ in our security analysis cannot find out the remaining parties in the route.) We remark that, in our security analysis, $\Functionality_{\LRM}$ is never invoked when both $\flrmSender$ and $\flrmGatewayentry$ are corrupt. When that is the case, our simulator inputs the packet to $\Functionality_{\LRM}$ through the forward interface when the packet is forwarded to an honest party.

   \end{itemize}

   $\Functionality_{\LRM}$ stores $\tid$ along with $\ppid$, $\flrmmessagerequest$, $\flrmroutelist$ and the index $\flrmindex'$ of the corrupt party $\flrmposition'$ that should use $\tid$ to make $\Functionality_{\LRM}$ process again the packet.

   Finally, $\Functionality_{\LRM}$ sends the message $\flrmsendend$ along with $\tid$, the sender identifier $\flrmSender$ and the first-layer node identifier $\flrmNodeArequest$ to the entry gateway $\flrmGatewayentry$. $\Functionality_{\LRM}$ also sends the list $\flrmlistleakage$, which is empty if $\flrmGatewayentry$ is honest. This implies that, in any construction $\mathrm{\Pi}_{\LRM}$ for $\Functionality_{\LRM}$, $\flrmGatewayentry$ learns $\flrmSender$ and $\flrmNodeArequest$, but cannot learn the remaining routing information, unless $\flrmGatewayentry$ and subsequent parties in the route of the packet are corrupt. The message $\flrmmessagerequest$ is leaked when $\flrmGatewayentry$ and all the subsequent parties in the route until and including the receiver $\flrmReceiver$ are corrupt. The global packet identifier $\ppid$ is only leaked when the sender is also corrupt.

   We remark that $\Functionality_{\LRM}$ does not abort if $\flrmGatewayentry$ is honest and $\flrmNodeArequest \allowbreak \notin \allowbreak \flrmSetA$. The reason is that this abortion is performed in the forward interface, i.e.\ if  an honest $\flrmGatewayentry$ forwards the packet to a party that is not a valid first-layer node, $\Functionality_{\LRM}$ aborts.

%% file: 4SecDefs4Definition4Reply.tex
\subsubsection{Reply Interface}
\label{sec:functionalityReply}

\paragraph{Definition of the reply interface.} $\Functionality_{\LRM}$ interacts with the parties described in~\S\ref{sec:functionalityParties} and uses the parameters, variables and algorithms defined in~\S\ref{sec:functionalityParameters}.

\begin{enumerate}
      \item[4.] On input $(\flrmreplyini, \allowbreak \sid, \allowbreak \tid, \allowbreak \flrmmessageresponse, \allowbreak \{\rpid, \allowbreak \flrmGatewayexitreply', \allowbreak \flrmNodeAresponse'\})$ from a receiver $\flrmReceiver$:

            \begin{itemize}

                \item Abort if a tuple $J \allowbreak = \allowbreak (\tid', \allowbreak \ppid, \allowbreak \flrmposition, \allowbreak \flrmindex, \flrmroutelist, \allowbreak \flrmmessagerequest)$ such that $\tid' \allowbreak = \allowbreak \tid$, $\tid \allowbreak \neq \allowbreak \bot$, $\flrmindex \allowbreak = \allowbreak 7$ and either $\flrmposition \allowbreak = \allowbreak \flrmReceiver$ or both $\flrmposition \allowbreak \in \allowbreak \flrmSetS$ and $\flrmReceiver \allowbreak \in \allowbreak \flrmSetS$ is not stored.
                
                \item Parse $\flrmroutelist$ as $[\flrmSender, \allowbreak \flrmGatewayentry, \allowbreak \flrmNodeArequest, \allowbreak \flrmNodeBrequest, \allowbreak \flrmNodeCrequest, \allowbreak \flrmGatewayexit, \allowbreak \flrmReceiver, \langle \flrmGatewayexitreply, \allowbreak \flrmNodeAresponse, \allowbreak \flrmNodeBresponse, \allowbreak \flrmNodeCresponse, \allowbreak  \flrmGatewayentryreply, \allowbreak \flrmSenderreply \rangle]$ and abort if $\langle \flrmGatewayexitreply, \allowbreak \flrmNodeAresponse, \allowbreak \flrmNodeBresponse, \allowbreak \flrmNodeCresponse, \allowbreak  \flrmGatewayentryreply, \allowbreak \flrmSenderreply \rangle = \bot$.


                \item If $\flrmReceiver \allowbreak \notin \allowbreak \flrmSetS$, abort if $\flrmmessageresponse \allowbreak \notin \flrmMessageSpace$, or if $\flrmGatewayexitreply \allowbreak \notin \allowbreak \flrmSetW$, or if $\flrmNodeAresponse \allowbreak \notin \allowbreak \flrmSetA$.

                \item Create a fresh $\qid$ and store the tuple $I \allowbreak = \allowbreak (\qid, \allowbreak \ppid, \allowbreak \flrmReceiver, \allowbreak \flrmindex, \allowbreak \flrmroutelist, \allowbreak \flrmmessageresponse)$, which takes all the values from the tuple $J$ and replaces $\tid$ by $\qid$,  $\flrmposition$ by $\flrmReceiver$, and $\flrmmessagerequest$ by $\flrmmessageresponse$.


                \item If $\flrmReceiver \allowbreak \in \allowbreak \flrmSetS$ and $\{\rpid, \allowbreak \flrmGatewayexitreply', \allowbreak \flrmNodeAresponse'\} \allowbreak \neq \allowbreak \bot$, do the following:
                \begin{itemize}

                    \item Replace $\flrmGatewayexitreply$ by $\flrmGatewayexitreply'$ in the route $\flrmroutelist$ in the tuple $I$.

                    \item Store $(\qid, \allowbreak \rpid, \allowbreak \flrmNodeAresponse')$.


                \end{itemize}


                \item Set $\flrmfirstmessage \allowbreak \gets \allowbreak 0$. 

                \item Send $(\flrmreplysim, \allowbreak \sid, \allowbreak \qid, \allowbreak \flrmReceiver, \allowbreak \flrmGatewayexitreply, \allowbreak \flrmfirstmessage)$ to $\Simulator$, where $\flrmGatewayexitreply$ is taken from the tuple $I$.

            \end{itemize}

        \item[S.] On input $(\flrmreplyrep, \allowbreak \sid, \allowbreak \qid)$ from $\Simulator$:

            \begin{itemize}

                \item Abort if a tuple $I = (\qid', \allowbreak \ppid, \allowbreak \flrmposition, \allowbreak \flrmindex, \flrmroutelist, \allowbreak \flrmmessagerequest)$ such that $\qid' \allowbreak = \allowbreak \qid$ is not stored.

                \item Parse $\flrmroutelist$ as $[\flrmSender, \allowbreak \flrmGatewayentry, \allowbreak \flrmNodeArequest, \allowbreak \flrmNodeBrequest, \allowbreak \flrmNodeCrequest, \allowbreak \flrmGatewayexit, \allowbreak \flrmReceiver, \langle \flrmGatewayexitreply, \allowbreak \flrmNodeAresponse, \allowbreak \flrmNodeBresponse, \allowbreak \flrmNodeCresponse, \allowbreak  \flrmGatewayentryreply, \allowbreak \flrmSenderreply \rangle]$.

                \item If $\flrmGatewayexitreply \allowbreak \notin \allowbreak \flrmSetS$, abort if $\flrmposition \allowbreak \notin \allowbreak \flrmSetU$ or if $\flrmGatewayexitreply \allowbreak \notin \allowbreak \flrmSetW$.

                \item If $\flrmGatewayexitreply \allowbreak \notin \allowbreak \flrmSetS$, abort if a tuple $(\sid, \allowbreak \flrmParty, \allowbreak 1)$ such that $\flrmParty \allowbreak = \allowbreak \flrmGatewayexitreply$ is not stored.

                \item If $\flrmGatewayexitreply \allowbreak \in \allowbreak \flrmSetS$, $\flrmposition \allowbreak \notin \allowbreak \flrmSetS$ and $\flrmSender \allowbreak \notin \allowbreak \flrmSetS$, do the following:
                \begin{itemize}
                
                    \item Set $(\flrmlistleakage, \flrmindexhonest) \gets \flrmCompLeakList(\flrmroutelist, 8, \flrmSetA, \flrmSetB, \flrmSetC, \flrmSetW, \allowbreak \flrmSetU, \allowbreak \flrmSetS)$.

                    \item Set $\flrmindex' \gets \flrmCompIndex(8, \flrmindexhonest)$.

                    \item Set $\flrmposition' \allowbreak \gets \allowbreak \flrmroutelist[\flrmindex']$ and create a fresh random $\tid$.

                    \item Set $\flrmmessageresponse' \gets \bot$ and set $\ppid' \gets \bot$.
                    
                \end{itemize}




                        

                        
                        
                    

                \item If  $\flrmGatewayexitreply \allowbreak \in \allowbreak \flrmSetS$, $\flrmposition \allowbreak \in \allowbreak \flrmSetS$ and $\flrmSender \allowbreak \notin \allowbreak \flrmSetS$, set $\flrmlistleakage \allowbreak \gets \allowbreak \bot$, set $\flrmposition' \allowbreak \gets \allowbreak \flrmGatewayexitreply$, set $\flrmindex' \allowbreak \gets \allowbreak 8$, set $\flrmmessageresponse' \gets \bot$,  set $\ppid' \gets \bot$ and create a fresh random $\tid$.

                \item If $\flrmGatewayexitreply \allowbreak \in \allowbreak \flrmSetS$ and $\flrmSender \allowbreak \in \allowbreak \flrmSetS$, set $\flrmlistleakage \allowbreak \gets \allowbreak \bot$, set $\flrmposition' \allowbreak \gets \allowbreak \bot$, set $\flrmindex' \allowbreak \gets \allowbreak \bot$, set $\tid \allowbreak \gets \allowbreak \bot$, set $\ppid' \allowbreak \gets \allowbreak \ppid$ and $\flrmmessageresponse' \allowbreak \gets \allowbreak \flrmmessagerequest$.

                \item If $\flrmGatewayexitreply \allowbreak \notin \allowbreak \flrmSetS$, set $\flrmlistleakage \allowbreak \gets \allowbreak \bot$, set $\flrmposition' \allowbreak \gets \allowbreak \flrmGatewayexitreply$, set  $\flrmindex' \allowbreak \gets \allowbreak 8$, set $\ppid' \allowbreak \gets \allowbreak \bot$, set $\flrmmessageresponse' \allowbreak \gets \allowbreak \bot$ and create a fresh random packet identifier $\tid$.

                \item Store the tuple $J = (\tid, \allowbreak \ppid, \allowbreak \flrmposition', \allowbreak \flrmindex', \flrmroutelist, \allowbreak \flrmmessagerequest)$, where $\flrmposition'$, $\flrmindex'$ and $\tid$ are set as described above and the remaining values are taken from the tuple $I$.

                \item If there is a tuple $(\qid', \allowbreak \rpid, \allowbreak \flrmNodeAresponse')$ such that $\qid' \allowbreak = \allowbreak \qid$, store $(\tid, \allowbreak \rpid, \allowbreak \flrmNodeAresponse')$ and delete $(\qid', \allowbreak \rpid, \allowbreak \flrmNodeAresponse')$.

                \item Delete the tuple $I$.


                \item Send the message $(\flrmreplyend, \allowbreak \sid, \allowbreak \tid, \allowbreak \ppid', \allowbreak \flrmmessageresponse', \allowbreak \flrmposition, \allowbreak \flrmNodeAresponse, \allowbreak \flrmlistleakage)$ to $\flrmGatewayexitreply$, where $\flrmNodeAresponse \allowbreak = \allowbreak \flrmNodeAresponse'$ in the tuple $(\tid', \allowbreak \rpid, \allowbreak \flrmNodeAresponse')$ such that $\tid' \allowbreak = \allowbreak \tid$ if such a tuple is stored, and otherwise $\flrmNodeAresponse$ is taken from the route $\flrmroutelist$ in the tuple $J$. $\flrmposition$ is taken from the tuple $I$. 

            \end{itemize}
\end{enumerate}

\paragraph{Explanation and discussion of the reply interface.} A receiver $\flrmReceiver$ sends the message $\flrmsendini$ along with a packet identifier $\tid$, a message $\flrmmessageresponse$ and the tuple $\{\rpid, \allowbreak \flrmGatewayexitreply', \allowbreak \flrmNodeAresponse'\}$. (If $\flrmReceiver$ is honest, then $\{\rpid, \flrmGatewayexitreply', \allowbreak \flrmNodeAresponse'\}$ is not used by $\Functionality_{\LRM}$.) We remark that $\Functionality_{\LRM}$ does not allow a corrupt receiver to send a reply packet whose routing information $\Functionality_{\LRM}$ does not store. I.e., $\Functionality_{\LRM}$ does not allow a corrupt receiver to input a global packet identifier $\ppid$ and routing information for the reply packet. In our construction, it can happen that a corrupt receiver $\flrmReceiver$ sends to an honest gateway $\flrmGatewayexitreply$ a reply packet associated with a request packet that was sent and routed by corrupt parties. When this happens, in our security analysis, the simulator $\Simulator$ receives a reply packet from the corrupt receiver $\flrmReceiver$ but is not able to tell whether this is a reply packet or a request packet produced by $\flrmReceiver$. In such a situation, $\Simulator$ will use the request interface of $\Functionality_{\LRM}$ to input  $\ppid$, the message and the routing information associated with this packet. Although $\Functionality_{\LRM}$ processes it as a request packet, in our security analysis we show that the ideal-world protocol is indistinguishable from the real-world protocol.  We recall that both interfaces use the same messages $\flrmsend$, and thus the output of the honest gateway is indistinguishable. Since $\Simulator$ uses the request interface, it is not needed that $\Functionality_{\LRM}$ allows a corrupt receiver to provide that input through the reply interface.

    $\Functionality_{\LRM}$  aborts if it does not store a tuple with the local packet identifier $\tid$ received as input such that $\flrmindex \allowbreak = \allowbreak 7$. $\Functionality_{\LRM}$ checks that $\flrmReceiver$ was the party supposed to use $\tid$, or that both $\flrmReceiver$ and the party that was supposed to use it are corrupt. (Adversarial parties collude with each other, and thus in our construction the packet could be produced by another corrupt party.) $\Functionality_{\LRM}$ also checks if there is routing information $\langle \flrmGatewayentryreply, \allowbreak \flrmGatewayexitreply,  \allowbreak \flrmNodeAresponse, \allowbreak \flrmNodeBresponse, \allowbreak \flrmNodeCresponse, \allowbreak \flrmSenderreply \rangle$ stored along with $\tid$ to send a reply packet. 
    $\Functionality_{\LRM}$ also aborts if the receiver $\flrmReceiver$ is honest and the message $\flrmmessageresponse$, or the identifier $\flrmGatewayexitreply$, or the identifier $\flrmNodeAresponse$, do not belong to the correct sets.

    If $\Functionality_{\LRM}$ does not abort, $\Functionality_{\LRM}$ creates a query identifier $\qid$ and stores $\qid$ along with $\ppid$,  the routing information $\flrmroutelist$, the receiver $\flrmReceiver$, the index $\flrmindex$ and the message $\flrmmessageresponse$.

   When the receiver $\flrmReceiver$ is corrupt, $\Functionality_{\LRM}$ may receive as input a reply packet identifier $\rpid$. This identifier is leaked by $\Functionality_{\LRM}$ to a corrupt first-layer node $\flrmNodeAresponse$. In our construction $\mathrm{\Pi}_{\LRM}$, if the receiver $\flrmReceiver$ and the node $\flrmNodeAresponse$ are corrupt, the adversary is able to trace a reply packet sent by $\flrmReceiver$ when it reaches $\flrmNodeAresponse$ because the gateway $\flrmGatewayexitreply$ does not process the packet and simply forwards it. This leakage occurs even if an honest sender computes the request packet that encrypts the single-use reply block used to compute the reply packet. We remark that the adversary cannot link the reply packet to the request packet computed by the honest sender if there is at least one honest party that processes the packet in the route of the request, and thus the global packet identifier $\ppid$ should not be leaked by $\Functionality_{\LRM}$.

   When the receiver $\flrmReceiver$ is corrupt, $\Functionality_{\LRM}$ allows the receiver to choose the exit gateway $\flrmGatewayexitreply'$ that should forward the packet. The reason is that exit gateways do not process reply packets, and thus in our construction a corrupt receiver is able to choose an exit gateway  $\flrmGatewayexitreply'$ different from the exit gateway $\flrmGatewayexitreply$ chosen by the sender. The corrupt receiver can also instruct  $\flrmGatewayexitreply'$ to send the packet to a different first-layer node $\flrmNodeAresponse'$. $\Functionality_{\LRM}$ stores $\flrmNodeAresponse'$ along with the reply packet identifier $\rpid$ and the query identifier $\qid$.

    $\Functionality_{\LRM}$ sets the bit $\flrmfirstmessage$ to $0$ because, to send a reply packet, a receiver does not need to retrieve any public keys in our construction. (In the request interface, the bit $\flrmfirstmessage$ is set to $1$ when a sender sends a packet to a receiver for the first time, because in that case the sender needs to retrieve the receiver's public key in our construction, and the simulator needs to know that information to simulate the retrieval of the public key towards the adversary.) The bit $\flrmfirstmessage$ needs to be included in the message $\flrmreplysim$ so that the messages of $\Functionality_{\LRM}$ in the request and in the reply interfaces cannot be distinguished.

    We remark that $\Functionality_{\LRM}$ does not delete the tuple that contains $\tid$. Therefore, a receiver $\flrmReceiver$ is allowed to invoke $\Functionality_{\LRM}$ using the same $\tid$ to send another reply packet. In other words, $\Functionality_{\LRM}$ does not prevent parties from reusing a single-use reply block or from sending duplicate packets. 
    

    $\Functionality_{\LRM}$ sends the message $\flrmreplysim$ along with the identifiers $\flrmReceiver$ and $\flrmGatewayexitreply$ and the bit $\flrmfirstmessage$ to the simulator $\Simulator$. This means that, in the real-world protocol $\mathrm{\Pi}_{\LRM}$, the adversary that eavesdrops the communication channel can learn $\flrmReceiver$ and $\flrmGatewayexitreply$, but cannot learn the message or the remaining routing information associated with the reply packet. 
 
    When the simulator sends the message $\flrmreplyrep$ along with a query identifier $\qid$, $\Functionality_{\LRM}$ finds a stored tuple with that $\qid$. $\Functionality_{\LRM}$ aborts if the party $\flrmGatewayexitreply$ is honest and $\flrmGatewayexitreply$ is not a gateway or the party $\flrmposition$ that acted as receiver does not belong to the set of users. We remark that, when the party that inputs the request packet to the functionality is corrupt, the routing information that $\Functionality_{\LRM}$ stores may be inadmissible. Therefore, $\Functionality_{\LRM}$ checks the admissibility of the routing information whenever the packet is received by an honest party. Additionally, a corrupt receiver may input an invalid party identifier $\flrmGatewayexitreply'$.

    Then $\Functionality_{\LRM}$ proceeds as follows:
    \begin{itemize}
    

        \item If $\flrmGatewayexitreply$ is corrupt, the receiver $\flrmposition$ is honest, and the sender $\flrmSender$ is honest, $\Functionality_{\LRM}$ uses the algorithms $\flrmCompLeakList$ and $\flrmCompIndex$ to compute the leakage list $\flrmlistleakage$ and the index $\flrmindex'$. $\flrmlistleakage$ contains all the parties in the route until and including the next honest party. The position $\flrmposition'$ is set to the corrupt party just before the next honest party, which is the one that will use $\tid$ to communicate to $\Functionality_{\LRM}$ that the packet should be forwarded to an honest party. The message and the global packet identifier are not disclosed. We recall that, since $\flrmSender$ is honest, it is guaranteed that there is an honest party in the route because the sender is the destination of the reply packet, and thus, unlike in the request interface, the index $\flrmindex'$ is never $\bot$.

        \item If $\flrmGatewayexitreply$ is corrupt, the receiver $\flrmposition$ is corrupt, and the sender $\flrmSender$ is honest,  the leakage list is empty, the position $\flrmposition'$ is set to $\flrmGatewayexitreply$ and the index $\flrmindex'$ is set to $8$.  When the receiver $\flrmposition$ is corrupt, the adversary already received the leakage list when  the request packet was received through the forward interface. The message and the global packet identifier are not disclosed.  We remark that it is possible that the simulator $\Simulator$ invokes the reply interface when both $\flrmGatewayexitreply$ and $\flrmposition$ are corrupt. This happens when the simulator receives from the adversary a reply packet that the adversary computed by using a single-use reply block created by an honest sender. In that case, the functionality expects $\flrmReceiver$ to input the reply packet, even if subsequent parties in the route are also corrupt.

        \item  If the gateway $\flrmGatewayexitreply$ and the sender $\flrmSender$ are corrupt, the leakage list is empty and $\flrmposition'$ and $\flrmindex'$ are set to $\bot$. (We recall that $\flrmSender$ refers to any corrupt party that created the request packet.) In this case, the functionality discloses the message and the global packet identifier $\ppid$, which already reveals all the routing information related to the packet. Additionally, the local packet identifier $\tid$ is set to $\bot$ because $\Functionality_{\LRM}$ will not process again this packet. If, subsequently, a corrupt party forwards the packet to an honest party, the functionality processes it as a new packet. In our security analysis, we show that the outputs of honest parties in the real and in the ideal world are indistinguishable even if the functionality processes the forwarded packet as a new packet. We remark that the functionality might not store routing information related to that packet beyond $\flrmGatewayexitreply$. The reason is that, when the party that computes the packet is corrupt, the simulator can only find out the route of the packet until and including the next corrupt party that processes the packet.  

        The global packet identifier $\ppid$ is leaked because, when both parties are corrupt, in our construction the adversary is able to link the packet received by $\flrmGatewayexitreply$ with the packet previously computed by another corrupt party. The message $\flrmmessageresponse$ is leaked because, when the request packet that enabled this reply was computed by a corrupt party, then the adversary can decrypt the reply packet whenever it is received by a corrupt party.  We remark that $\Functionality_{\LRM}$ checks if the sender $\flrmSender$ is corrupt, although the party that computed the packet is not necessarily $\flrmSender$. Corrupt nodes and gateways can inject packets into the network. When that happens, in our security analysis $\Simulator$ will use the forward interface of the functionality to input $\ppid$ and the routing information for that packet, and will mark the sender $\flrmSender$ as corrupt (although the packet is not necessarily computed by a sender). That is the reason why $\Functionality_{\LRM}$ checks $\flrmSender$ in the reply interface.


        \item If $\flrmGatewayexitreply$ is honest, $\flrmlistleakage$ is empty, $\flrmposition'$ is set to $\flrmGatewayexitreply$ and $\flrmindex'$ is set to $8$. The message and the global packet identifier are not disclosed. Therefore, when $\flrmGatewayexitreply$ is honest, $\Functionality_{\LRM}$ expects $\flrmGatewayexitreply$ to instruct the functionality to forward the packet.

    \end{itemize}
    

    We remark that the local packet identifier $\tid$ that $\Functionality_{\LRM}$ sends to $\flrmGatewayexitreply$ is different from the $\tid$ that $\Functionality_{\LRM}$ received from the receiver $\flrmposition$.  $\Functionality_{\LRM}$ changes $\tid$ so that  $\tid$ values cannot be used to trace the route of the packet throughout the mixnet.  In the forward interface, each time a packet is forwarded, $\Functionality_{\LRM}$ also generates a new $\tid$.
    
    
   $\Functionality_{\LRM}$  deletes the tuple with $\qid$. Therefore, if the simulator $\Simulator$ sends again a message $\flrmreplyrep$ with $\qid$, $\Functionality_{\LRM}$ sends an abortion message to $\flrmGatewayexitreply$. To take that into account, our construction $\mathrm{\Pi}_{\LRM}$ for $\Functionality_{\LRM}$ uses as building block the functionality $\Functionality_{\SMT}$, which sends an abortion message to $\flrmGatewayexitreply$ if the real-world adversary attempts to send again the message communicated by the receiver to the gateway through $\Functionality_{\SMT}$.

   Finally $\Functionality_{\LRM}$ sends the message $\flrmreplyend$ along with $\tid$ and the identifiers $\flrmposition$ and $\flrmNodeAresponse$ to $\flrmGatewayexitreply$.  This means that constructions  $\mathrm{\Pi}_{\LRM}$ for $\Functionality_{\LRM}$ can reveal $\flrmReceiver$ and $\flrmNodeAresponse$ to $\flrmGatewayexitreply$, but no further information regarding the message $\flrmmessageresponse$ or the routing information used to compute the reply packet. The global packet identifier $\ppid$, the message $\flrmmessageresponse$, and the list $\flrmlistleakage$ are leaked when the conditions discussed above are met.

%% file: 4SecDefs4Definition5Forward.tex
\subsubsection{Forward Interface}
\label{sec:functionalityForward}

\paragraph{Definition of the forward interface.}  $\Functionality_{\LRM}$ interacts with the parties described in~\S\ref{sec:functionalityParties} and uses the parameters, variables and algorithms defined in~\S\ref{sec:functionalityParameters}.

    \begin{enumerate}

   \item[5.] On input  $(\flrmforwardini, \allowbreak \sid, \allowbreak \tid, \allowbreak \flrmdestroymessage, \{\ppid, \allowbreak \flrmGatewayexit, \allowbreak \flrmReceiver, \allowbreak \flrmmessagerequest, \allowbreak \flrmNodeArequest, \allowbreak \flrmNodeBrequest, \allowbreak \flrmNodeCrequest, \allowbreak \langle   \flrmGatewayentryreply, \allowbreak \flrmGatewayexitreply, \allowbreak \flrmNodeAresponse, \allowbreak \flrmNodeBresponse, \allowbreak \flrmNodeCresponse, \allowbreak \flrmSenderreply \rangle\})$ from a node or a gateway  $\flrmParty$:

            \begin{itemize}

                \item If there is no tuple $J \allowbreak = \allowbreak (\tid', \allowbreak \ppid, \allowbreak \flrmposition, \allowbreak \flrmindex, \flrmroutelist, \allowbreak \flrmmessagerequest)$ such that $\tid' \allowbreak = \allowbreak \tid$, $\tid \allowbreak \neq \allowbreak \bot$, $\flrmindex \allowbreak \in \allowbreak [2,6] \cup [8,12]$, and either $\flrmroutelist[\flrmindex] \allowbreak = \allowbreak \flrmParty$ or both $\flrmroutelist[\flrmindex] \allowbreak \in \allowbreak \flrmSetS$ and $\flrmParty \allowbreak \in \allowbreak \flrmSetS$, do the following:
                \begin{itemize}
                    
                    \item If $\flrmParty \allowbreak \notin \allowbreak \flrmSetS$, or if $\flrmParty \allowbreak \in \allowbreak \flrmSetS$ and $\{\ppid, \allowbreak \flrmGatewayexit, \allowbreak \flrmReceiver, \allowbreak \flrmmessagerequest, \allowbreak \flrmNodeArequest, \allowbreak \flrmNodeBrequest, \allowbreak \flrmNodeCrequest, \allowbreak \langle   \flrmGatewayentryreply, \allowbreak \flrmGatewayexitreply, \allowbreak \flrmNodeAresponse, \allowbreak \flrmNodeBresponse, \allowbreak \flrmNodeCresponse, \allowbreak \flrmSenderreply \rangle\} = \bot$, abort.

                    \item Else, abort if a tuple of the form $(\tid, \allowbreak \ppid', \allowbreak \flrmposition, \allowbreak \flrmindex, \flrmroutelist, \allowbreak \flrmmessagerequest)$ with $\flrmroutelist \allowbreak = \allowbreak [\flrmSender', \allowbreak \flrmGatewayentry, \allowbreak \flrmNodeArequest, \allowbreak \flrmNodeBrequest, \allowbreak \flrmNodeCrequest, \allowbreak \flrmGatewayexit, \allowbreak \flrmReceiver, \langle \flrmGatewayexitreply, \allowbreak \flrmNodeAresponse, \allowbreak \flrmNodeBresponse, \allowbreak \flrmNodeCresponse, \allowbreak  \flrmGatewayentryreply, \allowbreak \flrmSenderreply \rangle]$ such that $\ppid' \allowbreak = \allowbreak \ppid$ and $\flrmSender' \allowbreak \in \allowbreak \flrmSetS$ is stored, or if a tuple of the form $(\qid, \allowbreak \ppid', \allowbreak \flrmposition,  \allowbreak \flrmindex, \allowbreak \flrmroutelist, \allowbreak \flrmmessagerequest)$ with $\flrmroutelist \allowbreak = \allowbreak [\flrmSender', \allowbreak \flrmGatewayentry, \allowbreak \flrmNodeArequest, \allowbreak \flrmNodeBrequest, \allowbreak \flrmNodeCrequest, \allowbreak \flrmGatewayexit, \allowbreak \flrmReceiver, \langle \flrmGatewayexitreply, \allowbreak \flrmNodeAresponse, \allowbreak \flrmNodeBresponse, \allowbreak \flrmNodeCresponse, \allowbreak  \flrmGatewayentryreply, \allowbreak \flrmSenderreply \rangle]$ such that $\ppid' \allowbreak = \allowbreak \ppid$ and $\flrmSender' \allowbreak \in \allowbreak \flrmSetS$ is stored.    

                    \item Else, store a tuple $J \allowbreak = \allowbreak (\tid, \allowbreak \ppid, \allowbreak \flrmposition, \allowbreak \flrmindex, \flrmroutelist, \allowbreak \flrmmessagerequest)$ with packet route $\flrmroutelist \allowbreak = \allowbreak [\flrmSender, \allowbreak \flrmGatewayentry, \allowbreak \flrmNodeArequest, \allowbreak \flrmNodeBrequest, \allowbreak \flrmNodeCrequest, \allowbreak \flrmGatewayexit, \allowbreak \flrmReceiver, \langle \flrmGatewayexitreply, \allowbreak \flrmNodeAresponse, \allowbreak \flrmNodeBresponse, \allowbreak \flrmNodeCresponse, \allowbreak  \flrmGatewayentryreply, \allowbreak \flrmSenderreply \rangle]$ as follows:
                    \begin{itemize}
                    
                        \item Set $\tid \allowbreak \gets \allowbreak \bot$.

                        \item Set $\flrmposition \allowbreak \gets \allowbreak \flrmParty$. 



                        \item Set $\flrmSender \allowbreak \gets \allowbreak \Adversary$.

                        \item If $\flrmParty \allowbreak = \allowbreak \flrmNodeArequest$, set $\flrmindex \allowbreak \gets \allowbreak 3$ and $\flrmGatewayentry \allowbreak \gets \allowbreak \bot$.

                        \item If $\flrmParty \allowbreak = \allowbreak \flrmNodeBrequest$, set $\flrmindex \allowbreak \gets \allowbreak 4$ and $\flrmGatewayentry \allowbreak \gets \allowbreak \bot$.

                        \item If $\flrmParty \allowbreak = \allowbreak \flrmNodeCrequest$, set $\flrmindex \allowbreak \gets \allowbreak 5$ and $\flrmGatewayentry \allowbreak \gets \allowbreak \bot$.

                        \item If $\flrmParty \allowbreak = \allowbreak \flrmGatewayexit$ and $\flrmNodeCrequest \allowbreak = \allowbreak \bot$, set $\flrmindex \allowbreak \gets \allowbreak 6$ and $\flrmGatewayentry \allowbreak \gets \allowbreak \bot$, else set $\flrmindex \allowbreak \gets \allowbreak 2$ and $\flrmGatewayentry \allowbreak \gets \allowbreak \flrmParty$.


                        \item The remaining values in $J$ are set to those in $\{\ppid, \allowbreak \flrmGatewayexit, \allowbreak \flrmReceiver, \allowbreak \flrmmessagerequest, \allowbreak \flrmNodeArequest, \allowbreak \flrmNodeBrequest, \allowbreak \flrmNodeCrequest, \allowbreak \langle   \flrmGatewayentryreply, \allowbreak \flrmGatewayexitreply, \allowbreak \flrmNodeAresponse, \allowbreak \flrmNodeBresponse, \allowbreak \flrmNodeCresponse, \allowbreak \flrmSenderreply \rangle\}$.
                    
                    \end{itemize}

                    \item If the following conditions are satisfied
                    \begin{itemize}
                    
                        \item $\flrmParty \allowbreak \in \allowbreak \flrmSetS$.

                        \item There is a tuple $(\tid, \allowbreak \ppid', \allowbreak \flrmposition',  \allowbreak \flrmindex', \allowbreak \flrmroutelist', \allowbreak \flrmmessagerequest')$ such that $\tid \allowbreak = \allowbreak \ppid$ and $\ppid \allowbreak \neq \allowbreak \bot$, where $\ppid$ is the global packet identifier received as input.


                        \item Given $\flrmroutelist' \allowbreak = \allowbreak [\flrmSender', \allowbreak \flrmGatewayentry', \allowbreak \flrmNodeArequest', \allowbreak \flrmNodeBrequest', \allowbreak \flrmNodeCrequest', \allowbreak \flrmGatewayexit', \allowbreak \flrmReceiver', \langle \flrmGatewayexitreply', \allowbreak \flrmNodeAresponse', \allowbreak \flrmNodeBresponse', \allowbreak \flrmNodeCresponse', \allowbreak  \flrmGatewayentryreply', \allowbreak \flrmSenderreply' \rangle]$ in that tuple,  $\flrmSender' \allowbreak \notin \allowbreak \flrmSetS$ and $\flrmposition' \allowbreak \in \allowbreak \flrmSetS$.

                        \item $\flrmindex \allowbreak < \allowbreak \flrmindex'$.
                    
                    \end{itemize}
                    do the following:
                    \begin{itemize}

                        \item If either (1) $2 \allowbreak < \allowbreak \flrmindex' \allowbreak \leq \allowbreak 7$,  or (2) $\flrmindex' \allowbreak > \allowbreak 7$, and $\langle \flrmGatewayexitreply, \allowbreak \flrmNodeAresponse, \allowbreak \flrmNodeBresponse, \allowbreak \flrmNodeCresponse, \allowbreak  \flrmGatewayentryreply, \allowbreak \flrmSenderreply \rangle \allowbreak \neq \allowbreak \bot$ in $\flrmroutelist$, and $\flrmroutelist'[7] \allowbreak \in \allowbreak \flrmSetS$, do the following:
                        \begin{itemize}

                            \item For $i \allowbreak \in \allowbreak [\flrmindex+1, \allowbreak \flrmindex']$, check that $\flrmroutelist[i] \allowbreak \notin \allowbreak \flrmSetS$ and that each party $\flrmroutelist[i]$ belongs to the correct set. Abort if that is not the case.

                            \item For $i \allowbreak \in \allowbreak [\flrmindex' + 1, \allowbreak 13]$, set $\flrmroutelist[i] \allowbreak \gets \allowbreak \flrmroutelist'[i]$.

                            \item Set $\flrmroutelist[1] \allowbreak \gets \allowbreak \flrmroutelist'[1]$.

                            \item Set $\ppid \allowbreak \gets \allowbreak \ppid'$.
                        
                            \item If $\flrmindex' \allowbreak \leq \allowbreak 6$ and $\flrmmessagerequest \allowbreak \neq \allowbreak \top$, set $\flrmmessagerequest \allowbreak \gets \allowbreak \flrmmessagerequest'$. If $\flrmindex' \allowbreak \leq \allowbreak 6$ and $\flrmmessagerequest \allowbreak = \allowbreak \top$,  set $\flrmmessagerequest \allowbreak \gets \allowbreak \bot$.

                        \end{itemize}
                        
                    \item If (3) $\flrmindex' \allowbreak > \allowbreak 8$ and $\langle \flrmGatewayexitreply, \allowbreak \flrmNodeAresponse, \allowbreak \flrmNodeBresponse, \allowbreak \flrmNodeCresponse, \allowbreak  \flrmGatewayentryreply, \allowbreak \flrmSenderreply \rangle \allowbreak = \allowbreak \bot$ in $\flrmroutelist$, do the following:
                    \begin{itemize}

                        \item For $i \allowbreak \in \allowbreak [\flrmindex+1, \allowbreak \flrmindex'-6]$, check that $\flrmroutelist[i] \allowbreak \notin \allowbreak \flrmSetS$ and that each party $\flrmroutelist[i]$ belongs to the correct set. Abort if that is not the case.

                        \item For $i \allowbreak \in \allowbreak [\flrmindex+7, \allowbreak \flrmindex']$, set $\flrmroutelist[i] \allowbreak \gets \allowbreak \flrmroutelist[i-6]$.

                        \item For $i \allowbreak \notin \allowbreak [\flrmindex+7, \allowbreak \flrmindex']$, set $\flrmroutelist[i] \allowbreak \gets \allowbreak \flrmroutelist'[i]$.

                        \item Set $\ppid \allowbreak \gets \allowbreak \ppid'$.
                        
                        \item If $\flrmmessagerequest \allowbreak \neq \allowbreak \top$, set $\flrmmessagerequest \allowbreak \gets \allowbreak \flrmmessagerequest'$, else set $\flrmmessagerequest \allowbreak \gets \allowbreak \bot$.

                        \item Set $\flrmindex \allowbreak \gets \allowbreak \flrmindex + 6$.

                        \item Mark $\qid$ as a query identifier for the reply interface.

                    \end{itemize}

                    \item Update the tuple $J$ with the newly computed values.
                    
                \end{itemize}

                \end{itemize}



                \item Create a fresh $\qid$ and store the tuple $I = (\qid, \allowbreak \ppid, \allowbreak \flrmposition, \allowbreak \flrmindex, \flrmroutelist, \allowbreak \flrmmessagerequest)$, which takes all the values from the tuple $J$, replaces $\tid$ by $\qid$, and sets $\flrmposition \allowbreak \gets \allowbreak \flrmParty$.

                \item If there is a tuple $(\tid', \allowbreak \rpid, \allowbreak \flrmNodeAresponse')$ such that $\tid' \allowbreak = \allowbreak \tid$, do the following:
                \begin{itemize}

                    \item If $\flrmNodeAresponse' \allowbreak = \allowbreak \flrmNodeAresponse$, where $\flrmNodeAresponse$ is stored in $\flrmroutelist$ in the tuple $I$, do the following:
                    \begin{itemize}

                        \item If both $\flrmNodeAresponse' \allowbreak \in \allowbreak \flrmSetS$ and $\flrmNodeAresponse \allowbreak \in \allowbreak \flrmSetS$, set $z \allowbreak \gets \allowbreak 2$.

                        \item If both $\flrmNodeAresponse' \allowbreak \notin \allowbreak \flrmSetS$ and $\flrmNodeAresponse \allowbreak \notin \allowbreak \flrmSetS$, set $z \allowbreak \gets \allowbreak 3$.

                    \end{itemize}

                    \item If $\flrmNodeAresponse' \allowbreak \neq \allowbreak \flrmNodeAresponse$, where $\flrmNodeAresponse$ is stored in $\flrmroutelist$ in the tuple $I$, do the following:
                    \begin{itemize}

                        \item If $\flrmNodeAresponse' \allowbreak \notin \allowbreak \flrmSetS$, set $z \allowbreak \gets \allowbreak 0$.
                    
                        \item If $\flrmNodeAresponse' \allowbreak \in \allowbreak \flrmSetS$ and $\flrmNodeAresponse \allowbreak \notin \allowbreak \flrmSetS$, set $z \allowbreak \gets \allowbreak 1$.

                        \item If $\flrmNodeAresponse' \allowbreak \in \allowbreak \flrmSetS$ and $\flrmNodeAresponse \allowbreak \in \allowbreak \flrmSetS$, set $z \allowbreak \gets \allowbreak 2$.

                        \item Replace $\flrmNodeAresponse$ by $\flrmNodeAresponse'$ in $\flrmroutelist$ in the tuple I.

                    \end{itemize}

                    \item Store $(\qid, \allowbreak \rpid, \allowbreak z)$.
                    
                \end{itemize}

                \item Abort if $\flrmParty \allowbreak \notin \allowbreak \flrmSetS$ and $\flrmverify(\flrmroutelist, \flrmindex, \flrmSetA, \flrmSetB, \flrmSetC, \flrmSetW, \flrmSetU) = 0$, where $\flrmSetA$, $\flrmSetB$, $\flrmSetC$, and $\flrmSetW$ are taken from $\sid$, and $\flrmroutelist$ and $\flrmindex$ are taken from the tuple $I$.

                \item If $\flrmParty \allowbreak \in \allowbreak \flrmSetS$ and $\flrmdestroymessage \allowbreak = \allowbreak 1$, do the following. 
                \begin{itemize}
                    
                    \item Set $\flrmmessagerequest \allowbreak \gets \allowbreak \bot$ in the tuple $I$.

                    \item If $\flrmindex \allowbreak \in \allowbreak [2,6]$, parse $\flrmroutelist$ in the tuple $I$ as $[\flrmSender, \allowbreak \flrmGatewayentry, \allowbreak \flrmNodeArequest, \allowbreak \flrmNodeBrequest, \allowbreak \flrmNodeCrequest, \allowbreak \flrmGatewayexit, \allowbreak \flrmReceiver, \langle \flrmGatewayexitreply, \allowbreak \flrmNodeAresponse, \allowbreak \flrmNodeBresponse, \allowbreak \flrmNodeCresponse, \allowbreak  \flrmGatewayentryreply, \allowbreak \flrmSenderreply \rangle]$ and set $\langle \flrmGatewayexitreply, \allowbreak \flrmNodeAresponse, \allowbreak \flrmNodeBresponse, \allowbreak \flrmNodeCresponse, \allowbreak  \flrmGatewayentryreply, \allowbreak \flrmSenderreply \rangle \allowbreak \gets \allowbreak \bot$ in $\flrmroutelist$ in the tuple $I$.
                
                \end{itemize}


                \item Send the message $(\flrmforwardsim, \allowbreak \sid, \allowbreak \qid, \allowbreak \flrmParty, \allowbreak \flrmroutelist[\flrmindex+1])$ to $\Simulator$, where $\flrmroutelist$ and $\flrmindex$ are taken from the tuple $I$.

            \end{itemize}

        \item[S.] On input $(\flrmforwardrep, \allowbreak \sid, \allowbreak \qid)$ from $\Simulator$:

            \begin{itemize}

                \item Abort if a tuple $I = (\qid', \allowbreak \ppid, \allowbreak \flrmposition, \allowbreak \flrmindex, \flrmroutelist, \allowbreak \flrmmessagerequest)$ such that $\qid' \allowbreak = \allowbreak \qid$ is not stored.

                \item Abort if $\flrmroutelist[\flrmindex+1] \allowbreak \notin \allowbreak \flrmSetS$ and $\flrmpriorvf(\flrmposition,\flrmindex+1) \allowbreak = \allowbreak 0$.

                \item If $\flrmroutelist[\flrmindex+1] \allowbreak \notin \allowbreak \flrmSetS$, abort if a tuple $(\sid, \allowbreak \flrmParty, \allowbreak 1)$ such that $\flrmParty \allowbreak = \allowbreak \flrmroutelist[\flrmindex+1]$ is not stored.

                \item Parse $\flrmroutelist$ as $[\flrmSender, \allowbreak \flrmGatewayentry, \allowbreak \flrmNodeArequest, \allowbreak \flrmNodeBrequest, \allowbreak \flrmNodeCrequest, \allowbreak \flrmGatewayexit, \allowbreak \flrmReceiver, \langle \flrmGatewayexitreply, \allowbreak \flrmNodeAresponse, \allowbreak \flrmNodeBresponse, \allowbreak \flrmNodeCresponse, \allowbreak  \flrmGatewayentryreply, \allowbreak \flrmSenderreply \rangle]$.

                \item If $\flrmindex +1 = 13$, do the following:

                \begin{itemize}


                    \item Abort if $\flrmSender \allowbreak \notin \allowbreak \flrmSetS$ and $\flrmposition \allowbreak \neq \allowbreak \flrmroutelist[\flrmindex]$.

                    \item If $\flrmSender \allowbreak = \allowbreak \flrmSenderreply$, or if both $\flrmSender \allowbreak \in \allowbreak \flrmSetS$ and $\flrmSenderreply \allowbreak \in \allowbreak \flrmSetS$, set $\ppid' \allowbreak \gets \allowbreak \ppid$ and $\flrmmessagerequest' \allowbreak \gets \allowbreak \flrmmessagerequest$, else set $\ppid' \allowbreak = \allowbreak \bot$ and $\flrmmessagerequest' \allowbreak = \allowbreak \bot$.
                    
                    \item Delete the tuple $I$.
                    
                    \item Send $(\flrmforwardend, \allowbreak \sid, \allowbreak \flrmposition, \allowbreak \flrmmessagerequest', \allowbreak \ppid')$ to $\flrmSenderreply$, where $\flrmposition$ is taken from the tuple $I$.
                    
                \end{itemize}

                \item If $\flrmindex +1 = 7$, do the following:

                \begin{itemize}

                    \item If $\flrmReceiver \allowbreak \in \allowbreak \flrmSetS$, $\flrmSender \allowbreak \notin \allowbreak \flrmSetS$, and $\bot \neq \langle \flrmGatewayexitreply, \allowbreak \flrmNodeAresponse, \allowbreak \flrmNodeBresponse, \allowbreak \flrmNodeCresponse, \allowbreak  \flrmGatewayentryreply, \allowbreak \flrmSenderreply \rangle$, do the following:
                    
                    \begin{itemize}

                        \item Set $(\flrmlistleakage, \flrmindexhonest) \gets \flrmCompLeakList(\flrmroutelist, 7, \flrmSetA, \flrmSetB, \flrmSetC, \flrmSetW, \allowbreak \flrmSetU, \allowbreak \flrmSetS)$. $\flrmSetA$, $\flrmSetB$, $\flrmSetC$ and $\flrmSetW$ are taken from $\sid$.

                        \item Set $\flrmindex' \gets \flrmCompIndex(7, \flrmindexhonest)$. ($\flrmindex' \allowbreak = 7$.)

                        \item Set $\flrmposition' \allowbreak \gets \allowbreak \flrmroutelist[\flrmindex']$ and create a fresh random $\tid$.

                        \item Set  $\ppid' \gets \bot$.


                    
                        

                        
                        
                    
                    \end{itemize}

                    \item If both $\flrmReceiver \allowbreak \in \allowbreak \flrmSetS$ and $\flrmSender \allowbreak \in \allowbreak \flrmSetS$, or if $\bot = \langle  \flrmGatewayentryreply, \allowbreak \flrmGatewayexitreply, \allowbreak \flrmNodeAresponse, \allowbreak \flrmNodeBresponse, \allowbreak \flrmNodeCresponse, \allowbreak \flrmSenderreply \rangle$, set $\flrmlistleakage \allowbreak = \allowbreak \bot$, $\flrmposition' \allowbreak = \allowbreak \bot$, $\flrmindex' \allowbreak \gets \allowbreak \bot$, and $\tid \allowbreak = \allowbreak \bot$. If both $\flrmReceiver \allowbreak \in \allowbreak \flrmSetS$ and $\flrmSender \allowbreak \in \allowbreak \flrmSetS$, set $\ppid' \allowbreak \gets \allowbreak \ppid$, else  set $\ppid' \allowbreak \gets \allowbreak \bot$.

                    \item If $\flrmReceiver \allowbreak \notin \allowbreak \flrmSetS$ and $\bot \neq \langle  \flrmGatewayentryreply, \allowbreak \flrmGatewayexitreply, \allowbreak \flrmNodeAresponse, \allowbreak \flrmNodeBresponse, \allowbreak \flrmNodeCresponse, \allowbreak \flrmSenderreply \rangle$,  set $\flrmlistleakage \allowbreak = \allowbreak \bot$, set $\flrmposition' \allowbreak \gets \allowbreak \flrmReceiver$, set $\flrmindex' \allowbreak \gets \allowbreak 7$, create a fresh random  $\tid$, and set $\ppid' \gets \bot$. 

                    \item Store the tuple $J = (\tid, \allowbreak \ppid, \allowbreak \flrmposition', \allowbreak \flrmindex', \flrmroutelist, \allowbreak \flrmmessagerequest)$, where $\flrmposition'$, $\flrmindex'$ and $\tid$ are set as described above and the remaining values are taken from the tuple $I$.


                    \item Delete the tuple $I$.


                    \item If $\bot \neq \langle  \flrmGatewayentryreply, \allowbreak \flrmGatewayexitreply, \allowbreak \flrmNodeAresponse, \allowbreak \flrmNodeBresponse, \allowbreak \flrmNodeCresponse, \allowbreak \flrmSenderreply \rangle$, set $\flrmNodeAresponse' \allowbreak \gets \allowbreak \flrmNodeAresponse$ and $\flrmGatewayexitreply' \allowbreak \gets \allowbreak \flrmGatewayexitreply$, else set $\flrmNodeAresponse' \allowbreak \gets \allowbreak \bot$ and $\flrmGatewayexitreply' \allowbreak \gets \allowbreak \bot$.

                    \item Send  $(\flrmforwardend, \allowbreak \sid, \allowbreak \flrmposition, \allowbreak \flrmmessagerequest, \allowbreak \tid, \allowbreak \flrmGatewayexitreply', \allowbreak \flrmNodeAresponse', \allowbreak \ppid', \allowbreak \flrmlistleakage)$ to $\flrmReceiver$, where $\flrmposition$ and $\flrmmessagerequest$ are taken from the tuple $I$.

                \end{itemize}

                \item If $\flrmindex + 1 \neq 13$ and $\flrmindex +1 \neq 7$, do the following:

                \begin{itemize}

                    \item If $\flrmroutelist[\flrmindex+1] \allowbreak \in \allowbreak \flrmSetS$, $\flrmposition \allowbreak \notin \allowbreak \flrmSetS$, and $\flrmSender \allowbreak \notin \allowbreak \flrmSetS$, do the following: 
                    \begin{itemize}

                        \item Set $(\flrmlistleakage, \flrmindexhonest) \allowbreak \gets \allowbreak \flrmCompLeakList(\flrmroutelist, \flrmindex + 1, \flrmSetA, \flrmSetB, \allowbreak \flrmSetC, \allowbreak \flrmSetW, \flrmSetU, \allowbreak \flrmSetS)$. $\flrmSetA$, $\flrmSetB$, $\flrmSetC$ and $\flrmSetW$ are taken from $\sid$.

                        \item Set $\flrmindex' \allowbreak \gets \allowbreak \flrmCompIndex(\flrmindex + 1, \flrmindexhonest)$.

                        \item If $\flrmindex' \allowbreak \neq \allowbreak \bot$, set $\flrmposition' \allowbreak \gets \allowbreak \flrmroutelist[\flrmindex']$ and create a fresh random $\tid$, else set $\flrmposition' \allowbreak \gets \allowbreak \bot$ and $\tid \allowbreak \gets \allowbreak \bot$.

                        \item If $\flrmindex' \allowbreak = \allowbreak \bot$ or $\flrmindex' \allowbreak = \allowbreak 7$, set $\flrmmessagerequest' \allowbreak \gets \allowbreak \flrmmessagerequest$, else set $\flrmmessagerequest' \gets \bot$.

                        \item Set  $\ppid' \gets \bot$.

                    \end{itemize}

                    \item If $\flrmroutelist[\flrmindex + 1] \allowbreak \in \allowbreak \flrmSetS$, $\flrmposition  \allowbreak \in \allowbreak \flrmSetS$ and $\flrmSender  \allowbreak \notin \allowbreak \flrmSetS$, set $\flrmlistleakage \allowbreak \gets \allowbreak \bot$, $\flrmposition' \allowbreak \gets \allowbreak \flrmroutelist[\flrmindex + 1]$, $\flrmindex' \allowbreak \gets \allowbreak \flrmindex + 1$, create a fresh local packet identifier $\tid$, set $\flrmmessagerequest' \allowbreak \gets \allowbreak \bot$, and set  $\ppid' \allowbreak \gets \allowbreak \bot$.
                    
                    \item If $\flrmroutelist[\flrmindex+1] \allowbreak \in \allowbreak \flrmSetS$ and $\flrmSender  \allowbreak \in \allowbreak \flrmSetS$, set $\flrmlistleakage \allowbreak \gets \allowbreak \bot$, $\flrmposition' \allowbreak \gets \allowbreak \bot$, $\flrmindex' \allowbreak \gets \allowbreak \bot$, $\tid \allowbreak \gets \allowbreak \bot$, $\flrmmessagerequest' \allowbreak \gets \allowbreak \flrmmessagerequest$ and  $\ppid' \allowbreak \gets \allowbreak \ppid$.

                    \item If $\flrmroutelist[\flrmindex + 1] \allowbreak \notin \allowbreak \flrmSetS$, set $\flrmlistleakage \allowbreak \gets \allowbreak \bot$, $\flrmposition' \allowbreak \gets \allowbreak \flrmroutelist[\flrmindex + 1]$, $\flrmindex' \allowbreak \gets \allowbreak \flrmindex + 1$, create a fresh local packet identifier $\tid$, set $\flrmmessagerequest' \allowbreak \gets \allowbreak \bot$, and set  $\ppid' \allowbreak \gets \allowbreak \bot$.
                    

                    \item Store the tuple $J = (\tid, \allowbreak \ppid, \allowbreak \flrmposition', \allowbreak \flrmindex', \flrmroutelist, \allowbreak \flrmmessagerequest)$, where $\flrmposition'$, $\flrmindex'$ and $\tid$ are set as described above and the remaining values are taken from the tuple $I$.




                    \item Delete the tuple $I$.
                
                    \item If there is a tuple $(\qid', \allowbreak \rpid, \allowbreak z)$ such that $\qid' \allowbreak = \allowbreak \qid$, do the following:
                    \begin{itemize}

                        \item If $z = 0$, set $\rpid' \allowbreak \gets \allowbreak \bot$ and set $\flrmroutelist[\flrmindex+2] \allowbreak \gets \allowbreak \bot$ in $\flrmroutelist$ in the tuple $J$.

                        \item If $z = 1$, set $\rpid' \allowbreak \gets \allowbreak \rpid$ and $\tid \allowbreak \gets \allowbreak \bot$, and replace $\tid$ by $\bot$ in the tuple $J$. Moreover, set $\flrmroutelist[\flrmindex+2] \allowbreak \gets \allowbreak \bot$ in $\flrmroutelist$ in the tuple $J$.

                        \item If $z = 2$, set $\rpid' \allowbreak \gets \allowbreak \rpid$.

                        \item If $z = 3$, set $\rpid' \allowbreak \gets \allowbreak \bot$.

                        \item Remove the tuple $(\qid, \allowbreak \rpid, \allowbreak z)$.

                    \end{itemize}
                    Else, set $\rpid' \allowbreak \gets \allowbreak \bot$.

                    \item Send the message $(\flrmforwardend, \allowbreak \sid, \allowbreak \tid, \allowbreak \ppid', \allowbreak \flrmmessagerequest', \allowbreak \allowbreak \flrmposition, \allowbreak \flrmroutelist[\flrmindex+2], \flrmlistleakage, \allowbreak \rpid')$ to the party 
                    $\flrmroutelist[\flrmindex+1]$.
                    
                \end{itemize}

            \end{itemize}

    \end{enumerate}

\paragraph{Explanation and discussion of the forward interface.} $\Functionality_{\LRM}$ receives the message $\flrmforwardini$ from a node or a gateway $\flrmParty$ along with a packet identifier $\tid$, a bit $\flrmdestroymessage$ and a tuple $\{\ppid, \allowbreak \flrmGatewayexit, \allowbreak \flrmReceiver, \allowbreak \flrmmessagerequest, \allowbreak \flrmNodeArequest, \allowbreak \flrmNodeBrequest, \allowbreak \flrmNodeCrequest, \allowbreak \langle   \flrmGatewayentryreply, \allowbreak \flrmGatewayexitreply, \allowbreak \flrmNodeAresponse, \allowbreak \flrmNodeBresponse, \allowbreak \flrmNodeCresponse, \allowbreak \flrmSenderreply \rangle\}$. The bit $\flrmdestroymessage$ and the aforementioned tuple are only used by $\Functionality_{\LRM}$ when $\flrmParty$ is corrupt.

$\Functionality_{\LRM}$ checks if it stores a tuple $J$ with $\tid$ and that $\tid$ is valid. $\Functionality_{\LRM}$ also checks that the index $\flrmindex$ of the party that is expected to forward the packet is the index of a node or a gateway. Moreover, $\Functionality_{\LRM}$ checks that the party that forwards the packet is the same party that previously received the packet, or that both parties are corrupt. If such a tuple $J$ is not stored, and either $\flrmParty$ is honest or the tuple $\{\ppid, \allowbreak \flrmGatewayexit, \allowbreak \flrmReceiver, \allowbreak \flrmmessagerequest, \allowbreak \flrmNodeArequest, \allowbreak \flrmNodeBrequest, \allowbreak \flrmNodeCrequest, \allowbreak \langle   \flrmGatewayentryreply, \allowbreak \flrmGatewayexitreply, \allowbreak \flrmNodeAresponse, \allowbreak \flrmNodeBresponse, \allowbreak \flrmNodeCresponse, \allowbreak \flrmSenderreply \rangle\}$ received as input is empty, $\Functionality_{\LRM}$ aborts. Therefore, $\Functionality_{\LRM}$ enforces that a party cannot forward a packet if that party could not have received the packet.
    
However, if $J$ is not stored, $\flrmParty$ is corrupt, the tuple $\{\ppid, \allowbreak \flrmGatewayexit, \allowbreak \flrmReceiver, \allowbreak \flrmmessagerequest, \allowbreak \flrmNodeArequest, \allowbreak \flrmNodeBrequest, \allowbreak \flrmNodeCrequest, \allowbreak \langle   \flrmGatewayentryreply, \allowbreak \flrmGatewayexitreply, \allowbreak \flrmNodeAresponse, \allowbreak \flrmNodeBresponse, \allowbreak \flrmNodeCresponse, \allowbreak \flrmSenderreply \rangle\}$ is not empty, and $\ppid$ in that tuple has not been used before by any corrupt party, then $\Functionality_{\LRM}$ creates $J$ with the routing information in that tuple. In our security analysis, when the simulator receives a packet from a corrupt node or gateway, and the packet was not seen before by the simulator (e.g., because it was sent by a corrupt sender to a corrupt gateway and only now reaches an honest party), the simulator uses the tuple $\{\ppid, \allowbreak \flrmGatewayexit, \allowbreak \flrmReceiver, \allowbreak \flrmmessagerequest, \allowbreak \flrmNodeArequest, \allowbreak \flrmNodeBrequest, \allowbreak \flrmNodeCrequest, \allowbreak \langle   \flrmGatewayentryreply, \allowbreak \flrmGatewayexitreply, \allowbreak \flrmNodeAresponse, \allowbreak \flrmNodeBresponse, \allowbreak \flrmNodeCresponse, \allowbreak \flrmSenderreply \rangle\}$ to send the routing information associated with that packet to the functionality. We make the following remarks:
    \begin{itemize}

        \item The simulator may set some of the input in the tuple to the symbol $\bot$. For example, if a request packet is sent to an honest node $\flrmNodeCrequest$, then the routing information for $\flrmNodeArequest$ is not needed and $\flrmNodeArequest = \bot$ in the tuple. 


        \item $\Functionality_{\LRM}$ stores the identity $\flrmSender$ of the sender as $\Adversary$ to record that the packet was computed by a corrupt party, although the packet was not necessarily computed by a sender (i.e., it could have been generated by a corrupt node or gateway).

        \item $\Functionality_{\LRM}$ always stores that the packet is a request packet. The reason is that the simulator is not able to distinguish request packets from reply packets when packets do not enable replies. Thanks to the fact that $\Functionality_{\LRM}$ uses the same forward interface for both requests and replies, it is not an issue that $\Functionality_{\LRM}$ processes a reply packet as a request packet that does not enable replies. In the security analysis, the simulator simply sets the tuple $\langle   \flrmGatewayentryreply, \allowbreak \flrmGatewayexitreply, \allowbreak \flrmNodeAresponse, \allowbreak \flrmNodeBresponse, \allowbreak \flrmNodeCresponse, \allowbreak \flrmSenderreply \rangle$ to $\bot$. If the packet is a request packet that enables a reply, the routing information for the reply is sent by the simulator in the tuple $\langle   \flrmGatewayentryreply, \allowbreak \flrmGatewayexitreply, \allowbreak \flrmNodeAresponse, \allowbreak \flrmNodeBresponse, \allowbreak \flrmNodeCresponse, \allowbreak \flrmSenderreply \rangle$.

    \end{itemize}
    Once $\Functionality_{\LRM}$ stores a tuple $J$ with routing information for the packet (either because it was stored beforehand or because it was created from the input received from a corrupt $\flrmParty$), $\Functionality_{\LRM}$ creates a tuple $I$ that stores the same information as $J$ and replaces the local packet identifier $\tid$ by a query identifier $\qid$. In $I$, the variable $\flrmposition$ stores the party identifier $\flrmParty$ of the party that forwards the packet, which may differ from the one contained in the packet route $\flrmroutelist$ when $\flrmParty$ is corrupt. $\Functionality_{\LRM}$ keeps the tuple $J$ unmodified because it does not perform detection of duplicates, so a party is allowed to forward the packet more than once.

    After creating $I$, $\Functionality_{\LRM}$ checks if there is a tuple $(\tid', \allowbreak \rpid, \allowbreak \flrmNodeAresponse')$ such that $\tid' \allowbreak = \allowbreak \tid$. That tuple could have been created in the reply interface when a corrupt receiver inputs a reply packet and instructs the gateway to forward it to a node $\flrmNodeAresponse'$. If a tuple is found, $\Functionality_{\LRM}$ compares the first-layer node $\flrmNodeAresponse'$ with the first-layer node $\flrmNodeAresponse$ stored in $I$ and sets a variable $z$ as follows:
    \begin{itemize}
    
        \item $z \allowbreak \gets \allowbreak 0$ when the two identifiers are different and $\flrmNodeAresponse'$ is honest. In this case, when the packet is forwarded to $\flrmNodeAresponse'$, $\Functionality_{\LRM}$ deletes $\flrmNodeBresponse$ from the packet route, so that $\flrmNodeAresponse'$ does not learn $\flrmNodeBresponse$ and  aborts if it tries to forward the packet.

        \item $z \allowbreak \gets \allowbreak 1$ when the two identifiers are different, $\flrmNodeAresponse'$ is corrupt and $\flrmNodeAresponse$ is honest. In this case, when the packet is forwarded to $\flrmNodeAresponse'$, $\Functionality_{\LRM}$ does not let $\flrmNodeAresponse'$ forward the packet later. The reason is that, because $\flrmNodeAresponse$ is honest, the adversary should not have the ability to process the packet and forward it. Any construction that realizes $\Functionality_{\LRM}$ should guarantee that. $\Functionality_{\LRM}$ does not disclose $\flrmNodeBresponse$ to  $\flrmNodeAresponse'$.

        \item $z \allowbreak \gets \allowbreak 2$ when the two parties are corrupt. In this case, because adversarial parties collude with each other, even if the packet is forwarded to a node $\flrmNodeAresponse'$ different from the node $\flrmNodeAresponse$ that has the ability to process the packet, $\Functionality_{\LRM}$ allows $\flrmNodeAresponse'$ or another adversarial party to forward the packet.

        \item $z \allowbreak \gets \allowbreak 3$ when the two parties are equal and honest. In this case, the adversary did not change the first-layer node that should process the packet.
        
    \end{itemize}

    Then $\Functionality_{\LRM}$ aborts if $\flrmParty$ is honest and the party that will receive the packet does not belong to the right set. To determine that, $\Functionality_{\LRM}$ uses the function $\flrmverify$.
    
    $\Functionality_{\LRM}$ sets the message $\flrmmessagerequest$ in the tuple $I$ to $\bot$ if $\flrmParty$ is corrupt and the bit $\flrmdestroymessage$ is set to $1$. Therefore,  $\Functionality_{\LRM}$ allows a corrupt node or gateway to destroy the message sent in a packet. This means that any construction $\mathrm{\Pi}_{\LRM}$ for $\Functionality_{\LRM}$ does not need to protect the integrity of $\flrmmessagerequest$ until a request packet reaches the receiver $\flrmReceiver$ or a reply packet reaches the sender. In~\S\ref{wpes:sec:previouswork}, we discussed an attack against sender anonymity derived from this issue.  
    
    We remark that $\Functionality_{\LRM}$ does not allow corrupt nodes or gateways to modify the routing information of request packets sent by honest senders or of reply packets generated by using a single-use reply block computed by an honest sender, and thus any construction $\mathrm{\Pi}_{\LRM}$ for $\Functionality_{\LRM}$ needs to protect the integrity of the routing information. Nevertheless, if $\flrmParty$ is corrupt, the bit $\flrmdestroymessage$ is set to $1$, and the packet is a request packet, $\Functionality_{\LRM}$ deletes the routing information $\langle \flrmGatewayexitreply, \allowbreak \flrmNodeAresponse, \allowbreak \flrmNodeBresponse, \allowbreak \flrmNodeCresponse, \allowbreak  \flrmGatewayentryreply, \allowbreak \flrmSenderreply \rangle$ for the reply packet. The reason is that, in our construction, the information needed to compute the reply packet is sent along with the message in the payload of the request packet, whose integrity is not protected. Therefore, if a corrupt party destroys the payload of a request packet, both the message and the information needed to compute the reply packet are destroyed.

    If $\Functionality_{\LRM}$ does not abort, $\Functionality_{\LRM}$ sends a message $\flrmforwardsim$ to the simulator along with $\qid$ and the party identifiers $\flrmParty$ and $\flrmroutelist[\flrmindex+1]$, which are the parties that forward and receive the packet respectively.

    Once the simulator sends the $\flrmforwardrep$ message on input $\qid$, $\Functionality_{\LRM}$ aborts if there is not a tuple $I$ stored for that $\qid$. $\Functionality_{\LRM}$ also aborts if the packet is received by an honest party $\flrmroutelist[\flrmindex+1]$ and the party that forwards the packet does not belong to the right set. The latter is checked by running the function $\flrmpriorvf$. Then $\Functionality_{\LRM}$ proceeds as follows:
    \begin{itemize}


        \item If $\flrmindex +1 = 13$, the party that receives the reply packet is the sender $\flrmSenderreply$. $\Functionality_{\LRM}$ proceeds as follows:
        \begin{itemize}

            \item $\Functionality_{\LRM}$ aborts when the sender $\flrmSender$ is honest and the party $\flrmposition$ that forwards the packet is not equal to the gateway $\flrmroutelist[\flrmindex]$. We recall that, when $\flrmSender$ is honest, $\flrmSender \allowbreak = \allowbreak \flrmSenderreply$. Therefore, $\flrmSenderreply$ knows the packet route and checks not only that the reply packet is received from a gateway, but also that the gateway is the one specified in the packet route.

            \item $\Functionality_{\LRM}$ checks if $\flrmSenderreply \allowbreak = \allowbreak \flrmSender$ or if both $\flrmSenderreply$ and $\flrmSender$ are corrupt. In that case, $\Functionality_{\LRM}$ reveals the message and the global packet identifier to $\flrmSenderreply$. Those are the conditions that enable $\flrmSenderreply$ to decrypt the reply packet, i.e., $\flrmSenderreply$ is able to decrypt the reply packet when it was created by using a single-use reply block that was computed by $\flrmSenderreply$, which happens when $\flrmSenderreply \allowbreak = \allowbreak \flrmSender$, or when both $\flrmSenderreply$ and $\flrmSender$ are corrupt, because corrupt parties collude with each other and thus $\flrmSenderreply$ could decrypt any reply packet created by using a single-use reply-block computed by another corrupt party.

            \item $\Functionality_{\LRM}$ also reveals to $\flrmSenderreply$ the party identifier $\flrmposition \allowbreak = \allowbreak \flrmParty$ of the party that forwards the reply packet to $\flrmSenderreply$. The global packet identifier $\ppid$ allows $\flrmSenderreply$ to link the received reply packet with the corresponding request packet.

        \end{itemize}


        \item If $\flrmindex + 1 = 7$, the party that receives the request packet is a receiver $\flrmReceiver$.  $\Functionality_{\LRM}$ proceeds as follows:
        \begin{itemize}
        
            \item When $\flrmReceiver$ is corrupt, the sender $\flrmSender$ is honest, and the request packet enables a reply, $\Functionality_{\LRM}$ runs the algorithm $\flrmCompLeakList$ to compute a leakage list $\flrmlistleakage$ that contains all the parties in the route of the reply packet until and including the first honest party that processes the packet. (We recall that exit gateways $\flrmGatewayexitreply$ do not process reply packets.) $\Functionality_{\LRM}$ runs the algorithm $\flrmCompIndex$ to compute the index $\flrmindex'$ of the next party that should interact with $\Functionality_{\LRM}$ regarding this packet. In this case, $\flrmindex' = 7$, since $\Functionality_{\LRM}$ requires the receiver $\flrmReceiver$ to input the reply message through the reply interface. The position $\flrmposition'$ is set to $\flrmReceiver$, and a local packet identifier $\tid$ is created to enable the receiver to input the reply packet. The global packet identifier is not disclosed to the receiver because the sender is honest.

            \item When both $\flrmReceiver$ and $\flrmSender$ are corrupt, or when the request packet does not enable a reply, then $\Functionality_{\LRM}$ sets $\flrmlistleakage$, $\flrmindex$, $\flrmposition$ and $\tid$ to $\bot$. If both $\flrmReceiver$ and $\flrmSender$ are corrupt, $\Functionality_{\LRM}$ discloses the global packet identifier to $\flrmReceiver$, but otherwise it does not do that. $\Functionality_{\LRM}$ sets $\flrmindex$, $\flrmposition$ and $\tid$ to $\bot$ because the packet will not be processed again. The reason is the following:
            \begin{itemize}

                \item When a reply is not enabled, the receiver is the last party that should receive the packet.

                \item When $\flrmReceiver$ and $\flrmSender$ are corrupt, and a reply is enabled, the reply packet input by the corrupt receiver will be processed  by $\Functionality_{\LRM}$ as a new packet. We remark that, because $\flrmSender$ and $\flrmReceiver$ are corrupt, $\Functionality_{\LRM}$ does not store information about the packet route beyond $\flrmReceiver$. In our security analysis, when the party  that generates the packet is corrupt, our simulator can only find the packet route until the next corrupt party that processes the packet. If the adversary submits a reply packet, the simulator will input it to the functionality as a new request packet because it cannot find out if it is a reply packet or a request packet that does not enable a reply.

            \end{itemize}

            The reason why the leakage list is set to $\bot$ is the following:
            \begin{itemize}
            
                \item When a reply is not enabled, the receiver is the last party in the route.

                \item When $\flrmReceiver$ and $\flrmSender$ are corrupt, $\Functionality_{\LRM}$ discloses the global packet identifier, which allows the adversary to link this packet with the packet generated by another corrupt party. Therefore, the whole packet route is revealed.
        
            \end{itemize}
            

            \item When $\flrmReceiver$ is honest and the request packet enables a reply, the position $\flrmposition'$ is set to $\flrmReceiver$ and the index $\flrmindex'$ is set to $7$. A local packet identifier $\tid$ is created to enable $\flrmReceiver$ to input the reply message through the reply interface. Both the leakage list $\flrmlistleakage$ and the global packet identifier $\ppid'$ are set to $\bot$ because $\flrmReceiver$ is honest.
            
        \end{itemize}

         $\Functionality_{\LRM}$ stores a tuple $J$ with the generated $\tid$, $\flrmposition'$ and $\flrmindex'$. The global packet identifier $\ppid$, the packet route $\flrmroutelist$ and the message $\flrmmessagerequest$ are taken from the tuple $I$. We remark that storing $\flrmmessagerequest$ is not needed here, but we do it so that the tuple $J$ has the same structure in all the interfaces. 
         
         Then $\Functionality_{\LRM}$ sends the party identifier $\flrmposition = \flrmParty$ and the message $\flrmmessagerequest$ stored in $I$ to $\flrmReceiver$. $\Functionality_{\LRM}$ also discloses $\tid$, $\ppid'$ and $\flrmlistleakage$, which are set as described above, and the party identifiers $\flrmGatewayexitreply'$ and $\flrmNodeAresponse'$, which are set to $\bot$ if a reply is not enabled. The receiver needs to learn both $\flrmGatewayexitreply$ and $\flrmNodeAresponse$ when the sender enables a reply because $\flrmGatewayexitreply$ does not process reply packets, and thus needs to learn $\flrmNodeAresponse$ from the receiver. We remark that, in our security analysis, the party identifiers $\flrmGatewayexitreply'$ and $\flrmNodeAresponse'$ are $\bot$ when $\flrmReceiver$ and $\flrmSender$ are corrupt because, as explained above, $\Functionality_{\LRM}$ does not receive from our simulator the packet route beyond $\flrmReceiver$.


        \item If $\flrmindex + 1 \neq 13$ and $\flrmindex +1 \neq 7$, the party that receives the packet is neither a sender nor a receiver. $\Functionality_{\LRM}$ proceeds as follows:
        \begin{itemize}
    

            \item If the party $\flrmroutelist[\flrmindex + 1]$ that receives the packet is corrupt, the party  $\flrmposition$ that forwards the packet is honest, and the sender $\flrmSender$ is honest, $\Functionality_{\LRM}$ uses the algorithms $\flrmCompLeakList$ and $\flrmCompIndex$ to compute the leakage list $\flrmlistleakage$ and the index $\flrmindex'$. We refer to~\S\ref{sec:leakagelist} for the explanation of how those algorithms work. We recall that, when the packet is a request packet that does not enable replies and all the parties after $\flrmroutelist[\flrmindex + 1]$ in the packet route are also corrupt, then $\flrmindex' \allowbreak = \allowbreak \bot$. If $\flrmindex' \allowbreak \neq \allowbreak \bot$, $\Functionality_{\LRM}$ creates a local packet identifier $\tid$ and sets $\flrmposition' \allowbreak \gets \allowbreak \flrmroutelist[\flrmindex']$ as the next party that should use $\tid$ to forward again the packet, else sets both $\tid$ and $\flrmposition'$ to $\bot$.
            The message is disclosed when all the remaining parties in the route of the packet are corrupt, which is the case when $\flrmindex' \allowbreak \neq \allowbreak \bot$ or when $\flrmindex' \allowbreak = \allowbreak 7$.
            The global packet identifier is not disclosed. 

            \item If the party $\flrmroutelist[\flrmindex + 1]$ that receives the packet is corrupt, the party  $\flrmposition$ that forwards the packet is corrupt, and the sender $\flrmSender$ is honest,  the leakage list is set to $\bot$, the position $\flrmposition'$ is set to $\flrmroutelist[\flrmindex + 1]$ and the index $\flrmindex'$ is set to $\flrmindex + 1$.  When $\flrmposition$ is corrupt, the adversary already received the leakage list when  the packet was received by $\flrmposition$ or another corrupt party before $\flrmposition$ in the packet route. The message and the global packet identifier are not disclosed.  We remark that it is possible that the simulator $\Simulator$ in our security analysis invokes the forward interface when both $\flrmroutelist[\flrmindex + 1]$ and $\flrmposition$ are corrupt. This happens when the simulator receives from the adversary a reply packet that the adversary computed by using a single-use reply block created by an honest sender. In that case, the functionality expects the receiver $\flrmReceiver$ to input the reply packet through the reply interface even if subsequent parties in the route are also corrupt. After using the reply interface, the simulator uses the forward interface for those corrupt parties in the route until the packet reaches an honest party.

            \item  If both the party $\flrmroutelist[\flrmindex + 1]$ that receives the packet and the sender $\flrmSender$ are corrupt, the leakage list, $\flrmposition'$ and $\flrmindex'$ are set to $\bot$. (We recall that $\flrmSender$ refers to any corrupt party that created the packet.) In this case, the functionality discloses the message and the global packet identifier $\ppid$, which already reveals all the routing information related to the packet to the adversary. Additionally, the local packet identifier $\tid$ is set to $\bot$ because $\Functionality_{\LRM}$ will not process again this packet. If, subsequently, a corrupt party forwards the packet to an honest party, the functionality processes it as a new packet. In our security analysis, we show that the outputs of honest parties in the real and in the ideal world are indistinguishable even if the functionality processes the forwarded packet as a new packet. We remark that the functionality might not store routing information related to that packet beyond $\flrmroutelist[\flrmindex + 1]$. The reason is that, when the party that computes the packet is corrupt, the simulator can only find out the route of the packet until and including the next corrupt party that processes the packet.  

            The global packet identifier $\ppid$ is leaked because, when both parties are corrupt, in our construction the adversary is able to link the packet received by $\flrmroutelist[\flrmindex + 1]$ with the packet previously computed by another corrupt party. The message $\flrmmessagerequest$ is leaked because the party was created by a corrupt party, so the adversary can decrypt it whenever it is received by a corrupt party.  
            


            \item If the party $\flrmroutelist[\flrmindex + 1]$ that receives the packet is honest, $\flrmlistleakage$ is set to $\bot$, a fresh local packet identifier $\tid$ is created, $\flrmposition'$ is set to $\flrmroutelist[\flrmindex + 1]$ and $\flrmindex'$ is set to $\flrmindex + 1$. The message and the global packet identifier are not disclosed. Therefore, when $\flrmroutelist[\flrmindex + 1]$ is honest, $\Functionality_{\LRM}$ expects $\flrmroutelist[\flrmindex + 1]$ to instruct the functionality to forward the packet by using $\tid$.

    \end{itemize}

    $\Functionality_{\LRM}$ stores the newly created $\tid$ along with $\flrmposition'$, $\flrmindex'$, $\ppid$, $\flrmmessagerequest$ and $\flrmroutelist$. The values $\ppid$, $\flrmmessagerequest$ and $\flrmroutelist$ are taken from the tuple $I$, which is subsequently deleted. We remark that $\tid$ does not provide any information about the packet route or the message to the adversary.

    $\Functionality_{\LRM}$ checks if there is a tuple $(\qid, \allowbreak \rpid, \allowbreak z)$ with a reply packet identifier $\rpid$. Such a tuple could exist if, in the reply interface, a corrupt receiver sends a reply packet and chooses an exit gateway or a first-layer node different from the ones specified by the honest sender that generated the request packet associated to that reply packet. As described above, the variable $z$ determines how $\Functionality_{\LRM}$ should proceed depending on whether the corrupt receiver chose a different first-layer node and on whether the node is corrupt or not. 

    Finally, $\Functionality_{\LRM}$ sends the $\flrmforwardend$ message to the party $\flrmroutelist[\flrmindex + 1]$. $\flrmroutelist[\flrmindex + 1]$ receives the identifier $\flrmposition$ of the party that forwarded the packet and the identifier $\flrmroutelist[\flrmindex + 2]$ of the next party in the route. It also receives $\tid$, $\ppid'$, $\flrmlistleakage$ and $\flrmmessagerequest'$, which are set as specified above. The reply packet identifier $\rpid'$ is disclosed to $\flrmroutelist[\flrmindex + 1]$ when $\flrmroutelist[\flrmindex + 1]$ is corrupt. We remark that $\rpid'$ is only disclosed to first-layer nodes $\flrmNodeAresponse$, as otherwise $\Functionality_{\LRM}$ does not store any tuple $(\qid, \allowbreak \rpid, \allowbreak z)$.
        

    \end{itemize}

%% file: 5Preliminaries.tex
\section{Preliminaries}
\label{sec:preliminaries}

In this section, we describe the building blocks used in our construction $\mathrm{\Pi}_{\LRM}$ in~\S\ref{sec:construction}. In~\S\ref{sec:funcREG}, we describe an ideal functionality for registration. 
In~\S\ref{sec:funcPREG}, we describe an ideal functionality for registration with privacy-preserving key retrieval.
In~\S\ref{sec:IdealFunctionalitySMT}, we describe an ideal functionality for secure message transmission.
In~\S\ref{sec:keyencapsulation}, we define the security properties of key encapsulation mechanisms.
In~\S\ref{sec:keyderivationfunction}, we describe key derivation functions.
In~\S\ref{sec:symmetrickey}, we define the security properties of symmetric key encryption schemes.
In~\S\ref{sec:authenticatedencryption}, we define the security properties of authenticated encryption schemes.


	

	


\subsection{Ideal Functionality for Registration}
\label{sec:funcREG}

Our construction $\mathrm{\Pi}_{\LRM}$ uses the ideal functionality $\Functionality_{\Freg}$ for registration by Canetti~\cite{DBLP:conf/focs/Canetti01}. $\Functionality_{\Freg}$ interacts with authorized parties $\FregT$ that register a message $\fregvalue$ and with any parties $\Party$ that retrieve the registered message. $\Functionality_{\Freg}$ consists of two interfaces $\fregregister$ and $\fregretrieve$. The interface $\fregregister$ is used by $\FregT$ to register a message $\fregvalue$ with $\Functionality_{\Freg}$. $\Functionality_{\Freg}$ checks whether $\FregT \allowbreak \in \allowbreak \mathbb{T}$, where  $\mathbb{T}$ is contained in the session identifier $\sid$.  A party $\Party$ uses $\fregretrieve$ on input a party identifier $\FregT$ to retrieve the message $\fregvalue$ associated with $\FregT$ from $\Functionality_{\Freg}$.



\paragraph{Ideal Functionality $\Functionality_{\Freg}$.} $\Functionality_{\Freg}$ is parameterized by a message space $\fregmessagespace$.
\begin{enumerate}
\item On input $(\fregregisterini, \allowbreak \sid, \allowbreak \fregvalue)$ from a party $\FpregT$: 

\begin{itemize}

\item Abort if $\sid \neq (\mathbb{T}, \sid')$, or if $\FpregT \notin \mathbb{T}$, or if $\fregvalue \notin \fregmessagespace$, or if there is a tuple $(\sid, \allowbreak \FpregT', \allowbreak \fregvalue, \allowbreak 0)$ stored such that $\FregT' \allowbreak = \allowbreak \FregT$.

\item Store $(\sid, \allowbreak \FpregT, \allowbreak \fregvalue, \allowbreak 0)$.

\item Send $(\fregregistersim, \allowbreak \sid, \allowbreak \FpregT, \allowbreak \fregvalue)$ to $\Simulator$.

\end{itemize}

\item[S.] On input $(\fregregisterrep, \allowbreak \sid, \allowbreak \FpregT)$ from the simulator $\Simulator$: 
 
\begin{itemize}

\item Abort if $(\sid, \allowbreak \FpregT', \fregvalue, 0)$ such that $\FpregT' = \FpregT$  is not stored, or if $(\sid, \allowbreak \FpregT', \allowbreak \fregvalue, \allowbreak 1)$ such that $\FpregT' = \FpregT$ is already stored.

\item Store $(\sid, \allowbreak \FpregT, \fregvalue, 1)$.

\item Send $(\fregregisterend, \allowbreak \sid)$ to $\FpregT$.

\end{itemize}

\item On input $(\fregretrieveini, \allowbreak \sid, \allowbreak \FpregT)$ from any party $\Party$:

\begin{itemize}

\item If $(\sid, \allowbreak \FpregT', \fregvalue, 1)$ such that $\FpregT' = \FpregT$  is stored, set $\fregvalue' \gets \fregvalue$; else set $\fregvalue' \gets \bot$.

\item Create a fresh $\ssid$ and store $(\ssid, \fregvalue', \Party)$.

\item Send $(\fregretrievesim, \allowbreak \sid, \allowbreak \ssid, \allowbreak \FpregT, \allowbreak \fregvalue')$ to $\Simulator$.

\end{itemize}

\item[S.] On input $(\fregretrieverep, \allowbreak \sid, \allowbreak \ssid)$ from  $\Simulator$:

\begin{itemize}

\item Abort if $(\ssid', \fregvalue', \Party)$ such that $\ssid' = \ssid$ is not stored.

\item Delete the record $(\ssid, \fregvalue', \Party)$.

\item Send $(\fregretrieveend, \sid, \fregvalue')$ to $\Party$.

\end{itemize}

\end{enumerate}

\subsection{Ideal Functionality for Registration with Privacy Preserving Key Retrieval}
\label{sec:funcPREG}

Our construction $\mathrm{\Pi}_{\LRM}$ uses an ideal functionality $\Functionality_{\Fpreg}$ for registration with privacy-preserving key retrieval. $\Functionality_{\Fpreg}$ is similar to $\Functionality_{\Freg}$ in~\S\ref{sec:funcREG}, but there are two differences. In the registration interface, $\Functionality_{\Fpreg}$ allows registrations from any party. In the retrieval interface, $\Functionality_{\Fpreg}$ does not leak the identity of the party whose registered message is being retrieved to the adversary. 



\paragraph{Ideal Functionality $\Functionality_{\Fpreg}$.} $\Functionality_{\Fpreg}$ is parameterized by a message space $\fpregmessagespace$.
\begin{enumerate}
\item On input $(\fpregregisterini, \allowbreak \sid, \allowbreak \fpregvalue)$ from a party $\FpregT$: 

\begin{itemize}

\item Abort if $\fpregvalue \notin \fpregmessagespace$ or if there is a tuple $(\sid, \allowbreak \FpregT', \allowbreak \fpregvalue, \allowbreak 0)$ stored such that $\FpregT' \allowbreak = \allowbreak \FpregT$.

\item Store $(\sid, \allowbreak \FpregT, \allowbreak \fpregvalue, \allowbreak 0)$.

\item Send $(\fpregregistersim, \allowbreak \sid, \allowbreak \FpregT, \allowbreak \fpregvalue)$ to $\Simulator$.

\end{itemize}

\item[S.] On input $(\fpregregisterrep, \allowbreak \sid, \allowbreak \FpregT)$ from the simulator $\Simulator$: 
 
\begin{itemize}

\item Abort if $(\sid, \allowbreak \FpregT', \fpregvalue, 0)$ such that $\FpregT' = \FpregT$  is not stored, or if $(\sid, \allowbreak \FpregT', \allowbreak \fpregvalue, \allowbreak 1)$ such that $\FpregT' = \FpregT$ is already stored.

\item Store $(\sid, \allowbreak \FpregT, \fpregvalue, 1)$.

\item Send $(\fpregregisterend, \allowbreak \sid)$ to $\FpregT$.

\end{itemize}

\item On input $(\fpregretrieveini, \allowbreak \sid, \allowbreak \FpregT)$ from any party $\Party$:

\begin{itemize}

\item If $(\sid, \allowbreak \FpregT', \fpregvalue, 1)$ such that $\FpregT' = \FpregT$  is stored, set $\fpregvalue' \gets \fpregvalue$; else set $\fpregvalue' \gets \bot$.

\item Create a fresh $\ssid$ and store $(\ssid, \fpregvalue', \Party)$.

\item Send $(\fpregretrievesim, \allowbreak \sid, \allowbreak \ssid)$ to $\Simulator$.

\end{itemize}

\item[S.] On input $(\fpregretrieverep, \allowbreak \sid, \allowbreak \ssid)$ from  $\Simulator$:

\begin{itemize}

\item Abort if $(\ssid', \fpregvalue', \Party)$ such that $\ssid' = \ssid$ is not stored.

\item Delete the record $(\ssid, \fpregvalue', \Party)$.

\item Send $(\fpregretrieveend, \sid, \fpregvalue')$ to $\Party$.

\end{itemize}

\end{enumerate}

\subsection{Ideal Functionality for Secure Message Transmission}
\label{sec:IdealFunctionalitySMT}

Our construction $\mathrm{\Pi}_{\LRM}$ uses the ideal functionality $\Functionality_{\SMT}$ for secure message transmission in~\cite{DBLP:conf/focs/Canetti01}. $\Functionality_{\SMT}$ interacts with a sender $\Sender$ and a receiver $\Receiver$, and consists of one interface $\fsmtsend$. $\Sender$ uses the $\fsmtsend$ interface to send a message $\SMTmessage$ to $\Receiver$. $\Functionality_{\SMT}$ leaks $\SMTfleakage(\SMTmessage)$ to the simulator $\Simulator$. $\SMTfleakage: \fsmtmessagespace \rightarrow \mathbb{N}$ is a function that leaks the message length. $\Simulator$ cannot modify $\SMTmessage$. The session identifier $\sid$ contains the identifiers of $\Sender$ and $\Receiver$. 



\paragraph{Ideal Functionality $\Functionality_{\SMT}$.} $\Functionality_{\SMT}$ is parameterized by a message space $\fsmtmessagespace$ and by a leakage function $\SMTfleakage: \fsmtmessagespace \rightarrow \mathbb{N}$, which leaks the message length.
\begin{enumerate}

\item On input $(\fsmtsendini, \allowbreak \sid, \allowbreak \SMTmessage)$ from a party $\Sender$:

\begin{itemize}

\item Abort if $\sid \neq (\Sender, \Receiver, \sid')$, or if $\SMTmessage \notin \fsmtmessagespace$.

\item Create a fresh $\ssid$ and store $(\ssid, \SMTmessage)$.

\item Send $(\fsmtsendsim, \allowbreak \sid, \allowbreak \ssid, \allowbreak \SMTfleakage(\SMTmessage))$ to $\Simulator$.

\end{itemize}

\item[S.] On input $(\fsmtsendrep, \allowbreak \sid, \allowbreak \ssid)$ from $\Simulator$:

\begin{itemize}

\item Abort if $(\ssid',  \SMTmessage)$ such that $\ssid' = \ssid$ is not stored.

\item Delete the record $(\ssid,  \SMTmessage)$.

\item Parse $\sid$ as $(\Sender, \Receiver, \sid')$ and send $(\fsmtsendend, \allowbreak \sid, \allowbreak \SMTmessage)$ to $\Receiver$.

\end{itemize}

\end{enumerate}

\subsection{Key Encapsulation Mechanism}
\label{sec:keyencapsulation}

Our construction $\mathrm{\Pi}_{\LRM}$ uses a key encapsulation mechanism. We define the algorithms of a key encapsulation mechanism in~\S\ref{sec:keyencapsulation:syntax}. In~\S\ref{sec:keyencapsulation:correctness}, we define correctness for a key encapsulation mechanism. In~\S\ref{sec:keyencapsulation:security}, we define the security properties of a key encapsulation mechanism. We follow the definitions in~\cite{cryptoeprint:2024/039}.

\subsubsection{Syntax}
\label{sec:keyencapsulation:syntax}

\begin{definition}[KEM.]  A key encapsulation mechanism is a triple of algorithms $\KEM \allowbreak = \allowbreak (\KEMKeyGen, \allowbreak \KEMEnc, \allowbreak \KEMDec)$ with public keyspace $\KEMpublickeyspace$, secret keyspace $\KEMsecretkeyspace$, ciphertext space $\KEMciphertextspace$, and shared keyspace $\KEMsharedkeyspace$. The triple of algorithms is defined as follows:
\begin{description}
    
    \item[$\KEMKeyGen(1^\securityparameter)$.] Randomized algorithm that, given the security parameter $1^\securityparameter$, outputs a secret key $\KEMsecretkey \allowbreak \in \KEMsecretkeyspace$, and a public key $\KEMpublickey \allowbreak \in \KEMpublickeyspace$.

    \item[$\KEMEnc(\KEMpublickey)$.] Randomized algorithm that, given a public key $\KEMpublickey \allowbreak \in \KEMpublickeyspace$, outputs a shared key $\KEMsharedkey \allowbreak \in \allowbreak \KEMsecretkeyspace$, and a ciphertext $\KEMciphertext \allowbreak \in \allowbreak \KEMciphertextspace$.

    \item[$\KEMDec(\KEMciphertext, \KEMsecretkey)$.] Deterministic algorithm that, given a secret key $\KEMsecretkey \allowbreak \in \allowbreak \KEMsecretkeyspace$  and a ciphertext $\KEMciphertext \allowbreak \in \allowbreak \KEMciphertextspace$, outputs the shared key $\KEMsharedkey \in \KEMsharedkeyspace$. In case of rejection, this algorithm outputs $\bot$.
    
\end{description}
\end{definition}

\subsubsection{Correctness}
\label{sec:keyencapsulation:correctness}

The correctness of a $\KEM$ imposes that, except with small probability drawn over the
random coin space of $\KEMKeyGen$ and $\KEMEnc$, we have that the algorithm $\KEMDec$ correctly
recovers the shared key produced by $\KEMEnc$. Formally, we say that a KEM is $\delta$-correct
if:
\begin{equation*}
\Prob \left[
\KEMsharedkey \neq \KEMsharedkey' \middle| \begin{array}{c} (\KEMsecretkey, \KEMpublickey) \getsdollar \KEMKeyGen(1^\securityparameter) \\ (\KEMsharedkey, \KEMciphertext) \getsdollar \KEMEnc(\KEMpublickey) \\ \KEMsharedkey' \gets \KEMDec(\KEMciphertext, \KEMsecretkey) \end{array}
\right] \leq \delta~.
\end{equation*}

\subsubsection{Security}
\label{sec:keyencapsulation:security}

The IND-CCA security game for a $\KEM$ is denoted as IND-CCA$^{b}_{\KEM,\A}$ and is shown in
Figure~\ref{fig:secKEM}.

\begin{definition}[IND-CCA advantage for a $\KEM$.]  The advantage of $\A$ in breaking the
IND-CCA security of a $\KEM$ is defined as
\begin{multline*}
    \Adv_{\textrm{IND-CCA},\A}^{\KEM} = \\ |\Pr[\textrm{IND-CCA}_{\KEM,\A}^{0} \Rightarrow 1] - \Pr[\textrm{IND-CCA}_{\KEM,\A}^{1} \Rightarrow 1]|  
\end{multline*}
\end{definition}

\begin{figure}

  \begin{tabular}{ll}
      \underline{\textbf{Game} IND-CCA$^{b}_{\KEM,\A}$} & \hspace{10mm} \underline{\textbf{Oracle} $\KEMOracleDec(\KEMciphertext)$}  \\
       $(\KEMsecretkey, \KEMpublickey) \getsdollar \KEMKeyGen(1^\securityparameter)$ &  \hspace{10mm} \textbf{if} $\KEMciphertext = \KEMciphertext^{\ast}$ \textbf{then}  \\
      $(\KEMsharedkey_0, \KEMciphertext^{\ast}) \getsdollar \KEMEnc(\KEMpublickey)$ & \hspace{13mm} \textbf{return} $\bot$ \\
      $\KEMsharedkey_1 \getsdollar \KEMsharedkeyspace$ & \hspace{10mm} $\KEMsharedkey \gets \KEMDec(\KEMciphertext, \KEMsecretkey)$ \\
      $b' \getsdollar \A^{\KEMOracleDec(\cdot)}(\KEMpublickey, \KEMciphertext^{\ast}, \KEMsharedkey_b)$ & \hspace{10mm} \textbf{return} $\KEMsharedkey$ \\ 
\textbf{return} $b'$ & 
      
  \end{tabular}
  
  \caption{IND-CCA security game for a $\KEM$}\label{fig:secKEM}
\end{figure}

\subsection{Key Derivation Function}
\label{sec:keyderivationfunction}

Our construction $\mathrm{\Pi}_{\LRM}$ uses a key derivation function $\KDFname$. In this section we include a definition for $\KDFname$ and a security definition for $\KDFname$, which depends on a source of keying material $\Sigma$. We follow the definitions in~\cite{cryptoeprint:2010/264}. Then we adapt the definition in~\cite{cryptoeprint:2010/264} for the case in which the source of keying material $\Sigma$ is the key encapsulation mechanism $\KEM$ in~\ref{sec:keyencapsulation}.

\begin{definition}[$\KDFname$.]  A key derivation function $\KDF$ accepts as input four arguments: a value $\sigma$ sampled from a source of keying material (see Definition~\ref{def:keyingmaterialsource}), a length value $\ell$, and two additional arguments, a salt value $r$ defined over a set of possible salt values and a context variable $ctx$, both of which are optional, i.e., can be set to the null string or to a constant. The $\KDF$ output is a string of $\ell$ bits.
\end{definition}

The security and quality of a $\KDF$ depends on the properties of the ``source of keying material'' from which the input $\sigma$ is chosen. 

\begin{definition}[Source of keying material $\Sigma$.] \label{def:keyingmaterialsource}  A source of keying material $\Sigma$ is a two-valued probability distribution $(\sigma, \allowbreak \alpha)$ generated by an efficient probabilistic algorithm. We will refer to both the probability distribution as well as the generating algorithm by $\Sigma$.
\end{definition}

The output $\sigma$ represents the (secret) source key material to be input to a $\KDF$, while $\alpha$ represents some auxiliary knowledge about $\sigma$ (or its distribution). We require a $\KDF$ to be secure on inputs $\sigma$ even when the knowledge value $\alpha$ is given to the adversary. We define the security of a $\KDF$ with respect to a specific source of keying material $\Sigma$. 

\begin{definition}[Secure $\KDF$.] \label{def:KDF:secure}
A key derivation function $\KDF$ is said to be $(t, \allowbreak q, \allowbreak \epsilon)$-secure with respect to a source
of keying material $\Sigma$ if no adversary $\Adversary$ running in time $t$ and making at most $q$ queries can win the
following distinguishing game with probability larger than $1/2 + \epsilon$:
\begin{enumerate}

    \item The algorithm $\Sigma$ is invoked to produce a pair $(\sigma, \allowbreak \alpha)$.

    \item A salt value $r$ is chosen at random from the set of possible salt values defined by KDF ($r$ may be set to a constant or a null value if so defined by KDF).

    \item The adversary $\Adversary$ is provided with $\alpha$ and $r$.

    \item For $i = 1, \ldots , q' \leq q$: $\Adversary$ chooses arbitrary values $ctx_i, \ell_i$ and receives the value $\KDF(\sigma, r, ctx_i, \ell_i)$. The queries by $\Adversary$ are adaptive, i.e., each query may depend on the responses to previous ones.

    \item $\Adversary$ chooses values $ctx$ and $\ell$ such that $ctx \notin \{ctx_1, \ldots, ctx_{q'}\}$.

    \item A bit $b \in \{0, 1\}$ is chosen at random. If $b = 0$, the adversary $\Adversary$ is provided with the output of
    $\KDF(\sigma, \allowbreak r, \allowbreak ctx, \allowbreak \ell)$, else $\Adversary$ is given a random string of $\ell$ bits.

    \item Step 4 is repeated for up to $q - q'$ queries (subject to the restriction $ctx_i \neq ctx$).

    \item $\Adversary$ outputs a bit $b' \in \{0,1\}$. It wins if $b' = b$.

\end{enumerate}

\end{definition}

We adapt Definition~\ref{def:KDF:secure} for the case in which the source of keying material $\Sigma$ is the key encapsulation mechanism $\KEM$ in~\ref{sec:keyencapsulation}. The input $\sigma$ of the $\KDF$ is the shared key $\KEMsharedkey$ output by $\KEMEnc$, while $\alpha$ consists of the values $\KEMpublickey$ and $\KEMciphertext$ output by $\KEMKeyGen$ and $\KEMEnc$ respectively. The secret key $\KEMsecretkey$ output by $\KEMKeyGen$ should not be given to the adversary, because otherwise the adversary is able to compute $\sigma$ on input $\alpha$. However, in $\mathrm{\Pi}_{\LRM}$ the adversary computes ciphertexts on input the public key of honest users that need be decrypted by honest users. Therefore, our reduction to the security of the $\KDF$ with respect to a source of keying material $\KEM$ needs a decryption oracle. To address this issue, we adapt the game described in Definition~\ref{def:KDF:secure}, so that the adversary can access a decryption oracle. This requires a $\KEM$ that is IND-CCA secure.


The context $ctx$ contains $\KEMpublickey$, $\KEMciphertext$ and the session identifier of the protocol. In the construction $\mathrm{\Pi}_{\LRM}$, we use a $\KDF$ that does not use a salt $r$.

\begin{definition}[Secure $\KDF$ with source IND-CCA $\KEM$.] \label{def:KDF:secure:KEM}
A  $\KDF$ is said to be $(t, \allowbreak q, \allowbreak \epsilon)$-secure with respect to a source
of keying material $\KEM$ defined by the algorithms $(\KEMKeyGen, \allowbreak \KEMEnc, \allowbreak \KEMDec)$ and that is IND-CCA secure if no adversary $\Adversary$ running in time $t$ and making at most $q$ queries can win the following distinguishing game with probability larger than $1/2 + \epsilon$:
\begin{enumerate}

    \item The algorithm $\KEMKeyGen(1^\securityparameter)$ is executed to produce a pair $(\KEMsecretkey, \allowbreak \KEMpublickey)$.

    \item The algorithm $\KEMEnc(\KEMpublickey)$ is executed to compute a pair $(\KEMciphertext^{\ast}, \allowbreak \KEMsharedkey)$.
    
    \item The value $\sigma$ is set as $\KEMsharedkey$ and the value $\alpha$ is set as $\KEMpublickey$ and $\KEMciphertext^{\ast}$. $\alpha$ is given to the adversary.

    \item For $i = 1, \ldots , q' \leq q$: $\Adversary$ chooses arbitrary values $ctx_i, \ell_i$ and receives the value $\KDF(\sigma, ctx_i, \ell_i)$. The queries by $\Adversary$ are adaptive, i.e., each query may depend on the responses to previous ones.

    \item $\Adversary$ chooses values $ctx$ and $\ell$ such that $ctx \notin \{ctx_1, \ldots, ctx_{q'}\}$.


    \item A bit $b \in \{0, 1\}$ is chosen at random. If $b \allowbreak = \allowbreak 0$, set $s_0 \allowbreak \gets \allowbreak \KDF(\KEMsharedkey, ctx, \ell)$, else set $s_1 \allowbreak \gets \allowbreak \{0,1\}^{\ell}$, where $\ell$ is the bit length of $s_0$.

    \item The adversary $\Adversary$ is provided with $s_b$.

    \item The queries to the $\KDF$ oracle  are repeated for up to $q - q'$ queries (subject to the restriction $ctx_i \neq ctx$).

     \item The adversary $\Adversary$ is provided with a decryption oracle $\KEMOracleDec$ that operates like the one described in Figure~\ref{fig:secKEM}. The oracle $\KEMOracleDec$ uses the secret key $\KEMsecretkey$ computed above as input to the algorithm $\KEMDec$ and the ciphertext $\KEMciphertext^{\ast}$ computed above as challenge ciphertext.




    \item $\Adversary$ outputs a bit $b' \in \{0,1\}$. It wins if $b' = b$.

\end{enumerate}

\end{definition}

\subsection{Authenticated Encryption with Associated Data}
\label{sec:authenticatedencryption}

Our construction $\mathrm{\Pi}_{\LRM}$ uses  an authenticated encryption  with associated data ($\AEAD$) scheme. In this section, we define an $\AEAD$ scheme and the security properties of an $\AEAD$ scheme. We follow the definitions in~\cite{DBLP:conf/ccs/Rogaway02}.

\begin{definition}[$\AEAD$.] \label{def:aead}
An authenticated encryption  with associated data $(\AEAD)$ scheme
is a tuple  $(\AEADKeyspace, \allowbreak \AEADE, \allowbreak \AEADD)$. Associated to $(\AEADKeyspace, \allowbreak \AEADE, \allowbreak \AEADD)$ are sets of strings $\AEADSetNonce \allowbreak = \allowbreak \{0, \allowbreak 1\}^{n}$ and $\AEADSetMessage \allowbreak \subseteq \allowbreak \{0, \allowbreak \allowbreak 1\}^{\ast}$, and also a set $\AEADSetHeader \allowbreak \subseteq \allowbreak \{0, \allowbreak 1\}^{\ast}$ that has a linear-time membership test. The key
space $\AEADKeyspace$ is a finite nonempty set of strings. The algorithms $(\AEADE, \allowbreak \AEADD)$ are defined as follows.
\begin{description}

\item[$\AEADE(\AEADkey,\AEADnonce, \AEADheader, \AEADmessage)$.] The encryption algorithm $\AEADE$ is a deterministic algorithm that takes strings $\AEADkey \allowbreak \in \allowbreak \AEADKeyspace$, $\AEADnonce \allowbreak \in \allowbreak \AEADSetNonce$, $\AEADheader \allowbreak \in \allowbreak \AEADSetHeader$, and $\AEADmessage \allowbreak \in \allowbreak \AEADSetMessage$. It outputs a string $\AEADciphertext$.

\item[$\AEADD(\AEADkey, \AEADnonce, \AEADheader, \AEADciphertext)$.] The decryption algorithm $\AEADD$ is a deterministic algorithm that takes strings $\AEADkey \allowbreak \in \allowbreak \AEADKeyspace$, $\AEADnonce \allowbreak \in \allowbreak \AEADSetNonce$, $\AEADheader \allowbreak \in \allowbreak \AEADSetHeader$, and $\AEADciphertext \allowbreak \in \allowbreak \{0,1\}^{\ast}$. It outputs either a string in $\AEADSetMessage$ or $\bot$.

\end{description}
\end{definition}

Correctness requires that $\AEADD(\AEADkey, \allowbreak \AEADnonce, \allowbreak \AEADheader, \allowbreak \AEADE(\AEADkey, \allowbreak \AEADnonce, \allowbreak \AEADheader, \allowbreak \AEADmessage)) \allowbreak = \allowbreak \AEADmessage$ for all $\AEADkey \allowbreak \in \allowbreak \AEADKeyspace$, $\AEADnonce \allowbreak \in \allowbreak \AEADSetNonce$, $\AEADheader \allowbreak \in \allowbreak \AEADSetHeader$ and $\AEADmessage \allowbreak \in \allowbreak \AEADSetMessage$. The length $|\AEADE(\AEADkey,\AEADnonce, \AEADheader, \AEADmessage)| = \ell(|\AEADmessage|)$ for some linear-time computable length function $\ell$.

\begin{definition}[IND\$-CPA.] \label{def:aead:ind-cpa}
Let $(\AEADKeyspace, \allowbreak \AEADE, \allowbreak \AEADD)$ be an $\AEAD$ scheme with length function $\ell$. Let $\$(\cdot, \allowbreak \cdot, \allowbreak \cdot)$ be an oracle that, on input $(\AEADnonce, \allowbreak \AEADheader, \allowbreak \AEADmessage)$ returns a random string of $\ell(|\AEADmessage|)$ bits. Let the advantage of the adversary $\Adversary$ be
\begin{multline*}
    \Adv_{\Adversary}^{\AEAD, \mathsf{IND\$-CPA}} = \\ |\Pr[\AEADkey \xleftarrow{\$} \AEADKeyspace: \Adversary^{\AEADE(\AEADkey,\cdot, \cdot, \cdot)}=1] - \Pr[\Adversary^{\$(\cdot, \cdot, \cdot)} = 1]|
\end{multline*}
This notion is named indistinguishability from random bits under a chosen-plaintext attack (IND\$-CPA).
\end{definition}

\begin{definition}[Unforgeability.] \label{def:aead:unforgeability}
Let the tuple $(\AEADKeyspace, \allowbreak \AEADE, \allowbreak \AEADD)$ be an $\AEAD$ scheme and let $\Adversary$ be an adversary with access to an oracle $\AEADE(\AEADkey,\cdot, \cdot, \cdot)$ for some key $\AEADkey$. We say that $\Adversary$ breaks the unforgeability property if, with non negligible probability, $\Adversary$ outputs $(\AEADnonce, \allowbreak \AEADheader, \allowbreak \AEADciphertext)$ such that $\AEADD(\AEADkey, \allowbreak \AEADnonce, \allowbreak \AEADheader, \allowbreak \AEADciphertext) \allowbreak \neq \allowbreak \bot$ and $\Adversary$ did not send a query $(\AEADnonce, \allowbreak \AEADheader, \allowbreak \AEADmessage)$ that resulted in a response $\AEADciphertext$ from the oracle.
\end{definition}

To prove security of construction $\mathrm{\Pi}_{\LRM}$, we need an $\AEAD$ scheme that is committing. We use the definition of committing $\AEAD$ in Section 3 in~\cite{cryptoeprint:2022/1260}.

In the description of Outfox algorithms in~\S\ref{sec:constructionalgorithms}, for simplicity, we omit the nonce and generate the associated data as an output of algorithm $\AEADE$. We refer to the $\AEAD$ ciphertext as $\beta$ and to the associated data as $\gamma$. (To avoid confusion with the header of a packet, we refer to the $\AEAD$ header as associated data.) The syntax of the algorithms $\AEAD$ algorithms is thus the following.
\begin{description}

\item[$\AEADE(k,m)$.] On input a key $k$ and a message $m$, output a ciphertext $\beta$ and associated data $\gamma$.

\item[$\AEADD(k, \beta, \gamma)$.] On input a key $k$, a ciphertext $\beta$ and  associated data $\gamma$, output a message $m$ or $\bot$.

\end{description}

\subsection{Symmetric-Key Encryption}
\label{sec:symmetrickey}

The implementation of Outfox uses the block cipher Lioness~\cite{fse-1996-2968} to encrypt the payload of a packet, and assumes that the message given as input is already encrypted by an encryption scheme that should provide both confidentiality and integrity. Such encryption scheme is chosen at the application level, and the security properties of combining it with Lioness depend on that choice.

For the purpose of analyzing the security of construction $\mathrm{\Pi}_{\LRM}$, we use a symmetric-key encryption scheme that provides indistinguishability from random bits under chosen-ciphertext attack (IND\$-CCA). 

\begin{definition}[$\blockcipher$.] \label{def:se}
A symmetric-key encryption $(\blockcipher)$ scheme is a tuple  $(\AEADKeyspace, \allowbreak \blockcipherenc, \allowbreak \blockcipherdec)$. Associated to $(\AEADKeyspace, \allowbreak \blockcipherenc, \allowbreak \blockcipherdec)$ is a message space $\blockciphermessagespace$ that consists of strings of bit length $\ell$. The ciphertext space is equal to the message space. The key space $\blockcipherkeyspace$ is a finite nonempty set of strings. The algorithms $(\blockcipherenc, \allowbreak \blockcipherdec)$ are defined as follows.
\begin{description}

\item[$\blockcipherenc(k,m)$.] On input a key $k \allowbreak \in \allowbreak \blockcipherkeyspace$ and a message $m \allowbreak \in \allowbreak \blockciphermessagespace$, output a ciphertext $c$.

\item[$\blockcipherdec(k,c)$.] On input a key $k \allowbreak \in \allowbreak \blockcipherkeyspace$ and a ciphertext $c$, output a message $m$.

\end{description}
\end{definition}

Correctness requires that $\blockcipherdec(k,\allowbreak\blockcipherenc(k, \allowbreak m)) \allowbreak = \allowbreak m$ and that $\blockcipherenc(k, \allowbreak \blockcipherdec(k, \allowbreak m)) \allowbreak = \allowbreak m$ for all $m \allowbreak \in \allowbreak \blockciphermessagespace$ and for all $k \allowbreak \in \allowbreak \blockcipherkeyspace$. I.e., we require both that decryption of a ciphertext outputs the encrypted message and that encryption of a decrypted message outputs the message.

\begin{definition}[IND\$-CCA.] \label{def:se:IND-CCA}
Let $(\blockcipherkeyspace, \allowbreak \blockcipherenc, \allowbreak \blockcipherdec)$ be a $\blockcipher$ scheme.  Let $\$(\cdot)$ be an oracle that returns a random string of bit length $\ell$. Let $\mathsf{Enc}$ and $\mathsf{Dec}$ be two oracles parameterized by a random key $k \allowbreak \in \allowbreak \blockcipherkeyspace$ and that store all their outputs in two sets $Q_{e}$ and $Q_{d}$ respectively.  The oracle $\mathsf{Enc}$ receives as input a member $m$ of $\blockciphermessagespace$ and, if $m \allowbreak \notin \allowbreak Q_{d}$, runs $c \allowbreak \gets \allowbreak \blockcipherenc(k,m)$, includes $c$ in $Q_{e}$, and outputs $c$. The oracle $\mathsf{Dec}$ receives as input a member $c$ of $\blockciphermessagespace$ (the ciphertext and the message spaces are equal) and, if $c \allowbreak \notin \allowbreak Q_{e}$, runs $m \allowbreak \gets \allowbreak \blockcipherdec(k,c)$, includes $m$ in $Q_{d}$, and outputs $m$.
Let the advantage of the adversary $\Adversary$ be
\begin{multline*}
    \Adv_{\Adversary}^{\blockcipher, \mathsf{IND\$-CCA}} = \\ |\Pr[k \xleftarrow{\$} \blockcipherkeyspace: \Adversary^{\mathsf{Enc}(k,\cdot), \mathsf{Dec}(k,\cdot)}=1] - \Pr[\Adversary^{\$(\cdot)} = 1]|
\end{multline*}
\end{definition}

%% file: 6OutfoxPacketFormat.tex
\section{Our Construction for a Layered Mixnet in the Client Model}
\label{sec:construction}

We describe our construction $\mathrm{\Pi}_{\LRM}$ for a layered mixnet in the client model.  
In~\S\ref{sec:descriptionconstruction}, we provide a formal description of $\mathrm{\Pi}_{\LRM}$ and we discuss our construction.

\subsection{Description of Our Construction}
\label{sec:descriptionconstruction}


We describe a construction $\mathrm{\Pi}_{\LRM}$ that realizes the functionality $\Functionality_{\LRM}$. $\mathrm{\Pi}_{\LRM}$ uses as building blocks the ideal functionality $\Functionality_{\Freg}$ for registration in~\S\ref{sec:funcREG}, the ideal functionality $\Functionality_{\Fpreg}$ for registration with privacy-preserving key retrieval in~\S\ref{sec:funcPREG}, the ideal functionality $\Functionality_{\SMT}$ for secure message transmission in~\S\ref{sec:IdealFunctionalitySMT}, and the key encapsulation mechanism  in~\S\ref{sec:keyencapsulation}. Additionally, $\mathrm{\Pi}_{\LRM}$ uses the algorithms for packet computation and processing that are described in~\S\ref{sec:definitionalgorithms} and~\S\ref{sec:constructionalgorithms}. 

In~\S\ref{sec:construction1setup}, we describe and discuss the setup interface.
In~\S\ref{sec:construction2registration}, we describe and discuss the registration interface.
In~\S\ref{sec:construction3request}, we describe and discuss the request interface.
In~\S\ref{sec:construction4reply}, we describe and discuss the reply interface.
Finally, in~\S\ref{sec:construction5forward}, we describe and discuss the forward interface.


\input{6OutfoxPacketFormat1Setup}
\input{6OutfoxPacketFormat2Registration}
\input{6OutfoxPacketFormat3Request}
\input{6OutfoxPacketFormat4Reply}
\input{6OutfoxPacketFormat5Forward}

%% file: 6OutfoxPacketFormat1Setup.tex
\subsubsection{Setup Interface of Our Construction}
\label{sec:construction1setup}

\paragraph{Construction of the setup interface.} $\mathrm{\Pi}_{\LRM}$ is parameterized by the parameters defined in~\S\ref{sec:functionalityParameters}, and by a security parameter $1^\securityparameter$.
\begin{enumerate}
    \item[1.] On input $(\flrmsetupini, \allowbreak \sid)$, a node or a gateway $\flrmParty$ does the following:

    \begin{itemize}

        \item $\flrmParty$ aborts if $\sid \neq (\flrmSetA, \allowbreak \flrmSetB, \allowbreak \flrmSetC, \allowbreak \flrmSetW, \allowbreak \sid')$, or if  $\flrmParty \notin \flrmSetA \cup \allowbreak \flrmSetB \cup \allowbreak \flrmSetC \cup \flrmSetW$, or if $|\flrmSetA| \allowbreak < \allowbreak \flrmNodesmin$, or if $|\flrmSetB| \allowbreak < \allowbreak \flrmNodesmin$, or if $|\flrmSetC| \allowbreak < \allowbreak \flrmNodesmin$, or if $|\flrmSetW| \allowbreak < \allowbreak \flrmGatewaysmin$, or if $\flrmSetA \allowbreak \cap \allowbreak \flrmSetB \neq \emptyset$, or if $\flrmSetA \allowbreak \cap \allowbreak \flrmSetC \neq \emptyset$, or if $\flrmSetA \allowbreak \cap \allowbreak \flrmSetW \neq \emptyset$, or if $\flrmSetB \allowbreak \cap \allowbreak \flrmSetC \neq \emptyset$, or if $\flrmSetB \allowbreak \cap \allowbreak \flrmSetW \neq \emptyset$, or if $\flrmSetC \allowbreak \cap \allowbreak \flrmSetW \neq \emptyset$, or if $(\sid, \allowbreak \KEMsecretkey, \allowbreak \KEMpublickey)$ is already stored.

        \item $\flrmParty$ runs $(\KEMsecretkey, \allowbreak \KEMpublickey) \allowbreak \gets \allowbreak \KEMKeyGen(1^\securityparameter)$.




        \item $\flrmParty$ sets $\sid_{\Freg} \allowbreak \gets \allowbreak (\flrmSetA \cup \flrmSetB \cup \flrmSetC \cup \flrmSetW, \allowbreak \sid)$, sends the message $(\fregregisterini, \allowbreak \sid_{\Freg}, \allowbreak  \KEMpublickey)$ to $\Functionality_{\Freg}$, and receives the message $(\fregregisterend, \allowbreak \sid_{\Freg})$ from $\Functionality_{\Freg}$.

        \item $\flrmParty$ stores $(\sid, \allowbreak \KEMsecretkey, \allowbreak \KEMpublickey)$.

        \item $\flrmParty$ outputs $(\flrmsetupend, \allowbreak \sid)$.

    \end{itemize}
\end{enumerate}

\paragraph{Discussion of the construction of the setup interface.} After receiving the $\flrmsetupini$ message, a node or a gateway $\flrmParty$ outputs an abortion message if the session identifier $\sid$ is malformed or if $\flrmParty$ already run the setup interface. If $\flrmParty$ does not abort, $\flrmParty$ computes a key pair $(\KEMsecretkey, \allowbreak \KEMpublickey)$ by using the key generation algorithm $\KEMKeyGen$ of the key encapsulation mechanism. 
    After that, $\flrmParty$ registers the public key $\KEMpublickey$ by using the interface $\fregregister$ of $\Functionality_{\Freg}$. The use of $\Functionality_{\Freg}$ guarantees that all honest users retrieve the same public keys for each of the nodes and gateways.


%% file: 6OutfoxPacketFormat2Registration.tex
\subsubsection{Registration Interface of Our Construction}
\label{sec:construction2registration}

\paragraph{Construction of the registration interface.} $\mathrm{\Pi}_{\LRM}$ is parameterized by the parameters defined in~\S\ref{sec:functionalityParameters}, and by a security parameter $1^\securityparameter$.
\begin{enumerate}

\item[2.] On input $(\flrmregisterini, \allowbreak \sid)$, a user $\flrmUser$ does the following:
        
    \begin{itemize}
        
        \item $\flrmUser$ aborts if $\flrmUser \allowbreak \notin \allowbreak \flrmSetU$, or if  a tuple $(\sid, \allowbreak \KEMsecretkey, \allowbreak \KEMpublickey)$ is stored.

        \item $\flrmUser$ parses $\sid$ as $(\flrmSetA, \allowbreak \flrmSetB, \allowbreak \flrmSetC, \allowbreak \flrmSetW, \allowbreak \sid')$ and sets $\sid_{\Freg} \allowbreak \gets \allowbreak (\flrmSetA \cup \flrmSetB \cup \flrmSetC \cup \flrmSetW, \allowbreak \sid)$. For every party $\flrmParty \allowbreak \in \allowbreak (\flrmSetA \cup \flrmSetB \cup \flrmSetC \cup \flrmSetW)$,  $\flrmUser$ does the following:
        \begin{itemize}

            \item $\flrmUser$ sends $(\fregretrieveini, \allowbreak \sid_{\Freg}, \allowbreak \flrmParty)$ to $\Functionality_{\Freg}$ and receives the message  $(\fregretrieveend, \allowbreak \sid_{\Freg}, \allowbreak \KEMpublickey)$ from $\Functionality_{\Freg}$.



            \item If $\KEMpublickey \allowbreak = \allowbreak \bot$, $\flrmUser$ outputs an abortion message, else $\flrmUser$ stores $(\sid, \allowbreak \flrmParty, \allowbreak \KEMpublickey)$.
        
        \end{itemize}

        \item $\flrmUser$ runs $(\KEMsecretkey, \allowbreak \KEMpublickey) \allowbreak \gets \allowbreak \KEMKeyGen(1^\securityparameter)$.




        \item $\flrmUser$ sends the message $(\fpregregisterini, \allowbreak \sid, \allowbreak \KEMpublickey)$ to $\Functionality_{\Fpreg}$, and receives $(\fpregregisterend, \allowbreak \sid)$ from $\Functionality_{\Fpreg}$.

        \item $\flrmUser$ stores $(\sid,  \allowbreak \KEMsecretkey, \allowbreak \KEMpublickey)$.

        \item $\flrmUser$ outputs $(\flrmregisterend, \allowbreak \sid)$.

    \end{itemize}

\end{enumerate}

\paragraph{Discussion of the construction of the registration interface.} After receiving the $\flrmregisterini$ message, a user $\flrmUser$  outputs an abortion message if $\flrmUser$ does not belong to the set of users, or if $\flrmUser$ already run the registration interface. If $\flrmUser$ does not abort, $\flrmUser$ retrieves the public keys of all the gateways and nodes by using the $\fregretrieve$ interface of $\Functionality_{\Freg}$. If the public key of any node or gateway is not available yet, $\flrmUser$ outputs an abortion message. Othewise $\flrmUser$ stores all the retrieved public keys.

    $\flrmUser$ computes a key pair $(\KEMsecretkey, \allowbreak \KEMpublickey)$ by running  $\KEMKeyGen$. After that, $\flrmUser$ registers $\KEMpublickey$ by using the interface $\fpregregister$ of $\Functionality_{\Fpreg}$. $\Functionality_{\Fpreg}$, like $\Functionality_{\Freg}$, guarantees that all honest users retrieve the same public key $\KEMpublickey$ for $\flrmUser$. However, $\Functionality_{\Fpreg}$, in contrast to $\Functionality_{\Freg}$, allows users to retrieve $\KEMpublickey$ without leaking $\KEMpublickey$ or $\flrmUser$ to the adversary, which is needed to avoid the leakage of the intended receiver of a packet in the request interface. Finally, $\flrmUser$ stores her key pair $(\KEMsecretkey, \allowbreak \KEMpublickey)$.

    It could be possible to use $\Functionality_{\Fpreg}$ to also register the public keys of nodes and gateways, so that users could retrieve those public keys in the request interface when they need them to compute a packet, instead of retrieving all of them in the register interface. We chose to use a different functionality $\Functionality_{\Freg}$ because it can be realized by more efficient protocols than $\Functionality_{\Fpreg}$. In a typical mixnet, there are less than 1000 nodes and gateways, so retrieving all their public keys is practical. In contrast, there are potentially millions of users, so using $\Functionality_{\Freg}$ to register user public keys is not practical because a user would need to retrieve all their keys in the registration phase.

%% file: 6OutfoxPacketFormat3Request.tex
\subsubsection{Request Interface of Our Construction}
\label{sec:construction3request}

\paragraph{Construction of the request interface.} $\mathrm{\Pi}_{\LRM}$ is parameterized by the parameters defined in~\S\ref{sec:functionalityParameters}, and by a security parameter $1^\securityparameter$.
\begin{enumerate}

\item[3.] On input the  message $(\flrmsendini, \allowbreak \sid, \allowbreak \ppid, \allowbreak \flrmGatewayentry, \allowbreak \flrmGatewayexit, \allowbreak \flrmReceiver, \allowbreak \flrmmessagerequest, \allowbreak \flrmNodeArequest, \allowbreak \flrmNodeBrequest, \allowbreak \flrmNodeCrequest, \allowbreak \langle    \flrmGatewayentryreply, \allowbreak \flrmGatewayexitreply, \allowbreak \flrmNodeAresponse, \allowbreak \flrmNodeBresponse, \allowbreak \flrmNodeCresponse, \allowbreak \flrmSenderreply \rangle)$, a sender $\flrmSender$ and a gateway $\flrmGatewayentry$ do the following:

    \begin{itemize}

        \item $\flrmSender$ parses $\sid$ as $(\flrmSetA, \allowbreak \flrmSetB, \allowbreak \flrmSetC, \allowbreak \flrmSetW, \allowbreak \sid')$ and aborts if $\flrmGatewayentry \notin \flrmSetW$, or if $\flrmGatewayexit \notin \flrmSetW$, or if $\flrmmessagerequest \notin \flrmMessageSpace$, or if $\flrmNodeArequest \notin \flrmSetA$, or if $\flrmNodeBrequest \notin \flrmSetB$, or if $\flrmNodeCrequest \notin \flrmSetC$.

        \item If $\langle  \allowbreak \flrmGatewayentryreply, \allowbreak \flrmGatewayexitreply, \flrmNodeAresponse, \allowbreak \flrmNodeBresponse, \allowbreak \flrmNodeCresponse, \allowbreak \flrmSenderreply \rangle \neq \bot$, $\flrmSender$ aborts if $\flrmGatewayentryreply \notin \flrmSetW$, or if $\flrmGatewayexitreply \notin \flrmSetW$, or if $\flrmNodeAresponse \notin \flrmSetA$, or if $\flrmNodeBresponse \notin \flrmSetB$, or if $\flrmNodeCresponse \notin \flrmSetC$, or if $\flrmSenderreply \neq \flrmSender$.

        \item $\flrmSender$ aborts if $(\sid,  \allowbreak \KEMsecretkey, \allowbreak \KEMpublickey)$ is not stored.

        \item $\flrmSender$ aborts if a tuple $(\sid, \allowbreak \ppid', \allowbreak \surbidentifier, \allowbreak \surbsecrets, \allowbreak \flrmGatewayentryreply)$ such that $\ppid' \allowbreak = \allowbreak \ppid$ is stored.

        \item $\flrmSender$ finds the  tuples $(\sid, \allowbreak \flrmGatewayentry, \allowbreak \KEMpublickey_{\flrmGatewayentry})$, $(\sid, \allowbreak \flrmGatewayexit, \allowbreak \KEMpublickey_{\flrmGatewayexit})$,  $(\sid, \allowbreak \flrmNodeArequest, \allowbreak \KEMpublickey_{\flrmNodeArequest})$, $(\sid, \allowbreak \flrmNodeBrequest, \allowbreak \KEMpublickey_{\flrmNodeBrequest})$ and $(\sid, \allowbreak \flrmNodeCrequest, \allowbreak \KEMpublickey_{\flrmNodeCrequest})$.

        \item If $\langle  \allowbreak \flrmGatewayentryreply, \allowbreak \flrmGatewayexitreply, \allowbreak \flrmNodeAresponse, \allowbreak \flrmNodeBresponse, \allowbreak \flrmNodeCresponse, \allowbreak \flrmSenderreply \rangle \neq \bot$, $\flrmSender$ retrieves the stored tuples $(\sid, \allowbreak \flrmGatewayentryreply, \allowbreak \KEMpublickey_{\flrmGatewayentryreply})$, $(\sid, \allowbreak \flrmGatewayexitreply, \allowbreak \KEMpublickey_{\flrmGatewayexitreply})$,  $(\sid, \allowbreak \flrmNodeAresponse, \allowbreak \KEMpublickey_{\flrmNodeAresponse})$, $(\sid, \allowbreak \flrmNodeBresponse, \allowbreak \KEMpublickey_{\flrmNodeBresponse})$ and finally $(\sid, \allowbreak \flrmNodeCresponse, \allowbreak \KEMpublickey_{\flrmNodeCresponse})$.

        \item If $(\sid, \allowbreak \flrmReceiver', \allowbreak \KEMpublickey)$ such that $\flrmReceiver' = \flrmReceiver$ is not stored, $\flrmSender$ does the following: 
        \begin{itemize}
        
             \item $\flrmSender$ sends the message $(\fpregretrieveini, \allowbreak \sid, \allowbreak \flrmReceiver)$ to $\Functionality_{\Fpreg}$ and receives $(\fpregretrieveend, \allowbreak \sid, \allowbreak \KEMpublickey)$ from $\Functionality_{\Fpreg}$.

            \item If $\KEMpublickey \allowbreak = \allowbreak \bot$, $\flrmSender$ outputs an abortion message, else $\flrmSender$ stores $(\sid, \allowbreak \flrmReceiver, \allowbreak \KEMpublickey)$.

        \end{itemize}
        
        \item If $\langle  \allowbreak \flrmGatewayentryreply, \allowbreak \flrmGatewayexitreply, \allowbreak \flrmNodeAresponse, \allowbreak \flrmNodeBresponse, \allowbreak \flrmNodeCresponse, \allowbreak \flrmSenderreply \rangle \neq \bot$, $\flrmSender$ does the following:

        \begin{itemize}


            \item $\flrmSender$ sets the routing information as follows:
            \begin{itemize}

            
                \item Set $\routinginformation_{\flrmNodeAresponse} \allowbreak \gets \allowbreak \flrmNodeBresponse$.

                \item Set $\routinginformation_{\flrmNodeBresponse} \allowbreak \gets \allowbreak \flrmNodeCresponse$.

                \item Set $\routinginformation_{\flrmNodeCresponse} \allowbreak \gets \allowbreak \flrmGatewayentryreply$.

                \item Set $\routinginformation_{\flrmGatewayentryreply} \allowbreak \gets \allowbreak \flrmSender$. (We recall that $\flrmSender \allowbreak = \allowbreak \flrmSenderreply$.)
                
            \end{itemize}

            \item Set the route $\routeresponse \allowbreak \gets \allowbreak (\langle \flrmNodeAresponse, \allowbreak \KEMpublickey_{\flrmNodeAresponse}, \allowbreak \routinginformation_{\flrmNodeAresponse} \rangle, \allowbreak \langle \flrmNodeBresponse, \KEMpublickey_{\flrmNodeBresponse}, \allowbreak \routinginformation_{\flrmNodeBresponse} \rangle, \allowbreak \langle \flrmNodeCresponse, \allowbreak \KEMpublickey_{\flrmNodeCresponse}, \allowbreak \routinginformation_{\flrmNodeCresponse} \rangle, \allowbreak \langle \flrmGatewayentryreply, \allowbreak \KEMpublickey_{\flrmGatewayentryreply}, \allowbreak \routinginformation_{\flrmGatewayentryreply} \rangle)$.

            \item Set the sender information $\senderinfo \allowbreak \gets \allowbreak \langle \flrmSender, \allowbreak \KEMpublickey, \allowbreak \bot \rangle$, where $\KEMpublickey$ is stored in the tuple $(\sid,  \allowbreak \KEMsecretkey, \allowbreak \KEMpublickey)$.

            \item Execute  $(\surb, \allowbreak \surbidentifier, \allowbreak \surbsecrets) \allowbreak \gets \allowbreak \lrmSurbCreate(\routeresponse, \allowbreak \senderinfo)$.

            \item Store $(\sid, \allowbreak \ppid, \allowbreak \surbidentifier, \allowbreak \surbsecrets, \allowbreak \flrmGatewayentryreply)$.


        \end{itemize}

        \item $\flrmSender$ sets the following routing information:

        \begin{itemize}

            \item Set $\routinginformation_{\flrmNodeArequest} \allowbreak \gets \allowbreak \flrmNodeBrequest$.

            \item Set $\routinginformation_{\flrmNodeBrequest} \allowbreak \gets \allowbreak \flrmNodeCrequest$.

            \item Set $\routinginformation_{\flrmNodeCrequest} \allowbreak \gets \allowbreak \flrmGatewayexit$.

            \item Set $\routinginformation_{\flrmGatewayexit} \allowbreak \gets \allowbreak \flrmReceiver$.

        \end{itemize}

        \item $\flrmSender$ sets the route $\route \allowbreak \gets \allowbreak (\langle \flrmNodeArequest, \allowbreak \KEMpublickey_{\flrmNodeArequest}, \allowbreak \routinginformation_{\flrmNodeArequest} \rangle, \allowbreak \langle \flrmNodeBrequest, \KEMpublickey_{\flrmNodeBrequest}, \allowbreak \routinginformation_{\flrmNodeBrequest} \rangle, \allowbreak \langle \flrmNodeCrequest, \allowbreak \KEMpublickey_{\flrmNodeCrequest}, \allowbreak \routinginformation_{\flrmNodeCrequest} \rangle, \allowbreak \langle \flrmGatewayexit, \allowbreak \KEMpublickey_{\flrmGatewayexit}, \allowbreak \routinginformation_{\flrmGatewayexit} \rangle)$.

        \item If $\langle  \allowbreak \flrmGatewayentryreply, \allowbreak \flrmGatewayexitreply, \allowbreak \flrmNodeAresponse, \allowbreak \flrmNodeBresponse, \allowbreak \flrmNodeCresponse, \allowbreak \flrmSenderreply \rangle \neq \bot$, $\flrmSender$ sets the receiver information $\receiverinfo \allowbreak \gets \allowbreak \langle \flrmReceiver, \allowbreak \KEMpublickey, \allowbreak (\flrmGatewayexitreply, \allowbreak  \flrmNodeAresponse) \rangle$, else $\flrmSender$ sets  $\receiverinfo \allowbreak \gets \allowbreak \langle \flrmReceiver, \allowbreak \KEMpublickey, \allowbreak \bot \rangle$ and $\surb \allowbreak \gets \allowbreak \bot$, where $\KEMpublickey$ is in the tuple  $(\sid, \allowbreak \flrmReceiver, \allowbreak \KEMpublickey)$.

        \item $\flrmSender$ runs $\packet \allowbreak \gets \allowbreak \lrmPacketCreate(\route, \flrmmessagerequest, \receiverinfo, \surb)$.

        \item $\flrmSender$ sets $\sid_{\SMT} \allowbreak \gets \allowbreak (\flrmSender, \allowbreak \flrmGatewayentry, \allowbreak \sid)$ and sends the message $(\fsmtsendini, \allowbreak \sid_{\SMT}, \allowbreak \langle \packet, \allowbreak \flrmNodeArequest \rangle)$ to  $\Functionality_{\SMT}$.

        \item $\flrmGatewayentry$ receives the message $(\fsmtsendend, \allowbreak \sid_{\SMT}, \allowbreak \langle \packet, \allowbreak \flrmNodeArequest \rangle)$ from the functionality $\Functionality_{\SMT}$.

        \item $\flrmGatewayentry$ parses $\sid_{\SMT}$ as $(\flrmSender, \allowbreak \flrmGatewayentry, \allowbreak \sid)$.

        \item $\flrmGatewayentry$ aborts if $(\sid',  \allowbreak \KEMsecretkey, \allowbreak \KEMpublickey)$ such that $\sid' \allowbreak = \allowbreak \sid$ is not stored, or if $\flrmSender \allowbreak \notin \allowbreak \flrmSetU$, or if $\flrmGatewayentry \allowbreak \notin \allowbreak \flrmSetW$.

        \item $\flrmGatewayentry$ creates a fresh random packet identifier $\tid$. If $\flrmNodeArequest \allowbreak \in \allowbreak \flrmSetA$,  $\flrmGatewayentry$ stores $(\sid, \allowbreak \tid, \allowbreak \packet, \allowbreak \flrmNodeArequest)$, else $\flrmGatewayentry$ stores $(\sid, \allowbreak \tid, \allowbreak \packet, \allowbreak \bot)$.

        \item $\flrmGatewayentry$ outputs $(\flrmsendend, \allowbreak \sid, \allowbreak \tid, \allowbreak \bot, \allowbreak \bot, \allowbreak \flrmSender, \allowbreak \flrmNodeArequest, \bot)$.

    \end{itemize}

\end{enumerate}

\paragraph{Discussion of the construction of the request interface.} After receiving the $\flrmsendini$ message, a sender $\flrmSender$ outputs an abortion message if any of the nodes and gateways received as input is not in the correct set, or if the message does not belong to the right space. $\flrmSender$ also outputs an abortion message if the input provides routing information to enable a reply but any of the nodes and gateways in that information is not in the correct set, or if the party $\flrmSenderreply$ that should receive the reply is different from $\flrmSender$. Additionally, $\flrmSender$ aborts if it has not run the registration interface. We recall that this guarantees that $\flrmSender$ already stores the public keys of all the nodes and gateways. Furthermore, $\flrmSender$ aborts if the global packet identifier $\ppid$ has already been used to identify another packet that enables replies.

    If $\flrmSender$ does not store the public key of the receiver $\flrmReceiver$, $\flrmSender$ uses the $\fpregretrieve$ interface of $\Functionality_{\Fpreg}$ to retrieve the registered public key. If the public key of $\flrmReceiver$ is not registered, $\flrmSender$ outputs an abortion message.

    If there is routing information to enable replies, $\flrmSender$ computes the single-use reply block $\surb$ for the reply. To this end,  $\flrmSender$ sets the route $\routeresponse$ to include, for each of the nodes and gateways in the routing information, its identifier, its public key, and the identifier of the party to whom they should forward the packet. We recall that, in a reply, the exit gateway $\flrmGatewayexitreply$ does not process the packet, so it is not included in $\routeresponse$. $\flrmSender$ also sets the sender information $\senderinfo$, which includes his identifier and public key. Then $\flrmSender$ runs the algorithm $\lrmSurbCreate$ on input $\routeresponse$ and $\senderinfo$ to compute the single-use reply block $\surb$, the header identifier $\surbidentifier$ and the secret information $\surbsecrets$ that will enable $\flrmSender$ to retrieve the reply message once the reply is received. Finally, $\flrmSender$ stores $\surbidentifier$ and $\surbsecrets$ along with the global packet identifier $\ppid$ and the entry gateway $\flrmGatewayentryreply$.

    To compute a packet, $\flrmSender$ sets the route $\route$ in a way similar to how $\routeresponse$ was set to compute $\surb$. We recall that, in a request, the entry gateway $\flrmGatewayentry$ does not process the packet, so it is not included in $\route$. To set the receiver information $\receiverinfo$, $\flrmSender$ includes the identifiers of the exit gateway $\flrmGatewayexitreply$ and of the first-layer node $\flrmNodeAresponse$ if a reply is enabled, whereas no party identifiers are included if a reply is not enabled. ($\flrmNodeAresponse$ is included because exit gateways do not process reply packets, so the receiver needs to tell the exit gateway the identifier of the first-layer node to which the reply packet should be forwarded.)  Then $\flrmSender$ runs the algorithm $\lrmPacketCreate$ on input $\route$, $\receiverinfo$, the request message $\flrmmessagerequest$ and $\surb$ to obtain a packet $\packet$. (If replies are not enabled, $\surb \allowbreak = \allowbreak \bot$.) After that, $\flrmSender$ sends $\packet$ and the first-layer node identifier $\flrmNodeArequest$ to $\flrmGatewayentry$ by using the $\fsmtsend$ interface of $\Functionality_{\SMT}$. $\flrmNodeArequest$ is included so that $\flrmGatewayentry$ knows to whom the packet should be forwarded.  $\Functionality_{\SMT}$ guarantees both confidentiality and integrity. Therefore, the adversary only learns that $\flrmSender$ sends a message to $\flrmGatewayentry$, but learns neither $\packet$ nor $\flrmNodeArequest$ and cannot modify them.

    When $\flrmGatewayentry$ receives $\packet$ and $\flrmNodeArequest$ from $\Functionality_{\SMT}$, $\flrmGatewayentry$ outputs an abortion message if $\flrmGatewayentry$ did not run the setup interface for the protocol identified by the session identifier $\sid$, or if $\flrmSender$ or $\flrmGatewayentry$ do not belong to the correct sets. Otherwise $\flrmGatewayentry$ creates a random fresh packet identifier $\tid$ and stores it along with $\packet$ and either $\flrmNodeArequest$, if $\flrmNodeArequest$ belongs to the correct set of nodes, or $\bot$, if $\flrmNodeArequest$ does not belong to the correct set of nodes. The latter indicates that the packet should not be forwarded. Finally,  $\flrmGatewayentry$ outputs $\tid$, $\flrmSender$ and $\flrmNodeArequest$. The global packet identifier    $\ppid'$, the message $\flrmmessagerequest'$, and the leakage list $\flrmlistleakage$ are set to $\bot$ in the message output by $\flrmGatewayentry$ because $\Functionality_{\LRM}$ sets them to $\bot$ when $\flrmGatewayentry$ is honest.

%% file: 6OutfoxPacketFormat4Reply.tex
\subsubsection{Reply Interface of Our Construction}
\label{sec:construction4reply}

\paragraph{Construction of the reply interface.} $\mathrm{\Pi}_{\LRM}$ is parameterized by the parameters defined in~\S\ref{sec:functionalityParameters}, and by a security parameter $1^\securityparameter$.
\begin{enumerate}

\item[4.] On input the message $(\flrmreplyini, \allowbreak \sid, \allowbreak \tid, \allowbreak \flrmmessageresponse, \allowbreak \{\rpid, \allowbreak \flrmGatewayexitreply', \allowbreak \flrmNodeAresponse'\})$, a receiver $\flrmReceiver$ and a gateway $\flrmGatewayexitreply$ do the following:

    \begin{itemize}

        \item $\flrmReceiver$ aborts if $(\sid, \allowbreak \tid', \allowbreak \surb, \allowbreak \flrmGatewayexitreply, \allowbreak \flrmNodeAresponse)$ such that $\tid' = \tid$ is not stored, or if $\flrmmessageresponse \allowbreak \notin \allowbreak \flrmMessageSpace$.

        \item $\flrmReceiver$ parses $\sid$ as $(\flrmSetA, \allowbreak \flrmSetB, \allowbreak \flrmSetC, \allowbreak \flrmSetW, \allowbreak \sid')$, and aborts if $\flrmGatewayexitreply \allowbreak \notin \allowbreak \flrmSetW$ or if $\flrmNodeAresponse \allowbreak \notin \allowbreak \flrmSetA$.

        \item $\flrmReceiver$ runs $\packet \allowbreak \gets \allowbreak \lrmSurbUse(\surb, \allowbreak \flrmmessageresponse)$.

        
        \item $\flrmReceiver$ sets $\sid_{\SMT} \allowbreak \gets \allowbreak (\flrmReceiver, \allowbreak \flrmGatewayexitreply, \allowbreak \sid)$ and sends the message $(\fsmtsendini, \allowbreak \sid_{\SMT}, \allowbreak \langle \packet, \allowbreak \flrmNodeAresponse \rangle)$ to $\Functionality_{\SMT}$.

        \item $\flrmGatewayexitreply$ receives the message $(\fsmtsendend, \allowbreak \sid_{\SMT}, \allowbreak  \langle \packet, \allowbreak \flrmNodeAresponse \rangle)$ from $\Functionality_{\SMT}$.

        \item $\flrmGatewayexitreply$ parses $\sid_{\SMT}$ as $(\flrmReceiver, \allowbreak \flrmGatewayexitreply, \allowbreak \sid)$.

        \item $\flrmGatewayexitreply$ aborts if $(\sid', \allowbreak \KEMsecretkey, \allowbreak \KEMpublickey)$ such that $\sid' \allowbreak = \allowbreak \sid$ is not stored, or if $\flrmReceiver \allowbreak \notin \allowbreak \flrmSetU$, or if $\flrmGatewayexitreply \allowbreak \notin \allowbreak \flrmSetW$.



        \item $\flrmGatewayexitreply$ creates a fresh random packet identifier $\tid$. If $\flrmNodeAresponse \allowbreak \in \allowbreak \flrmSetA$, $\flrmGatewayexitreply$ stores $(\sid, \allowbreak \tid, \allowbreak \packet, \allowbreak \flrmNodeAresponse)$, else $\flrmGatewayexitreply$ stores $(\sid, \allowbreak \tid, \allowbreak \packet, \allowbreak \bot)$.


        \item $\flrmGatewayexitreply$ outputs $(\flrmreplyend, \allowbreak \sid, \allowbreak \tid, \allowbreak \bot, \allowbreak \bot, \allowbreak \flrmReceiver, \allowbreak \flrmNodeAresponse, \allowbreak \bot)$.

    \end{itemize}

\end{enumerate}

\paragraph{Discussion of the construction of the reply interface.} A receiver $\flrmReceiver$ receives the $\flrmsendini$ message along with a packet identifier $\tid$, a response message $\flrmmessageresponse$, and the tuple $\{\rpid, \allowbreak \flrmGatewayexitreply', \allowbreak \flrmNodeAresponse'\}$. That tuple is not used by $\mathrm{\Pi}_{\LRM}$. It is only used in the security analysis by the simulator $\Simulator$ when a corrupt receiver sends a reply packet that is forwarded to a corrupt first-layer node, when the corrupt receiver uses an exit gateway different from the one chosen by the party that computed the packet, or when it instructs the gateway to forward the packet to a first-layer node different from the one chosen by the party that computed the packet.
    
    The receiver $\flrmReceiver$ outputs an abortion message if the message $\flrmmessageresponse$ does not belong to the correct space, or if $\flrmReceiver$ does not store any single-use reply block $\surb$ associated with the local packet identifier $\tid$. (We remark that the latter implies that $\flrmReceiver$ has run the registration interface.) $\flrmReceiver$ also aborts if the party identifiers $\flrmGatewayexitreply$ and $\flrmNodeAresponse$ associated with $\tid$ do not belong to the correct sets.
    
    If $\flrmReceiver$ does not abort, $\flrmReceiver$ runs the algorithm $\lrmSurbUse$ on input the single-use reply block $\surb$ and the message $\flrmmessageresponse$ to create a reply packet $\packet$. Then $\flrmReceiver$ uses the $\fsmtsend$ interface of the ideal functionality $\Functionality_{\SMT}$ to send $\packet$ and  $\flrmNodeAresponse$ to the exit gateway $\flrmGatewayexitreply$. We recall that, thanks to the use of $\Functionality_{\SMT}$, an adversary that controls the communication channel cannot learn or modify $\packet$ or $\flrmNodeAresponse$.

    When $\flrmGatewayexitreply$ receives $\packet$ and $\flrmNodeAresponse$ from $\Functionality_{\SMT}$, $\flrmGatewayexitreply$ outputs an abortion message if $\flrmGatewayexitreply$ did not run the setup interface for the protocol identified by the session identifier $\sid$, or if $\flrmReceiver$ or $\flrmGatewayexitreply$ do not belong to the correct sets. Otherwise $\flrmGatewayexitreply$ creates a random fresh packet identifier $\tid$ and stores it along with $\packet$ and either $\flrmNodeAresponse$, if $\flrmNodeAresponse$ belongs to the correct set of nodes, or $\bot$, if $\flrmNodeAresponse$ does not belong to the correct set of nodes. The latter indicates that the packet should not be forwarded. Finally, $\flrmGatewayexitreply$ outputs $\tid$, $\flrmReceiver$ and $\flrmNodeAresponse$. The global packet identifier    $\ppid'$, the message $\flrmmessageresponse'$, and the leakage list $\flrmlistleakage$ are set to $\bot$ in the message output by $\flrmGatewayexitreply$ because $\Functionality_{\LRM}$ sets them to $\bot$ when $\flrmGatewayexitreply$ is honest.

%% file: 6OutfoxPacketFormat5Forward.tex
\subsubsection{Forward Interface of Our Construction}
\label{sec:construction5forward}

\paragraph{Construction of the forward interface.} $\mathrm{\Pi}_{\LRM}$ is parameterized by the parameters defined in~\S\ref{sec:functionalityParameters}, and by a security parameter $1^\securityparameter$.
\begin{enumerate}

 \item[5.] On input  $(\flrmforwardini, \allowbreak \sid, \allowbreak \tid, \allowbreak \flrmdestroymessage, \{\ppid, \flrmGatewayexit, \allowbreak \flrmReceiver, \allowbreak \flrmmessagerequest, \allowbreak \flrmNodeArequest, \allowbreak \flrmNodeBrequest, \allowbreak \flrmNodeCrequest, \allowbreak \langle   \flrmGatewayentryreply, \allowbreak \flrmGatewayexitreply, \allowbreak \flrmNodeAresponse, \allowbreak \flrmNodeBresponse, \allowbreak \flrmNodeCresponse, \allowbreak \flrmSender \rangle\})$, a gateway or a node $\flrmParty$ and a gateway, node or user $\flrmParty'$ do the following:

    \begin{itemize}

        \item $\flrmParty$ aborts if a tuple $(\sid, \allowbreak \tid', \allowbreak \packet, \allowbreak \flrmParty')$ such that $\tid' \allowbreak = \allowbreak \tid$ is not stored. 

        \item $\flrmParty$ aborts if $\flrmParty' \allowbreak = \allowbreak \bot$ in the tuple $(\sid, \allowbreak \tid', \allowbreak \packet, \allowbreak \flrmParty')$ such that $\tid' \allowbreak = \allowbreak \tid$.


        \item $\flrmParty$ sets $\sid_{\SMT} \allowbreak \gets \allowbreak (\flrmParty, \allowbreak \flrmParty', \allowbreak \sid)$ and sends $(\fsmtsendini, \allowbreak \sid_{\SMT}, \allowbreak \packet)$ to $\Functionality_{\SMT}$.

        \item $\flrmParty'$ receives $(\fsmtsendend, \allowbreak \sid_{\SMT}, \allowbreak \packet)$ from $\Functionality_{\SMT}$.

        \item $\flrmParty'$ parses $\sid_{\SMT}$ as $(\flrmParty, \allowbreak \flrmParty', \allowbreak \sid)$.

        \item $\flrmParty'$ aborts if $(\sid', \allowbreak \KEMsecretkey, \allowbreak \KEMpublickey)$ such that $\sid' \allowbreak = \allowbreak \sid$ is not stored.


        \item If $\flrmParty' \allowbreak \in \allowbreak \flrmSetU$, $\flrmParty'$ does the following:

        \begin{itemize}

            \item Abort if $\flrmParty \allowbreak \notin \allowbreak \flrmSetW$.

            \item Set $b \gets 0$.

            \item For all the stored tuples  $(\sid', \allowbreak \ppid, \allowbreak \surbidentifier, \allowbreak \surbsecrets, \allowbreak \flrmGatewayentryreply)$ such that $\sid' \allowbreak = \allowbreak \sid$, do the following:

            \begin{itemize}

                \item Run $b \allowbreak \gets \allowbreak \lrmSurbCheck(\packet, \allowbreak \surbidentifier)$.

                \item If $b = 1$, do the following:

                \begin{itemize}

                    \item Abort if $\flrmParty \allowbreak \neq \allowbreak \flrmGatewayentryreply$.

                    \item Run $\flrmmessageresponse \allowbreak \gets \allowbreak \lrmSurbRecover(\packet, \surbsecrets)$.


                    \item Output $(\flrmforwardend, \allowbreak \sid, \allowbreak \flrmParty, \allowbreak \flrmmessageresponse, \allowbreak \ppid)$.

                \end{itemize}

            \end{itemize}

            \item If $b = 0$, do the following:

            \begin{itemize} 
            
                \item Run $(\routing{k}, \surb, \flrmmessagerequest)  \allowbreak \gets \allowbreak \lrmPacketProcess(\KEMsecretkey, \allowbreak \packet, \allowbreak 1)$.

                \item If $(\routing{k}, \surb, \flrmmessagerequest) \allowbreak = \allowbreak \top$, abort.

                \item Else, if $(\routing{k}, \surb, \flrmmessagerequest) \allowbreak = \allowbreak \bot$, set  $\tid \allowbreak \gets \allowbreak \bot$, $\flrmGatewayexitreply' \allowbreak \gets \allowbreak \bot$, $\flrmNodeAresponse' \allowbreak \gets \allowbreak \bot$, and $\flrmmessagerequest \allowbreak \gets \allowbreak \bot$.

                \item Else, if $\routing{k} = \surb = \bot$, set $\tid \allowbreak \gets \allowbreak \bot$, $\flrmGatewayexitreply' \allowbreak \gets \allowbreak \bot$ and $\flrmNodeAresponse' \allowbreak \gets \allowbreak \bot$.
                
                \item Else, do the following:
                \begin{itemize}
                    
                    \item Create a fresh random packet identifier $\tid$.

                    \item Parse $\routing{k}$ as $(\flrmGatewayexitreply', \allowbreak \flrmNodeAresponse')$. If $\flrmGatewayexitreply' \allowbreak \notin \allowbreak \flrmSetW$ or $\flrmNodeAresponse' \allowbreak \notin \allowbreak \flrmSetA$, set $\flrmGatewayexitreply \allowbreak \gets \allowbreak \bot$ and $\flrmNodeAresponse \allowbreak \gets \allowbreak \bot$, else set $\flrmGatewayexitreply \allowbreak \gets \allowbreak \flrmGatewayexitreply'$ and $\flrmNodeAresponse \allowbreak \gets \allowbreak \flrmNodeAresponse'$.

                    \item Store $(\sid, \allowbreak \tid, \allowbreak \surb, \allowbreak \flrmGatewayexitreply, \allowbreak \flrmNodeAresponse)$.
                    
                \end{itemize}

                \item Output $(\flrmforwardend, \allowbreak \sid, \allowbreak \flrmParty, \allowbreak \flrmmessagerequest, \allowbreak \tid, \allowbreak \flrmGatewayexitreply', \allowbreak \flrmNodeAresponse', \allowbreak \bot, \allowbreak \bot)$.
                
            \end{itemize}
            
        \end{itemize}

        \item If $\flrmParty' \allowbreak \in \allowbreak \{\flrmSetW, \flrmSetA, \flrmSetB, \flrmSetC\}$, $\flrmParty'$ does the following:
        
        \begin{itemize}

            \item If $\flrmParty' \in \flrmSetA$, abort if $\flrmParty \notin \flrmSetW$.

            \item If $\flrmParty' \in \flrmSetB$, abort if $\flrmParty \notin \flrmSetA$.

            \item If $\flrmParty' \in \flrmSetC$, abort if $\flrmParty \notin \flrmSetB$.

            \item If $\flrmParty' \in \flrmSetW$, abort if $\flrmParty \notin \flrmSetC$.

            \item Run $(\packet', \allowbreak \routinginformation) \allowbreak \gets \allowbreak \lrmPacketProcess(\KEMsecretkey, \allowbreak \packet, \allowbreak 0)$.

            \item If $(\packet', \allowbreak \routinginformation) \allowbreak = \allowbreak \top$, set $\packet' \allowbreak \gets \allowbreak \bot$ and $\routinginformation \allowbreak \gets \allowbreak \bot$.

            \item Else, do the following:
            \begin{itemize}
                
                \item If $\flrmParty' \in \flrmSetA$, set $\routinginformation \gets \bot$ if $\routinginformation \notin \flrmSetB$.
 
                \item If $\flrmParty' \in \flrmSetB$, set $\routinginformation \gets \bot$ if $\routinginformation \notin \flrmSetC$.

                \item If $\flrmParty' \in \flrmSetC$, set $\routinginformation \gets \bot$ if $\routinginformation \notin \flrmSetW$.

                \item If $\flrmParty' \in \flrmSetW$, set $\routinginformation \gets \bot$ if $\routinginformation  \notin \flrmSetU$.

                \item Create a fresh local packet identifier $\tid$ and store $(\sid, \allowbreak \tid, \allowbreak \packet', \allowbreak \routinginformation)$.

                \item Output  $(\flrmforwardend, \allowbreak \sid, \allowbreak \tid, \allowbreak \bot, \allowbreak \bot, \allowbreak \allowbreak \flrmParty, \allowbreak \routinginformation, \bot, \allowbreak \bot)$.

            \end{itemize}

        \end{itemize}
               
    \end{itemize}

\end{enumerate}

\paragraph{Discussion of the construction of the forward interface.} In this interface, a node or a gateway $\flrmParty$ receives the $\flrmforwardini$ message along with a packet identifier $\tid$. We remark that the remaining input sent along with the $\flrmforwardini$ message is only used by $\Functionality_{\LRM}$ when $\flrmParty$ is corrupt, and thus is ignored by an honest node or gateway $\flrmParty$ in our construction. 

$\flrmParty$ outputs an abortion message if $\tid$ is not stored. $\flrmParty$ also aborts if the party identifier $\flrmParty'$ stored along with $\tid$ is set to $\bot$, which indicates that the packet should not be forwarded because the intended recipient does not belong to the correct set of parties. If $\flrmParty$ does not abort,  $\flrmParty$ uses the $\fsmtsend$ interface of $\Functionality_{\SMT}$ to send the packet $\packet$ to $\flrmParty'$.

    
$\flrmParty'$ outputs an abortion message if it did not run the setup interface, in the case that $\flrmParty'$ is a node or a gateway, or the registration interface, in the case that $\flrmParty'$ is a user. If $\flrmParty'$ does not abort, $\flrmParty'$ proceeds as follows:
    \begin{description}

        \item[$\flrmParty'$ is a user.] If $\flrmParty'$ is a user, first $\flrmParty'$ outputs an abortion message if $\flrmParty$ is not a gateway. If $\flrmParty'$ does not abort, $\flrmParty'$ checks whether the received packet is a reply packet. To this end, $\flrmParty'$ runs the algorithm $\lrmSurbCheck$ on input all the single-use reply block identifiers $\surbidentifier$ that it stores. If a match is found, $\flrmParty'$ runs the algorithm $\lrmSurbRecover$ on input the secrets $\surbsecrets$ stored along with $\surbidentifier$ in order to obtain the reply message $\flrmmessageresponse$ encrypted in $\packet$. We remark that $\flrmmessageresponse$ may equal $\bot$, which could happen e.g.\ when a corrupt party destroys the payload. $\flrmParty'$ outputs the party identifier $\flrmParty$, the reply message $\flrmmessageresponse$, and the global packet identifier $\ppid$, which allows $\flrmParty'$ to link the reply packet to the corresponding request packet.

        If the algorithm $\lrmSurbCheck$ does not find a match, then $\flrmParty'$ treats the packet $\packet$ as a request packet and runs  $\lrmPacketProcess$ on input her secret key $\KEMsecretkey$ and the packet $\packet$. Then $\flrmParty'$ proceeds as follows:
        \begin{itemize}
            
            \item If $\lrmPacketProcess$ outputs $\top$, it means that the processing of the header was not successful, which could happen e.g.\ because the packet was not intended to be processed by $\flrmParty'$, and so $\flrmParty'$ aborts.

            \item If $\lrmPacketProcess$ outputs $\bot$, it means that the processing of the payload was not successful. Then $\flrmParty'$ sets the message $\flrmmessagerequest$ as $\bot$ and does not store any information to send a reply packet.

            \item If $\lrmPacketProcess$ outputs $\routing{k} = \surb = \bot$, i.e. it outputs only a message, it means that the request packet does not enable a reply, and thus $\flrmParty'$ does not store any information to send a reply packet.

            \item If $\lrmPacketProcess$ outputs $\routing{k}$, $\surb$ and a message $\flrmmessagerequest$, $\flrmParty'$ retrieves $\flrmGatewayexitreply'$ and  $\flrmNodeAresponse$, checks if they are valid and stores them along with $\surb$ to be able to send a reply.
            
        \end{itemize}
         $\flrmParty'$ outputs $\tid$, $\flrmGatewayexitreply'$ and $\flrmNodeAresponse'$ in addition to $\flrmParty$ and  $\flrmmessagerequest$. The leakage list $\flrmlistleakage$ and the global packet identifier $\ppid$ are set to $\bot$. $\Functionality_{\LRM}$ sets them to $\bot$ when $\flrmParty'$ is honest.

        \item[$\flrmParty'$ is a node or a gateway.] $\flrmParty'$ aborts if the packet is received from a party $\flrmParty$ that is not in the correct set. If $\flrmParty'$ does not abort, $\flrmParty'$ runs the algorithm $\lrmPacketProcess$ on input the packet $\packet$ and a secret key $\KEMsecretkey$ to obtain a processed packet $\packet'$ and the identifier $\routinginformation$ of the next party that should receive the packet. $\flrmParty'$ aborts if the processing of the packet by $\lrmPacketProcess$ fails.
        

         $\flrmParty'$ sets the the identifier $\routinginformation$ to $\bot$ if $\routinginformation$ does not belong to the correct set, so that the packet is not forwarded later. Then $\flrmParty'$ creates a packet identifier $\tid$ and stores it along with $\packet'$ and the party identifier $\routinginformation$.  
         
         Finally, $\flrmParty'$ outputs $\tid$, $\flrmParty$ and $\routinginformation$. The message $\flrmmessagerequest'$, the global packet identifier $\ppid'$, the leakage list $\flrmlistleakage$ and the reply packet identifier $\rpid'$ are set to $\bot$ as the functionality $\Functionality_{\LRM}$ sets them to $\bot$ when $\flrmParty'$ is honest.
         

    \end{description}

%% file: 7SecAnalysis.tex
\section{Security Analysis of Our Construction}\label{sec:securityanalysis}

To prove that the construction $\mathrm{\Pi}_{\LRM}$ in~\S\ref{sec:descriptionconstruction} securely realizes the ideal functionality $\Functionality_{\LRM}$ in~\S\ref{sec:functionalityDefinition}, we have to show that for any environment $\Environment$ and any adversary $\Adversary$ there exists a simulator $\Simulator$ such that $\Environment$ cannot distinguish between whether it is interacting with $\Adversary$ and the construction $\mathrm{\Pi}_{\LRM}$ in the real world or with $\Simulator$ and the ideal functionality $\Functionality_{\LRM}$ in the ideal world. The simulator thereby plays the role of all honest parties and interacts with $\Functionality_{\LRM}$ for all corrupt parties in the ideal world.

$\Simulator$ runs a copy of any adversary $\Adversary$, which is used to provide to $\Environment$ a view that is indistinguishable from the view given by $\Adversary$ in the real world. To achieve that, $\Simulator$ must simulate the real-world protocol towards the copy of $\Adversary$, in such a way that $\Adversary$ cannot distinguish an interaction with $\Simulator$ from an interaction with the real-world protocol. $\Simulator$ uses the information provided by $\Functionality_{\LRM}$ to provide a simulation of the real-world protocol.

Our simulator $\Simulator$ runs copies of the functionalities $\Functionality_{\Freg}$, $\Functionality_{\Fpreg}$, and $\Functionality_{\SMT}$. When any of the copies of those functionalities aborts, $\Simulator$ implicitly forwards the abortion message to the adversary if the functionality sends the abortion message to a corrupt party. 


\subsection{Simulator} 
\label{sec:simulator}

We describe the simulator $\Simulator$ for the case in which a subset of users $\flrmUser$, a subset of gateways $\flrmGateway$, a subset of first-layer nodes $\flrmNodeArequest$, a subset of second-layer nodes $\flrmNodeBrequest$, and a subset of third-layer nodes $\flrmNodeCrequest$ are corrupt. $\Simulator$ simulates the honest parties in the protocol $\mathrm{\Pi_{\LRM}}$ and runs copies of the ideal functionalities involved.

\input{7SecAnalysis11SimulatorSetup}
\input{7SecAnalysis12SimulatorRegistration}

\input{7SecAnalysis13SimulatorRequest}
\input{7SecAnalysis14SimulatorReply}
\input{7SecAnalysis15SimulatorForward}

\input{7SecAnalysis20SecurityProof}

%% file: 7SecAnalysis11SimulatorSetup.tex
\subsubsection{Simulation of the Setup Interface} 
\label{sec:simulatorSetup}

We describe the simulator $\Simulator$ for the setup interface.
\begin{description}


    \item[Honest node or gateway $\flrmParty$ starts setup.] When $\Functionality_{\LRM}$ sends the message $(\flrmsetupsim, \allowbreak \sid, \allowbreak \flrmParty)$, the simulator $\Simulator$ proceeds with the same computations of honest party $\flrmParty$ in $\mathrm{\Pi}_{\LRM}$:
    \begin{itemize}
        
        \item $\Simulator$ runs $(\KEMsecretkey, \allowbreak \KEMpublickey) \allowbreak \gets \allowbreak \KEMKeyGen(1^\securityparameter)$ and stores $(\flrmParty, \allowbreak \KEMsecretkey, \allowbreak \KEMpublickey)$.
        


        \item $\Simulator$ parses $\sid$ as $(\flrmSetA, \allowbreak \flrmSetB, \allowbreak \flrmSetC, \allowbreak \flrmSetW, \allowbreak \sid')$, sets $\sid_{\Freg} \allowbreak \gets \allowbreak (\flrmSetA \cup \flrmSetB \cup \flrmSetC \cup \flrmSetW, \allowbreak \sid)$ and sends $(\fregregisterini, \allowbreak \sid_{\Freg}, \allowbreak \KEMpublickey)$ to the functionality $\Functionality_{\Freg}$.
        
    \end{itemize}
    When $\Functionality_{\Freg}$ sends $(\fregregistersim, \allowbreak \sid_{\Freg}, \allowbreak \flrmParty, \allowbreak \KEMpublickey)$, $\Simulator$ forwards that message to $\Adversary$.
    
    \item[Honest node or gateway $\flrmParty$ ends setup.] When the adversary $\Adversary$ sends the message $(\fregregisterrep, \allowbreak \sid_{\Freg}, \allowbreak \flrmParty)$, $\Simulator$ runs a copy of $\Functionality_{\Freg}$ on input that message. When $\Functionality_{\Freg}$ sends $(\fregregisterend, \allowbreak \sid_{\Freg})$, $\Simulator$ sends  $(\flrmsetuprep, \allowbreak \sid, \allowbreak \flrmParty)$ to $\Functionality_{\LRM}$.

    \item[Corrupt node or gateway $\flrmParty$ starts setup.]  When the adversary $\Adversary$ sends the message $(\fregregisterini, \allowbreak \sid_{\Freg}, \allowbreak \KEMpublickey)$, $\Simulator$ runs the copy of $\Functionality_{\Freg}$ on input that message.  When $\Functionality_{\Freg}$ sends $(\fregregistersim, \allowbreak \sid_{\Freg}, \allowbreak \flrmParty, \allowbreak  \KEMpublickey)$, $\Simulator$ forwards that message to $\Adversary$.

    \item[Corrupt node or gateway $\flrmParty$ ends setup.] As  $\Adversary$ sends the message $(\fregregisterrep, \allowbreak \sid_{\Freg}, \allowbreak \flrmParty)$, $\Simulator$ runs a copy of $\Functionality_{\Freg}$ on input that message. When $\Functionality_{\Freg}$ sends $(\fregregisterend, \allowbreak \sid_{\Freg})$, $\Simulator$ forwards that message to the adversary $\Adversary$.

\end{description}

%% file: 7SecAnalysis12SimulatorRegistration.tex
\subsubsection{Simulation of the Registration Interface} 
\label{sec:simulatorRegistration}

We describe the simulator $\Simulator$ for the registration interface.
\begin{description}

    \item[Honest user $\flrmUser$ starts registration.] When  $\Functionality_{\LRM}$ sends the message $(\flrmregistersim, \allowbreak \sid, \allowbreak \flrmUser)$, $\Simulator$ parses $\sid$ as $(\flrmSetA, \allowbreak \flrmSetB, \allowbreak \flrmSetC, \allowbreak \flrmSetW, \allowbreak \sid')$ and sets $\sid_{\Freg} \allowbreak \gets \allowbreak (\flrmSetA \cup \flrmSetB \cup \flrmSetC \cup \flrmSetW, \allowbreak \sid)$. For every party $\flrmParty \allowbreak \in \allowbreak (\flrmSetA, \allowbreak \flrmSetB, \allowbreak \flrmSetC, \allowbreak \flrmSetW)$, $\Simulator$ sends $(\fregretrieveini, \allowbreak \sid_{\Freg}, \allowbreak \flrmParty)$ to the copy of $\Functionality_{\Freg}$. When $\Functionality_{\Freg}$ sends the message $(\fregretrievesim, \sid_{\Freg}, \ssid, \allowbreak \flrmParty, \KEMpublickey)$, $\Simulator$ forwards that message to $\Adversary$. ($\Simulator$ runs $\Functionality_{\Freg}$ for each party $\flrmParty$ sequentially, i.e., $\Adversary$ receives one $(\fregretrievesim, \sid_{\Freg}, \ssid, \allowbreak \flrmParty, \KEMpublickey)$ message at a time.) 
        
    \item[Honest user $\flrmUser$ continues registration.] When the adversary $\Adversary$ sends the message $(\fregretrieverep, \allowbreak \sid_{\Freg}, \allowbreak \ssid)$, $\Simulator$ runs a copy of $\Functionality_{\Freg}$ on input that message. When $\Functionality_{\Freg}$ sends $(\fregretrieveend, \allowbreak \sid_{\Freg}, \allowbreak  \KEMpublickey)$, $\Simulator$ includes $\flrmParty$ in a set $\mathbb{P}$, which is initially empty, and stores $(\flrmUser, \allowbreak \mathbb{P})$. (We remark that it is guaranteed that $\KEMpublickey \allowbreak \neq \allowbreak \bot$ because $\Functionality_{\LRM}$ outputs an abortion message when an honest user invokes the registration interface and not all nodes and gateways have run the setup interface.) If $\mathbb{P}$ in the tuple $(\flrmUser', \allowbreak \mathbb{P})$ such that $\flrmUser' \allowbreak = \allowbreak \flrmUser$ contains all the parties in $(\flrmSetA, \allowbreak \flrmSetB, \allowbreak \flrmSetC, \allowbreak \flrmSetW)$, $\Simulator$ proceeds with ``Honest user $\flrmUser$ continues registration (2)'', else $\Simulator$ goes back to ``Honest user $\flrmUser$ starts registration''to simulate the retrieval of the key of the next party $\flrmParty \allowbreak \in \allowbreak (\flrmSetA, \allowbreak \flrmSetB, \allowbreak \flrmSetC, \allowbreak \flrmSetW)$.

    \item[Honest user $\flrmUser$ continues registration (2).]  $\Simulator$  runs $(\KEMsecretkey, \allowbreak \KEMpublickey) \allowbreak \gets \allowbreak \KEMKeyGen(1^\securityparameter)$. $\Simulator$  sends the message $(\fpregregisterini, \allowbreak \sid, \allowbreak \KEMpublickey)$ to $\Functionality_{\Fpreg}$. When $\Functionality_{\Fpreg}$ sends $(\fpregregistersim, \allowbreak \sid, \allowbreak \flrmUser, \allowbreak  \KEMpublickey)$, $\Simulator$ forwards that message to $\Adversary$.

    \item[Honest user $\flrmUser$ ends registration.] As the adversary $\Adversary$ sends the message $(\fpregregisterrep, \allowbreak \sid, \allowbreak \flrmUser)$, $\Simulator$ runs a copy of $\Functionality_{\Fpreg}$ on input that message. When $\Functionality_{\Fpreg}$ sends the message $(\fpregregisterend, \allowbreak \sid)$, $\Simulator$ stores $(\flrmUser, \allowbreak \KEMsecretkey, \allowbreak \KEMpublickey)$ and sends the message $(\flrmregisterrep, \allowbreak \sid, \allowbreak \flrmUser)$ to $\Functionality_{\LRM}$.

    \item[Corrupt user $\flrmUser$ starts registration.] As the adversary $\Adversary$ sends the message  $(\fregretrieveini, \allowbreak \sid_{\Freg}, \allowbreak \flrmParty)$, $\Simulator$ runs $\Functionality_{\Freg}$ on input that message. When $\Functionality_{\Freg}$ sends $(\fregretrievesim, \allowbreak \sid_{\Freg}, \allowbreak \ssid, \allowbreak \flrmParty, \allowbreak \KEMpublickey)$, $\Simulator$ forwards that message to $\Adversary$.

    \item[Corrupt user $\flrmUser$ continues registration.] When the adversary $\Adversary$ sends the message $(\fregretrieverep, \allowbreak \sid_{\Freg}, \allowbreak \ssid)$, $\Simulator$ runs the functionality $\Functionality_{\Freg}$ on input that message. When $\Functionality_{\Freg}$ sends $(\fregretrieveend, \allowbreak \sid_{\Freg}, \allowbreak \KEMpublickey)$, $\Simulator$ forwards that message to $\Adversary$.

    \item[Corrupt user $\flrmUser$ continues registration (2).] As the adversary $\Adversary$ sends the message $(\fpregregisterini, \allowbreak \sid, \allowbreak  \KEMpublickey)$, $\Simulator$ runs a copy of $\Functionality_{\Fpreg}$ on input that message. When the functionality $\Functionality_{\Fpreg}$ sends $(\fpregregistersim, \allowbreak \sid, \allowbreak \flrmUser, \allowbreak \KEMpublickey)$, $\Simulator$ forwards that message to $\Adversary$.

    \item[Corrupt user $\flrmUser$ ends registration.] When the adversary $\Adversary$ sends the message $(\fpregregisterrep, \allowbreak \sid, \allowbreak \flrmUser)$,  $\Simulator$ runs $\Functionality_{\Fpreg}$ on input that message. When $\Functionality_{\Fpreg}$ sends $(\fpregregisterend, \allowbreak \sid)$, $\Simulator$ forwards that message to $\Adversary$.

    \end{description}

%% file: 7SecAnalysis13SimulatorRequest.tex
\subsubsection{Simulation of the Request Interface} 
\label{sec:simulatorRequest}

We describe the simulator $\Simulator$ for the request interface.
\begin{description}

    \item[Honest sender $\flrmSender$ sends a packet.]  When the functionality $\Functionality_{\LRM}$ sends the message $(\flrmsendsim, \allowbreak \sid, \allowbreak \qid, \allowbreak \flrmSender, \allowbreak \flrmGatewayentry, \allowbreak \flrmfirstmessage)$, if $\flrmfirstmessage \allowbreak = \allowbreak 0$, $\Simulator$ proceeds with the case ``Honest sender $\flrmSender$ sends a packet (3)'', else $\Simulator$ chooses randomly an honest user $\flrmUser$ from one of the stored tuples $(\flrmUser, \allowbreak \KEMsecretkey, \allowbreak \KEMpublickey)$, runs the copy of $\Functionality_{\Fpreg}$ on input $(\fpregretrieveini, \allowbreak \sid, \allowbreak \flrmUser)$ and, when $\Functionality_{\Fpreg}$ sends $(\fpregretrievesim, \allowbreak \sid, \allowbreak \ssid)$, $\Simulator$ forwards that message to the adversary $\Adversary$.

    \item[Honest sender $\flrmSender$ sends a packet (2).] As the adversary $\Adversary$ sends  $(\fpregretrieverep, \allowbreak \sid, \allowbreak \ssid)$, $\Simulator$ runs the copy of $\Functionality_{\Fpreg}$ on input that message. When $\Functionality_{\Fpreg}$ sends $(\fpregretrieveend, \allowbreak \sid, \allowbreak \KEMpublickey)$, $\Simulator$ proceeds with the case ``Honest sender $\flrmSender$ sends a packet (3).''
    
    \item[Honest sender $\flrmSender$ sends a packet (3).]  $\Simulator$ sets a random string $\flrmstring$ of the same length as the message $\langle \packet, \allowbreak \flrmNodeArequest \rangle$. $\Simulator$ sets $\sid_{\SMT} \allowbreak \gets \allowbreak (\flrmSender, \allowbreak \flrmGatewayentry, \allowbreak \sid)$ and runs a copy of the functionality $\Functionality_{\SMT}$ on input  $(\fsmtsendini, \allowbreak \sid_{\SMT}, \allowbreak \flrmstring)$. When $\Functionality_{\SMT}$ sends the message $(\fsmtsendsim, \allowbreak \sid_{\SMT}, \allowbreak \ssid, \allowbreak \SMTfleakage(\flrmstring))$, $\Simulator$ forwards that message to $\Adversary$.

    \item[Honest gateway $\flrmGatewayentry$ gets a packet from an honest sender $\flrmSender$.] As  $\Adversary$ sends the message $(\fsmtsendrep, \allowbreak \sid_{\SMT}, \allowbreak \ssid)$, $\Simulator$ runs $\Functionality_{\SMT}$ on input that message. When $\Functionality_{\SMT}$ outputs the message $(\fsmtsendend, \allowbreak \sid_{\SMT}, \allowbreak \flrmstring)$, $\Simulator$ sends  $(\flrmsendrep, \allowbreak \sid, \allowbreak \qid)$ to $\Functionality_{\LRM}$.

    \item[Corrupt gateway $\flrmGatewayentry$ receives a packet.] When the functionality $\Functionality_{\LRM}$ sends the message $(\flrmsendend, \allowbreak \sid, \allowbreak \tid, \allowbreak  \ppid', \allowbreak \flrmmessagerequest', \allowbreak \flrmSender, \allowbreak \flrmNodeArequest, \allowbreak \flrmlistleakage)$, the simulator  $\Simulator$ checks if $\ppid' \allowbreak = \allowbreak \bot$. If $\ppid' \allowbreak \neq \allowbreak \bot$, $\Simulator$ proceeds with ``Corrupt $\flrmGatewayexitreply$ receives a reply packet'' in~\S\ref{sec:simulatorReply}. If $\ppid' \allowbreak = \allowbreak \bot$, $\Simulator$ continues the execution described below.
    
   The simulator $\Simulator$ runs a copy of the honest sender in construction $\mathrm{\Pi}_{\LRM}$ on input the message $(\flrmsendini, \allowbreak \sid, \allowbreak \ppid, \allowbreak \flrmGatewayentry, \allowbreak \flrmGatewayexit, \allowbreak \flrmReceiver, \allowbreak \flrmmessagerequest, \allowbreak \flrmNodeArequest, \allowbreak \flrmNodeBrequest, \allowbreak \flrmNodeCrequest, \allowbreak \langle    \flrmGatewayentryreply, \allowbreak \flrmGatewayexitreply, \flrmNodeAresponse, \allowbreak \flrmNodeBresponse, \allowbreak \flrmNodeCresponse, \allowbreak \flrmSenderreply \rangle)$. To set the input given to the copy of the honest sender, $\Simulator$ proceeds as follows:
    \begin{itemize}

        \item The simulator $\Simulator$ takes the party identifier $\flrmGatewayentry$  from the message $(\flrmsendsim, \allowbreak \sid, \allowbreak \qid, \allowbreak \flrmSender, \allowbreak \flrmGatewayentry, \allowbreak \flrmfirstmessage)$ sent by  $\Functionality_{\LRM}$ in ``Honest sender $\flrmSender$ sends a packet.'' 

        \item $\Simulator$ takes the party identifier $\flrmNodeArequest$ from  $(\flrmsendend, \allowbreak \sid, \allowbreak \tid, \allowbreak  \ppid', \allowbreak \flrmmessagerequest', \allowbreak \flrmSender, \allowbreak \flrmNodeArequest, \allowbreak \flrmlistleakage)$ sent by $\Functionality_{\LRM}$.

        \item If any of the input values $(\flrmNodeBrequest, \allowbreak \flrmNodeCrequest, \allowbreak \flrmGatewayexit, \allowbreak \flrmReceiver)$ is contained in $\flrmlistleakage$, $\Simulator$ sets it to the corresponding value in $\flrmlistleakage$, else $\Simulator$ sets it to a random honest party from the correct domain, i.e. $\flrmNodeBrequest \allowbreak \gets \allowbreak \flrmSetB$, $\flrmNodeCrequest \allowbreak \gets \allowbreak \flrmSetC$, $\flrmGatewayexit \allowbreak \gets \allowbreak \flrmSetW$, and $\flrmReceiver$ to any registered honest user in the set $\flrmSetU$.

        \item If $\flrmmessagerequest' \allowbreak \neq \allowbreak \bot$, $\Simulator$ sets $\flrmmessagerequest \allowbreak \gets \allowbreak \flrmmessagerequest'$, else $\Simulator$ chooses a random message $\flrmmessagerequest \allowbreak \gets \allowbreak \flrmMessageSpace$.

        \item If the party identifiers $\langle    \flrmGatewayentryreply, \allowbreak \flrmGatewayexitreply, \flrmNodeAresponse, \allowbreak \flrmNodeBresponse, \allowbreak \flrmNodeCresponse, \allowbreak \flrmSenderreply \rangle$ are not present in $\flrmlistleakage$ and $\flrmmessagerequest' \allowbreak \neq \allowbreak \bot$, then $\Simulator$ sets the input values as $\langle    \flrmGatewayentryreply, \allowbreak \flrmGatewayexitreply, \flrmNodeAresponse, \allowbreak \flrmNodeBresponse, \allowbreak \flrmNodeCresponse, \allowbreak \flrmSenderreply \rangle \allowbreak \gets \allowbreak \bot$. When the party identifiers $\langle    \flrmGatewayentryreply, \allowbreak \flrmGatewayexitreply, \flrmNodeAresponse, \allowbreak \flrmNodeBresponse, \allowbreak \flrmNodeCresponse, \allowbreak \flrmSenderreply \rangle$ are not present in $\flrmlistleakage$ and $\flrmmessagerequest' \allowbreak = \allowbreak \bot$, $\Simulator$ sets  $\langle    \flrmGatewayentryreply, \allowbreak \flrmGatewayexitreply, \flrmNodeAresponse, \allowbreak \flrmNodeBresponse, \allowbreak \flrmNodeCresponse, \allowbreak \flrmSenderreply  \rangle$  to random honest parties from the correct sets. Else, $\Simulator$ sets $\flrmSenderreply \allowbreak \gets \allowbreak \flrmSender$, $\Simulator$ sets the input values $\langle    \flrmGatewayentryreply, \allowbreak \flrmGatewayexitreply, \flrmNodeAresponse, \allowbreak \flrmNodeBresponse, \allowbreak \flrmNodeCresponse \rangle$ that are included in $\flrmlistleakage$ to the values in $\flrmlistleakage$, and $\Simulator$ sets the input values $\langle    \flrmGatewayentryreply, \allowbreak \flrmGatewayexitreply, \flrmNodeAresponse, \allowbreak \flrmNodeBresponse, \allowbreak \flrmNodeCresponse \rangle$ that are not included in $\flrmlistleakage$ to random honest parties from the correct domains, i.e., $\flrmGatewayexitreply \allowbreak \gets \allowbreak \flrmSetW$, $\flrmNodeAresponse \allowbreak \gets \allowbreak \flrmSetA$, $\flrmNodeBresponse \allowbreak \gets \allowbreak \flrmSetB$, $\flrmNodeCresponse \allowbreak \gets \allowbreak \flrmSetC$, $\flrmGatewayentryreply \allowbreak \gets \allowbreak \flrmSetW$.

        We remark that, when the party identifiers $\langle    \flrmGatewayentryreply, \allowbreak \flrmGatewayexitreply, \flrmNodeAresponse, \allowbreak \flrmNodeBresponse, \allowbreak \flrmNodeCresponse, \allowbreak \flrmSenderreply \rangle$ are not present in $\flrmlistleakage$ and the message $\flrmmessagerequest'$ equals $\bot$,  $\Simulator$ does not know if the packet sent by the honest sender enables a reply or not. To simulate the packet towards the adversary, $\Simulator$ computes a packet that enables replies. In~\S\ref{sec:securityproof}, we show that in that situation the adversary cannot distinguish a request packet that enables replies from a request packet that does not enable replies. The reason is that there are parties in the route of the  packet that are honest, and thus the adversary is not able to fully process the packet to retrieve the message $\flrmmessagerequest$ and check whether replies are enabled.


    \end{itemize}
    To run the copy of the honest sender on that input, $\Simulator$ proceeds as follows:
    \begin{itemize}

        \item $\Simulator$ stores the tuple $(\sid, \allowbreak \KEMsecretkey, \allowbreak \KEMpublickey)$ in the copy of the honest sender, where  $(\KEMsecretkey, \allowbreak \KEMpublickey)$ are taken from the stored tuple $(\flrmUser, \allowbreak \KEMsecretkey, \allowbreak \KEMpublickey)$ such that $\flrmUser \allowbreak = \allowbreak \flrmSender$. The tuple $(\flrmUser, \allowbreak \KEMsecretkey, \allowbreak \KEMpublickey)$ is stored by $\Simulator$ in ``Honest user $\flrmUser$ ends registration'', and the sender identifier $\flrmSender$ is taken from the message $(\flrmsendend, \allowbreak \sid, \allowbreak \tid, \allowbreak  \ppid', \allowbreak \flrmmessagerequest', \allowbreak \flrmSender, \allowbreak \flrmNodeArequest, \allowbreak \flrmlistleakage)$ sent by the functionality $\Functionality_{\LRM}$.

        \item If the copy of the sender sends  $(\fpregretrieveini, \allowbreak \sid, \allowbreak \flrmReceiver)$, $\Simulator$ replies with the message $(\fpregretrieveend, \allowbreak \sid, \allowbreak \KEMpublickey)$, where $\KEMpublickey$  is stored by the copy of $\Functionality_{\Fpreg}$ in the tuple $(\sid, \allowbreak \FpregT', \allowbreak \KEMpublickey, \allowbreak 1)$ such that $\FpregT' \allowbreak = \allowbreak \flrmReceiver$. We remark that the message $(\fpregretrievesim, \allowbreak \sid, \allowbreak \ssid)$ was already sent to $\Adversary$ in ``Honest sender $\flrmSender$ sends a packet.''



        \item If the last party $\flrmParty$ in $\flrmlistleakage$ is honest, $\Simulator$ proceeds as follows:
        \begin{itemize}
            
            \item Let $\flrmParty \allowbreak \in \allowbreak \langle      \flrmNodeAresponse, \allowbreak \flrmNodeBresponse, \allowbreak \flrmNodeCresponse, \allowbreak \flrmGatewayentryreply, \allowbreak \flrmSenderreply \rangle$. This means that the list $\flrmlistleakage$ ends with a party that is in the route of the reply packet. (We recall that $\flrmGatewayexitreply$ is never the last party in $\flrmlistleakage$). In that case, when the copy of the honest sender runs the algorithm $\lrmSurbCreate$, $\Simulator$ picks the layer $i$ associated with $\flrmParty$ and stores the tuple $(\tid, \allowbreak \flrmParty, \allowbreak h_{i}, \allowbreak \surb, \allowbreak \surbsecrets, \allowbreak \flrmlistleakage)$. In that tuple, $h_{i}$ is the header computed by $\lrmSurbCreate$ for that layer, and $\surb$ and $\surbsecrets$ are part of the output of $\lrmSurbCreate$.   $\Simulator$ stores $(\tid, \allowbreak \flrmParty, \allowbreak h_{i}, \allowbreak \surb, \allowbreak \surbsecrets, \allowbreak \flrmlistleakage)$ so that it can associate a reply packet sent by the adversary with the $\surb$ sent to $\Adversary$. There are two possibilities regarding how $\Adversary$ processes the packet:
            \begin{itemize}
                
                \item $\Adversary$ processes the request packet, computes a reply packet by using the single-use reply block $\surb$, and processes the reply packet until forwarding it to $\flrmParty$. In this case, $\Simulator$ will use the header $h_{i}$ to associate the reply packet to the corresponding request packet and retrieve $\tid$ from the tuple
                $(\tid, \allowbreak \flrmParty, \allowbreak h_{i}, \allowbreak \surb, \allowbreak \surbsecrets, \allowbreak \flrmlistleakage)$.

                \item $\Adversary$ processes the request packet, computes a reply packet by using the single-use reply block $\surb$, and sends it to an honest gateway $\flrmGatewayexitreply'$. (We recall that exit gateways do not process reply packets, and thus $\Adversary$ can choose a gateway  $\flrmGatewayexitreply'$ different from the gateway $\flrmGatewayexitreply$ contained in $\flrmlistleakage$.) In that case, $\Simulator$ will use the header $h_{0}$ contained in $\surb$ to associate the reply packet to the corresponding request packet and retrieve $\tid$ from the tuple $(\tid, \allowbreak \flrmParty, \allowbreak h_{i}, \allowbreak \surb, \allowbreak \surbsecrets, \allowbreak \flrmlistleakage)$.
                
            \end{itemize}

            The secrets $\surbsecrets$ are used to decrypt the response message encrypted in the reply packet.

            Additionally, for each encryption layer $j \in [i+1,k]$ after the party $\flrmParty$, $\Simulator$ stores the tuples $(\flrmParty_j, \allowbreak h_j)$, where $\flrmParty_j$ is the identifier of an honest party and $h_j$ is the header for that layer. $\Simulator$ stores those tuples in order to check in the forward interface whether the adversary is able to decrypt layers that he is not supposed to be able to decrypt. $\Simulator$ outputs failure if the adversary is able to do that.
            
            

            \item Let $\flrmParty \allowbreak \in \allowbreak \langle \flrmNodeArequest, \flrmNodeBrequest, \allowbreak \flrmNodeCrequest, \allowbreak \flrmGatewayexit, \allowbreak \flrmReceiver \rangle$. This means that the list $\flrmlistleakage$ ends with a party that is in the route of the request packet, or that it is a reply packet. In those cases, when the copy of the honest sender runs the algorithm $\lrmPacketCreate$, $\Simulator$ picks the layer $i$ associated with $\flrmParty$ and stores $(\tid, \allowbreak \flrmParty, \allowbreak h_{i}, \allowbreak \payload{i}, \allowbreak \payloadsecrets)$, where $h_{i}$ and $\payload{i}$ are the header and the payload computed by the algorithm $\lrmPacketCreate$ for that layer. $\Simulator$ stores $(\tid, \allowbreak \flrmParty, \allowbreak h_{i}, \allowbreak \payload{i}, \allowbreak \payloadsecrets)$ so that, when $\Adversary$ processes and forwards the packet to the party $\flrmParty$, $\Simulator$ can compare the header of the packet sent by $\Adversary$ with $h_{i}$ and then retrieve the corresponding local packet identifier $\tid$, which is used to communicate with the functionality $\Functionality_{\LRM}$. The payload $\payload{i}$ and the secret keys $\payloadsecrets$ used to encrypt the payload are stored to find out whether the adversary has destroyed it or not.

            Additionally, for each encryption layer $j \in [i+1,k]$ after the party $\flrmParty$, $\Simulator$ stores the tuples $(\flrmParty_j, \allowbreak h_j, \allowbreak \payload{j})$, where $\flrmParty_j$ is the identifier of an honest party, $h_j$ is the header for that layer, and $\payload{j}$ is the payload for that layer. $\Simulator$ also stores $(\flrmNodeAresponse, \allowbreak \flrmmessagerequest, \allowbreak \surb)$ when the receiver $\flrmReceiver$ is the party $\flrmParty$ or is situated after $\flrmParty$ in the packet route. $\flrmNodeAresponse$ is the identifier of the first-layer node that should process $\surb$. $\Simulator$ stores those tuples in order to check  whether the adversary is able to decrypt layers that he is not supposed to be able to decrypt. $\Simulator$ outputs failure if the adversary is able to do that.
            
        \end{itemize}

        \item When the copy of the honest sender sends the message $(\fsmtsendini, \allowbreak \sid_{\SMT}, \allowbreak \langle \packet, \allowbreak \flrmNodeArequest \rangle)$, $\Simulator$ stores $(\tid, \allowbreak \packet, \allowbreak \flrmNodeArequest)$ and sends the message $(\fsmtsendend, \allowbreak \sid_{\SMT}, \allowbreak \langle \packet, \allowbreak \flrmNodeArequest \rangle)$ to the adversary $\Adversary$. We remark that the message $(\fsmtsendsim, \allowbreak \sid_{\SMT}, \allowbreak \ssid, \allowbreak \SMTfleakage(\flrmstring))$ was already sent to $\Adversary$ in ``Honest sender $\flrmSender$ sends a packet (3).''

    \end{itemize}

    \item[Corrupt sender $\flrmSender$ requests receiver's public key.] As the adversary $\Adversary$ sends the message  $(\fpregretrieveini, \allowbreak \sid, \allowbreak \flrmReceiver)$, $\Simulator$ runs the copy of $\Functionality_{\Fpreg}$ on input that message. When the copy of $\Functionality_{\Fpreg}$ outputs $(\fpregretrievesim, \allowbreak \sid, \allowbreak \ssid)$, $\Simulator$ forwards that message to $\Adversary$.
    
    \item[Corrupt sender $\flrmSender$ retrieves receiver's public key.] As soon as the adversary $\Adversary$ sends the message $(\fpregretrieverep, \allowbreak \sid, \allowbreak \ssid)$, $\Simulator$ runs the copy of $\Functionality_{\Fpreg}$ on input that message. When $\Functionality_{\Fpreg}$ sends $(\fpregretrieveend, \allowbreak \sid, \allowbreak \fpregvalue')$, $\Simulator$ forwards that message to $\Adversary$. We remark that, if $\flrmReceiver$ is honest,  $\fpregvalue' \allowbreak = \allowbreak \bot$ if $\Simulator$ did not receive a message $(\flrmregistersim, \allowbreak \sid, \allowbreak \flrmUser)$ such that $\flrmUser \allowbreak = \allowbreak \flrmReceiver$ in ``Honest user $\flrmUser$ starts registration.''  Otherwise $\fpregvalue' \allowbreak = \allowbreak \KEMpublickey$, where $\KEMpublickey$ was computed by $\Simulator$ in ``Honest user $\flrmUser$ continues registration (2).'' If  $\flrmReceiver$ is corrupt, $\fpregvalue' \allowbreak = \allowbreak \bot$ if $\Simulator$ did not receive a message $(\fpregregisterini, \allowbreak \sid, \allowbreak  \KEMpublickey)$ from the corrupt user $\flrmReceiver$ in ``Corrupt user $\flrmUser$ continues registration (2).''  


    \item[Corrupt sender $\flrmSender$ sends a packet.] When the adversary $\Adversary$ sends the message $(\fsmtsendini, \allowbreak \sid_{\SMT}, \allowbreak \langle \packet, \allowbreak \flrmNodeArequest \rangle)$, $\Simulator$ runs the copy of $\Functionality_{\SMT}$ on input that message.  When $\Functionality_{\SMT}$ sends the message $(\fsmtsendsim, \allowbreak \sid_{\SMT}, \allowbreak \ssid, \allowbreak \SMTfleakage(\langle \packet, \allowbreak \flrmNodeArequest \rangle))$, $\Simulator$ forwards that message to $\Adversary$.

    \item[Honest gateway $\flrmGatewayentry$ gets a packet from a corrupt sender $\flrmSender$.] As $\Adversary$ sends the message $(\fsmtsendrep, \allowbreak \sid_{\SMT}, \allowbreak \ssid)$, $\Simulator$ runs the copy of $\Functionality_{\SMT}$ on input that message.   When $\Functionality_{\SMT}$ sends the message $(\fsmtsendend, \allowbreak \sid_{\SMT}, \allowbreak \langle \packet, \allowbreak \flrmNodeArequest \rangle)$, the simulator  $\Simulator$ forwards that message to $\Adversary$, $\Simulator$ runs a copy of the honest gateway $\flrmGatewayentry$ on input $(\fsmtsendend, \allowbreak \sid_{\SMT}, \allowbreak \langle \packet, \allowbreak \flrmNodeArequest \rangle)$ as follows:
    \begin{itemize}
        
        \item To set up the copy of the honest gateway $\flrmGatewayentry$, the adversary stores the tuple $(\sid, \allowbreak \KEMsecretkey, \allowbreak \KEMpublickey)$ in that copy. The values $(\KEMsecretkey, \allowbreak \KEMpublickey)$ are taken from the tuple $(\flrmParty, \allowbreak \KEMsecretkey, \allowbreak \KEMpublickey)$ such that $\flrmParty \allowbreak = \allowbreak \flrmGatewayentry$, which is stored by $\Simulator$ in ``Honest node or gateway $\flrmParty$ starts setup.''


        \item If the copy of the honest gateway $\flrmGatewayentry$ outputs an abortion message, $\Simulator$ chooses a random fresh $\ppid$ and  sends $(\flrmsendini, \allowbreak \sid, \allowbreak \ppid, \allowbreak \flrmGatewayentry, \allowbreak \bot, \allowbreak \bot, \allowbreak \bot, \allowbreak \flrmNodeArequest, \allowbreak \bot, \allowbreak \bot, \allowbreak \langle \bot \rangle)$ to the functionality $\Functionality_{\LRM}$. We remark that the copy of the honest gateway outputs an abortion message when a tuple $(\flrmParty, \allowbreak \KEMsecretkey, \allowbreak \KEMpublickey)$ such that $\flrmParty \allowbreak = \allowbreak \flrmGatewayentry$ is not stored (because the honest gateway $\flrmGatewayentry$ did not run the setup interface), or because the corrupt user that sends the message  $(\fsmtsendrep, \allowbreak \sid_{\SMT}, \allowbreak \ssid)$ or the honest party that receives it, whose identifiers are included in $\sid_{\SMT}$,  do not belong to correct sets. In the ideal world, after sending the message $(\flrmsendsim, \allowbreak \sid, \allowbreak \qid, \allowbreak \flrmSender, \allowbreak \flrmGatewayentry, \allowbreak \flrmfirstmessage)$ to $\Simulator$ and receiving the message $(\flrmsendini, \allowbreak \sid, \allowbreak \qid)$ from $\Simulator$, $\Functionality_{\LRM}$ sends an abortion message to the honest gateway $\flrmGatewayentry$ for any of those reasons. 
        


        \item Else, when the copy of the honest gateway $\flrmGatewayentry$ outputs $(\flrmsendend, \allowbreak \sid, \allowbreak \tid, \allowbreak  \bot, \allowbreak \bot, \allowbreak \flrmSender, \allowbreak \flrmNodeArequest, \allowbreak \bot)$, $\Simulator$ proceeds with the procedure ``$\Simulator$ processes a packet''.

    \end{itemize}

    \item[$\Simulator$ processes a packet.] $\Simulator$ processes a packet received from a corrupt sender in order to find out the honest parties in the route of the packet. If all the parties in the route are honest, $\Simulator$ can send the encrypted message and all the party identifiers that define the route to the functionality $\Functionality_{\LRM}$. Moreover, if the packet contains a single-use reply block $\surb$, $\Simulator$ can send all the party identifiers that define the route for the reply. However, if at least one party in the route of the packet is corrupt, $\Simulator$ finds out all the party identifiers that define the route until and including the next corrupt party. If the next corrupt party is in the route of the reply, $\Simulator$ also finds out the encrypted message.

    We remark that, if the only corrupt party in the route is the exit gateway $\flrmGatewayexitreply$, $\Simulator$ could find all the parties in the route because $\flrmGatewayexitreply$ does not process the packet. However, $\Simulator$ stops processing the packet when it reaches $\flrmGatewayexitreply$, in the same way as it does with corrupt parties that process the packet.

    To process a packet, $\Simulator$ runs sequentially copies of the next honest parties that should process the packet  as follows:
    \begin{description}

        \item[Honest entry gateway $\flrmGatewayentry$ forwards the packet.] $\Simulator$ runs the copy of the honest entry gateway $\flrmGatewayentry$ on input the message $(\flrmforwardini, \allowbreak \sid, \allowbreak \tid, \allowbreak 0, \allowbreak \bot)$. ($\tid$ was output by the copy of $\flrmGatewayentry$ in ``Honest gateway $\flrmGatewayentry$ receives a packet from a corrupt sender $\flrmSender$.'') Then $\Simulator$ proceeds as follows:
        \begin{itemize}

            \item If the copy of $\flrmGatewayentry$ sends an abortion message (which happens when $\flrmNodeArequest \notin \flrmSetA$), $\Simulator$ ends the packet processing, chooses a random fresh $\ppid$,  and sends the message $(\flrmsendini, \allowbreak \sid, \allowbreak \ppid, \allowbreak \flrmGatewayentry, \allowbreak \bot, \allowbreak \bot, \allowbreak \bot, \allowbreak \flrmNodeArequest, \allowbreak \bot, \allowbreak \bot, \allowbreak \langle \bot \rangle)$ to the functionality $\Functionality_{\LRM}$. When the functionality $\Functionality_{\LRM}$ sends the message $(\flrmsendsim, \allowbreak \sid, \allowbreak \qid, \allowbreak \flrmSender, \allowbreak \flrmGatewayentry, \allowbreak \flrmfirstmessage)$, $\Simulator$ sends $(\flrmsendrep, \allowbreak \sid, \allowbreak \qid)$ to $\Functionality_{\LRM}$. We remark that, in this case, if the environment instructs the honest gateway $\flrmGatewayentry$ in the ideal world to forward the packet by using the forward interface, $\Functionality_{\LRM}$ aborts because   $\flrmNodeArequest \notin \flrmSetA$ and sends an abortion message to $\flrmGatewayentry$. 

            \item If the copy of the honest gateway $\flrmGatewayentry$ does not abort and sends $(\fsmtsendini, \allowbreak \sid_{\SMT}, \allowbreak \packet)$, and $\flrmNodeArequest$ is corrupt, $\Simulator$ ends the processing of the packet, chooses a random fresh $\ppid$, stores $(\ppid, \allowbreak \packet, \allowbreak \flrmNodeArequest)$, and sends the message $(\flrmsendini, \allowbreak \sid, \allowbreak \ppid, \allowbreak \flrmGatewayentry, \allowbreak \bot, \allowbreak \bot, \allowbreak \bot, \allowbreak \flrmNodeArequest, \allowbreak \bot, \allowbreak \bot, \allowbreak \langle \bot \rangle)$ to $\Functionality_{\LRM}$. When $\Functionality_{\LRM}$ sends the message $(\flrmsendsim, \allowbreak \sid, \allowbreak \qid, \allowbreak \flrmSender, \allowbreak \flrmGatewayentry, \allowbreak \flrmfirstmessage)$, $\Simulator$ sends $(\flrmsendrep, \allowbreak \sid, \allowbreak \qid)$ to the functionality $\Functionality_{\LRM}$. We remark that, in this case, if the environment instructs the gateway $\flrmGatewayentry$ in the ideal world to forward the packet by using the forward interface, $\Functionality_{\LRM}$ will leak $\ppid$ to $\Simulator$, which allows $\Simulator$ to retrieve the stored packet $\packet$ and send it to the corrupt $\flrmNodeArequest$. 

            \item If the copy of the honest gateway $\flrmGatewayentry$ does not abort and $\flrmNodeArequest$ is honest,  the simulator $\Simulator$ parses $\packet$ as $(h, \allowbreak \payloa)$. $\Simulator$ does the following:
            \begin{itemize}


                \item $\Simulator$ outputs failure if there is a tuple $(\flrmNodeAresponse, \allowbreak \flrmmessagerequest, \allowbreak \surb)$ such that $\flrmNodeAresponse \allowbreak = \allowbreak \flrmNodeArequest$ and such that $\surb \allowbreak = \allowbreak (h', \allowbreak s_k^p)$ with $h' \allowbreak = \allowbreak h$. The tuples $(\flrmNodeAresponse, \allowbreak \flrmmessagerequest, \allowbreak \surb)$ are stored by the simulator when computing a request packet that encrypts a single-use reply block $\surb$, and the adversary should not be able to retrieve the header $h'$ in $\surb$ because it was encrypted under a layer of encryption associated to an honest party. $\Simulator$ outputs failure because the adversary has computed a  packet that uses such a header.

                \item $\Simulator$ parses $h$ as $(\KEMciphertext \ || \  \beta \ || \ \gamma)$. $\Simulator$ checks whether there is a tuple $(\tid, \allowbreak \flrmParty, \allowbreak \allowbreak h', \allowbreak \payloa', \allowbreak \payloadsecrets)$ such that $\flrmParty \allowbreak = \allowbreak \flrmNodeArequest$ and $h' \allowbreak = \allowbreak  (\KEMciphertext' \ || \  \beta' \ || \ \gamma')$ such that either $\KEMciphertext' \allowbreak = \allowbreak \KEMciphertext$ and $(\beta' \ || \ \gamma') \allowbreak \neq \allowbreak (\beta \ || \ \gamma)$, or $\KEMciphertext' \allowbreak \neq \allowbreak \KEMciphertext$ and $(\beta' \ || \ \gamma') \allowbreak = \allowbreak (\beta \ || \ \gamma)$. If that is the case, $\Simulator$ outputs failure. The reason why $\Simulator$ outputs failure is that the adversary modified the header of a packet computed by an honest sender and the honest party that processed the packet did not abort.

                \item $\Simulator$ parses $h$ as $(\KEMciphertext \ || \  \beta \ || \ \gamma)$. $\Simulator$ checks whether there is a tuple $(\tid, \allowbreak \flrmParty, \allowbreak h', \allowbreak \surb, \allowbreak \surbsecrets, \allowbreak \flrmlistleakage)$ such that $\flrmParty \allowbreak = \allowbreak \flrmNodeArequest$ and $h' \allowbreak = \allowbreak  (\KEMciphertext' \ || \  \beta' \ || \ \gamma')$ such that either $\KEMciphertext' \allowbreak = \allowbreak \KEMciphertext$ and $(\beta' \ || \ \gamma') \allowbreak \neq \allowbreak (\beta \ || \ \gamma)$, or $\KEMciphertext' \allowbreak \neq \allowbreak \KEMciphertext$ and $(\beta' \ || \ \gamma') \allowbreak = \allowbreak (\beta \ || \ \gamma)$. If that is the case, $\Simulator$ outputs failure. The reason why $\Simulator$ outputs failure is that the adversary modified the header of a $\surb$ computed by an honest sender and the honest party that processed the packet did not abort.

                \item $\Simulator$ checks whether there is a tuple $(\tid, \allowbreak \flrmParty, \allowbreak \allowbreak h', \allowbreak \payloa', \allowbreak \payloadsecrets)$ such that $h \allowbreak = \allowbreak h'$ and $\flrmParty \allowbreak = \allowbreak \flrmNodeArequest$. If that is the case, it means that the adversary is resending a packet that previously the simulator sent to a corrupt gateway or to a corrupt first-layer node. If $\payloa \allowbreak = \allowbreak \payloa'$, $\Simulator$ sets $\flrmmessagerequest \allowbreak \gets \allowbreak \bot$, else sets  $\flrmmessagerequest \allowbreak \gets \allowbreak \top$ to indicate that the payload has been destroyed. ($\Simulator$ uses $\payloadsecrets$ to decrypt the payload. If the payload was modified but the result of decryption is not $\bot$, $\Simulator$ outputs failure.) Then the simulator sets $\ppid \allowbreak \gets \allowbreak \tid$ and sends $(\flrmsendini, \allowbreak \sid, \allowbreak \ppid, \allowbreak \flrmGatewayentry, \allowbreak \bot, \allowbreak \bot, \allowbreak \flrmmessagerequest, \allowbreak \bot, \allowbreak \bot, \allowbreak \bot, \allowbreak \langle \bot \rangle)$ to $\Functionality_{\LRM}$. When $\Functionality_{\LRM}$ sends $(\flrmsendsim, \allowbreak \sid, \allowbreak \qid, \allowbreak \flrmSender, \allowbreak \flrmGatewayentry, \allowbreak \flrmfirstmessage)$, $\Simulator$ sends the message $(\flrmsendini, \allowbreak \sid, \allowbreak \qid)$ to $\Functionality_{\LRM}$.

                \item $\Simulator$ checks whether there is a tuple $(\tid, \allowbreak \flrmParty, \allowbreak h', \allowbreak \surb, \allowbreak \surbsecrets, \allowbreak \flrmlistleakage)$ such that $h \allowbreak = \allowbreak h'$ and $\flrmParty \allowbreak = \allowbreak \flrmNodeArequest$. If that is the case, it means that the adversary is sending a reply packet computed on input a single-use reply block $\surb$ that was generated by the simulator. In that case, $\Simulator$ executes the procedure ``Single-use reply block computed by an honest sender'' in~\S\ref{sec:simulatorReply}.
                

            \end{itemize}

            \item If the copy of the honest gateway $\flrmGatewayentry$ does not abort and  $\flrmNodeArequest$ is honest, and $\Simulator$ neither outputs failure nor ends processing the packet, $\Simulator$ proceeds with ``Honest first-layer node $\flrmNodeArequest$ receives packet.''

        \end{itemize}
        
        \item[Honest first-layer node $\flrmNodeArequest$ receives packet.] $\Simulator$ runs the copy of the honest first-layer node $\flrmNodeArequest$ on input the message $(\fsmtsendend, \allowbreak \sid_{\SMT}, \allowbreak \packet)$ as follows:
        \begin{itemize}
            
            \item To set up the copy of the honest first-layer node $\flrmNodeArequest$, the adversary stores the tuple $(\sid, \allowbreak \KEMsecretkey, \allowbreak \KEMpublickey)$ in that copy. The values $(\KEMsecretkey, \allowbreak \KEMpublickey)$ are taken from the tuple $(\flrmParty, \allowbreak \KEMsecretkey, \allowbreak \KEMpublickey)$ such that $\flrmParty \allowbreak = \allowbreak \flrmNodeArequest$, which is stored by $\Simulator$ in ``Honest node or gateway $\flrmParty$ starts setup.''

            \item If the copy of the honest  first-layer node $\flrmNodeArequest$ outputs an abortion message, $\Simulator$ chooses a random fresh $\ppid$ and  sends the message $(\flrmsendini, \allowbreak \sid, \allowbreak \ppid, \allowbreak \flrmGatewayentry, \allowbreak \bot, \allowbreak \bot, \allowbreak \bot, \allowbreak \flrmNodeArequest, \allowbreak \bot, \allowbreak \bot, \allowbreak \langle \bot \rangle)$ to $\Functionality_{\LRM}$. We remark that the copy of the first-layer node $\flrmNodeArequest$ outputs an abortion message when a tuple $(\flrmParty, \allowbreak \KEMsecretkey, \allowbreak \KEMpublickey)$ such that $\flrmParty \allowbreak = \allowbreak \flrmNodeArequest$ is not stored because the honest first-layer node $\flrmNodeArequest$ did not run the setup interface. (The copy of the first-layer node $\flrmNodeArequest$ could also abort if the party identifier $\flrmGatewayentry$ included in $\sid_{\SMT}$ does not belong to the set $\flrmSetW$, but in this case $\Simulator$ has already checked the correctness of $\flrmGatewayentry$.) When $\Functionality_{\LRM}$ sends the message $(\flrmsendsim, \allowbreak \sid, \allowbreak \qid, \allowbreak \flrmSender, \allowbreak \flrmGatewayentry, \allowbreak \flrmfirstmessage)$, $\Simulator$ sends the message $(\flrmsendini, \allowbreak \sid, \allowbreak \qid)$ to $\Functionality_{\LRM}$. We remark that, if the environment instructs $\flrmGatewayentry$ in the ideal world to forward the packet, then $\Functionality_{\LRM}$ will detect that $\flrmNodeArequest$ has not run the setup interface and will send an abortion message to $\flrmNodeArequest$ in the ideal world. 

            \item Else, when the copy of the first-layer node $\flrmNodeArequest$ outputs the message $(\flrmforwardend, \allowbreak \sid, \allowbreak \tid, \allowbreak \bot, \allowbreak \bot, \allowbreak \allowbreak \flrmGatewayentry, \allowbreak \flrmNodeBrequest, \bot, \allowbreak \bot)$, $\Simulator$ continues with ``Honest first-layer node $\flrmNodeArequest$ forwards the packet''. We remark that the copy of the first-layer node $\flrmNodeArequest$ stores a tuple $(\sid, \allowbreak \tid, \allowbreak \packet', \allowbreak \flrmNodeBrequest)$, where $\packet'$ is the processed packet after removing one layer of encryption.
            
        \end{itemize}

    \item[Honest first-layer node $\flrmNodeArequest$ forwards the packet.] $\Simulator$ runs the copy of the honest first-layer node $\flrmNodeArequest$ on input  the message $(\flrmforwardini, \allowbreak \sid, \allowbreak \tid, \allowbreak 0, \allowbreak \bot)$. ($\tid$ was output by the copy of $\flrmNodeArequest$ in ``Honest first-layer node $\flrmNodeArequest$ receives packet.'') Then $\Simulator$ proceeds as follows:
        \begin{itemize}

            \item If the copy of $\flrmNodeArequest$ sends an abortion message (which happens when $\flrmNodeBrequest \notin \flrmSetB$, where $\flrmNodeBrequest$ is stored in the tuple $(\sid, \allowbreak \tid, \allowbreak \packet', \allowbreak \flrmNodeBrequest)$), the simulator $\Simulator$ ends the packet processing, chooses a random fresh $\ppid$,  and sends the message $(\flrmsendini, \allowbreak \sid, \allowbreak \ppid, \allowbreak \flrmGatewayentry, \allowbreak \bot, \allowbreak \bot, \allowbreak \bot, \allowbreak \flrmNodeArequest, \allowbreak \flrmNodeBrequest, \allowbreak \bot, \allowbreak \langle \bot \rangle)$ to the functionality $\Functionality_{\LRM}$. When  $\Functionality_{\LRM}$ sends $(\flrmsendsim, \allowbreak \sid, \allowbreak \qid, \allowbreak \flrmSender, \allowbreak \flrmGatewayentry, \allowbreak \flrmfirstmessage)$, the simulator $\Simulator$ sends $(\flrmsendrep, \allowbreak \sid, \allowbreak \qid)$ to  $\Functionality_{\LRM}$. We remark that, in this case, if the environment instructs the honest gateway $\flrmGatewayentry$ and then the honest first-layer node $\flrmNodeArequest$ in the ideal world to forward the packet by using the forward interface, $\Functionality_{\LRM}$ aborts because   $\flrmNodeBrequest \allowbreak \notin \allowbreak \flrmSetB$ and sends an abortion message to $\flrmNodeArequest$. 

            \item If the copy of the  first-layer node $\flrmNodeArequest$ does not abort and sends $(\fsmtsendini, \allowbreak \sid_{\SMT}, \allowbreak \packet')$,  and $\flrmNodeBrequest$ is corrupt, the simulator $\Simulator$ ends the processing of the packet, chooses a random fresh $\ppid$, stores $(\ppid, \allowbreak \packet', \allowbreak \flrmNodeBrequest)$, and sends the message $(\flrmsendini, \allowbreak \sid, \allowbreak \ppid, \allowbreak \flrmGatewayentry, \allowbreak \bot, \allowbreak \bot, \allowbreak \bot, \allowbreak \flrmNodeArequest, \allowbreak \flrmNodeBrequest, \allowbreak \bot, \allowbreak \langle \bot \rangle)$ to the functionality $\Functionality_{\LRM}$. When $\Functionality_{\LRM}$ sends the message $(\flrmsendsim, \allowbreak \sid, \allowbreak \qid, \allowbreak \flrmSender, \allowbreak \flrmGatewayentry, \allowbreak \flrmfirstmessage)$, $\Simulator$ sends $(\flrmsendrep, \allowbreak \sid, \allowbreak \qid)$ to $\Functionality_{\LRM}$. We remark that, in this case, if the environment instructs the gateway $\flrmGatewayentry$  and then the honest first-layer node $\flrmNodeArequest$ in the ideal world to forward the packet by using the forward interface, $\Functionality_{\LRM}$ will leak $\ppid$ to $\Simulator$, which allows $\Simulator$ to retrieve the stored packet $\packet'$ and send it to the corrupt $\flrmNodeBrequest$.

            \item If the copy of the honest first-layer node $\flrmNodeArequest$ does not abort and  $\flrmNodeBrequest$ is honest,  the simulator $\Simulator$ parses $\packet'$ as $(h, \allowbreak \payloa)$. $\Simulator$ does the following:
            \begin{itemize}

                \item $\Simulator$ outputs failure if there is a tuple $(\flrmParty_j, \allowbreak h_j, \allowbreak \payload{j})$ such that $\flrmNodeBrequest \allowbreak = \allowbreak \flrmParty_j$ and $h \allowbreak = \allowbreak h_j$. We recall that the tuples   $(\flrmParty_j, \allowbreak h_j, \allowbreak \payload{j})$ store packet headers and  payloads computed by the simulator that the adversary should not be able to retrieve because they were encrypted under a layer of encryption associated to an honest party. $\Simulator$ outputs failure because the adversary has computed a  packet that uses such a header.

                \item $\Simulator$ also outputs failure if there is a tuple $(\flrmParty_j, \allowbreak h_j)$ such that $\flrmNodeBrequest \allowbreak = \allowbreak \flrmParty_j$ and $h \allowbreak = \allowbreak h_j$. The tuples $(\flrmParty_j, \allowbreak h_j)$ were stored by the simulator when computing a single-use reply block, and the adversary should not be able to retrieve the header $h_j$ because it was encrypted under a layer of encryption associated to an honest party. $\Simulator$ outputs failure because the adversary has computed a  packet that uses such a header.

                \item $\Simulator$ parses $h$ as $(\KEMciphertext \ || \  \beta \ || \ \gamma)$. $\Simulator$ checks whether there is a tuple $(\tid, \allowbreak \flrmParty, \allowbreak \allowbreak h', \allowbreak \payloa', \allowbreak \payloadsecrets)$ such that $\flrmParty \allowbreak = \allowbreak \flrmNodeBrequest$ and $h' \allowbreak = \allowbreak  (\KEMciphertext' \ || \  \beta' \ || \ \gamma')$ such that either $\KEMciphertext' \allowbreak = \allowbreak \KEMciphertext$ and $(\beta' \ || \ \gamma') \allowbreak \neq \allowbreak (\beta \ || \ \gamma)$, or $\KEMciphertext' \allowbreak \neq \allowbreak \KEMciphertext$ and $(\beta' \ || \ \gamma') \allowbreak = \allowbreak (\beta \ || \ \gamma)$. If that is the case, $\Simulator$ outputs failure. The reason why $\Simulator$ outputs failure is that the adversary modified the header of a packet computed by an honest sender and the honest party that processed the packet did not abort.

                \item $\Simulator$ parses $h$ as $(\KEMciphertext \ || \  \beta \ || \ \gamma)$. $\Simulator$ checks whether there is a tuple $(\tid, \allowbreak \flrmParty, \allowbreak h', \allowbreak \surb, \allowbreak \surbsecrets, \allowbreak \flrmlistleakage)$ such that $\flrmParty \allowbreak = \allowbreak \flrmNodeBrequest$ and $h' \allowbreak = \allowbreak  (\KEMciphertext' \ || \  \beta' \ || \ \gamma')$ such that either $\KEMciphertext' \allowbreak = \allowbreak \KEMciphertext$ and $(\beta' \ || \ \gamma') \allowbreak \neq \allowbreak (\beta \ || \ \gamma)$, or $\KEMciphertext' \allowbreak \neq \allowbreak \KEMciphertext$ and $(\beta' \ || \ \gamma') \allowbreak = \allowbreak (\beta \ || \ \gamma)$. If that is the case, $\Simulator$ outputs failure. The reason why $\Simulator$ outputs failure is that the adversary modified the header of a $\surb$ computed by an honest sender and the honest party that processed the packet did not abort.

                \item $\Simulator$ checks whether there is a tuple $(\tid, \allowbreak \flrmParty, \allowbreak \allowbreak h', \allowbreak \payloa', \allowbreak \payloadsecrets)$ such that $h \allowbreak = \allowbreak h'$ and $\flrmParty \allowbreak = \allowbreak \flrmNodeBrequest$. If that is the case, it means that the adversary is resending a packet that previously the simulator sent to a corrupt first-layer node after removing the layer of encryption for that corrupt first-layer node and adding a layer of encryption for an honest first-layer node. If $\payloa \allowbreak = \allowbreak \payloa'$, $\Simulator$ sets $\flrmmessagerequest \allowbreak \gets \allowbreak \bot$, else sets  $\flrmmessagerequest \allowbreak \gets \allowbreak \top$ to indicate that the payload has been destroyed. ($\Simulator$ uses $\payloadsecrets$ to decrypt the payload. If the payload was modified but the result of decryption is not $\bot$, $\Simulator$ outputs failure.) Then the simulator sets $\ppid \allowbreak \gets \allowbreak \tid$ and sends $(\flrmsendini, \allowbreak \sid, \allowbreak \ppid, \allowbreak \flrmGatewayentry, \allowbreak \bot, \allowbreak \bot, \allowbreak \flrmmessagerequest, \allowbreak \flrmNodeArequest, \allowbreak \bot, \allowbreak \bot, \allowbreak \langle \bot \rangle)$ to $\Functionality_{\LRM}$. When $\Functionality_{\LRM}$ sends $(\flrmsendsim, \allowbreak \sid, \allowbreak \qid, \allowbreak \flrmSender, \allowbreak \flrmGatewayentry, \allowbreak \flrmfirstmessage)$, $\Simulator$ sends the message $(\flrmsendini, \allowbreak \sid, \allowbreak \qid)$ to $\Functionality_{\LRM}$.

                \item $\Simulator$ checks whether there is a tuple $(\tid, \allowbreak \flrmParty, \allowbreak h', \allowbreak \surb, \allowbreak \surbsecrets, \allowbreak \flrmlistleakage)$ such that $h \allowbreak = \allowbreak h'$ and $\flrmParty \allowbreak = \allowbreak \flrmNodeBrequest$. If that is the case, it means that the adversary is sending a reply packet computed on input a single-use reply block $\surb$ that was sent to the adversary by the simulator. The adversary removed the layer of encryption corresponding to a corrupt first-layer node from $\surb$ and added one layer of encryption associated to an honest first-layer node. $\Simulator$ uses $\surbsecrets$ to decrypt the payload $\payloa$ and obtain the message $\flrmmessagerequest$.\footnote{To decrypt the payload of a reply packet, the user uses secret keys for all the layers. However, it is still possible that an honest user decrypts the payload successfully even if the adversary adds and removes layers of encryption. For that, the adversary can encrypt the payload by using the secret key included in the $\surb$ (as usual), and additionally decrypt it by using the payload secret keys for the layers on encryption that were removed, and encrypt it by using the payload secret keys for the layers of encryption that were added.} The simulator executes the procedure ``Use reply interface'' in~\S\ref{sec:simulatorForward} on input $\tid$ and gets a local packet identifier $\tid'$. Then the simulator sets $\ppid \allowbreak \gets \allowbreak \tid'$ and sends $(\flrmsendini, \allowbreak \sid, \allowbreak \ppid, \allowbreak \flrmGatewayentry, \allowbreak \bot, \allowbreak \bot, \allowbreak \flrmmessagerequest, \allowbreak \flrmNodeArequest, \allowbreak \bot, \allowbreak \bot, \allowbreak \langle \bot \rangle)$ to $\Functionality_{\LRM}$. When $\Functionality_{\LRM}$ sends $(\flrmsendsim, \allowbreak \sid, \allowbreak \qid, \allowbreak \flrmSender, \allowbreak \flrmGatewayentry, \allowbreak \flrmfirstmessage)$, $\Simulator$ sends the message $(\flrmsendini, \allowbreak \sid, \allowbreak \qid)$ to $\Functionality_{\LRM}$. 

            \end{itemize}

            \item If the copy of the honest first-layer node $\flrmNodeArequest$ does not abort and  $\flrmNodeBrequest$ is honest, and $\Simulator$ neither outputs failure nor ends processing the packet, $\Simulator$ proceeds with ``Honest second-layer node $\flrmNodeBrequest$ receives packet.''

        \end{itemize}

        \item[Honest second-layer node $\flrmNodeBrequest$ receives packet.] The simulator $\Simulator$ runs the copy of the honest second-layer node $\flrmNodeBrequest$ on input the message $(\fsmtsendend, \allowbreak \sid_{\SMT}, \allowbreak \packet')$ as follows:
        \begin{itemize}
            
            \item To set up the copy of the honest second-layer node $\flrmNodeBrequest$, the adversary stores the tuple $(\sid, \allowbreak \KEMsecretkey, \allowbreak \KEMpublickey)$ in that copy. The values $(\KEMsecretkey, \allowbreak \KEMpublickey)$ are taken from the tuple $(\flrmParty, \allowbreak \KEMsecretkey, \allowbreak \KEMpublickey)$ such that $\flrmParty \allowbreak = \allowbreak \flrmNodeBrequest$, which is stored by $\Simulator$ in ``Honest node or gateway $\flrmParty$ starts setup.''

            \item If the copy of the honest second-layer node $\flrmNodeBrequest$ outputs an abortion message, $\Simulator$ chooses a random fresh $\ppid$ and  sends the message $(\flrmsendini, \allowbreak \sid, \allowbreak \ppid, \allowbreak \flrmGatewayentry, \allowbreak \bot, \allowbreak \bot, \allowbreak \bot, \allowbreak \flrmNodeArequest, \allowbreak \flrmNodeBrequest, \allowbreak \bot, \allowbreak \langle \bot \rangle)$ to $\Functionality_{\LRM}$. We remark that the copy of the second-layer node $\flrmNodeBrequest$ outputs an abortion message when a tuple $(\flrmParty, \allowbreak \KEMsecretkey, \allowbreak \KEMpublickey)$ such that $\flrmParty \allowbreak = \allowbreak \flrmNodeBrequest$ is not stored because the honest second-layer node $\flrmNodeBrequest$ did not run the setup interface. (The copy of the second-layer node $\flrmNodeBrequest$ could also abort if the party identifier $\flrmNodeArequest$ included in $\sid_{\SMT}$ does not belong to the set $\flrmSetA$, but in this case $\Simulator$ has already checked the correctness of $\flrmNodeArequest$.) When $\Functionality_{\LRM}$ sends the message $(\flrmsendsim, \allowbreak \sid, \allowbreak \qid, \allowbreak \flrmSender, \allowbreak \flrmGatewayentry, \allowbreak \flrmfirstmessage)$, $\Simulator$ sends the message $(\flrmsendini, \allowbreak \sid, \allowbreak \qid)$ to $\Functionality_{\LRM}$. We remark that, if the environment instructs $\flrmGatewayentry$ and then $\flrmNodeArequest$ in the ideal world to forward the packet, then $\Functionality_{\LRM}$ will detect that $\flrmNodeBrequest$ has not run the setup interface and will send an abortion message to $\flrmNodeBrequest$ in the ideal world. 

            \item Else, when the copy of the second-layer node $\flrmNodeBrequest$ outputs the message $(\flrmforwardend, \allowbreak \sid, \allowbreak \tid, \allowbreak \bot, \allowbreak \bot, \allowbreak \allowbreak \flrmNodeArequest, \allowbreak \flrmNodeCrequest, \bot, \allowbreak \bot)$, $\Simulator$ continues with ``Honest second-layer node $\flrmNodeBrequest$ forwards the packet''. We remark that the copy of the second-layer node $\flrmNodeBrequest$ stores a tuple $(\sid, \allowbreak \tid, \allowbreak \packet', \allowbreak \flrmNodeCrequest)$, where $\packet'$ is the processed packet after removing the layer of encryption associated with $\flrmNodeBrequest$.
            
        \end{itemize}

        \item[Honest second-layer node $\flrmNodeBrequest$ forwards the packet.] The simulator $\Simulator$ runs the copy of the honest second-layer node $\flrmNodeBrequest$ on input the message $(\flrmforwardini, \allowbreak \sid, \allowbreak \tid, \allowbreak 0, \allowbreak \bot)$. ($\tid$ was output by the copy of $\flrmNodeBrequest$ in ``Honest second-layer node $\flrmNodeBrequest$ receives packet.'') Then $\Simulator$ proceeds as follows:
        \begin{itemize}

            \item If the copy of $\flrmNodeBrequest$ sends an abortion message (which happens when $\flrmNodeCrequest \allowbreak \notin \allowbreak \flrmSetC$, where $\flrmNodeCrequest$ is stored in the tuple $(\sid, \allowbreak \tid, \allowbreak \packet', \allowbreak \flrmNodeCrequest)$), $\Simulator$ ends the packet processing, chooses a random fresh $\ppid$,  and sends the message $(\flrmsendini, \allowbreak \sid, \allowbreak \ppid, \allowbreak \flrmGatewayentry, \allowbreak \bot, \allowbreak \bot, \allowbreak \bot, \allowbreak \flrmNodeArequest, \allowbreak \flrmNodeBrequest, \allowbreak \flrmNodeCrequest, \allowbreak \langle \bot \rangle)$ to the functionality $\Functionality_{\LRM}$. When  the functionality $\Functionality_{\LRM}$ sends $(\flrmsendsim, \allowbreak \sid, \allowbreak \qid, \allowbreak \flrmSender, \allowbreak \flrmGatewayentry, \allowbreak \flrmfirstmessage)$, the simulator $\Simulator$ sends $(\flrmsendrep, \allowbreak \sid, \allowbreak \qid)$ to $\Functionality_{\LRM}$. We remark that, in this case, if the environment instructs the honest gateway $\flrmGatewayentry$, then the honest first-layer node $\flrmNodeArequest$, and then the honest second-layer node $\flrmNodeBrequest$  in the ideal world to forward the packet by using the forward interface, $\Functionality_{\LRM}$ aborts because   $\flrmNodeCrequest \allowbreak \notin \allowbreak \flrmSetC$ and sends an abortion message to $\flrmNodeBrequest$. 

            \item If the copy of the  second-layer node $\flrmNodeBrequest$ does not abort and sends $(\fsmtsendini, \allowbreak \sid_{\SMT}, \allowbreak \packet')$, and $\flrmNodeCrequest$ is corrupt, $\Simulator$ ends the processing of the packet, chooses a random fresh $\ppid$, stores $(\ppid, \allowbreak \packet', \allowbreak \flrmNodeCrequest)$, and sends the message $(\flrmsendini, \allowbreak \sid, \allowbreak \ppid, \allowbreak \flrmGatewayentry, \allowbreak \bot, \allowbreak \bot, \allowbreak \bot, \allowbreak \flrmNodeArequest, \allowbreak \flrmNodeBrequest, \allowbreak \flrmNodeCrequest, \allowbreak \langle \bot \rangle)$ to $\Functionality_{\LRM}$. When  the functionality $\Functionality_{\LRM}$ sends the message $(\flrmsendsim, \allowbreak \sid, \allowbreak \qid, \allowbreak \flrmSender, \allowbreak \flrmGatewayentry, \allowbreak \flrmfirstmessage)$, $\Simulator$ sends the message $(\flrmsendrep, \allowbreak \sid, \allowbreak \qid)$ to $\Functionality_{\LRM}$. We remark that, in this case, if the environment instructs the gateway $\flrmGatewayentry$, then the honest first-layer node $\flrmNodeArequest$, and then the second-layer node $\flrmNodeBrequest$ in the ideal world to forward the packet by using the forward interface, $\Functionality_{\LRM}$ will leak $\ppid$ to $\Simulator$, which allows $\Simulator$ to retrieve the stored packet $\packet'$ and send it to the corrupt $\flrmNodeCrequest$. 

            \item If the copy of the honest second-layer node $\flrmNodeBrequest$ does not abort and  $\flrmNodeCrequest$ is honest,  the simulator $\Simulator$ parses $\packet'$ as $(h, \allowbreak \payloa)$. $\Simulator$ does the following:
            \begin{itemize}

                \item $\Simulator$ outputs failure if there is a tuple $(\flrmParty_j, \allowbreak h_j, \allowbreak \payload{j})$ such that $\flrmNodeCrequest \allowbreak = \allowbreak \flrmParty_j$ and $h \allowbreak = \allowbreak h_j$. We recall that the tuples   $(\flrmParty_j, \allowbreak h_j, \allowbreak \payload{j})$ store packet headers and  payloads computed by the simulator that the adversary should not be able to retrieve because they were encrypted under a layer of encryption associated to an honest party. $\Simulator$ outputs failure because the adversary has computed a  packet that uses such a header.

                \item $\Simulator$ also outputs failure if there is a tuple $(\flrmParty_j, \allowbreak h_j)$ such that $\flrmNodeCrequest \allowbreak = \allowbreak \flrmParty_j$ and $h \allowbreak = \allowbreak h_j$. The tuples $(\flrmParty_j, \allowbreak h_j)$ were stored by the simulator when computing a single-use reply block, and the adversary should not be able to retrieve the header $h_j$ because it was encrypted under a layer of encryption associated to an honest party. $\Simulator$ outputs failure because the adversary has computed a  packet that uses such a header.

                \item $\Simulator$ parses $h$ as $(\KEMciphertext \ || \  \beta \ || \ \gamma)$. $\Simulator$ checks whether there is a tuple $(\tid, \allowbreak \flrmParty, \allowbreak \allowbreak h', \allowbreak \payloa', \allowbreak \payloadsecrets)$ such that $\flrmParty \allowbreak = \allowbreak \flrmNodeCrequest$ and $h' \allowbreak = \allowbreak  (\KEMciphertext' \ || \  \beta' \ || \ \gamma')$ such that either $\KEMciphertext' \allowbreak = \allowbreak \KEMciphertext$ and $(\beta' \ || \ \gamma') \allowbreak \neq \allowbreak (\beta \ || \ \gamma)$, or $\KEMciphertext' \allowbreak \neq \allowbreak \KEMciphertext$ and $(\beta' \ || \ \gamma') \allowbreak = \allowbreak (\beta \ || \ \gamma)$. If that is the case, $\Simulator$ outputs failure. The reason why $\Simulator$ outputs failure is that the adversary modified the header of a packet computed by an honest sender and the honest party that processed the packet did not abort.

                \item $\Simulator$ parses $h$ as $(\KEMciphertext \ || \  \beta \ || \ \gamma)$. $\Simulator$ checks whether there is a tuple $(\tid, \allowbreak \flrmParty, \allowbreak h', \allowbreak \surb, \allowbreak \surbsecrets, \allowbreak \flrmlistleakage)$ such that $\flrmParty \allowbreak = \allowbreak \flrmNodeCrequest$ and $h' \allowbreak = \allowbreak  (\KEMciphertext' \ || \  \beta' \ || \ \gamma')$ such that either $\KEMciphertext' \allowbreak = \allowbreak \KEMciphertext$ and $(\beta' \ || \ \gamma') \allowbreak \neq \allowbreak (\beta \ || \ \gamma)$, or $\KEMciphertext' \allowbreak \neq \allowbreak \KEMciphertext$ and $(\beta' \ || \ \gamma') \allowbreak = \allowbreak (\beta \ || \ \gamma)$. If that is the case, $\Simulator$ outputs failure. The reason why $\Simulator$ outputs failure is that the adversary modified the header of a $\surb$ computed by an honest sender and the honest party that processed the packet did not abort.

                \item $\Simulator$ checks whether there is a tuple $(\tid, \allowbreak \flrmParty, \allowbreak \allowbreak h', \allowbreak \payloa', \allowbreak \payloadsecrets)$ such that $h \allowbreak = \allowbreak h'$ and $\flrmParty \allowbreak = \allowbreak \flrmNodeCrequest$. If that is the case, it means that the adversary is resending a packet that previously the simulator sent to a corrupt second-layer node after removing the layer of encryption for that corrupt second-layer node and adding two layers of encryption. If $\payloa \allowbreak = \allowbreak \payloa'$, $\Simulator$ sets $\flrmmessagerequest \allowbreak \gets \allowbreak \bot$, else sets  $\flrmmessagerequest \allowbreak \gets \allowbreak \top$ to indicate that the payload has been destroyed. ($\Simulator$ uses $\payloadsecrets$ to decrypt the payload. If the payload was modified but the result of decryption is not $\bot$, $\Simulator$ outputs failure.) Then the simulator sets $\ppid \allowbreak \gets \allowbreak \tid$ and sends $(\flrmsendini, \allowbreak \sid, \allowbreak \ppid, \allowbreak \flrmGatewayentry, \allowbreak \bot, \allowbreak \bot, \allowbreak \flrmmessagerequest, \allowbreak \flrmNodeArequest, \allowbreak \flrmNodeBrequest, \allowbreak \bot, \allowbreak \langle \bot \rangle)$ to $\Functionality_{\LRM}$. When $\Functionality_{\LRM}$ sends $(\flrmsendsim, \allowbreak \sid, \allowbreak \qid, \allowbreak \flrmSender, \allowbreak \flrmGatewayentry, \allowbreak \flrmfirstmessage)$, $\Simulator$ sends the message $(\flrmsendini, \allowbreak \sid, \allowbreak \qid)$ to $\Functionality_{\LRM}$.

                \item $\Simulator$ checks whether there is a tuple $(\tid, \allowbreak \flrmParty, \allowbreak h', \allowbreak \surb, \allowbreak \surbsecrets, \allowbreak \flrmlistleakage)$ such that $h \allowbreak = \allowbreak h'$ and $\flrmParty \allowbreak = \allowbreak \flrmNodeCrequest$. If that is the case, it means that the adversary is sending a reply packet computed on input a single-use reply block $\surb$ that was sent to the adversary by the simulator. The adversary removed two layers of encryption from $\surb$ and added  two layers of encryption associated with an honest first-layer node and an honest second-layer node. $\Simulator$ uses $\surbsecrets$ to decrypt the payload $\payloa$ and obtain the message $\flrmmessagerequest$. The simulator executes the procedure ``Use reply interface'' in~\S\ref{sec:simulatorForward} on input $\tid$ and gets a local packet identifier $\tid'$. Then the simulator sets $\ppid \allowbreak \gets \allowbreak \tid'$ and sends $(\flrmsendini, \allowbreak \sid, \allowbreak \ppid, \allowbreak \flrmGatewayentry, \allowbreak \bot, \allowbreak \bot, \allowbreak \flrmmessagerequest, \allowbreak \flrmNodeArequest, \allowbreak \flrmNodeBrequest, \allowbreak \bot, \allowbreak \langle \bot \rangle)$ to $\Functionality_{\LRM}$. When $\Functionality_{\LRM}$ sends $(\flrmsendsim, \allowbreak \sid, \allowbreak \qid, \allowbreak \flrmSender, \allowbreak \flrmGatewayentry, \allowbreak \flrmfirstmessage)$, $\Simulator$ sends the message $(\flrmsendini, \allowbreak \sid, \allowbreak \qid)$ to $\Functionality_{\LRM}$. 

            \end{itemize}

            \item If the copy of the honest second-layer node $\flrmNodeBrequest$ does not abort and  $\flrmNodeCrequest$ is honest, and $\Simulator$ neither outputs failure nor ends processing the packet, $\Simulator$ proceeds with ``Honest third-layer node $\flrmNodeCrequest$ receives packet.''

        \end{itemize}

        \item[Honest third-layer node $\flrmNodeCrequest$ receives packet.]  The simulator $\Simulator$ runs the copy of the honest third-layer node $\flrmNodeCrequest$ on input the message $(\fsmtsendend, \allowbreak \sid_{\SMT}, \allowbreak \packet')$ as follows:
        \begin{itemize}
            
            \item To set up the copy of the third-layer node $\flrmNodeCrequest$, the adversary stores the tuple $(\sid, \allowbreak \KEMsecretkey, \allowbreak \KEMpublickey)$ in that copy. The values $(\KEMsecretkey, \allowbreak \KEMpublickey)$ are taken from the tuple $(\flrmParty, \allowbreak \KEMsecretkey, \allowbreak \KEMpublickey)$ such that $\flrmParty \allowbreak = \allowbreak \flrmNodeCrequest$, which is stored by $\Simulator$ in ``Honest node or gateway $\flrmParty$ starts setup.''

            \item If the copy of the honest third-layer node $\flrmNodeCrequest$ outputs an abortion message, $\Simulator$ chooses a random fresh $\ppid$ and sends the message $(\flrmsendini, \allowbreak \sid, \allowbreak \ppid, \allowbreak \flrmGatewayentry, \allowbreak \bot, \allowbreak \bot, \allowbreak \bot, \allowbreak \flrmNodeArequest, \allowbreak \flrmNodeBrequest, \allowbreak \flrmNodeCrequest, \allowbreak \langle \bot \rangle)$ to $\Functionality_{\LRM}$. We remark that the copy of the third-layer node $\flrmNodeCrequest$ outputs an abortion message when a tuple $(\flrmParty, \allowbreak \KEMsecretkey, \allowbreak \KEMpublickey)$ such that $\flrmParty \allowbreak = \allowbreak \flrmNodeCrequest$ is not stored because the honest third-layer node $\flrmNodeCrequest$ did not run the setup interface. (The copy of the third-layer node $\flrmNodeCrequest$ could also abort if the party identifier $\flrmNodeBrequest$ included in $\sid_{\SMT}$ does not belong to the set $\flrmSetB$, but in this case $\Simulator$ has already checked the correctness of $\flrmNodeBrequest$.) When $\Functionality_{\LRM}$ sends the message $(\flrmsendsim, \allowbreak \sid, \allowbreak \qid, \allowbreak \flrmSender, \allowbreak \flrmGatewayentry, \allowbreak \flrmfirstmessage)$, $\Simulator$ sends the message $(\flrmsendini, \allowbreak \sid, \allowbreak \qid)$ to $\Functionality_{\LRM}$. We remark that, if the environment instructs $\flrmGatewayentry$, then $\flrmNodeArequest$, and then $\flrmNodeBrequest$  in the ideal world to forward the packet, then $\Functionality_{\LRM}$ will detect that $\flrmNodeCrequest$ has not run the setup interface and will send an abortion message to $\flrmNodeCrequest$ in the ideal world. 

            \item Else, when the copy of the third-layer node $\flrmNodeCrequest$ outputs the message $(\flrmforwardend, \allowbreak \sid, \allowbreak \tid, \allowbreak \bot, \allowbreak \bot, \allowbreak \allowbreak \flrmNodeBrequest, \allowbreak \flrmGatewayexit, \bot, \allowbreak \bot)$, $\Simulator$ continues with ``Honest third-layer node $\flrmNodeCrequest$ forwards the packet''. We remark that the copy of the third-layer node $\flrmNodeCrequest$  stores a tuple $(\sid, \allowbreak \tid, \allowbreak \packet', \allowbreak \flrmGatewayexit)$, where $\packet'$ is the processed packet after removing the layer of encryption associated with $\flrmNodeCrequest$.
            
        \end{itemize}

        \item[Honest third-layer node $\flrmNodeCrequest$ forwards the packet.] The \newline simulator $\Simulator$ runs the copy of the honest third-layer node $\flrmNodeCrequest$ on input the message $(\flrmforwardini, \allowbreak \sid, \allowbreak \tid, \allowbreak 0, \allowbreak \bot)$. ($\tid$ was output by the copy of $\flrmNodeCrequest$ in ``Honest third-layer node $\flrmNodeCrequest$ receives packet.'') Then $\Simulator$ proceeds as follows:
        \begin{itemize}

            \item If the copy of $\flrmNodeCrequest$ sends an abortion message (which happens when $\flrmGatewayexit \allowbreak \notin \allowbreak \flrmSetW$, where $\flrmGatewayexit$ is stored in the tuple $(\sid, \allowbreak \tid, \allowbreak \packet', \allowbreak \flrmGatewayexit)$), $\Simulator$ ends the packet processing, chooses a random fresh $\ppid$,  and sends the message $(\flrmsendini, \allowbreak \sid, \allowbreak \ppid, \allowbreak \flrmGatewayentry, \allowbreak \flrmGatewayexit, \allowbreak \bot, \allowbreak \bot, \allowbreak \flrmNodeArequest, \allowbreak \flrmNodeBrequest, \allowbreak \flrmNodeCrequest, \allowbreak \langle \bot \rangle)$ to the functionality $\Functionality_{\LRM}$. When  $\Functionality_{\LRM}$ sends the message $(\flrmsendsim, \allowbreak \sid, \allowbreak \qid, \allowbreak \flrmSender, \allowbreak \flrmGatewayentry, \allowbreak \flrmfirstmessage)$, $\Simulator$ sends $(\flrmsendrep, \allowbreak \sid, \allowbreak \qid)$ to $\Functionality_{\LRM}$. We remark that, in this case, if the environment instructs $\flrmGatewayentry$, then  $\flrmNodeArequest$, then  $\flrmNodeBrequest$, and then $\flrmNodeCrequest$  in the ideal world to forward the packet by using the forward interface, $\Functionality_{\LRM}$ aborts because   $\flrmGatewayexit \allowbreak \notin \allowbreak \flrmSetW$ and sends an abortion message to $\flrmNodeCrequest$. 

            \item If the copy of the  third-layer node $\flrmNodeCrequest$ does not abort and sends $(\fsmtsendini, \allowbreak \sid_{\SMT}, \allowbreak \packet')$, and $\flrmGatewayexit$ is corrupt, $\Simulator$ ends the processing of the packet, chooses a random fresh $\ppid$, stores $(\ppid, \allowbreak \packet', \allowbreak \flrmGatewayexit)$, and sends the message  $(\flrmsendini, \allowbreak \sid, \allowbreak \ppid, \allowbreak \flrmGatewayentry, \allowbreak \flrmGatewayexit, \allowbreak \bot, \allowbreak \bot, \allowbreak \flrmNodeArequest, \allowbreak \flrmNodeBrequest, \allowbreak \flrmNodeCrequest, \allowbreak \langle \bot \rangle)$ to $\Functionality_{\LRM}$. When the ideal functionality $\Functionality_{\LRM}$ sends the message  $(\flrmsendsim, \allowbreak \sid, \allowbreak \qid, \allowbreak \flrmSender, \allowbreak \flrmGatewayentry, \allowbreak \flrmfirstmessage)$,  $\Simulator$ sends the message $(\flrmsendrep, \allowbreak \sid, \allowbreak \qid)$ to $\Functionality_{\LRM}$. We remark that, in this case, if the environment instructs $\flrmGatewayentry$, then $\flrmNodeArequest$, then  $\flrmNodeBrequest$, and then $\flrmNodeCrequest$ in the ideal world to forward the packet by using the forward interface, $\Functionality_{\LRM}$ will leak $\ppid$ to $\Simulator$, which allows $\Simulator$ to retrieve the stored packet $\packet'$ and send it to the corrupt $\flrmGatewayexit$. 

            \item If the copy of the honest third-layer node $\flrmNodeCrequest$ does not abort and $\flrmGatewayexit$ is honest, the simulator $\Simulator$ parses $\packet'$ as $(h, \allowbreak \payloa)$. $\Simulator$ does the following:
            \begin{itemize}

                \item $\Simulator$ outputs failure if there is a tuple $(\flrmParty_j, \allowbreak h_j, \allowbreak \payload{j})$ such that $\flrmGatewayexit \allowbreak = \allowbreak \flrmParty_j$ and $h \allowbreak = \allowbreak h_j$. We recall that the tuples   $(\flrmParty_j, \allowbreak h_j, \allowbreak \payload{j})$ store packet headers and  payloads computed by the simulator that the adversary should not be able to retrieve because they were encrypted under a layer of encryption associated to an honest party. $\Simulator$ outputs failure because the adversary has computed a  packet that uses such a header.

                \item $\Simulator$ also outputs failure if there is a tuple $(\flrmParty_j, \allowbreak h_j)$ such that $\flrmGatewayexit \allowbreak = \allowbreak \flrmParty_j$ and $h \allowbreak = \allowbreak h_j$. The tuples $(\flrmParty_j, \allowbreak h_j)$ were stored by the simulator when computing a single-use reply block, and the adversary should not be able to retrieve the header $h_j$ because it was encrypted under a layer of encryption associated to an honest party. $\Simulator$ outputs failure because the adversary has computed a  packet that uses such a header.

                \item $\Simulator$ parses $h$ as $(\KEMciphertext \ || \  \beta \ || \ \gamma)$. $\Simulator$ checks whether there is a tuple $(\tid, \allowbreak \flrmParty, \allowbreak \allowbreak h', \allowbreak \payloa', \allowbreak \payloadsecrets)$ such that $\flrmParty \allowbreak = \allowbreak \flrmGatewayexit$ and $h' \allowbreak = \allowbreak  (\KEMciphertext' \ || \  \beta' \ || \ \gamma')$ such that either $\KEMciphertext' \allowbreak = \allowbreak \KEMciphertext$ and $(\beta' \ || \ \gamma') \allowbreak \neq \allowbreak (\beta \ || \ \gamma)$, or $\KEMciphertext' \allowbreak \neq \allowbreak \KEMciphertext$ and $(\beta' \ || \ \gamma') \allowbreak = \allowbreak (\beta \ || \ \gamma)$. If that is the case, $\Simulator$ outputs failure. The reason why $\Simulator$ outputs failure is that the adversary modified the header of a packet computed by an honest sender and the honest party that processed the packet did not abort.

                \item $\Simulator$ parses $h$ as $(\KEMciphertext \ || \  \beta \ || \ \gamma)$. $\Simulator$ checks whether there is a tuple $(\tid, \allowbreak \flrmParty, \allowbreak h', \allowbreak \surb, \allowbreak \surbsecrets, \allowbreak \flrmlistleakage)$ such that $\flrmParty \allowbreak = \allowbreak \flrmGatewayexit$ and $h' \allowbreak = \allowbreak  (\KEMciphertext' \ || \  \beta' \ || \ \gamma')$ such that either $\KEMciphertext' \allowbreak = \allowbreak \KEMciphertext$ and $(\beta' \ || \ \gamma') \allowbreak \neq \allowbreak (\beta \ || \ \gamma)$, or $\KEMciphertext' \allowbreak \neq \allowbreak \KEMciphertext$ and $(\beta' \ || \ \gamma') \allowbreak = \allowbreak (\beta \ || \ \gamma)$. If that is the case, $\Simulator$ outputs failure. The reason why $\Simulator$ outputs failure is that the adversary modified the header of a $\surb$ computed by an honest sender and the honest party that processed the packet did not abort.

                \item $\Simulator$ checks whether there is a tuple $(\tid, \allowbreak \flrmParty, \allowbreak \allowbreak h', \allowbreak \payloa', \allowbreak \payloadsecrets)$ such that $h \allowbreak = \allowbreak h'$ and $\flrmParty \allowbreak = \allowbreak \flrmGatewayexit$. If that is the case, it means that the adversary is resending a packet that previously the simulator sent to a corrupt third-layer node after removing the layer of encryption for that corrupt third-layer node and adding three layers of encryption. If $\payloa \allowbreak = \allowbreak \payloa'$, $\Simulator$ sets $\flrmmessagerequest \allowbreak \gets \allowbreak \bot$, else sets  $\flrmmessagerequest \allowbreak \gets \allowbreak \top$ to indicate that the payload has been destroyed. ($\Simulator$ uses $\payloadsecrets$ to decrypt the payload. If the payload was modified but the result of decryption is not $\bot$, $\Simulator$ outputs failure.) Then the simulator sets $\ppid \allowbreak \gets \allowbreak \tid$ and sends $(\flrmsendini, \allowbreak \sid, \allowbreak \ppid, \allowbreak \flrmGatewayentry, \allowbreak \bot, \allowbreak \bot, \allowbreak \flrmmessagerequest, \allowbreak \flrmNodeArequest, \allowbreak \flrmNodeBrequest, \allowbreak \flrmNodeCrequest, \allowbreak \langle \bot \rangle)$ to $\Functionality_{\LRM}$. When $\Functionality_{\LRM}$ sends $(\flrmsendsim, \allowbreak \sid, \allowbreak \qid, \allowbreak \flrmSender, \allowbreak \flrmGatewayentry, \allowbreak \flrmfirstmessage)$, $\Simulator$ sends the message $(\flrmsendini, \allowbreak \sid, \allowbreak \qid)$ to $\Functionality_{\LRM}$.

                \item $\Simulator$ checks whether there is a tuple $(\tid, \allowbreak \flrmParty, \allowbreak h', \allowbreak \surb, \allowbreak \surbsecrets, \allowbreak \flrmlistleakage)$ such that $h \allowbreak = \allowbreak h'$ and $\flrmParty \allowbreak = \allowbreak \flrmGatewayexit$. If that is the case, it means that the adversary is sending a reply packet computed on input a single-use reply block $\surb$ that was sent to the adversary by the simulator. The adversary removed three layers of encryption from $\surb$ and added  three layers of encryption associated with an honest first-layer node, an honest second-layer node and an honest third-layer node. $\Simulator$ uses $\surbsecrets$ to decrypt the payload $\payloa$ and obtain the message $\flrmmessagerequest$. The simulator executes the procedure ``Use reply interface'' in~\S\ref{sec:simulatorForward} on input $\tid$ and gets a local packet identifier $\tid'$. Then the simulator sets $\ppid \allowbreak \gets \allowbreak \tid'$ and sends $(\flrmsendini, \allowbreak \sid, \allowbreak \ppid, \allowbreak \flrmGatewayentry, \allowbreak \bot, \allowbreak \bot, \allowbreak \flrmmessagerequest, \allowbreak \flrmNodeArequest, \allowbreak \flrmNodeBrequest, \allowbreak \flrmNodeCrequest, \allowbreak \langle \bot \rangle)$ to $\Functionality_{\LRM}$. When $\Functionality_{\LRM}$ sends $(\flrmsendsim, \allowbreak \sid, \allowbreak \qid, \allowbreak \flrmSender, \allowbreak \flrmGatewayentry, \allowbreak \flrmfirstmessage)$, $\Simulator$ sends the message $(\flrmsendini, \allowbreak \sid, \allowbreak \qid)$ to $\Functionality_{\LRM}$. 

            \end{itemize}

            \item If the copy of the third-layer node $\flrmNodeCrequest$ does not abort and  $\flrmGatewayexit$ is honest, and $\Simulator$ neither outputs failure nor ends processing the packet, $\Simulator$ proceeds with ``Honest exit gateway $\flrmGatewayexit$ receives packet.''

        \end{itemize}

        \item[Honest exit gateway $\flrmGatewayexit$ receives packet.]  The simulator $\Simulator$ runs the copy of the honest exit gateway $\flrmGatewayexit$ on input $(\fsmtsendend, \allowbreak \sid_{\SMT}, \allowbreak \packet')$ as follows:
        \begin{itemize}
            
            \item To set up the copy of the exit gateway $\flrmGatewayexit$, the adversary stores the tuple $(\sid, \allowbreak \KEMsecretkey, \allowbreak \KEMpublickey)$ in that copy. The values $(\KEMsecretkey, \allowbreak \KEMpublickey)$ are taken from the tuple $(\flrmParty, \allowbreak \KEMsecretkey, \allowbreak \KEMpublickey)$ such that $\flrmParty \allowbreak = \allowbreak \flrmGatewayexit$, which is stored by $\Simulator$ in ``Honest node or gateway $\flrmParty$ starts setup.''

            \item If the copy of the honest exit gateway $\flrmGatewayexit$ outputs an abortion message, $\Simulator$ chooses a random fresh $\ppid$ and sends the message $(\flrmsendini, \allowbreak \sid, \allowbreak \ppid, \allowbreak \flrmGatewayentry, \allowbreak \flrmGatewayexit, \allowbreak \bot, \allowbreak \bot, \allowbreak \flrmNodeArequest, \allowbreak \flrmNodeBrequest, \allowbreak \flrmNodeCrequest, \allowbreak \langle \bot \rangle)$ to $\Functionality_{\LRM}$. We remark that the copy of the exit gateway $\flrmGatewayexit$ outputs an abortion message when a tuple $(\flrmParty, \allowbreak \KEMsecretkey, \allowbreak \KEMpublickey)$ such that $\flrmParty \allowbreak = \allowbreak \flrmGatewayexit$ is not stored because the honest exit gateway $\flrmGatewayexit$ did not run the setup interface. (The copy of the exit gateway $\flrmGatewayexit$ could also abort if the party identifier $\flrmNodeCrequest$ included in $\sid_{\SMT}$ does not belong to the set $\flrmSetC$, but in this case $\Simulator$ has already checked the correctness of $\flrmNodeCrequest$.) When $\Functionality_{\LRM}$ sends the message $(\flrmsendsim, \allowbreak \sid, \allowbreak \qid, \allowbreak \flrmSender, \allowbreak \flrmGatewayentry, \allowbreak \flrmfirstmessage)$, $\Simulator$ sends the message $(\flrmsendini, \allowbreak \sid, \allowbreak \qid)$ to $\Functionality_{\LRM}$. We remark that, if the environment instructs $\flrmGatewayentry$, then $\flrmNodeArequest$, then $\flrmNodeBrequest$, and then $\flrmNodeCrequest$ in the ideal world to forward the packet, then $\Functionality_{\LRM}$ will detect that $\flrmGatewayexit$ has not run the setup interface and will send an abortion message to $\flrmGatewayexit$ in the ideal world. 

            \item Else, when the copy of the exit gateway $\flrmGatewayexit$ outputs the message $(\flrmforwardend, \allowbreak \sid, \allowbreak \tid, \allowbreak \bot, \allowbreak \bot, \allowbreak \allowbreak \flrmNodeCrequest, \allowbreak \flrmReceiver, \bot, \allowbreak \bot)$, the simulator $\Simulator$ continues with ``Honest exit gateway $\flrmGatewayexit$ forwards the packet''. We remark that the copy of the exit gateway $\flrmGatewayexit$  stores a tuple $(\sid, \allowbreak \tid, \allowbreak \packet', \allowbreak \flrmReceiver)$, where $\packet'$ is the processed packet after removing the layer of encryption associated with $\flrmGatewayexit$.
            
        \end{itemize}

        \item[Honest exit gateway $\flrmGatewayexit$ forwards the packet.] The simulator  $\Simulator$ runs the copy of the honest exit gateway $\flrmGatewayexit$ on input  $(\flrmforwardini, \allowbreak \sid, \allowbreak \tid, \allowbreak 0, \allowbreak \bot)$. ($\tid$ was output by the copy of $\flrmGatewayexit$ in ``Honest exit gateway $\flrmGatewayexit$ receives packet.'') Then $\Simulator$ proceeds as follows:
        \begin{itemize}

            \item If the copy of $\flrmGatewayexit$ sends an abortion message (which happens when $\flrmReceiver \allowbreak \notin \allowbreak \flrmSetU$, where $\flrmReceiver$ is stored in the tuple $(\sid, \allowbreak \tid, \allowbreak \packet', \allowbreak \flrmReceiver)$), $\Simulator$ ends the packet processing, chooses a random fresh $\ppid$,  and sends the message $(\flrmsendini, \allowbreak \sid, \allowbreak \ppid, \allowbreak \flrmGatewayentry, \allowbreak \flrmGatewayexit, \allowbreak \flrmReceiver, \allowbreak \bot, \allowbreak \flrmNodeArequest, \allowbreak \flrmNodeBrequest, \allowbreak \flrmNodeCrequest, \allowbreak \langle \bot \rangle)$ to the functionality $\Functionality_{\LRM}$. When  $\Functionality_{\LRM}$ sends the message $(\flrmsendsim, \allowbreak \sid, \allowbreak \qid, \allowbreak \flrmSender, \allowbreak \flrmGatewayentry, \allowbreak \flrmfirstmessage)$, $\Simulator$ sends the message $(\flrmsendrep, \allowbreak \sid, \allowbreak \qid)$ to $\Functionality_{\LRM}$. We remark that, in this case, if the environment instructs $\flrmGatewayentry$, then  $\flrmNodeArequest$, then  $\flrmNodeBrequest$, then $\flrmNodeCrequest$, and then $\flrmGatewayexit$  in the ideal world to forward the packet by using the forward interface, $\Functionality_{\LRM}$ aborts because   $\flrmReceiver \allowbreak \notin \allowbreak \flrmSetU$ and sends an abortion message to $\flrmGatewayexit$. 

            \item If the copy of the  exit gateway $\flrmGatewayexit$ does not abort and sends $(\fsmtsendini, \allowbreak \sid_{\SMT}, \allowbreak \packet')$, and $\flrmReceiver$ is corrupt, $\Simulator$ ends the processing of the packet, chooses a random fresh $\ppid$, stores $(\ppid, \allowbreak \packet', \allowbreak \flrmReceiver)$, and sends the message  $(\flrmsendini, \allowbreak \sid, \allowbreak \ppid, \allowbreak \flrmGatewayentry, \allowbreak \flrmGatewayexit, \allowbreak \flrmReceiver, \allowbreak \bot, \allowbreak \flrmNodeArequest, \allowbreak \flrmNodeBrequest, \allowbreak \flrmNodeCrequest, \allowbreak \langle \bot \rangle)$ to $\Functionality_{\LRM}$. When the ideal functionality $\Functionality_{\LRM}$ sends the message  $(\flrmsendsim, \allowbreak \sid, \allowbreak \qid, \allowbreak \flrmSender, \allowbreak \flrmGatewayentry, \allowbreak \flrmfirstmessage)$,  $\Simulator$ sends the message $(\flrmsendrep, \allowbreak \sid, \allowbreak \qid)$ to $\Functionality_{\LRM}$. We remark that, in this case, if the environment instructs $\flrmGatewayentry$, then $\flrmNodeArequest$, then  $\flrmNodeBrequest$, then $\flrmNodeCrequest$, and then $\flrmGatewayexit$ in the ideal world to forward the packet by using the forward interface, $\Functionality_{\LRM}$ will leak $\ppid$ to $\Simulator$, which allows $\Simulator$ to retrieve the stored packet $\packet'$ and send it to the corrupt $\flrmReceiver$. 

            \item If the copy of the honest gateway $\flrmGatewayexit$ does not abort and $\flrmReceiver$ is honest, the simulator $\Simulator$ parses $\packet'$ as $(h, \allowbreak \payloa)$. $\Simulator$ does the following:
            \begin{itemize}

                \item $\Simulator$ outputs failure if there is a tuple $(\flrmParty_j, \allowbreak h_j, \allowbreak \payload{j})$ such that $\flrmReceiver \allowbreak = \allowbreak \flrmParty_j$ and $h \allowbreak = \allowbreak h_j$. We recall that the tuples   $(\flrmParty_j, \allowbreak h_j, \allowbreak \payload{j})$ store packet headers and  payloads computed by the simulator that the adversary should not be able to retrieve because they were encrypted under a layer of encryption associated to an honest party. $\Simulator$ outputs failure because the adversary has computed a  packet that uses such a header.

                \item $\Simulator$ also outputs failure if there is a tuple $(\flrmParty_j, \allowbreak h_j)$ such that $\flrmReceiver \allowbreak = \allowbreak \flrmParty_j$ and $h \allowbreak = \allowbreak h_j$. The tuples $(\flrmParty_j, \allowbreak h_j)$ were stored by the simulator when computing a single-use reply block, and the adversary should not be able to retrieve the header $h_j$ because it was encrypted under a layer of encryption associated to an honest party. $\Simulator$ outputs failure because the adversary has computed a  packet that uses such a header.

                \item $\Simulator$ parses $h$ as $(\KEMciphertext \ || \  \beta \ || \ \gamma)$. $\Simulator$ checks whether there is a tuple $(\tid, \allowbreak \flrmParty, \allowbreak \allowbreak h', \allowbreak \payloa', \allowbreak \payloadsecrets)$ such that $\flrmParty \allowbreak = \allowbreak \flrmReceiver$ and $h' \allowbreak = \allowbreak  (\KEMciphertext' \ || \  \beta' \ || \ \gamma')$ such that either $\KEMciphertext' \allowbreak = \allowbreak \KEMciphertext$ and $(\beta' \ || \ \gamma') \allowbreak \neq \allowbreak (\beta \ || \ \gamma)$, or $\KEMciphertext' \allowbreak \neq \allowbreak \KEMciphertext$ and $(\beta' \ || \ \gamma') \allowbreak = \allowbreak (\beta \ || \ \gamma)$. If that is the case, $\Simulator$ outputs failure. The reason why $\Simulator$ outputs failure is that the adversary modified the header of a packet computed by an honest sender and the honest party that processed the packet did not abort.

                \item $\Simulator$ parses $h$ as $(\KEMciphertext \ || \  \beta \ || \ \gamma)$. $\Simulator$ checks whether there is a tuple $(\tid, \allowbreak \flrmParty, \allowbreak h', \allowbreak \surb, \allowbreak \surbsecrets, \allowbreak \flrmlistleakage)$ such that $\flrmParty \allowbreak = \allowbreak \flrmReceiver$ and $h' \allowbreak = \allowbreak  (\KEMciphertext' \ || \  \beta' \ || \ \gamma')$ such that either $\KEMciphertext' \allowbreak = \allowbreak \KEMciphertext$ and $(\beta' \ || \ \gamma') \allowbreak \neq \allowbreak (\beta \ || \ \gamma)$, or $\KEMciphertext' \allowbreak \neq \allowbreak \KEMciphertext$ and $(\beta' \ || \ \gamma') \allowbreak = \allowbreak (\beta \ || \ \gamma)$. If that is the case, $\Simulator$ outputs failure. The reason why $\Simulator$ outputs failure is that the adversary modified the header of a $\surb$ computed by an honest sender and the honest party that processed the packet did not abort.

                \item $\Simulator$ checks whether there is a tuple $(\tid, \allowbreak \flrmParty, \allowbreak \allowbreak h', \allowbreak \payloa', \allowbreak \payloadsecrets)$ such that $h \allowbreak = \allowbreak h'$ and $\flrmParty \allowbreak = \allowbreak \flrmReceiver$. If that is the case, it means that the adversary is resending a packet that previously the simulator sent to a corrupt gateway after removing the layer of encryption for that corrupt gateway and adding four layers of encryption. If $\payloa \allowbreak = \allowbreak \payloa'$, $\Simulator$ sets $\flrmmessagerequest \allowbreak \gets \allowbreak \bot$, else sets  $\flrmmessagerequest \allowbreak \gets \allowbreak \top$ to indicate that the payload has been destroyed. ($\Simulator$ uses $\payloadsecrets$ to decrypt the payload. If the payload was modified but the result of decryption is not $\bot$, $\Simulator$ outputs failure.) Then the simulator sets $\ppid \allowbreak \gets \allowbreak \tid$ and sends $(\flrmsendini, \allowbreak \sid, \allowbreak \ppid, \allowbreak \flrmGatewayentry, \allowbreak \bot, \allowbreak \bot, \allowbreak \flrmmessagerequest, \allowbreak \flrmNodeArequest, \allowbreak \flrmNodeBrequest, \allowbreak \flrmNodeCrequest, \allowbreak \langle \bot \rangle)$ to $\Functionality_{\LRM}$. When $\Functionality_{\LRM}$ sends $(\flrmsendsim, \allowbreak \sid, \allowbreak \qid, \allowbreak \flrmSender, \allowbreak \flrmGatewayentry, \allowbreak \flrmfirstmessage)$, $\Simulator$ sends  $(\flrmsendini, \allowbreak \sid, \allowbreak \qid)$ to $\Functionality_{\LRM}$.

                \item $\Simulator$ checks whether there is a tuple $(\tid, \allowbreak \flrmParty, \allowbreak h', \allowbreak \surb, \allowbreak \surbsecrets, \allowbreak \flrmlistleakage)$ such that $h \allowbreak = \allowbreak h'$ and $\flrmParty \allowbreak = \allowbreak \flrmReceiver$. If that is the case, it means that the adversary is sending a reply packet computed on input a single-use reply block $\surb$ that was sent to the adversary by the simulator. The adversary removed four layers of encryption from $\surb$ and added  four layers of encryption associated with an honest first-layer node, an honest second-layer node, an honest third-layer node, and an honest gateway. $\Simulator$ uses $\surbsecrets$ to decrypt the payload $\payloa$ and obtain the message $\flrmmessagerequest$. The simulator executes the procedure ``Use reply interface'' in~\S\ref{sec:simulatorForward} on input $\tid$ and gets a local packet identifier $\tid'$. Then the simulator sets $\ppid \allowbreak \gets \allowbreak \tid'$ and sends the message $(\flrmsendini, \allowbreak \sid, \allowbreak \ppid, \allowbreak \flrmGatewayentry, \allowbreak \bot, \allowbreak \bot, \allowbreak \flrmmessagerequest, \allowbreak \flrmNodeArequest, \allowbreak \flrmNodeBrequest, \allowbreak \flrmNodeCrequest, \allowbreak \langle \bot \rangle)$ to $\Functionality_{\LRM}$. When the functionality $\Functionality_{\LRM}$ sends $(\flrmsendsim, \allowbreak \sid, \allowbreak \qid, \allowbreak \flrmSender, \allowbreak \flrmGatewayentry, \allowbreak \flrmfirstmessage)$, $\Simulator$ sends the message $(\flrmsendini, \allowbreak \sid, \allowbreak \qid)$ to $\Functionality_{\LRM}$. 

            \end{itemize}

            \item If the copy of the exit gateway $\flrmGatewayexit$ does not abort and  $\flrmReceiver$ is honest, and $\Simulator$ neither outputs failure nor ends processing the packet, $\Simulator$ proceeds with ``Honest receiver $\flrmReceiver$ receives packet.''

        \end{itemize}

        \item[Honest receiver $\flrmReceiver$ receives packet.]   $\Simulator$ runs the copy of the honest receiver $\flrmReceiver$ on input $(\fsmtsendend, \allowbreak \sid_{\SMT}, \allowbreak \packet')$ as follows:
        \begin{itemize}
            
            \item To set up the copy of the receiver $\flrmReceiver$, the adversary stores the tuple $(\sid, \allowbreak \KEMsecretkey, \allowbreak \KEMpublickey)$ in that copy. The values $(\KEMsecretkey, \allowbreak \KEMpublickey)$ are taken from the tuple $(\flrmUser, \allowbreak \KEMsecretkey, \allowbreak \KEMpublickey)$ such that $\flrmUser \allowbreak = \allowbreak \flrmReceiver$, which is stored by $\Simulator$ in ``Honest user $\flrmUser$ ends registration.''

            \item If the copy of the honest receiver $\flrmReceiver$ outputs an abortion message, $\Simulator$ chooses a random fresh $\ppid$ and sends the message $(\flrmsendini, \allowbreak \sid, \allowbreak \ppid, \allowbreak \flrmGatewayentry, \allowbreak \flrmGatewayexit, \allowbreak \flrmReceiver, \allowbreak \bot, \allowbreak \flrmNodeArequest, \allowbreak \flrmNodeBrequest, \allowbreak \flrmNodeCrequest, \allowbreak \langle \bot \rangle)$ to $\Functionality_{\LRM}$. We remark that the copy of the honest receiver $\flrmReceiver$ outputs an abortion message when a tuple $(\flrmUser, \allowbreak \KEMsecretkey, \allowbreak \KEMpublickey)$ such that $\flrmUser \allowbreak = \allowbreak \flrmReceiver$ is not stored because the honest receiver $\flrmReceiver$ did not run the registration interface. (The copy of the receiver $\flrmReceiver$ could also abort if the party identifier $\flrmGatewayexit$ included in $\sid_{\SMT}$ does not belong to the set $\flrmSetW$, but in this case $\Simulator$ has already checked the correctness of $\flrmGatewayexit$.) When $\Functionality_{\LRM}$ sends the message $(\flrmsendsim, \allowbreak \sid, \allowbreak \qid, \allowbreak \flrmSender, \allowbreak \flrmGatewayentry, \allowbreak \flrmfirstmessage)$, $\Simulator$ sends the message $(\flrmsendini, \allowbreak \sid, \allowbreak \qid)$ to $\Functionality_{\LRM}$. We remark that, if the environment instructs $\flrmGatewayentry$, then $\flrmNodeArequest$, then $\flrmNodeBrequest$, then $\flrmNodeCrequest$, and then $\flrmGatewayexit$  in the ideal world to forward the packet, then $\Functionality_{\LRM}$ will detect that $\flrmReceiver$ has not run the registration interface and will send an abortion message to $\flrmReceiver$ in the ideal world. 

            \item Else, if  the copy of the receiver $\flrmReceiver$ outputs the message $(\flrmforwardend, \allowbreak \sid, \allowbreak \flrmParty, \allowbreak \flrmmessageresponse, \allowbreak \ppid)$, $\Simulator$ outputs failure. We remark that, if $\flrmReceiver$ outputs that message, it is because it has identified $\packet'$ as a reply packet when running 

            \item Else, when the copy of the exit gateway $\flrmGatewayexit$ outputs the message $(\flrmforwardend, \allowbreak \sid, \allowbreak \tid, \allowbreak \bot, \allowbreak \bot, \allowbreak \allowbreak \flrmNodeCrequest, \allowbreak \flrmReceiver, \bot, \allowbreak \bot)$, the simulator $\Simulator$ continues with ``Honest exit gateway $\flrmGatewayexit$ forwards the packet''. We remark that the copy of the exit gateway $\flrmGatewayexit$  stores a tuple $(\sid, \allowbreak \tid, \allowbreak \packet', \allowbreak \flrmReceiver)$, where $\packet'$ is the processed packet after removing the layer of encryption associated with $\flrmGatewayexit$.
            
        \end{itemize}

  \end{description}

\end{description}

%% file: 7SecAnalysis14SimulatorReply.tex
\subsubsection{Simulation of the Reply Interface} 
\label{sec:simulatorReply}

We describe the simulator $\Simulator$ for the reply interface.
\begin{description}

    \item[Honest receiver $\flrmReceiver$ sends a reply packet.] On input the message  $(\flrmreplysim, \allowbreak \sid, \allowbreak \qid, \allowbreak \flrmReceiver, \allowbreak \flrmGatewayexitreply, \allowbreak \flrmfirstmessage)$ from $\Functionality_{\LRM}$, $\Simulator$ proceeds with the case ``Honest sender $\flrmSender$ sends a packet (3)'' in~\S\ref{sec:simulatorRequest}.


    \item[Honest $\flrmGatewayexitreply$ receives a reply packet from an honest $\flrmReceiver$.] When the adversary $\Adversary$ sends the message $(\fsmtsendrep, \allowbreak \sid_{\SMT}, \allowbreak \ssid)$, $\Simulator$ proceeds with the case ``Honest gateway $\flrmGatewayentry$ receives a packet from an honest sender $\flrmSender$'' in~\S\ref{sec:simulatorRequest}.

    \item[Corrupt $\flrmGatewayexitreply$ receives a reply packet.] On input  $(\flrmreplyend, \allowbreak \sid, \allowbreak \tid, \allowbreak \ppid', \allowbreak \flrmmessageresponse', \allowbreak \flrmReceiver, \allowbreak \flrmNodeAresponse, \allowbreak \flrmlistleakage)$ from $\Functionality_{\LRM}$,  the simulator $\Simulator$ proceeds as follows:
    \begin{itemize}

        \item If $\ppid' \allowbreak = \allowbreak \bot$, $\Simulator$ proceeds with the case ``Corrupt gateway $\flrmGatewayentry$ receives a packet'' in~\S\ref{sec:simulatorRequest}. The reason why $\Simulator$ proceeds that way is that, in general, it is not able to distinguish whether the message $(\flrmreplyend, \allowbreak \sid, \allowbreak \tid, \allowbreak \ppid',  \allowbreak \flrmmessageresponse', \allowbreak \flrmReceiver, \allowbreak \flrmNodeAresponse, \allowbreak \flrmlistleakage)$ received from the functionality $\Functionality_{\LRM}$ corresponds to a request packet computed by an honest sender or to a reply packet computed by an honest receiver. More concretely:
        \begin{itemize}
        
            \item If the leakage list $\flrmlistleakage$ includes party identifiers for a reply packet route, then $\Simulator$ knows that this is a request packet that enables a reply and therefore should proceed with ``Corrupt gateway $\flrmGatewayentry$ receives a packet'' in~\S\ref{sec:simulatorRequest}.

            \item If the leakage list $\flrmlistleakage$ does not include party identifiers for a reply packet route, and $\flrmlistleakage$ ends with the party identifier of a user, and the message $\flrmmessagerequest$ is not $\bot$, then $\Simulator$ knows that this is a request packet that does not enable a reply and therefore should proceed with ``Corrupt gateway $\flrmGatewayentry$ receives a packet'' in~\S\ref{sec:simulatorRequest}.  We remark that, since $\flrmGatewayexitreply$ is corrupt and $\ppid' \allowbreak = \allowbreak \bot$, $\Simulator$ knows that the packet was computed by an honest sender. Therefore, if this was a reply packet, the message $\flrmmessagerequest$ would be $\bot$ because the sender that receives the reply is known to be honest and thus the functionality does not disclose $\flrmmessagerequest$.

            \item If the leakage list $\flrmlistleakage$ does not include party identifiers for a reply packet route, and the message $\flrmmessagerequest$ is equal to $\bot$, then $\Simulator$ does not know if this is a reply packet or a request packet. In that case, $\Simulator$ also proceeds with ``Corrupt gateway $\flrmGatewayentry$ receives a packet'' in~\S\ref{sec:simulatorRequest}. In our security proof in~\S\ref{sec:securityproof}, we show that, when the sender that computes a single-use reply block is honest, a request packet computed by an honest sender is indistinguishable from a reply packet computed by an honest receiver from such a single-use reply block.

        \end{itemize}

        \item If $\ppid' \allowbreak \neq \allowbreak \bot$, $\Simulator$ knows that the sender (or the party that computed the packet) is corrupt. Thanks to that, $\Simulator$ knows that the message $(\flrmreplyend, \allowbreak \sid, \allowbreak \tid, \allowbreak \ppid', \allowbreak \flrmmessageresponse', \allowbreak \flrmReceiver, \allowbreak \flrmNodeAresponse, \allowbreak \flrmlistleakage)$ received from $\Functionality_{\LRM}$ corresponds to the reply interface. The reason is that, if it corresponded to the request interface, then $\Simulator$ itself would have prompted the execution of the request interface with a message $\flrmsendini$. However, the case of a corrupt sender sending a packet to a corrupt gateway only involves corrupt parties, and thus $\Simulator$ never inputs the packet to the functionality through the request interface in that case. (If the corrupt gateway or a subsequent party in the route sends the packet to an honest party, then $\Simulator$ will input the packet to the functionality by using the forward interface.) Therefore, $\Simulator$ knows that the message from the functionality corresponds to the reply interface. 

        $\Simulator$ also knows that the receiver is honest if $\Simulator$ was not the one that invoked the reply interface with a  message $\flrmreplyini$. In the simulator of the forward interface in~\S\ref{sec:simulatorForward}, there is a case in which the simulator invokes the reply interface when the receiver is corrupt and the gateway is also corrupt. Here we deal with the case in which the simulator did not invoke the reply interface.


        Therefore, after receiving a message $(\flrmreplyend, \allowbreak \sid, \allowbreak \tid, \allowbreak \ppid', \allowbreak \flrmmessageresponse', \allowbreak \flrmReceiver, \allowbreak \flrmNodeAresponse, \allowbreak \flrmlistleakage)$ such that $\ppid' \allowbreak \neq \allowbreak \bot$, $\Simulator$ should compute a reply packet and send it to the corrupt gateway. To this end, $\Simulator$ retrieves the tuple $(\ppid, \allowbreak \surb)$ such that $\ppid \allowbreak = \allowbreak \ppid'$. (This tuple was stored by $\Simulator$ when it received a request packet that enables a reply from a corrupt sender in the request interface, or from a corrupt node or gateway in the forward interface.) Then $\Simulator$ runs a copy of a receiver in the construction $\mathrm{\Pi}_{\LRM}$ on input  $(\flrmreplyini, \allowbreak \sid, \allowbreak \tid, \allowbreak \flrmmessageresponse, \allowbreak \{\rpid, \allowbreak \flrmGatewayexitreply', \allowbreak \flrmNodeAresponse'\})$. To set the copy of the receiver, $\Simulator$ proceeds as follows:
        \begin{itemize}

            \item $\Simulator$ stores in the copy of the receiver a tuple $(\sid, \allowbreak \KEMsecretkey, \allowbreak \KEMpublickey)$, where $(\KEMsecretkey, \allowbreak \KEMpublickey)$ are taken from the tuple $(\flrmUser, \allowbreak \KEMsecretkey, \allowbreak \KEMpublickey)$ such that  $\flrmUser \allowbreak = \allowbreak  \flrmReceiver$, which was stored by $\Simulator$ when it simulated the registration phase for the honest user $\flrmUser$.

            \item $\Simulator$ stores in the copy of the receiver a tuple $(\sid, \allowbreak \tid, \allowbreak \surb, \allowbreak \flrmGatewayexitreply, \allowbreak \flrmNodeAresponse)$, where $\surb$ is taken from the tuple $(\ppid, \allowbreak \surb)$ mentioned above, and $\tid$, $\flrmGatewayexitreply$ and $\flrmNodeAresponse$ are taken from the messages received from $\Functionality_{\LRM}$.

            \item The simulator $\Simulator$ sets the input $(\tid, \allowbreak \flrmmessageresponse, \allowbreak \{\rpid, \allowbreak \flrmGatewayexitreply', \allowbreak \flrmNodeAresponse'\})$ of the  $\flrmreplyini$ message as follows: $\tid$ is taken from the message $\flrmreplyend$ sent by the functionality $\Functionality_{\LRM}$;  $\flrmmessageresponse \allowbreak \gets \allowbreak \flrmmessageresponse'$, where $\flrmmessageresponse'$ is in the $\flrmreplyend$ message  sent by the functionality $\Functionality_{\LRM}$; and $\{\rpid, \allowbreak \flrmGatewayexitreply', \allowbreak \flrmNodeAresponse'\} \allowbreak = \allowbreak \bot$. We recall that $\{\rpid, \allowbreak \flrmGatewayexitreply', \allowbreak \flrmNodeAresponse'\}$ is not used by the receiver in $\mathrm{\Pi}_{\LRM}$.

        \end{itemize}
        When $\Simulator$ runs the copy of the receiver, the copy of the receiver sends $(\fsmtsendini, \allowbreak \sid_{\SMT}, \allowbreak \langle \packet, \allowbreak \flrmNodeAresponse \rangle)$. Then the simulator $\Simulator$ sends the message $(\fsmtsendend, \allowbreak \sid_{\SMT}, \allowbreak  \langle \packet, \allowbreak \flrmNodeAresponse \rangle)$ to the adversary. We remark that the message $(\fsmtsendsim, \allowbreak \sid, \allowbreak \ssid, \allowbreak \SMTfleakage(\flrmstring))$ was already sent to $\Adversary$ in ``Honest receiver $\flrmReceiver$ sends a reply packet.'' 

    \end{itemize}

    \item[Corrupt $\flrmReceiver$ sends a packet.] Once $\Adversary$ sends $(\fsmtsendini, \allowbreak \sid_{\SMT}, \allowbreak \langle \packet, \allowbreak \flrmNodeAresponse \rangle)$, $\Simulator$ runs the copy of $\Functionality_{\SMT}$ on input that message. When $\Functionality_{\SMT}$ sends the message $(\fsmtsendsim, \allowbreak \sid_{\SMT}, \allowbreak \ssid, \allowbreak \SMTfleakage(\langle \packet, \allowbreak \flrmNodeAresponse \rangle))$,  $\Simulator$ forwards that message to $\Adversary$.

    \item[Honest $\flrmGatewayexitreply$ receives a packet from corrupt $\flrmReceiver$.] As $\Adversary$ sends the message $(\fsmtsendrep, \allowbreak \sid_{\SMT}, \allowbreak \ssid)$, $\Simulator$ runs the copy of $\Functionality_{\SMT}$ on input that message. When $\Functionality_{\SMT}$ sends  $(\fsmtsendend, \allowbreak \sid_{\SMT}, \allowbreak \langle \packet, \allowbreak \flrmNodeAresponse \rangle)$, $\Simulator$ proceeds with the case ``Honest gateway $\flrmGatewayentry$ receives a packet from a corrupt sender $\flrmSender$'' in~\S\ref{sec:simulatorRequest}. When executing ``Honest gateway $\flrmGatewayentry$ receives a packet from a corrupt sender $\flrmSender$'', $\Simulator$ checks whether the packet sent by the adversary is a reply packet computed on input a single-use reply block generated by an honest sender. In that case, $\Simulator$ executes the procedure ``Single-use reply block computed by an honest sender'' described below.

    \begin{description}
        
    \item[Single-use reply block computed by an honest sender.] The simulator $\Simulator$ parses $\packet$ as $(h_0, \allowbreak \payload{0})$. When there is a tuple $(\tid, \allowbreak \flrmParty, \allowbreak h_{i}, \allowbreak \surb, \allowbreak \surbsecrets, \allowbreak \flrmlistleakage)$  such that $\surb \allowbreak = \allowbreak (h'_{0}, \allowbreak s_k^{p})$ and $h'_0 \allowbreak = \allowbreak h_0$, $\Simulator$ knows that this is a reply packet computed by using a single-use reply block that was previously sent by $\Simulator$ to the adversary encrypted in a request packet. The request packet was computed by $\Simulator$ when $\Functionality_{\LRM}$ informed $\Simulator$ that a packet computed by an honest sender  was sent or forwarded to a corrupt party. As described in ``Honest gateway $\flrmGatewayentry$ receives a packet from a corrupt sender $\flrmSender$'', $\Simulator$ runs a copy of the honest gateway $\flrmGatewayexitreply$ to receive and forward the packet. If the copy of honest gateway does not abort and $\Simulator$ does not output failure, $\Simulator$ proceeds as described below.





        $\Simulator$ uses $\surbsecrets$ to decrypt the payload of the packet $\packet$ and obtain either a response message $\flrmmessageresponse$ or $\bot$. Then $\Simulator$ proceeds as follows:
        \begin{itemize}

            \item If $\flrmNodeAresponse \allowbreak \in \allowbreak \flrmlistleakage$ is honest, and both  $\flrmGatewayexitreply \allowbreak \in \allowbreak \flrmlistleakage$ and $\flrmNodeAresponse \allowbreak \in \allowbreak \flrmlistleakage$  are equal to $\flrmGatewayexitreply$ in $\sid_{\SMT}$ and $\flrmNodeAresponse$ in $\langle \packet, \allowbreak \flrmNodeAresponse \rangle$, then $\Simulator$ sends  $(\flrmreplyini, \allowbreak \sid, \allowbreak \tid, \allowbreak \flrmmessageresponse, \allowbreak \bot)$ to $\Functionality_{\LRM}$. In this case, the adversary does not change the gateway $\flrmGatewayexitreply$ and the node $\flrmNodeAresponse$ that were specified in the request packet sent by the simulator to the adversary.

            \item If $\flrmNodeAresponse \allowbreak \in \allowbreak \flrmlistleakage$ is corrupt, or if $\flrmGatewayexitreply \allowbreak \in \allowbreak \flrmlistleakage$ is not equal to   $\flrmGatewayexitreply$ in $\sid_{\SMT}$, or if  $\flrmNodeAresponse \allowbreak \in \allowbreak \flrmlistleakage$ is not equal to $\flrmNodeAresponse$ in $\langle \packet, \allowbreak \flrmNodeAresponse \rangle$, $\Simulator$ creates a random fresh reply packet identifier $\rpid$, stores $(\rpid, \allowbreak \packet)$, and sends $(\flrmreplyini, \allowbreak \sid, \allowbreak \tid, \allowbreak \flrmmessageresponse, \allowbreak \{\rpid, \allowbreak \flrmGatewayexitreply', \allowbreak \flrmNodeAresponse'\})$ to $\Functionality_{\LRM}$, where $\tid$ is in the tuple $(\tid, \allowbreak \flrmParty, \allowbreak h_{i}, \allowbreak \surb, \allowbreak \surbsecrets, \allowbreak \flrmlistleakage)$,   $\flrmGatewayexitreply'$ is in $\sid_{\SMT}$ and $\flrmNodeAresponse'$ is equal to $ \flrmNodeAresponse$ in $\langle \packet, \allowbreak \flrmNodeAresponse \rangle$. In the forward interface, when the functionality $\Functionality_{\LRM}$ informs $\Simulator$ that the honest gateway $\flrmGatewayexitreply'$ forwards the packet to the corrupt first-layer node $\flrmNodeAresponse$, $\Functionality_{\LRM}$ discloses $\rpid$ and the simulator $\Simulator$  uses the stored tuple $(\rpid, \allowbreak \packet)$ to retrieve the packet $\packet$ that should be sent to the adversary $\Adversary$.

        \end{itemize}

        \item[Single-use reply block computed by a corrupt party.] If the simulator  $\Simulator$ does not store a tuple $(\tid, \allowbreak \flrmParty, \allowbreak h_{i}, \allowbreak \surb, \allowbreak \surbsecrets, \allowbreak \flrmlistleakage)$ such that $\surb \allowbreak = \allowbreak (h'_{0}, \allowbreak s_k^{p})$ and $h'_{0} \allowbreak = \allowbreak h_0$, $\Simulator$ continues with the case ``Honest gateway $\flrmGatewayentry$ receives a packet from a corrupt sender $\flrmSender$''. We note that, if such a tuple is not stored, then the packet was computed by the adversary by using a single-use reply block generated by a corrupt party. Then $\Simulator$ cannot distinguish whether this is a reply packet computed by a corrupt receiver on input a single-use reply block, or whether this is a request packet computed a corrupt sender. (When running ``Honest gateway $\flrmGatewayentry$ receives a packet from a corrupt sender $\flrmSender$'', the simulator $\Simulator$ could find out that the packet encrypts a single-use reply block, and thus is a request packet. However, $\Simulator$ cannot distinguish a request packet that does not enable replies from a reply packet.) In our security proof in~\S\ref{sec:securityproof}, we show indistinguishability between request packets and reply packets. We remark that, in our construction, the processing of request packets and reply packets by gateways and nodes is the same, so in~\S\ref{sec:securityproof} we can prove indistinguishability between the real-world protocol and our simulation.
        \end{description}

\end{description}

%% file: 7SecAnalysis15SimulatorForward.tex
\subsubsection{Simulation of the Forward Interface} 
\label{sec:simulatorForward}

We describe the simulator $\Simulator$ for the forward interface.
\begin{description}

    \item[Honest node or gateway $\flrmParty$ forwards a packet.]   When the ideal functionality $\Functionality_{\LRM}$ sends the message $(\flrmforwardsim, \allowbreak \sid, \allowbreak \qid, \allowbreak \flrmParty, \allowbreak \flrmroutelist[\flrmindex+1])$, $\Simulator$ sets $\flrmParty' \allowbreak \gets \allowbreak \flrmroutelist[\flrmindex+1]$ and creates a random string $\flrmstring$ of the same length as the message $\packet$. $\Simulator$ sets $\sid_{\SMT} \allowbreak \gets \allowbreak (\flrmParty, \allowbreak \flrmParty', \allowbreak \sid)$ and runs a copy of the functionality $\Functionality_{\SMT}$ on input  $(\fsmtsendini, \allowbreak \sid_{\SMT}, \allowbreak \flrmstring)$. When $\Functionality_{\SMT}$ sends $(\fsmtsendsim, \allowbreak \sid_{\SMT}, \allowbreak \ssid, \allowbreak \SMTfleakage(\flrmstring))$, $\Simulator$ forwards that message to $\Adversary$.

    \item[Honest gateway, node or user $\flrmParty'$ receives a packet.] When   $\Adversary$ sends the message $(\fsmtsendrep, \allowbreak \sid_{\SMT}, \allowbreak \ssid)$, $\Simulator$ runs $\Functionality_{\SMT}$ on input that message. When $\Functionality_{\SMT}$ outputs $(\fsmtsendend, \allowbreak \sid_{\SMT}, \allowbreak \flrmstring)$,  $\Simulator$ sends  $(\flrmforwardrep, \allowbreak \sid, \allowbreak \qid)$ to $\Functionality_{\LRM}$.

    \item[Corrupt sender $\flrmParty'$ receives a reply packet.] When  $\Functionality_{\LRM}$ sends the message $(\flrmforwardend, \allowbreak \sid, \allowbreak \flrmposition, \allowbreak \flrmmessagerequest', \allowbreak \ppid')$, the simulator $\Simulator$ retrieves the stored tuple $(\ppid, \allowbreak \surb)$ such that $\ppid \allowbreak = \allowbreak \ppid'$. Then $\Simulator$ runs the procedure ``$\Simulator$ creates a reply packet'' in~\S\ref{sec:simulatorRequest} on input the single-use reply block $\surb$ and the message $\flrmmessagerequest'$ to obtain a reply packet $\packet$. After that, $\Simulator$ runs the procedure ``$\Simulator$ processes a reply packet'' in~\S\ref{sec:simulatorRequest} until reaching the step ``Honest entry gateway $\flrmGatewayentryreply$ forwards the packet''. When, in that step, the copy of $\flrmGatewayentryreply$ sends the message $(\fsmtsendini, \allowbreak \sid_{\SMT}, \allowbreak \packet')$, the simulator $\Simulator$     sends $(\fsmtsendend, \allowbreak \sid_{\SMT}, \allowbreak \packet')$ to the adversary $\Adversary$.

    Basically, the simulator retrieves a single-use reply block $\surb$ previously stored when a corrupt party inputs a request packet that enables a reply. Then $\Simulator$ uses $\surb$ and the message $\flrmmessagerequest'$ to simulate the reply packet computed by an honest receiver using $\surb$ and $\flrmmessagerequest'$. After that,  $\Simulator$ processes the reply packet by removing all the layers of encryption until leaving only the one corresponding to the corrupt sender $\flrmParty'$.

    We remark that $\Functionality_{\LRM}$ only sends  $(\flrmforwardend, \allowbreak \sid, \allowbreak \flrmposition, \allowbreak \flrmmessagerequest', \allowbreak \ppid')$ to a corrupt user when a corrupt party sends a request packet to an honest receiver, that request packet enables a reply, the honest receiver sends a reply packet associated to that request packet, and the reply packet is received by a corrupt user. The reason is the following:
    \begin{itemize}

        \item For  request packets that enable a reply computed by honest senders, the destination of the reply packet is an honest sender. Therefore, reply packets computed by corrupt receivers by using single-use reply blocks created by honest senders are not received by corrupt users.

        \item Reply packets computed by corrupt parties on input single-use reply blocks generated by corrupt parties are processed as request packets by the functionality. We recall that $\Simulator$ cannot distinguish those packets from request packets that do not enable replies.

    \end{itemize}

    \item[Corrupt receiver $\flrmParty'$ receives a request packet.] When the ideal functionality $\Functionality_{\LRM}$ sends the message $(\flrmforwardend, \allowbreak \sid, \allowbreak \flrmposition, \allowbreak \flrmmessagerequest, \allowbreak \tid, \allowbreak \flrmGatewayexitreply', \allowbreak \flrmNodeAresponse', \allowbreak \ppid', \allowbreak \flrmlistleakage)$, $\Simulator$ proceeds as follows:
    \begin{itemize}

        \item If $\ppid' \allowbreak \neq \allowbreak \bot$, then $\Simulator$ finds the tuple $(\ppid, \allowbreak \packet, \allowbreak \flrmParty')$ such that $\ppid \allowbreak = \allowbreak \ppid'$, sets $\sid_{\SMT} \allowbreak \gets \allowbreak (\flrmposition, \allowbreak \flrmParty', \allowbreak \sid)$ and  sends  $(\fsmtsendend, \allowbreak \sid_{\SMT}, \allowbreak \packet)$ to the adversary $\Adversary$. We recall that $\Functionality_{\LRM}$ outputs $\ppid$ to a corrupt receiver when the packet was input to the functionality by a corrupt party. In that case, the simulator stores a tuple $(\ppid, \allowbreak \packet, \allowbreak \flrmParty')$ when running the procedure ``$\Simulator$ processes a packet'' in~\S\ref{sec:simulatorRequest}, where $\packet$ is the packet that should be forwarded to the corrupt receiver and is obtained by running copies of the honest parties that should process the packet that was input by the adversary.

        \item If $\ppid' \allowbreak = \allowbreak \bot$, $\Simulator$ computes a packet $\packet$ by running a procedure similar to the one described in ``Corrupt gateway $\flrmGatewayentry$ receives a packet'' in~\S\ref{sec:simulatorRequest}. In that procedure, $\Simulator$ runs a copy of an honest sender to compute a packet. $\Simulator$ uses the data in $\flrmlistleakage$ to construct part of the packet route, whereas the parties in the route that are not contained in $\flrmlistleakage$ are set to random honest parties by the simulator. Moreover, if $\Functionality_{\LRM}$ discloses the message, $\Simulator$ gives it as input to the copy of the honest sender, and otherwise uses a random message.

        Here, $\Simulator$ computes a packet $\packet$ with just one layer of encryption, which is the layer that should be removed by the corrupt receiver. To compute this packet, $\Simulator$ uses as input the message $\flrmmessagerequest$ sent by $\Functionality_{\LRM}$. Additionally, if $\flrmGatewayexitreply' \allowbreak \neq \allowbreak \bot$ and  $\flrmNodeAresponse' \allowbreak \neq \allowbreak \bot$, $\Simulator$ also computes a single-use reply block $\surb$ and encrypts it along the message $\flrmmessagerequest$. $\surb$ is computed on input a packet route the contains $\flrmNodeAresponse'$ and any other party identifiers in  $\flrmlistleakage$, whereas the part of the route not contained in  $\flrmlistleakage$ is filled with random honest parties from the correct sets. Once $\Simulator$ computes the packet $\packet$, $\Simulator$ sets $\sid_{\SMT} \allowbreak \gets \allowbreak (\flrmposition, \allowbreak \flrmParty', \allowbreak \sid)$ and sends $(\fsmtsendend, \allowbreak \sid_{\SMT}, \allowbreak \packet)$ to the adversary $\Adversary$.

        We recall that, when $\ppid' \allowbreak = \allowbreak \bot$, the packet was computed by an honest sender. $\Simulator$ uses the information in $\flrmlistleakage$ and the message $\flrmmessagerequest$  to compute a packet $\packet$ that is indistinguishable from the one computed by the honest sender.

    \end{itemize}

    \item[Corrupt node or gateway $\flrmParty'$ receives a packet.]  When the ideal functionality $\Functionality_{\LRM}$ sends   $(\flrmforwardend, \allowbreak \sid, \allowbreak \tid, \allowbreak \ppid', \allowbreak \flrmmessagerequest', \allowbreak \allowbreak \flrmposition, \allowbreak \flrmroutelist[\flrmindex+2], \flrmlistleakage, \allowbreak \rpid')$, $\Simulator$ proceeds as follows:
    \begin{itemize}

        \item If $\rpid' \allowbreak \neq \allowbreak \bot$, $\Simulator$ finds the tuple $(\rpid, \allowbreak \packet)$ such that $\rpid \allowbreak = \allowbreak \rpid'$, sets $\sid_{\SMT} \allowbreak \gets \allowbreak (\flrmposition, \allowbreak \flrmParty', \allowbreak \sid)$ and sends the message $(\fsmtsendend, \allowbreak \sid_{\SMT}, \allowbreak \packet)$ to the adversary $\Adversary$. 

        The tuple $(\rpid, \allowbreak \packet)$ is stored by $\Simulator$ in the reply interface when a corrupt receiver sends to an honest gateway $\flrmGatewayexitreply$ a reply packet computed by using a single-use reply block from an honest sender. $\Functionality_{\LRM}$ reveals $\rpid$ to a corrupt first-layer node $\flrmNodeAresponse$.

        \item If $\ppid \allowbreak \neq \allowbreak \bot$, $\Simulator$ proceeds as follows:
        \begin{itemize}
            
            \item If there is a tuple $(\ppid, \allowbreak \surb)$ such that $\ppid \allowbreak = \allowbreak \ppid'$, $\Simulator$ runs  the procedure ``$\Simulator$ creates a reply packet'' in~\S\ref{sec:simulatorRequest} on input the single-use reply block $\surb$ and the message $\flrmmessagerequest'$ to obtain a reply packet $\packet$. After that, $\Simulator$ runs the procedure ``$\Simulator$ processes a reply packet'' in~\S\ref{sec:simulatorRequest} until reaching the step in which a message $(\fsmtsendini, \allowbreak \sid_{\SMT}, \allowbreak \packet')$ is sent to the corrupt node or gateway $\flrmParty'$. Then the simulator $\Simulator$ sends $(\fsmtsendend, \allowbreak \sid_{\SMT}, \allowbreak \packet')$ to the adversary $\Adversary$.

             Here the simulator retrieves a single-use reply block $\surb$ previously stored when a corrupt party inputs a request packet that enables a reply. Then $\Simulator$ uses $\surb$ and the message $\flrmmessagerequest'$ to simulate the reply packet computed by an honest receiver using $\surb$ and $\flrmmessagerequest'$. After that,  $\Simulator$ processes the reply packet by removing the layers of encryption until the one corresponding to the corrupt party $\flrmParty'$.

             We remark that, if a tuple $(\ppid, \allowbreak \surb)$ such that $\ppid \allowbreak = \allowbreak \ppid'$ is stored, the corrupt node or gateway $\flrmParty'$ is receiving a reply packet computed by an honest receiver to reply to a request packet generated by a corrupt party. $\Simulator$ only stores  $(\ppid, \allowbreak \surb)$ when it can retrieve a single-use reply block from a request packet sent by a corrupt party to an honest receiver. This implies that parties in the route of the request packet (after the corrupt party that sent it) are all honest, and thus $\Functionality_{\LRM}$ does not interact with the simulator regarding this packet until the corresponding reply packet is produced (if there are corrupt parties in the route of the reply).

             \item Else, $\Simulator$ finds the tuple $(\ppid, \allowbreak \packet, \allowbreak \flrmParty')$ such that $\ppid \allowbreak = \allowbreak \ppid'$, sets $\sid_{\SMT} \allowbreak \gets \allowbreak (\flrmposition, \allowbreak \flrmParty', \allowbreak \sid)$ and  sends the message $(\fsmtsendend, \allowbreak \sid_{\SMT}, \allowbreak \packet)$ to the adversary $\Adversary$. 

             In this case, the party $\flrmParty'$ receives a request or reply packet computed by another corrupt party after being forwarded by honest parties in the route. (In the case of a reply packet, it is computed on input a single-use reply block created by a corrupt party, so $\Simulator$ cannot distinguish it from a request packet that does not enable replies, and  $\Functionality_{\LRM}$ processes it as a request packet.) The tuple $(\ppid, \allowbreak \packet, \allowbreak \flrmParty')$ was stored by $\Simulator$ when processing the packet sent by the corrupt party, and $\packet$ is the packet that should be sent to the corrupt party $\flrmParty'$ after the processing done by honest parties.

        \end{itemize}

           \item If $\ppid \allowbreak = \allowbreak \bot$, $\Simulator$ computes a packet $\packet$ by running a procedure similar to the one described in ``Corrupt gateway $\flrmGatewayentry$ receives a packet'' in~\S\ref{sec:simulatorRequest}. In that procedure, $\Simulator$ runs a copy of an honest sender to compute a packet. $\Simulator$ uses the data in $\flrmlistleakage$ to construct part of the packet route, whereas the parties in the route that are not contained in $\flrmlistleakage$ are set to random honest parties by the simulator. Moreover, if $\Functionality_{\LRM}$ discloses the message, $\Simulator$ gives it as input to the copy of the honest sender, and otherwise uses a random message.

             Here, $\Simulator$ computes a packet $\packet$ with a number of layers of encryption that corresponds to the position of $\flrmParty'$ in a packet route, i.e. $2$ layers if $\flrmParty' \allowbreak \in \allowbreak \flrmSetW$, $3$ layers if $\flrmParty' \allowbreak \in \allowbreak \flrmSetC$, $4$ layers if $\flrmParty' \allowbreak \in \allowbreak \flrmSetB$, and $5$ layers if $\flrmParty' \allowbreak \in \allowbreak \flrmSetA$.  To compute this packet, $\Simulator$ uses as packet route the information contained in the leakage list $\flrmlistleakage$. If $\flrmlistleakage$ contains the identifier of a corrupt user, then $\flrmmessagerequest' \allowbreak \neq \allowbreak \bot$ in the message sent by the functionality and $\Simulator$ uses it as input to compute the packet, else $\Simulator$ uses a random message. When there are party identifiers in $\flrmlistleakage$ after the corrupt user, $\Simulator$ also computes a single-use reply block $\surb$ and encrypts it along the message $\flrmmessagerequest$. $\surb$ is computed on input a packet route the contains the party identifiers in  $\flrmlistleakage$, whereas the part of the route not contained in  $\flrmlistleakage$ is filled with random honest parties from the correct sets. Once $\Simulator$ computes the packet $\packet$, $\Simulator$ sets $\sid_{\SMT} \allowbreak \gets \allowbreak (\flrmposition, \allowbreak \flrmParty', \allowbreak \sid)$ and sends $(\fsmtsendend, \allowbreak \sid_{\SMT}, \allowbreak \packet)$ to the adversary $\Adversary$.

             We recall that, when $\ppid' \allowbreak = \allowbreak \bot$, the packet was computed by an honest sender. $\Simulator$ uses the information in $\flrmlistleakage$ and, if given, the message $\flrmmessagerequest$  to compute a packet $\packet$ that is indistinguishable from the one computed by the honest sender. We recall that $\flrmlistleakage$ ends either with a corrupt user, or with the identifier of an honest party that processes the packet. The adversary is not able to find out the random parties (if any) that $\Simulator$ chooses to complete the packet route because, for that, the adversary would need to remove a layer of encryption associated with an honest party. 
 
    \end{itemize}

    \item[Corrupt node or gateway $\flrmParty$ forwards a packet.] When the adversary $\Adversary$ sends the message $(\fsmtsendini, \allowbreak \sid_{\SMT}, \allowbreak \packet)$, $\Simulator$ runs the copy of $\Functionality_{\SMT}$ on input that message. When $\Functionality_{\SMT}$ sends the message $(\fsmtsendsim, \allowbreak \sid_{\SMT}, \allowbreak \ssid, \allowbreak \SMTfleakage(\packet))$,  $\Simulator$ forwards that message to $\Adversary$.
    
    \item[Honest node, gateway or user $\flrmParty$ receives a packet.] When $\Adversary$ \newline sends the message $(\fsmtsendrep, \allowbreak \sid_{\SMT}, \allowbreak \ssid)$, $\Simulator$ runs the copy of $\Functionality_{\SMT}$ on input that message. When $\Functionality_{\SMT}$ sends the message $(\fsmtsendend, \allowbreak \sid_{\SMT}, \allowbreak \langle \packet, \allowbreak \flrmNodeArequest \rangle)$, $\Simulator$ proceeds as in the procedure ``$\Simulator$ processes a packet'' in~\S\ref{sec:simulatorRequest}.

\end{description}

%% file: 7SecAnalysis20SecurityProof.tex
\subsection{Security Proof}
\label{sec:securityproof}

\newcounter{gameKDF}
\setcounter{gameKDF}{1}
\newcounter{gameKDFsurb}
\setcounter{gameKDFsurb}{2}
\newcounter{gameAEADunf}
\setcounter{gameAEADunf}{3}
\newcounter{gameAEADcommitting}
\setcounter{gameAEADcommitting}{4}
\newcounter{gamepayloadintegrity}
\setcounter{gamepayloadintegrity}{5}
\newcounter{gamepayloadintegrityreply}
\setcounter{gamepayloadintegrityreply}{6}
\newcounter{gameAEADindcpa}
\setcounter{gameAEADindcpa}{7}
\newcounter{gameAEADindcpasurb}
\setcounter{gameAEADindcpasurb}{8}
\newcounter{gameSEindcca}
\setcounter{gameSEindcca}{9}
\newcounter{gameSEindccasurb}
\setcounter{gameSEindccasurb}{10}
\newcounter{gameSEindccasurbprocess}
\setcounter{gameSEindccasurbprocess}{11}
\newcounter{gamefailureheader}
\setcounter{gamefailureheader}{12}
\newcounter{gamerequestreplacement}
\setcounter{gamerequestreplacement}{13}

\newcounter{gamereverse}
\setcounter{gamereverse}{16}

\paragraph{Intuition.} The security proof shows that the environment cannot distinguish construction $\mathrm{\Pi}_{\LRM}$  from the ideal-world protocol defined by $\Functionality_{\LRM}$. The view of the environment consists fundamentally of two parts: (1) the outputs of honest parties, and (2) the messages received from the adversary. Therefore, to prove indistinguishability between $\mathrm{\Pi}_{\LRM}$ and the ideal-world protocol defined by $\Functionality_{\LRM}$ we need to prove two things: (1) that the outputs of honest parties in the real and in the ideal world are indistinguishable, and (2) that the messages received by the environment from the adversary in the real and in the ideal world are indistinguishable.

Our simulator $\Simulator$ runs a copy of any real-world adversary $\Adversary$. To guarantee (1), when the copy of the adversary $\Adversary$ sends a packet to honest parties, our simulator processes it to extract the input needed by the functionality. Extraction is successful unless in the cases in which our simulator outputs failure. Once the functionality receives that input, it sends to honest parties in the ideal world the same outputs that honest parties compute in construction $\mathrm{\Pi}_{\LRM}$. We remark that the outputs of honest parties consist also of abortion messages, and our functionality and simulator are designed in such a way that the abortion messages output by honest parties are the same both in construction $\mathrm{\Pi}_{\LRM}$  and in the ideal-world protocol defined by $\Functionality_{\LRM}$. The cases in which our simulator outputs failure imply that our simulator is not able to provide correct input to $\Functionality_{\LRM}$.

To guarantee (2),  the simulator $\Simulator$ forwards the messages sent by the copy of $\Adversary$ to the environment. To ensure that those messages are indistinguishable from those that the copy of $\Adversary$ sends in construction $\mathrm{\Pi}_{\LRM}$, our simulator, given the input received from the ideal functionality $\Functionality_{\LRM}$, needs to produce a view towards the adversary that is indistinguishable from the view that the adversary gets when construction $\mathrm{\Pi}_{\LRM}$ is run.

Consequently, to prove (1), we need to prove that our simulator fails with negligible probability given any message sent by the adversary to honest parties. To prove (2), we need to prove that our simulator produces a view towards the adversary that is indistinguishable from the view the adversary gets when construction $\mathrm{\Pi}_{\LRM}$ is run.

In the proof, we use a list of games. The first game ($\Gam 0$) is $\mathrm{\Pi}_{\LRM}$. The next games modify $\mathrm{\Pi}_{\LRM}$ step by step, until the last game ($\Gam \thegamereverse$), which is indistinguishable from our simulator. For each game $\Gam i$ such that $i \allowbreak \in \allowbreak [1, \allowbreak 16]$, we prove that it is indistinguishable from $\Gam i-1$. 

In some of the games, some modifications are performed to the layers of encryption of request packets and single-use reply blocks that fulfill the following conditions:
    \begin{enumerate}

        \item The request packet or single-use reply block in which that layer of encryption is included is computed by an honest sender.

        \item The layer of encryption should be processed by an honest party.

        \item The layer of encryption is received by a corrupt party from an honest party or obtained by a corrupt party after processing a packet or single-use reply block.
        
    \end{enumerate}  
We refer to those conditions as conditions (1), (2) and (3).

The list of games can be summarized as follows:
\begin{itemize}

    \item In $\Gam \thegameKDF$, when a layer of encryption that fulfills conditions (1), (2) and (3) in a request packet is computed, the output of the key derivation function $\KDF$ is replaced by a random string of the same length. In $\Gam \thegameKDFsurb$, the same is done when a layer of encryption that fulfills conditions (1), (2) and (3) in a single-use reply block is computed. Indistinguishability between those games and their previous games is based on the fact that our construction uses a $\KDF$ that is secure with respect to the source of keying material IND-CCA $\KEM$.

    $\Gam \thegameKDF$ and $\Gam \thegameKDFsurb$ allow us to prove that the $\KEM$ ciphertext included in the header of every layer of encryption that fulfills conditions (1), (2) and (3) does not reveal any information about the secret keys used to compute the $\AEAD$ ciphertext and the payload. Thanks to that, subsequently we can use the security properties of the $\AEAD$ scheme and of the symmetric-key encryption scheme used to compute the payload in order to prove indistinguishability between the next games. 

    \item $\Gam \thegameAEADunf$ outputs failure when the adversary is able to replace the $\AEAD$ ciphertext by another valid $\AEAD$ ciphertext in the header of a layer of encryption that fulfills conditions (1), (2) and (3). Indistinguishablity between this game and the previous one is guaranteed by the unforgeability property of the $\AEAD$ scheme.


    We recall that the $\AEAD$ scheme encrypts the header of the next layer of encryption and the next party in the packet route. Therefore, this game is useful to show that the adversary cannot modify the header of a packet computed by an honest sender in such a way that the packet route is modified.

    \item $\Gam \thegameAEADcommitting$ outputs failure when the adversary is able to replace the $\KEM$ ciphertext and/or the associated data in the header of a layer of encryption that fulfills conditions (1), (2) and (3) in such a way that the decryption of the $\AEAD$ ciphertext does not fail. Indistinguishability between this game and the previous one is guaranteed by the committing property of the $\AEAD$ scheme.


    $\Gam \thegameAEADcommitting$, together with $\Gam \thegameAEADunf$, guarantee that, if the adversary modifies the header of a packet computed by an honest user, the packet processing fails. Additionally, it guarantees that, if the packet is processed by an honest party different from the one that should process it, then processing fails.
    
    \item $\Gam \thegamepayloadintegrity$ outputs failure when the payload of a request packet computed by an honest sender is modified by the adversary, and yet the payload processing by an honest receiver does not output $\bot$. $\Gam \thegamepayloadintegrityreply$ operates similarly for reply packets computed by honest receivers on input single-use reply blocks generated by honest senders.  $\Gam \thegamepayloadintegrity$ and  $\Gam \thegamepayloadintegrityreply$ also output failure when the payload processing does not output $\bot$ but the honest user that processes the payload is not the one that should process it. Indistinguishability between those games and the previous ones are guaranteed by the integrity property of the encryption scheme used to encrypt the request or the reply messages.

    $\Gam \thegamepayloadintegrity$ and $\Gam \thegamepayloadintegrityreply$ guarantee that, if the adversary modifies or replaces the payload of a packet, or if the processing of the payload is done by the wrong user, then decryption fails. This ensures that the adversary cannot modify the request or reply messages encrypted in a packet.

    \item In $\Gam \thegameAEADindcpa$, when a layer of encryption that fulfills conditions (1), (2) and (3) in a request packet is computed, the output of the algorithm $\AEADE$ is replaced by a random string of the same length. In $\Gam \thegameAEADindcpasurb$, the same is done when a layer of encryption that fulfills conditions (1), (2) and (3) in a single-use reply block is computed. Indistinguishablity between those games and their previous games is based on the fact that our construction uses an IND\$-CPA $\AEAD$ scheme.

    $\Gam \thegameAEADindcpa$ and $\Gam \thegameAEADindcpasurb$ allow us to prove that the $\AEAD$ ciphertext included in the header of every layer of encryption that fulfills conditions (1), (2) and (3) does not reveal any information about the header of the next layer of encryption or about the packet route.

    \item In $\Gam \thegameSEindcca$, when a layer of encryption that fulfills conditions (1), (2) and (3) in a request packet is computed, the output of the algorithm $\blockcipherenc$ is replaced by a random string of the same length. In $\Gam \thegameSEindccasurb$, the same is done when an honest receiver computes a reply packet on input a single-use reply block computed by an honest sender. In $\Gam \thegameSEindccasurbprocess$, when an honest party processes a layer of encryption of a reply packet computed on input a single-use reply block generated by an honest sender, and that layer is received by the adversary, the output of the algorithm $\blockcipherdec$ is replaced by a random string of the same length.  Indistinguishablity between those games and their previous games is based on the IND\$-CCA property of the $\blockcipher$ scheme.

    $\Gam \thegameSEindcca$, $\Gam \thegameSEindccasurb$ and $\Gam \thegameSEindccasurbprocess$ allow us to prove that the payloads generated by honest parties do not reveal any information about their encrypted content. 

    \item In $\Gam \thegamefailureheader$, all the headers of layers of encryption in a request packet or single-use reply block that are encrypted under a layer of encryption that fulfills conditions (1), (2) and (3) are stored. Then, if the adversary sends a packet that contains one of those headers, and the adversary did not receive it from an honest party, $\Gam \thegamefailureheader$ outputs failure.

    The probability that $\Gam \thegamefailureheader$ outputs failure is negligible because, in $\Gam \thegamefailureheader$, a layer of encryption that fulfills conditions (1), (2) and (3) does not reveal any information about the layers of encryption under it. The reason is that both the $\AEAD$ ciphertext and the payload have been replaced by random strings.

    \item  In $\Gam \thegamerequestreplacement$, when an honest party processes and forwards to a corrupt party a packet computed by an honest sender, the packet is replaced by a new packet. This new packet is computed on input the same data used by the honest sender. However, the positions in the packet route before the corrupt party are disregarded, i.e.\ the outermost layer of encryption in the new packet is the one associated with that corrupt party. Indistinguishability is again based on the fact that a layer of encryption that fulfills conditions (1), (2) and (3) does not reveal any information about the layers of encryption under it, which allows us to replace the packet obtained after processing one of those layers of encryption by a new packet. Additionally, we note that a layer of encryption does not contain any information about layers of encryption above it, and thus a packet computed by disregarding the positions in the packet route before the corrupt party is indistinguishable from the packet obtained after processing the packet computed by the sender.
    
\end{itemize}

\begin{theorem} \label{th:all}

    When a subset of users in the set $\flrmSetU$, a subset of first-layer nodes in the set $\flrmSetA$, a subset of second-layer nodes in the set $\flrmSetB$, a subset of third-layer nodes in the set $\flrmSetC$, and a subset of gateways in the set $\flrmSetW$ are corrupt, $\mathrm{\Pi}_{\LRM}$ securely realizes $\Functionality_{\LRM}$ in the $(\Functionality_{\SMT}, \allowbreak \Functionality_{\Freg}, \allowbreak \Functionality_{\Fpreg})$-hybrid model if the key derivation function $\KDF$ is secure with respect to the source of keying material IND-CCA $\KEM$, if the symmetric-key encryption scheme $\blockcipher$ is IND\$-CCA secure, if the authenticated encryption with associated data $\AEAD$ scheme is IND\$-CPA secure, unforgeable and committing, and if the encryption scheme used to encrypt request and reply messages at the application level provides integrity.

\end{theorem}

\paragraph{Proof of Theorem~\ref{th:all}.}
We show by means of a series of hybrid games that the environment $\Environment$ cannot distinguish between the ensemble $\RealEnsemble_{\mathrm{\Pi}_{\LRM},\Adversary,\Environment}$ and the ensemble $\IdealEnsemble_{\Functionality_{\LRM},\Simulator,\Environment}$ with non-negligible probability. We denote by $\Prob[\Gam i]$ the probability that the environment distinguishes $\Gam i$ from the real-world protocol.

\begin{description}

\newcounter{gamect}
\setcounter{gamect}{0}

    \item[$\Gam \thegamect$:] This game corresponds to the execution of the real-world protocol. Therefore, $\Prob[\Gam 0] = 0$.

    \stepcounter{gamect}
    \newcounter{prgamect}
    \setcounter{prgamect}{0}

    \item[$\Gam \thegamect$:] This game proceeds as $\Gam \theprgamect$, except that, in $\Gam \thegamect$, when a request packet $\packet$ is computed, for each layer $i$ that fulfills conditions (1), (2) and (3), the computation of the packet with algorithm $\lrmPacketCreate$ is modified as follows. The output $(s^{h}, \allowbreak s^{p})$ of the $\KDF$ is replaced by two random strings $s^{h} \allowbreak \gets \allowbreak \{0,1\}^{l_h}$ and $s^{p} \allowbreak \gets \allowbreak \{0,1\}^{l_p}$, where $l_h$ and $l_p$ are the bit lengths of $s^{h}$ and $s^{p}$ respectively. Moreover, when a packet is processed, for each layer $i$ that fulfills conditions (1), (2) and (3), the processing with algorithm $\lrmPacketProcess$ is modified as follows: The output of the $\KDF$ is replaced by the strings $(s^{h}, \allowbreak s^{p})$ that were used to compute the layer of encryption. The probability that the environment distinguishes between $\Gam \thegamect$ and $\Gam \theprgamect$ is bounded by the following theorem.
    \begin{theorem} \label{th:KDF:secure}

        Under the security of the $\KDF$ with respect to a source of keying material IND-CCA $\KEM$, we have that $|\Pr[\Gam \thegamect]-\Pr[\Gam \theprgamect]| \allowbreak \leq \allowbreak \Adv_{\Adversary}^{\KDF, \mathsf{Sec\ wrt\ IND-CCA\ \KEM}} \cdot N_{r}$, where $N_r$ is the number of layers of encryption of request packets that fulfill conditions (1), (2) and (3).

    \end{theorem}

    \paragraph{Proof of Theorem~\ref{th:KDF:secure}.}  The proof uses a sequence of games $\Gam \theprgamect.j.i$, for $j=1$ to $N_h$, and for $i=0$ to $N_j$, where $N_h$ is the number of honest parties, and $N_j$ is the number of layers of encryption of request packets that fulfill conditions (1), (2) and (3) that should be processed by the honest party number $j$. (Therefore, $N_r \allowbreak = \allowbreak \sum_{j=0}^{N_h} N_j$.) $\Gam \theprgamect.1.0$ is equal to $\Gam \theprgamect$, whereas $\Gam \theprgamect.N_h.N_{N_h}$ is equal to $\Gam \thegamect$. In $\Gam \theprgamect.j.i$, the first $i$ layers of encryption of request packets that fulfill conditions (1), (2) and (3) and should be processed by honest party $j$ use two random strings to set the keys $(s^{h}, \allowbreak s^{p})$, whereas the $N_j-i$ remaining ones are set as in $\Gam \theprgamect$, i.e., the output of $\KDF$ is used to set $(s^{h}, \allowbreak s^{p})$. Moreover, in $\Gam \theprgamect.j.i$, in the processing of the first $i$ layers of encryption that fulfill conditions (1), (2) and (3) and should be processed by honest party $j$, the output of the $\KDF$ is replaced  by the strings $(s^{h}, \allowbreak s^{p})$ that were used to compute the layer of encryption, whereas it is not replaced in the processing of the remaining $N_j-i$ layers of encryption.
    
    Given an adversary that is able to distinguish the game $\Gam \theprgamect.j.i$ from the game $\Gam \theprgamect.j.(i+1)$ with non-negligible probability $\tau$, we construct an algorithm $B$ that breaks the security of the $\KDF$ with respect to a source of keying material IND-CCA $\KEM$  with non-negligible probability $\tau$. 
    
    $B$ receives from the challenger of the secure $\KDF$ w.r.t.\ IND-CCA $\KEM$ game the tuple $(\KEMpublickey, \allowbreak \KEMciphertext^{\ast}, \allowbreak s_b)$. The algorithm $B$ proceeds as in $\Gam \theprgamect.j.i$ to compute and process the first $i$ layers of encryption that fulfill conditions (1), (2) and (3), and as in $\Gam \theprgamect.j.(i+1)$ to compute and process the last $N-i-1$ layers. $B$ simulates the honest parties towards the adversary as in $\Gam \theprgamect.j.i$, with the following changes.
    \begin{itemize}

        \item $B$ uses the public key $\KEMpublickey$ to set the public key of the honest party  number $j$, whose identifier is $\flrmParty$. When the adversary requests the public key of that party through $\Functionality_{\Freg}$ (if $\flrmParty$ is a node or a gateway) or $\Functionality_{\Fpreg}$ (if $\flrmParty$ is a user), $B$ replies by sending $\KEMpublickey$.

        \item When the honest sender computes the packet in which the layer of encryption $i+1$ that should be processed by $\flrmParty$ is included, $B$ uses the challenge ciphertext $\KEMciphertext^{\ast}$ to set the $\KEM$ ciphertext of the header for that layer. Moreover, $B$ parses $s_b \allowbreak \in \allowbreak \{0,1\}^{l_h+l_p}$ as $(s^{h}, \allowbreak s^{p})$ and uses $(s^{h}, \allowbreak s^{p})$ as output of $\KDF$. After that, the computation of the packet proceeds as specified in algorithm $\lrmPacketCreate$.

        \item When the adversary sends a packet that includes a layer of encryption associated with $\flrmParty$, and that layer was not computed by $B$, to process that packet $B$ sends the $\KEM$ ciphertext of the header of that layer to the decryption oracle $\KEMOracleDec$ provided by the challenger of the secure $\KDF$ w.r.t. IND-CCA $\KEM$ game and obtains the key $\KEMsharedkey$ from the oracle, which allows $B$ to process the layer. 

        \item For any layers of encryption associated with $\flrmParty$ that are computed by $B$ and that subsequently need to be processed by $B$, $B$ does not need to use the decryption oracle $\KEMOracleDec$ because $B$ already knows the encrypted content. Therefore, if the packet that includes the layer $i+1$ in which $\KEMciphertext^{\ast}$ was used is sent by the adversary to $B$, $B$ can process and forward the packet without needing to use the decryption oracle $\KEMOracleDec$. 

    \end{itemize}
    When the adversary outputs its guess $b'$, the algorithm $B$ forwards $b'$ to the challenger of the secure $\KDF$ w.r.t. IND-CCA $\KEM$ game. As can be seen, when  the key  $s_b$ is computed by the challenger by running the algorithm $\KDF$ on input the shared key $\KEMsharedkey$ output by $\KEMEnc(\KEMpublickey)$, the view of the adversary is that of $\Gam \theprgamect.j.i$. When the key  $s_b$  is a random string in $\{0,1\}^{l_h+l_p}$ sampled randomly by the challenger, the view of the adversary is that of $\Gam \theprgamect.j.(i+1)$. Therefore, if the adversary distinguishes between $\Gam \theprgamect.j.i$ and $\Gam \theprgamect.j.(i+1)$ with non-negligible probability $\tau$, $B$ has non-negligible probability $\tau$ in winning the secure $\KDF$ w.r.t. IND-CCA $\KEM$ game.

    \stepcounter{gamect}
    \stepcounter{prgamect}

    \item[$\Gam \thegamect$:] This game proceeds as $\Gam \theprgamect$, except that, in $\Gam \thegamect$, when a single-use reply block $\surb$ is computed, for each layer $i$ that fulfills conditions (1), (2) and (3), the computation of the single-use reply block with algorithm $\lrmSurbCreate$ is modified as follows. The output $(s^{h}, \allowbreak s^{p})$ of the $\KDF$ is replaced by two random strings $s^{h} \allowbreak \gets \allowbreak \{0,1\}^{l_h}$ and $s^{p} \allowbreak \gets \allowbreak \{0,1\}^{l_p}$, where $l_h$ and $l_p$ are the bit lengths of $s^{h}$ and $s^{p}$ respectively.  Moreover, when a reply packet is processed, for each layer $i$ that fulfills conditions (1), (2) and (3), the processing with algorithm $\lrmPacketProcess$ is modified as follows. The output of the $\KDF$ is replaced by the strings $(s^{h}, \allowbreak s^{p})$ that were used to compute the layer of encryption. The probability that the environment distinguishes between $\Gam \thegamect$ and $\Gam \theprgamect$ is bounded by the following theorem.
    \begin{theorem} \label{th:KDF:secure:surb}

        Under the security of the $\KDF$ with respect to a source of keying material IND-CCA $\KEM$, we have that $|\Pr[\Gam \thegamect]-\Pr[\Gam \theprgamect]| \allowbreak \leq \allowbreak \Adv_{\Adversary}^{\KDF, \mathsf{Sec\ wrt\ IND-CCA\ \KEM}} \cdot N_s$, where $N_s$ is the number of layers of encryption of single-use reply blocks that fulfill conditions (1), (2) and (3).

    \end{theorem}

    \paragraph{Proof of Theorem~\ref{th:KDF:secure:surb}.} The proof is very similar to the proof of Theorem~\ref{th:KDF:secure}. We omit it.

    \stepcounter{gamect}
    \stepcounter{prgamect}

    \item[$\Gam \thegamect$:] This game proceeds as $\Gam \theprgamect$, except that, in $\Gam \thegamect$, when the adversary sends a packet with a layer of encryption $i$ that needs to be processed by an honest party $\flrmParty$, $\Gam \thegamect$ proceeds as follows. First, $\Gam \thegamect$ processes the layer of encryption $i$ by running $\lrmPacketProcess$. Then, if  $\lrmPacketProcess$ does not fail, $\Gam \thegamect$ checks whether it computed a layer of encryption that fulfills conditions (1), (2) and (3) and that should be processed by the same party $\flrmParty$. In that case, let $h_{i} \gets (\KEMciphertext_i \ || \  \beta_{i} \ || \ \gamma_{i})$ be the header of the layer of encryption computed by $\Gam \thegamect$, and let $h'_{i} \gets (\KEMciphertext'_i \ || \  \beta'_{i} \ || \ \gamma'_{i})$ be the header of the layer of encryption submitted by the adversary. If $\KEMciphertext_i \allowbreak = \allowbreak \KEMciphertext'_i$ but $\beta_{i} \allowbreak \neq \allowbreak \beta'_{i}$, $\Gam \thegamect$ outputs failure. The probability that $\Gam \thegamect$ outputs failure, which enables the environment to distinguish between  $\Gam \thegamect$ and $\Gam \theprgamect$, is bounded by the following claim.
    \begin{theorem} \label{th:AEAD:unforgeability}

        Under the unforgeability property of the $\AEAD$ scheme,  $|\Pr[\Gam \thegamect]-\Pr[\Gam \theprgamect]| \allowbreak \leq \allowbreak \Adv_{\Adversary}^{\AEAD, \mathsf{Unforgeability}} \cdot (N_r+N_s)$, where $N_r+N_s$ is the number of layers of encryption of request packets and single-use reply blocks that fulfill conditions (1), (2) and (3).

    \end{theorem}
    
    \paragraph{Proof of Theorem~\ref{th:AEAD:unforgeability}.} Given an adversary that makes $\Gam \thegamect$ abort with non-negligible probability $\tau$, we construct an algorithm $B$ that breaks the unforgeability property of the $\AEAD$ scheme with probability $\tau/(N_r+N_s)$. $B$ picks randomly  $i \allowbreak \in \allowbreak [1, \allowbreak N_r+N_s]$. When computing the layer of encryption $i$ by running $\lrmPacketCreate$ or $\lrmSurbCreate$, to set the $\AEAD$ ciphertext for the header of layer $i$, $B$ sets the message $\AEADmessage$, the nonce $\AEADnonce$ and the associated data $\AEADheader$ as specified in those algorithms and sends $(\AEADmessage, \allowbreak \AEADnonce, \allowbreak \AEADheader)$ to the encryption oracle $\AEADE(\AEADkey,\cdot, \cdot, \cdot)$ provided by the challenger of the $\AEAD$ unforgeability game. (We recall that, in the description of Outfox in~\ref{sec:constructionalgorithms}, the nonce is omitted.) $B$ receives a ciphertext $\AEADciphertext$, which $B$ uses as part of the header of layer $i$. When the adversary $\Adversary$ sends a packet with a layer of encryption that makes $\Gam \thegamect$ abort, $B$ checks whether that layer of encryption is $i$. If not, $B$ fails. Otherwise $B$ parses the header of the layer of encryption submitted by the adversary as $h'_{i} \gets (\KEMciphertext'_i \ || \  \beta'_{i} \ || \ \gamma'_{i})$ and sends $(\AEADnonce', \allowbreak \gamma'_{i}, \allowbreak \beta'_{i})$ to break the unforgeability property of the $\AEAD$ scheme, where $\AEADnonce'$ is the nonce for that header. 

    \stepcounter{gamect}
    \stepcounter{prgamect}

    \item[$\Gam \thegamect$:] This game proceeds as $\Gam \theprgamect$, except that, in $\Gam \thegamect$, when the adversary sends a packet with a layer of encryption $i$ that needs to be processed by an honest party $\flrmParty$, $\Gam \thegamect$ proceeds as follows. First, $\Gam \thegamect$ processes the layer of encryption $i$ by running $\lrmPacketProcess$. Then, if  $\lrmPacketProcess$ does not fail, $\Gam \thegamect$ checks if there is any computed layer of encryption that fulfills conditions (1), (2) and (3) such that the following holds. Let $h_{i} \gets (\KEMciphertext_i \ || \  \beta_{i} \ || \ \gamma_{i})$ be the header of the layer of encryption computed by $\Gam \thegamect$ and that should be processed by an honest party $\flrmParty'$, and let $h'_{i} \allowbreak \gets \allowbreak (\KEMciphertext'_i \ || \  \beta'_{i} \ || \ \gamma'_{i})$ be the header of the layer of encryption submitted by the adversary. If $\beta_{i} \allowbreak = \allowbreak \beta'_{i}$ but either $(\KEMciphertext_i, \allowbreak \gamma_{i}) \allowbreak \neq \allowbreak (\KEMciphertext'_i, \allowbreak \gamma'_{i})$ or $\flrmParty \allowbreak \neq \allowbreak \flrmParty'$, $\Gam \thegamect$ outputs failure. The probability that $\Gam \thegamect$ outputs failure, which enables the environment to distinguish between  $\Gam \thegamect$ and $\Gam \theprgamect$, is bounded by the following claim.
    \begin{theorem} \label{th:AEAD:committing}

        Under the committing property of the $\AEAD$ scheme,  $|\Pr[\Gam \thegamect]-\Pr[\Gam \theprgamect]| \allowbreak \leq \allowbreak \Adv_{\Adversary}^{\AEAD, \mathsf{Committing}}$.

    \end{theorem}

    \paragraph{Proof of Theorem~\ref{th:AEAD:committing}.}  Given an adversary that makes $\Gam \thegamect$ abort with non-negligible probability $\tau$, we construct an algorithm $B$ that breaks the committing property of the $\AEAD$ scheme with probability $\tau$. We follow the game described in Section 3 in~\cite{cryptoeprint:2022/1260}. In that game, first the procedure $\mathsf{Initialize}$ is run to generate multiple keys and initialize the sets of queries, of corrupted keys and of revealed keys. Then $B$ works as follows.
    \begin{itemize}

        \item $B$ initializes to $0$ a counter $i$ of the layers of encryption that fulfill conditions (1), (2) and (3). Let $N$ be the maximum number of such layers of encryption. $B$ initializes to $N$ a counter $j$ to count the number of keys retrieved from the adversary. 

        \item When running  $\lrmPacketCreate$ or $\lrmSurbCreate$, if a layer of encryption fulfills conditions (1), (2) and (3), $B$ proceeds as follows:
        \begin{itemize}
            
            \item $B$ increments the counter $i$.

            \item $B$ sets the message $\AEADmessage$, the nonce $\AEADnonce$ and the associated data $\AEADheader$ as specified in the algorithms $\lrmPacketCreate$ or $\lrmSurbCreate$. 

            \item $B$ sends $(i, \allowbreak \AEADnonce, \allowbreak \AEADheader, \allowbreak  \AEADmessage)$ to the procedure $\mathsf{Enc}(i, \allowbreak \AEADnonce, \allowbreak \AEADheader, \allowbreak  \AEADmessage)$. 

            \item $B$ receives a ciphertext $\AEADciphertext$ as output of $\mathsf{Enc}$ and uses $\AEADciphertext$ to set the ciphertext $\beta$ in the layer of encryption $i$.    

            \item $B$ stores the tuple $(i, \allowbreak \KEMciphertext, \allowbreak \beta, \allowbreak \gamma, \allowbreak \AEADnonce, \allowbreak \flrmParty)$, where $\KEMciphertext$ is the $\KEM$ ciphertext, $\gamma \allowbreak = \AEADheader$ and $\flrmParty$ is the honest party associated with the layer of encryption.
            
        \end{itemize}

        \item When running algorithm $\lrmPacketProcess$ for an honest party $\flrmParty$, if the layer of encryption with header $(\KEMciphertext, \allowbreak \beta, \allowbreak \gamma)$ and nonce $\AEADnonce$ contains a ciphertext $\beta$ that is equal to one of the outputs of the procedure $\mathsf{Enc}$, $B$ proceeds as follows:
        \begin{itemize}

            \item $B$ retrieves the tuple $(i, \allowbreak \KEMciphertext', \allowbreak \beta', \allowbreak \gamma', \allowbreak \AEADnonce', \allowbreak \flrmParty')$ such that $\beta' \allowbreak = \allowbreak \beta$.

            \item If $\KEMciphertext' \allowbreak \neq \allowbreak \KEMciphertext$, or if $\flrmParty' \allowbreak \neq \allowbreak \flrmParty$,  then $B$ does the following:
            \begin{itemize}

                \item $B$ decrypts $\KEMciphertext$ to obtain a shared key $\KEMsharedkey$.

                \item $B$ runs  $(s^{h}, \allowbreak s^{p}) \allowbreak \gets \allowbreak \KDF(\KEMsharedkey)$.

                \item $B$ increments $j$ and sends $(j, \allowbreak s^{h})$ to the procedure $\mathsf{Cor}$ to indicate that the key with index $j$ is corrupt.

                \item $B$ sends $(j, \allowbreak \AEADnonce, \allowbreak \gamma, \allowbreak \beta)$ to the procedure $\mathsf{Dec}$.

                \item $B$ uses the output $\AEADmessage$ of  $\mathsf{Dec}$ as the output of algorithm $\AEADD$ and continues the processing of the layer according to $\lrmPacketProcess$.

            \end{itemize}
            \item If $\KEMciphertext' \allowbreak = \allowbreak \KEMciphertext$ and $\flrmParty \allowbreak = \allowbreak \flrmParty'$, then $B$ does the following:
            \begin{itemize}

                \item $B$ sends $(i, \allowbreak \AEADnonce, \allowbreak \gamma, \allowbreak \beta)$ to the procedure $\mathsf{Dec}$. In this case, the key that should be used to decrypt the ciphertext $\beta$ is the same as the one used to encrypt it.

                \item $B$ uses the output $\AEADmessage$ of  $\mathsf{Dec}$ as the output of algorithm $\AEADD$ and continues the processing of the layer according to $\lrmPacketProcess$.

            \end{itemize}

        \end{itemize}

        \item When the execution of the protocol ends, the procedure $\mathsf{Finalize}$ is run. That procedure compares the tuples $(\AEADkey, \allowbreak \AEADnonce, \allowbreak \AEADheader, \allowbreak \AEADmessage, \allowbreak \AEADciphertext)$ that were stored when running the procedures $\mathsf{Enc}$ and $\mathsf{Dec}$. $\AEADkey$ is the key, $\AEADnonce$ is the nonce, $\AEADheader$ is the associated data, $\AEADmessage$ is the message, and $\AEADciphertext$ is the ciphertext.   If there are two tuples $(\AEADkey, \allowbreak \AEADnonce, \allowbreak \AEADheader, \allowbreak \AEADmessage, \allowbreak \AEADciphertext)$ and $(\AEADkey', \allowbreak \AEADnonce', \allowbreak \AEADheader', \allowbreak \AEADmessage', \allowbreak \AEADciphertext')$ such that $\AEADmessage \allowbreak \neq \allowbreak \bot$, $\AEADmessage' \allowbreak \neq \allowbreak \bot$, $\AEADciphertext \allowbreak = \allowbreak \AEADciphertext'$, and $(\AEADkey, \allowbreak \AEADnonce, \allowbreak \AEADheader, \allowbreak \AEADmessage) \allowbreak \neq \allowbreak (\AEADkey', \allowbreak \AEADnonce', \allowbreak \AEADheader', \allowbreak \AEADmessage')$, then a pair of tuples that violates the committing property of $\AEAD$ scheme is found. 

        \item If a pair of tuples that violates the committing property of the $\AEAD$ scheme is found, the procedure $\mathsf{chk}$ is run on input the keys $\AEADkey$ and $\AEADkey'$ to check if the keys fulfill the specified level of corruption. In our case, the level of corruption should be set to ``0X'' in the notation of~\cite{cryptoeprint:2022/1260}, i.e., our construction should use an $\AEAD$ scheme such that the committing property holds against attacks in which the adversary creates one of the keys, whereas the other key remains hidden from the adversary. As can be seen, the keys used by the procedure $\mathsf{Enc}$ remain honest. The keys that are submitted to $\mathsf{Dec}$ are honest when the $\KEM$ ciphertext in the layer of encryption of the packet sent by the adversary is equal to the $\KEM$ ciphertext in the header of the layer of encryption computed by $B$ and the honest party associated to both layers is the same. In that case, the same key should be used to decrypt the $\AEAD$ ciphertext. The keys that are submitted to $\mathsf{Dec}$ are corrupt when the $\KEM$ ciphertext is different, or when the honest party that processes the packet is not the honest party associated to the layer of encryption. We remark that the procedure $\mathsf{Finalize}$ could find two tuples  $(\AEADkey, \allowbreak \AEADnonce, \allowbreak \AEADheader, \allowbreak \AEADmessage, \allowbreak \AEADciphertext)$ and $(\AEADkey', \allowbreak \AEADnonce', \allowbreak \AEADheader', \allowbreak \AEADmessage', \allowbreak \AEADciphertext')$ such that $\AEADmessage \allowbreak \neq \allowbreak \bot$, $\AEADmessage' \allowbreak \neq \allowbreak \bot$, $\AEADciphertext \allowbreak = \allowbreak \AEADciphertext'$, and $(\AEADkey, \allowbreak \AEADnonce, \allowbreak \AEADheader, \allowbreak \AEADmessage) \allowbreak \neq \allowbreak (\AEADkey', \allowbreak \AEADnonce', \allowbreak \AEADheader', \allowbreak \AEADmessage')$ and such that  both keys are corrupt when the adversary violates the committing property of the $\AEAD$ scheme more than once for the same ciphertext, but this always implies that there is also a tuple with an honestly generated key in which the violation occurs.

    \end{itemize}
    Given an adversary that makes  $\Gam \thegamect$ abort with non-negligible probability $\tau$, $B$ breaks the committed property with probability $\tau$. In comparison to the reduction of the unforgeability property of the $\AEAD$ scheme, $B$ does not fail here because the committing $\AEAD$ game in~\cite{cryptoeprint:2022/1260} is defined for multiple keys, so $B$ does not have to guess the layer of encryption that will be used by the adversary to make $\Gam \thegamect$ abort.

    \stepcounter{gamect}
    \stepcounter{prgamect}

    \item[$\Gam \thegamect$:] This game proceeds as $\Gam \theprgamect$, except that, in $\Gam \thegamect$, when an honest receiver $\flrmReceiver$ receives a request packet computed by an honest sender $\flrmSender$ for a receiver $\flrmReceiver'$, $\Gam \thegamect$ proceeds as follows. When the honest sender $\flrmSender$ computes the request packet, $\Gam \thegamect$ stores the payload $\payload{k}$ of the innermost layer of encryption, i.e. the payload that encrypts the request message and possibly a single-use reply block. When the honest receiver $\flrmReceiver$ receives  the request packet, $\Gam \thegamect$ compares the received payload $\payload{k}'$ with $\payload{k}$. If $\payload{k}' \allowbreak \neq \allowbreak \payload{k}$, or if $\flrmReceiver' \allowbreak \neq \allowbreak \flrmReceiver$, and algorithm $\lrmPacketProcess$ does not output $\bot$, $\Gam \thegamect$ outputs failure. The probability that $\Gam \thegamect$ outputs failure, which enables the environment to distinguish between  $\Gam \thegamect$ and $\Gam \theprgamect$, is bounded by the following claim.
    \begin{theorem} \label{th:integrity}

        Under the integrity property of the method used to encrypt the request message,  $|\Pr[\Gam \thegamect]-\Pr[\Gam \theprgamect]| \allowbreak \leq \allowbreak \Adv_{\Adversary}^{\mathsf{Integrity}}$.

    \end{theorem}
    
    \paragraph{Proof of Theorem~\ref{th:integrity}.} The implementation of Outfox  assumes that the message given as input is already encrypted by an encryption scheme that should provide both confidentiality and integrity. Such encryption scheme is chosen at the application level. The proof of Theorem~\ref{th:integrity} depends on the integrity property provided by this encryption scheme.

    If the encrypted message is prefixed with a string of $k$ zeroes, as shown in~\S\ref{sec:constructionalgorithms}, and considering that the output of the decryption algorithm of the symmetric-key encryption scheme used to encrypt the payload is indistinguishable from a random string of the same length, as required by the IND\$-CCA property in Definition~\ref{def:se:IND-CCA}, decryption of a modified payload yields a random string. Similarly, if a payload is decrypted by using an incorrect key, the result is a random string. Therefore, the probability that algorithm $\lrmPacketProcess$ does not output $\bot$ when decrypting a modified payload is $1/2^k$, which is negligible for a sufficiently large $k$.

    \stepcounter{gamect}
    \stepcounter{prgamect}

    \item[$\Gam \thegamect$:] This game proceeds as $\Gam \theprgamect$, except that, in $\Gam \thegamect$, when an honest  sender $\flrmSender$ receives a reply packet computed by an honest receiver $\flrmReceiver$ on input a single-use reply block produced by an honest sender $\flrmSender'$, $\Gam \thegamect$ proceeds as follows. When the honest receiver computes the reply packet, $\Gam \thegamect$ stores the payload $\payload{k}$ of the innermost layer of encryption, i.e. the payload that encrypts the reply message. When the honest sender receives the reply packet, if $\flrmSender' \allowbreak = \allowbreak \flrmSender$, after the payload is processed with algorithm $\lrmSurbRecover$ to yield the payload $\payload{k}'$ of the innermost layer of encryption, $\Gam \thegamect$ compares  $\payload{k}'$ with $\payload{k}$. If $\payload{k}' \allowbreak \neq \allowbreak \payload{k}$, and algorithm $\lrmSurbRecover$ does not output $\bot$, $\Gam \thegamect$ outputs failure. If $\flrmSender' \allowbreak \neq \allowbreak \flrmSender$, after the payload is processed with algorithm $\lrmPacketProcess$ to yield the reply message, if $\lrmPacketProcess$ does not output $\bot$, $\Gam \thegamect$ outputs failure. The probability that $\Gam \thegamect$ outputs failure, which enables the environment to distinguish between  $\Gam \thegamect$ and $\Gam \theprgamect$, is bounded by the following claim.
    \begin{theorem} \label{th:integrity:surb}

        Under the integrity property of the method used to encrypt the reply message,  $|\Pr[\Gam \thegamect]-\Pr[\Gam \theprgamect]| \allowbreak \leq \allowbreak \Adv_{\Adversary}^{\mathsf{Integrity}}$.

    \end{theorem}
    
    \paragraph{Proof of Theorem~\ref{th:integrity:surb}.} The proof of this theorem is very similar to the proof of Theorem~\ref{th:integrity}. We omit it.

    \stepcounter{gamect}
    \stepcounter{prgamect}

    \item[$\Gam \thegamect$:] This game proceeds as $\Gam \theprgamect$, except that, in $\Gam \thegamect$, when a request packet $\packet$ is computed, for each layer $i$ that fulfills conditions (1), (2) and (3), the computation of the packet with algorithm $\lrmPacketCreate$ is modified as follows. In the header $h$, the ciphertext $\beta$ of the $\AEAD$ scheme is replaced by a random string of the same length as $\beta$, and the message $\AEADmessage$ that should be encrypted by $\beta$ is stored. Moreover, when a packet is processed, for each layer $i$ that fulfills conditions (1), (2) and (3), the processing with algorithm $\lrmPacketProcess$ is modified as follows. The output of $\AEADD$ is replaced by the message $\AEADmessage$ that was stored when computing the layer of encryption.  The probability that the environment distinguishes between $\Gam \thegamect$ and $\Gam \theprgamect$ is bounded by the following theorem.
    \begin{theorem} \label{th:AEAD:IND-CPA}

        Under the IND\$-CPA property of the $\AEAD$ scheme,  $|\Pr[\Gam \thegamect]-\Pr[\Gam \theprgamect]| \allowbreak \leq \allowbreak \Adv_{\Adversary}^{\AEAD, \mathsf{IND\$-CPA}} \cdot N_r$, where $N_r$ is the number of layers of encryption of request packets that fulfill conditions (1), (2) and (3).

    \end{theorem}

    \paragraph{Proof of Theorem~\ref{th:AEAD:IND-CPA}.}  The proof uses a sequence of games $\Gam \theprgamect.i$, for $i=0$ to $N_r$. $\Gam \theprgamect.0$ is equal to $\Gam \theprgamect$, whereas $\Gam \theprgamect.N_r$ is equal to $\Gam \thegamect$. In $\Gam \theprgamect.i$, in the first $i$ layers of encryption of request packets that fulfill conditions (1), (2) and (3), the ciphertext $\beta$ of the $\AEAD$ scheme is replaced by a random string of the same length as $\beta$, and the message $\AEADmessage$ that should be encrypted is stored. The $N_r-i$ remaining layers of encryption are set as in $\Gam \theprgamect$, i.e., the ciphertext $\beta$ is given by the output of the algorithm $\AEADE$. Moreover, in $\Gam \theprgamect.i$, in the processing of the first $i$ layers of encryption that fulfill conditions (1), (2) and (3), the output of $\AEADD$ is replaced by the message $\AEADmessage$ that was stored when computing the layer of encryption, whereas it is not replaced in the processing of the remaining $N_r-i$ layers of encryption.
    
    Given an adversary that is able to distinguish $\Gam \theprgamect.i$ from $\Gam \theprgamect.(i+1)$ with non-negligible probability $\tau$, we construct an algorithm $B$ that breaks the IND\$-CPA property of the $\AEAD$ scheme with non-negligible probability $\tau$. $B$ proceeds as in $\Gam \theprgamect.i$ to compute and process the first $i$ layers of encryption that fulfill conditions (1), (2) and (3), and as in $\Gam \theprgamect.(i+1)$ to compute and process the $N_r-i-1$ remaining ones. $B$ simulates the honest parties towards the adversary as in $\Gam \theprgamect.i$, with the following changes.
    \begin{itemize}

        \item In the computation of the layer of encryption $i+1$ that fulfills conditions (1), (2) and (3), to set the ciphertext $\beta$, $B$ sets the message $\AEADmessage$, the header $\AEADheader$ and the nonce $\AEADnonce$ as done in $\Gam \theprgamect.i$ and calls the oracle provided by the challenger of the IND\$-CPA property of the $\AEAD$ scheme on input $(\AEADnonce, \allowbreak \AEADheader, \allowbreak \AEADmessage)$. When the challenger outputs a value $\KEMciphertext$, $B$ uses that value to set $\beta$.

        \item When $B$ needs to process the header of layer of encryption $i+1$, $B$ retrieves the message $\AEADmessage$ given as input to the oracle to compute the layer of encryption $i+1$ instead of running the decryption algorithm of the $\AEAD$ scheme because $B$ does not know the secret key used to compute $\beta$.  

    \end{itemize}
    As can be seen, when  the challenger of the IND\$-CPA property of the $\AEAD$ scheme uses the algorithm $\AEADE$ to compute $\KEMciphertext$, the view of the adversary is that of $\Gam \theprgamect.i$. When the challenger uses the oracle $\$(\cdot, \allowbreak \cdot, \allowbreak \cdot)$, $\KEMciphertext$ is a random string and thus the view of the adversary is that of $\Gam \theprgamect.(i+1)$. Therefore, if the adversary distinguishes between $\Gam \theprgamect.i$ and $\Gam \theprgamect.(i+1)$ with non-negligible advantage $\tau$, $B$ has non-negligible advantage $\tau$ against the IND\$-CPA property of the $\AEAD$ scheme.

    \stepcounter{gamect}
    \stepcounter{prgamect}
    
    \item[$\Gam \thegamect$:] This game proceeds as $\Gam \theprgamect$, except that, in $\Gam \thegamect$, when a single-use reply block $\surb$ is computed, for each layer $i$ that fulfills conditions (1), (2) and (3), the computation of the single-use reply block with algorithm $\lrmSurbCreate$ is modified as follows. In the header $h$, the ciphertext $\beta$ of the $\AEAD$ scheme is replaced by a random string of the same length as $\beta$, and the message $\AEADmessage$ that should be encrypted by $\beta$ is stored.  Moreover, when a reply packet is processed, for each layer $i$ that fulfills conditions (1), (2) and (3), the processing with algorithm $\lrmPacketProcess$ is modified as follows. The output of $\AEADD$ is replaced by the message $\AEADmessage$ that was stored when computing the layer of encryption. The probability that the environment distinguishes between $\Gam \thegamect$ and $\Gam \theprgamect$ is bounded by the following theorem.
    \begin{theorem} \label{th:AEAD:IND-CPA:surb}

        Under the IND\$-CPA property of the $\AEAD$ scheme, $|\Pr[\Gam \thegamect]-\Pr[\Gam \theprgamect]| \allowbreak \leq \allowbreak \Adv_{\Adversary}^{\AEAD, \mathsf{IND\$-CPA}} \cdot N_s$, where $N_s$ is the number of layers of encryption of single-use reply blocks that fulfill conditions (1), (2) and (3).

    \end{theorem}
    
    \paragraph{Proof of Theorem~\ref{th:AEAD:IND-CPA:surb}.} The proof is very similar to the proof of Theorem~\ref{th:AEAD:IND-CPA}. We omit it.
    
    \stepcounter{gamect}
    \stepcounter{prgamect}

    \item[$\Gam \thegamect$:] This game proceeds as $\Gam \theprgamect$, except that, in $\Gam \thegamect$, when a request packet $\packet$ is computed, for each layer $i$ that fulfills conditions (1), (2) and (3), the computation of the packet with algorithm $\lrmPacketCreate$ is modified as follows. The payload $\payloa$ computed by algorithm $\blockcipherenc$ is replaced by a random string of the same length as $\payloa$, and the message $m$ that should be encrypted by $\payloa$ is stored. Moreover, when a packet is processed, for each layer $i$ that fulfills conditions (1), (2) and (3), the processing with algorithm $\lrmPacketProcess$ is modified as follows. If the payload $\payloa$ is not modified, the output of $\blockcipherdec$ is replaced by the message $m$ that was stored when computing the layer of encryption, else the output of $\blockcipherdec$ is replaced by a random string of the same length as $m$. The probability that the environment distinguishes between $\Gam \thegamect$ and $\Gam \theprgamect$ is bounded by the following theorem.
    \begin{theorem} \label{th:SE:IND-CCA}

        Under the IND\$-CCA property of the $\blockcipher$ scheme, we have that $|\Pr[\Gam \thegamect]-\Pr[\Gam \theprgamect]| \allowbreak \leq \allowbreak \Adv_{\Adversary}^{\blockcipher, \mathsf{IND\$-CCA}} \cdot N_r$, where $N_r$ is the number of layers of encryption of request packets that fulfill conditions (1), (2) and (3).

    \end{theorem}

    \paragraph{Proof of Theorem~\ref{th:SE:IND-CCA}.} The proof uses a sequence of games $\Gam \theprgamect.i$, for $i=0$ to $N_r$. $\Gam \theprgamect.0$ is equal to $\Gam \theprgamect$, whereas $\Gam \theprgamect.N_r$ is equal to $\Gam \thegamect$. In $\Gam \theprgamect.i$, in the first $i$ layers of encryption of request packets that fulfill conditions (1), (2) and (3), the payload $\payloa$ is replaced by a random string of the same length as $\payloa$ and the message $m$ that should be encrypted is stored. The $N_r-i$ remaining layers of encryption are set as in $\Gam \theprgamect$, i.e., the payload $\payloa$ is given by the output of the algorithm $\blockcipherenc$.  Moreover, in $\Gam \theprgamect.i$, in the processing of the first $i$ layers of encryption that fulfill conditions (1), (2) and (3), the output of $\blockcipherdec$ is replaced by the message $m$ that was stored when computing the layer of encryption if the payload was not modified, else the output is replaced by a random string of the same length as $m$. For the $N_r-i$ remaining layers of encryption, the output of $\blockcipherdec$ is used.
    
    Given an adversary that is able to distinguish $\Gam \theprgamect.i$ from $\Gam \theprgamect.(i+1)$ with non-negligible probability $\tau$, we construct an algorithm $B$ that breaks the IND\$-CCA property of the $\blockcipher$ scheme with non-negligible probability $\tau$. $B$ proceeds as in $\Gam \theprgamect.i$ to compute the first $i$ layers of encryption that fulfill conditions (1), (2) and (3), and as in $\Gam \theprgamect.(i+1)$ to compute the $N_r-i-1$ remaining ones. $B$ simulates the honest parties towards the adversary as in $\Gam \theprgamect.i$, with the following changes.
    \begin{itemize}

        \item In the computation of the layer of encryption $i+1$ that fulfills conditions (1), (2) and (3), to set the payload $\payloa$, $B$ calls the oracle $\mathsf{Enc}$ provided by the challenger of the IND\$-CCA property of the $\blockcipher$ scheme on input the message to be encrypted (i.e.\ the request message and possibly a single-use reply block if this is layer $k$ in the packet, or the payload of the next layer otherwise). When the challenger outputs a value $c$, $B$ uses that value to set $\payloa$.

        \item When $B$ needs to process the layer of encryption $i+1$, $B$ proceeds as follows:
        \begin{itemize}

            \item If the payload is the same value $c$ output by the oracle $\mathsf{Enc}$, $B$ retrieves the message given as input to the oracle $\mathsf{Enc}$. We remark that $B$ cannot call $\mathsf{Dec}$ on input $c$.

            \item If the payload $c'$ is different from $c$ (because the adversary has modified the payload), $B$ calls the decryption oracle $\mathsf{Dec}$ on input $c'$ in order to process the packet.

        \end{itemize}

    \end{itemize}
    As can be seen, when  the challenger of the IND\$-CCA property of the $\blockcipher$ scheme uses the oracle $\mathsf{Enc}$ to compute $c$, and the oracle $\mathsf{Dec}$ on input payloads $c'$,  the view of the adversary is that of $\Gam \theprgamect.i$. When the challenger uses the oracle $\$(\cdot)$, its outputs are  random strings and thus the view of the adversary is that of $\Gam \theprgamect.(i+1)$. Therefore, if the adversary distinguishes between $\Gam \theprgamect.i$ and $\Gam \theprgamect.(i+1)$ with non-negligible advantage $\tau$, $B$ has non-negligible advantage $\tau$ against the IND\$-CCA property of the $\blockcipher$ scheme.

    \stepcounter{gamect}
    \stepcounter{prgamect}

    \item[$\Gam \thegamect$:] This game proceeds as $\Gam \theprgamect$, except that, in $\Gam \thegamect$, when a reply packet $\packet$ is computed, if (4) the single-use reply block $\surb$ used as input was computed by an honest sender, and (5) the user that computes the reply packet is honest, then the computation of the packet with algorithm $\lrmSurbUse$ is modified as follows. The payload $\payloa$ computed by algorithm $\blockcipherenc$ is replaced by a random string $c$ of the same length as $\payloa$, and both $c$ and the message $m$ that should be encrypted by $\payloa$ are stored. Moreover, when a reply packet that fulfills conditions (4) and (5) is processed by the honest sender that computed $\surb$, the processing with algorithm $\lrmSurbRecover$ is modified as follows. If the input to the algorithm $\blockcipherdec$ is equal to $c$, then $\blockcipherdec$ is not run and the message $m$ is used as its output, else the output is replaced by $\bot$. The probability that the environment distinguishes between $\Gam \thegamect$ and $\Gam \theprgamect$ is bounded by the following theorem.
    \begin{theorem} \label{th:SE:IND-CCA:reply}

        Under the IND\$-CCA property of the $\blockcipher$ scheme, we have that $|\Pr[\Gam \thegamect]-\Pr[\Gam \theprgamect]| \allowbreak \leq \allowbreak \Adv_{\Adversary}^{\blockcipher, \mathsf{IND\$-CCA}} \cdot N_u$, where $N_u$ is the number of reply packets that fulfill conditions (4) and (5).

    \end{theorem}
    \paragraph{Proof of Theorem~\ref{th:SE:IND-CCA:reply}.} The proof uses a sequence of games $\Gam \theprgamect.i$, for $i=0$ to $N_u$. $\Gam \theprgamect.0$ is equal to $\Gam \theprgamect$, whereas $\Gam \theprgamect.N_u$ is equal to $\Gam \thegamect$. In $\Gam \theprgamect.i$, in the first $i$ reply packets that fulfill conditions (4) and (5), the payload $\payloa$ is replaced by a random string $c$ of the same length as $\payloa$ and the message $m$ that should be encrypted is stored. The $N_u-i$ remaining layers of encryption are set as in $\Gam \theprgamect$, i.e., the payload $\payloa$ is given by the output of the algorithm $\blockcipherenc$.  Moreover, in $\Gam \theprgamect.i$, when a reply packet that fulfills conditions (4) and (5) is processed by the honest sender who computed the single-use reply block $\surb$ used to compute the reply packet, if the input to the algorithm $\blockcipherdec$ is $c$, then $\blockcipherdec$ is not run and its output is replaced by the message $m$ that was stored when computing the reply packet, else its output is replaced by $\bot$. For the $N_u-i$ remaining reply packets, the output of $\blockcipherdec$ is used.
    
    Given an adversary that is able to distinguish $\Gam \theprgamect.i$ from $\Gam \theprgamect.(i+1)$ with non-negligible probability $\tau$, we construct an algorithm $B$ that breaks the IND\$-CCA property of the $\blockcipher$ scheme with non-negligible probability $\tau$. $B$ proceeds as in $\Gam \theprgamect.i$ to compute the first $i$ reply packets that fulfill conditions (4) and (5), and as in $\Gam \theprgamect.(i+1)$ to compute the $N_u-i-1$ remaining ones. $B$ simulates the honest parties towards the adversary as in $\Gam \theprgamect.i$, with the following changes.
    \begin{itemize}

        \item In the computation of the reply packet $i+1$ that fulfills conditions (4) and (5), to set the payload $\payloa$, $B$ calls the oracle $\mathsf{Enc}$ provided by the challenger of the IND\$-CCA property of the $\blockcipher$ scheme on input the reply message $\flrmmessagerequest$. When the challenger outputs a value $c$, $B$ uses that value to set $\payloa$.

        \item When the reply packet $i+1$ is processed  by the honest sender who computed the single-use reply block $\surb$ used to compute the reply packet, $B$ proceeds as follows:
        \begin{itemize}

            \item If the input to the algorithm $\blockcipherdec$ is $c$, then $\blockcipherdec$ is not run and its output is replaced by the message $\flrmmessagerequest$ that was sent as input to the oracle $\mathsf{Enc}$.

            \item If the payload $c'$ is different from $c$ (because the adversary has modified the payload), $B$ sets the output of $\lrmSurbRecover$ to $\bot$.

        \end{itemize}

    \end{itemize}
    As can be seen, when the challenger of the IND\$-CCA property of the $\blockcipher$ scheme uses the oracle $\mathsf{Enc}$ to compute $c$,  the view of the adversary is that of $\Gam \theprgamect.i$. When the challenger uses the oracle $\$(\cdot)$, the view of the adversary is that of $\Gam \theprgamect.(i+1)$. Therefore, if the adversary distinguishes between $\Gam \theprgamect.i$ and $\Gam \theprgamect.(i+1)$ with non-negligible advantage $\tau$, $B$ has non-negligible advantage $\tau$ against the IND\$-CCA property of the $\blockcipher$ scheme.

    \stepcounter{gamect}
    \stepcounter{prgamect}
    \newcounter{gamereplypayloadprocessing}
    \setcounter{gamereplypayloadprocessing}{\value{gamect}}
 
    \item[$\Gam \thegamect$:] This game proceeds as $\Gam \theprgamect$, except that, in $\Gam \thegamect$, when the layer of a reply packet $\packet$ is processed, if (4) the single-use reply block $\surb$ used as input to compute the reply packet was computed by an honest sender, and (6) the party that processes the layer of encryption is an honest node or gateway, and (7) the layer of encryption is received or obtained by a corrupt party, then the processing of the packet with algorithm $\lrmPacketProcess$ is modified as follows. The output $\payloa$ computed by algorithm $\blockcipherdec$ is replaced by a random string $c$ of the same length as $\payloa$, and both $c$ and the payload $\payloa'$ that is input to $\blockcipherdec$ is stored. Moreover, when the layer of a reply packet that fulfills conditions (4), (6) and (7) is processed by the honest sender that computed $\surb$, the processing with algorithm $\lrmSurbRecover$ is modified as follows. If, for a layer that fulfills the conditions (4), (6) and (7), the input to the algorithm $\blockcipherenc$ is equal to the payload $\payloa'$ stored when processing that layer, then $\blockcipherenc$ is not run and $c$ is used as its output, else the output of $\lrmSurbRecover$ is $\bot$. The probability that the environment distinguishes between $\Gam \thegamect$ and $\Gam \theprgamect$ is bounded by the following theorem.
    \begin{theorem} \label{th:SE:IND-CCA:replylayer}

        Under the IND\$-CCA property of the $\blockcipher$ scheme, we have that $|\Pr[\Gam \thegamect]-\Pr[\Gam \theprgamect]| \allowbreak \leq \allowbreak \Adv_{\Adversary}^{\blockcipher, \mathsf{IND\$-CCA}} \cdot N_t$, where $N_t$ is the number of layers of encryption of reply packets that fulfill conditions (4), (6) and (7).

    \end{theorem}
    
     \paragraph{Proof of Theorem~\ref{th:SE:IND-CCA:replylayer}.} The proof uses a sequence of games $\Gam \theprgamect.i$, for $i=0$ to $N_t$. $\Gam \theprgamect.0$ is equal to $\Gam \theprgamect$, whereas $\Gam \theprgamect.N_t$ is equal to $\Gam \thegamect$. In $\Gam \theprgamect.i$, in the first $i$ layers of reply packets that fulfill conditions (4), (6) and (7), when processing the layer the output $\payloa$ computed by algorithm $\blockcipherdec$ is replaced by a random string $c$ of the same length as $\payloa$, and both $c$ and the payload $\payloa'$ that is input to $\blockcipherdec$ is stored. The $N_t-i$ remaining layers of encryption are set as in $\Gam \theprgamect$, i.e., the payload $\payloa$ is given by the output of the algorithm $\blockcipherdec$.  Moreover, in $\Gam \theprgamect.i$, when the layer of a reply packet that fulfills conditions (4), (6) and (7) is processed by the honest sender who computed the single-use reply block $\surb$ used to compute the reply packet, if  the input to the algorithm $\blockcipherenc$ is equal to the payload $\payloa'$ stored when processing that layer, then $\blockcipherenc$ is not run and $c$ is used as its output, else the output of $\lrmSurbRecover$ is $\bot$. For the $N_t-i$ remaining reply packets, the output of $\blockcipherenc$ is used.
    
    Given an adversary that is able to distinguish the game $\Gam \theprgamect.i$ from the game $\Gam \theprgamect.(i+1)$ with non-negligible probability $\tau$, we construct an algorithm $B$ that breaks the IND\$-CCA property of the $\blockcipher$ scheme with non-negligible probability $\tau$. $B$ proceeds as in $\Gam \theprgamect.i$ to compute the first $i$ layers of reply packets that fulfill conditions (4), (6) and (7), and as in $\Gam \theprgamect.(i+1)$ to compute the $N_t-i-1$ remaining ones. $B$ simulates the honest parties towards the adversary as in $\Gam \theprgamect.i$, with the following changes.
    \begin{itemize}

        \item In the processing of the layer $i+1$ that fulfills conditions (4), (6) and (7), to set the payload $\payloa$, $B$ calls the oracle $\mathsf{Dec}$ provided by the challenger of the IND\$-CCA property of the $\blockcipher$ scheme on input the payload $\payloa'$. When the challenger outputs a value $c$, $B$ uses that value to set $\payloa$.

        \item When the layer $i+1$ is processed by the honest sender who computed the single-use reply block $\surb$ used to compute the reply packet for that layer, $B$ proceeds as follows:
        \begin{itemize}

            \item If the input to the algorithm $\blockcipherenc$ is $c$, then $\blockcipherenc$ is not run and its output is replaced by the payload $\payloa'$ that was sent as input to the oracle $\mathsf{Dec}$.

            \item If the payload $c'$ is different from $c$ (because the adversary has modified the payload), $B$ sets the output of $\lrmSurbRecover$ to $\bot$.

        \end{itemize}

    \end{itemize}
    As can be seen, when the challenger of the IND\$-CCA property of the $\blockcipher$ scheme uses the oracle $\mathsf{Dec}$ to compute $c$,  the view of the adversary is that of $\Gam \theprgamect.i$. When the challenger uses the oracle $\$(\cdot)$, the view of the adversary is that of $\Gam \theprgamect.(i+1)$. Therefore, if the adversary distinguishes between $\Gam \theprgamect.i$ and $\Gam \theprgamect.(i+1)$ with non-negligible advantage $\tau$, $B$ has non-negligible advantage $\tau$ against the IND\$-CCA property of the $\blockcipher$ scheme.

    \stepcounter{gamect}
    \stepcounter{prgamect}
    \newcounter{gameStoreHeadersHonestSender}
    \setcounter{gameStoreHeadersHonestSender}{\value{gamect}}
 
    \item[$\Gam \thegamect$:] This game proceeds as $\Gam \theprgamect$, except that, in $\Gam \thegamect$, when an honest sender computes a request packet, if the entry gateway is corrupt, $\Gam \thegamect$ checks if there is a layer of encryption that fulfills conditions (1), (2) and (3) after the entry gateway. For that purpose, $\Gam \thegamect$ checks the packet route of the request packet and, if a layer of encryption that fulfills conditions (1), (2) and (3) is not found there, $\Gam \thegamect$ continues with the route of the reply packet (if it exists). If there are layers of encryption that fulfills conditions (1), (2) and (3),  $\Gam \thegamect$ takes the first one that fulfills them (following the packet route from the sender) and stores the following:
    \begin{itemize}
        
        \item If the first layer of encryption that fulfills conditions (1), (2) and (3) is found in the route of the request packet, $\Gam \thegamect$ stores all the headers and payloads computed by algorithm $\lrmPacketCreate$ of the layers of the request packet that are situated after the first layer that fulfills the conditions (1), (2) and (3) in the packet route, along with the identifier of the party that should process them. If the route of the reply packet is not empty, $\Gam \thegamect$ also stores the headers for all the layers of the single-use reply block computed by algorithm $\lrmSurbCreate$, along with the identifier of the party that should process them. Moreover, it stores the single-use reply block $\surb$.

        \item If the first layer of encryption that fulfills conditions (1), (2) and (3) is found in the route of the reply packet, $\Gam \thegamect$  stores the headers for all the layers of the single-use reply block computed by algorithm $\lrmSurbCreate$ after that first layer of encryption that fulfills conditions (1), (2) and (3), along with the identifier of the party that should process them.
        
    \end{itemize}
     Then, if the adversary sends a packet that contains one of those stored headers to the party that should process it, or a request packet that encrypts the single-use reply block $\surb$, or a reply packet that is computed on input $\surb$, and the adversary did not receive the header from an honest party, $\Gam \thegamect$ outputs failure. 
    
    The probability that the environment distinguishes between $\Gam \thegamect$ and $\Gam \theprgamect$ is negligible. The reason is that, in $\Gam \theprgamect$, a layer of encryption that fulfills conditions (1), (2) and (3) is computed in such a way that it is independent of the packet route and of the encrypted message. As can be seen, the $\AEAD$ ciphertext and the payload are replaced by random strings, and thus they cannot contain any information about the next layers of encryption, or the encrypted message and (possibly) single-use reply block. Therefore, if the adversary makes  $\Gam \thegamect$ output failure, the only possibility is that, by chance, it computed the same header, and the probability of that happening is negligible.

    \stepcounter{gamect}
    \stepcounter{prgamect}

    \item[$\Gam \thegamect$:] This game proceeds as $\Gam \theprgamect$, except that, in $\Gam \thegamect$, when an honest party processes and forwards a request packet to a corrupt party, if the request packet was computed by an honest sender,  $\Gam \thegamect$ computes a new packet that replaces the packet computed by the honest sender. To compute this packet, $\Gam \thegamect$ follows the way honest senders compute packets in $\Gam \theprgamect$, except that $\Gam \thegamect$ does not compute the layers of encryption associated with positions in the packet route that are situated before the corrupt party that receives the packet.

    The probability that the environment distinguishes between $\Gam \thegamect$ and $\Gam \theprgamect$ is 0. The reason is again that, in $\Gam \thegamect$, a layer of encryption that fulfills conditions (1), (2) and (3) is computed in such a way that it is independent of the layers of encryption under it. Therefore, changing the layers of encryption computed under the layer that fulfills conditions (1), (2) and (3) does not alter the view of the environment. 
    
    Furthermore, a layer of encryption only encrypts information about the next layers of encryption, i.e., it does not contain any information about previous layers of encryption. Therefore, a request packet computed by the sender and processed until the layer of encryption associated to the corrupt party that receives the packet is indistinguishable from a newly computed packet that uses the same packet route as the request packet starting from the position of the corrupt party in the packet route. 
    
    We remark that the above holds even if there were layers of encryption associated to corrupt parties before the layer associated with the honest party that forwards the packet. The reason is that the layer associated to the honest party that forwards the packet, or a previous layer associated to the first honest party that receives the packet from a corrupt party, fulfills conditions (1), (2) and (3), and is thus computed in such a way that does not reveal any information about the next layers of encryption. Therefore, a corrupt party that receives the packet before that honest party does not obtain any information about those next layers of encryption. Consequently, when the corrupt party after the honest party receives the packet, the adversary cannot know whether they are the same packet, i.e.\ they are unlinkable.

    \stepcounter{gamect}
    \stepcounter{prgamect}

    \item[$\Gam \thegamect$:] This game proceeds as $\Gam \theprgamect$, except that, in $\Gam \thegamect$, when an honest party (except for an exit gateway $\flrmGatewayexitreply$) forwards a reply packet computed on input a single-use reply block $\surb$ generated by an honest sender, $\Gam \theprgamect$ computes a new packet that replaces that reply packet. To compute this packet, $\Gam \theprgamect$ follows the way honest parties replace request packets by newly computed packets packets when forwarding to corrupt parties in $\Gam \theprgamect$, for the case in which the first layer of encryption that fulfills conditions (1), (2) and (3) is found in the route of the request packet. 
    
    The probability that the environment distinguishes between $\Gam \thegamect$ and $\Gam \theprgamect$ is 0. The reason is again that the layer of encryption associated to the honest party that forwards the reply packet, or a previous layer associated to the first honest party that receives the reply packet from a corrupt party, fulfills conditions (1), (2) and (3), and thus, when computing a $\surb$, the header was created in a way that does not reveal any information about headers encrypted under it. Moreover in $\Gam \thegamereplypayloadprocessing$ we have shown that processing the payload by an honest party yields a payload that is indistinguishable from random.

    We remark that, because the single-use reply block $\surb$ was computed by an honest sender, it is guaranteed that the last layer of encryption is associated to an honest sender. Therefore, even if the reply packet was computed by the adversary on input a single-use reply block generated by an honest sender, the reply packet can be simulated in the same way as a request packet because, once a packet is processed by an honest party, the adversary cannot link it to the packet it computed.

    We stress that the above does not apply to the case in which an honest exit gateway $\flrmGatewayexitreply$ forwards a reply packet. The reason is that exit gateways do not process reply packets.

    \stepcounter{gamect}
    \stepcounter{prgamect}

    \item[$\Gam \thegamect$:] This game proceeds as $\Gam \theprgamect$, except that, in $\Gam \thegamect$, for all the layers of encryption computed by honest senders whose headers are stored in $\Gam \theprgamect$, $\Gam \thegamect$ replaces those layers of encryption by layers of encryption associated with random honest parties from the correct sets. Additionally, if the first layer of encryption that fulfills conditions (1), (2) and (3) is found in the route of the request packet, $\Gam \theprgamect$ picks a random request message and decides randomly whether to encrypt a single-use reply block or not next to the request message.
    
    The probability that the environment distinguishes between $\Gam \thegamect$ and $\Gam \theprgamect$ is 0. The reason is again that, in $\Gam \theprgamect$, a layer of encryption that fulfills conditions (1), (2) and (3) is computed in such a way that it is independent of the packet route and of the encrypted message. Therefore, changing the layers of encryption computed under the layer that fulfills conditions (1), (2) and (3) does not alter the view of the environment.

    \stepcounter{gamect}
    \stepcounter{prgamect}

    \item[$\Gam \thegamect$:] This game proceeds as $\Gam \theprgamect$, except that, in $\Gam \thegamect$, when an honest sender computes a packet, the computation of layers of encryption that fulfill conditions (1), (2) and (3) is done as in the real protocol. 

    The probability that the environment distinguishes between $\Gam \thegamect$ and $\Gam \theprgamect$ is negligible. In the previous games, we have shown step by step that the environment distinguishes with negligible probability the computation of the layers of encryption that fulfill conditions (1), (2) and (3) in the real protocol from their computation in a way that does not depend on the layers of encryption under them. Therefore, when doing the reverse change, the environment can distinguish with negligible probability.

\end{description}
The distribution of $\Gam \thegamect$ is indistinguishable from our simulation. Therefore, the environment distinguishes the real protocol from our simulation with negligible probability, as stated in Theorem~\ref{th:all}.

%% file: 8Efficiency.tex
\section{Efficiency Analysis}
\label{sec:implementationefficiency}

\subsection{Efficiency Analysis of Outfox}

\paragraph{Computation cost.} For an algorithm $B$, let $|B|$ denote the average running time in seconds of that algorithm for a given security parameter $1^k$. Let $l+1$ denote the number of layers of encryption of a packet or a single-use reply block. In Table~\ref{tab:computationcostOutfox}, we show the computation cost of each of the algorithms of Outfox.

\begin{table}
    \centering
    \begin{tabular}{|l|l|} \hline 
        Algorithm    & Running time \\ \hline
        $\lrmPacketCreate$ & $(l+1) \cdot (|\KEMEnc| + |\KDF| + |\AEADE| + |\blockcipherenc|)$  \\ \hline
        $\lrmPacketProcess$ & $|\KEMDec| + |\KDF| + |\AEADD| + |\blockcipherdec|$ \\ \hline
        $\lrmSurbCreate$ & $(l+1) \cdot (|\KEMEnc| + |\KDF| + |\AEADE|)$ \\ \hline
        $\lrmSurbUse$ & $|\blockcipherenc|$ \\ \hline
        $\lrmSurbCheck$ & Negligible \\ \hline
        $\lrmSurbRecover$ & $l \cdot |\blockcipherenc| + |\blockcipherdec|$  \\ \hline
    \end{tabular}
    \caption{Computation cost of Outfox}
    \label{tab:computationcostOutfox}
\end{table}

The computation cost of Outfox is dominated by the KEM scheme used. To compute a packet $\packet$ or a single-use reply block $\surb$ of $l+1$ layers, algorithm $\KEMEnc$ is run $l+1$ times. To process a packet, algorithm $\KEMDec$ is run once. The computation costs of the key derivation function $\KDF$, of the symmetric-key encryption scheme $\blockcipher$,  or of the $\AEAD$ scheme are small compared to the $\KEM$, which requires public-key encryption operations. 

In Table~\ref{tab:kem_performance} and in Table~\ref{tab:kem_performance2}, we show performance measurements for the implementation of algorithms $\KEMKeyGen$, $\KEMEnc$ and $\KEMDec$ for the following $\KEM$ schemes. 
\begin{itemize}

    \item X25519~\cite{langley2016elliptic,bernstein2006curve25519}, which is based on elliptic curve Diffie-Hellman. It is not post-quantum secure. The algorithm $\KEMEnc$ requires 2 exponentiations, whereas $\KEMDec$ requires 1 exponentiation. 

    \item ML-KEM-768~\cite{nistmlkem}, a lattice-based post-quantum $\KEM$.

    \item Classic McEliece~\cite{josefsson-mceliece-02}, a code-based post-quantum $\KEM$ with parameter set mceliece460896f.

    \item X-Wing $\KEM$ Draft 02~\cite{connolly-cfrg-xwing-kem-02}, a hybrid post-quantum $\KEM$ that combines the $\KEM$ schemes based on X25519 and ML-KEM-768.

\end{itemize}

\begin{table}
    \centering
    \begin{tabular}{|l|c|c|c|c|}
        \hline
        & x25519 & ML-KEM-768 & X-Wing & Classic McEliece \\
        \hline
        $\KEMKeyGen$ & \SI{34.428}{\micro\second}  & \SI{13.732}{\micro\second}  & \SI{48.438}{\micro\second}  & \SI{264.65}{\milli\second}  \\
        $\KEMEnc$ & \SI{69.587}{\micro\second}  & \SI{14.485}{\micro\second}  & \SI{84.337}{\micro\second}  & \SI{50.280}{\milli\second}  \\
        $\KEMDec$ & \SI{34.331}{\micro\second}  & \SI{15.896}{\micro\second}  & \SI{50.218}{\micro\second}  & \SI{30.881}{\milli\second}  \\
        \hline
    \end{tabular}
    \caption{Performance comparison of $\KEM$ schemes on server}
    \label{tab:kem_performance}
\end{table}

\begin{table}
    \centering
    \begin{tabular}{|l|c|c|c|c|}
        \hline
        & x25519 & ML-KEM-768 & X-Wing & Classic McEliece \\
        \hline
        $\KEMKeyGen$ & \SI{29.080}{\micro\second} & \SI{10.985}{\micro\second} & \SI{40.647}{\micro\second} & \SI{114.77}{\milli\second} \\
        $\KEMEnc$ & \SI{58.189}{\micro\second} & \SI{11.406}{\micro\second} & \SI{71.844}{\micro\second} & \SI{54.287}{\milli\second} \\
        $\KEMDec$ & \SI{30.054}{\micro\second} & \SI{12.670}{\micro\second} & \SI{42.628}{\micro\second} & \SI{24.067}{\milli\second} \\
        \hline
    \end{tabular}
    \caption{Performance comparison of $\KEM$ schemes on server on laptop}
    \label{tab:kem_performance2}
\end{table}

In Table~\ref{tab:kem_performance}, the performance measurements were obtained by using a server running Ubuntu Linux on an AMD Ryzen 5 3600 6-Core processor. In Table~\ref{tab:kem_performance2}, the measurements were obtained by using a Lenovo
T14s laptop running Arch Linux on an AMD Ryzen 7 PRO 5850U.

\paragraph{Communication cost.} In terms of communication cost, each packet consists of a header and a payload. The payload length is equal to the length of a ciphertext of the symmetric-key encryption scheme. The header consists of three elements: a ciphertext and associated data for the $\AEAD$ scheme, and a ciphertext of the $\KEM$ scheme. 

The bit length of a header is given by the bit lengths of the $\KEM$ ciphertext, of the $\AEAD$ ciphertext and of the associated data. Let $p$ denote the bit length of the $\KEM$ ciphertext. For the security parameter $1^k$, the bit length of the associated data of the $\AEAD$ scheme is $k$. (This allows a fair comparison with Sphinx in~\S\ref{sec:sphinxcomparison}.) The bit length of the $\AEADE$ ciphertext depends on the layer in which the ciphertext is located. It can be computed as follows:
\begin{itemize}
    
    \item For layer $l$, the ciphertext encrypts the party identifiers of the exit gateway $\flrmGatewayexitreply$ and of the first-layer node $\flrmNodeAresponse$. In general, we consider that the bit length of a party identifier is $k$, and thus the bit length of this ciphertext is $2k$. However, to allow a fair comparison with Sphinx, we use $3k$ as size.\footnote{Sphinx operates in the service model and considers that destination addresses are of size $2k$, whereas other party identifiers are of size $k$.}

    \item For layer $l-1$, the ciphertext encrypts the identifier of the receiver $\flrmReceiver$, of size $k$, and the header of the next layer, which consists of a $\KEM$ ciphertext of size $p$, an $\AEAD$ ciphertext of size $3k$, and associated data of size $k$. The bit length is thus $k + p + 3k + k = p + 5k$.

    \item In general, for the layer $l-i$ (for $i \allowbreak \in \allowbreak [0,l]$), the bit length is $3k + ip + 2ik$.

    \item Therefore, for layer $0$, the bit length is $3k + lp + 2lk$.
    
\end{itemize}
Consequently, the bit length of the header for layer $l-i$ is $4k + (i+1)p + 2ik$, which is obtained by adding the bit length $p$ of the $\KEM$ ciphertext and the bit length $k$ of the associated data to the bit length $3k + ip + 2ik$ of the $\AEAD$ ciphertext.

The bit length of a payload ciphertext $\payloa$ is given by the bit length of the values that need to be encrypted in the innermost layer of encryption. For the security parameter $1^k$, those elements are a padding of $k$ zeroes, a message $\flrmmessagerequest$ of bit length $|\flrmmessagerequest|$, and a single-use reply block $\surb$, which consists of a header of bit length $4k + (l+1)p + 2lk$ and a key of length $k$. Consequently, the bit length of a payload is $k + |\flrmmessagerequest| + 4k + (l+1)p + 2lk + k = 6k + |\flrmmessagerequest| + (l+1)p + 2lk$.

In Table~\ref{tab:communicationcostOutfox}, we summarize the communication cost of Outfox. In Table~\ref{tab:public_keys_ciphertexts}, we show the bit length $p$ of $\KEM$ ciphertexts for the $\KEM$ schemes mentioned above.

\begin{table}
    \centering
    \begin{tabular}{|l|l|} \hline 
        Value    & Bit length \\ \hline
        $\KEM$ ciphertext & p  \\ \hline
        Associated data & k \\ \hline
        $\AEAD$ ciphertext at layer $l-i$ & $3k + ip + 2ik$ \\ \hline
        Header at layer $l-i$ & $4k + (i+1)p + 2ik$ \\ \hline
        Header at layer $0$ & $4k + (l+1)p + 2lk$ \\ \hline
        Single-use reply block &  $5k + (l+1)p + 2lk$  \\ \hline
        Payload & $6k + |\flrmmessagerequest| + (l+1)p + 2lk$ \\ \hline
        Packet at layer $l-i$ & $10k + |\flrmmessagerequest| + 2(i+1)p + 2(i+l)k$ \\ \hline
        Packet at layer $0$ & $10k + |\flrmmessagerequest| + 2(l+1)p + 4lk$ \\ \hline
    \end{tabular}
    \caption{Communication cost of Outfox. The security parameter is $1^k$ and the number of layers is $l+1$.}
    \label{tab:communicationcostOutfox}
\end{table}

\begin{table}
    \centering
    \begin{tabular}{|l|c|c|c|c|}
        \hline 
        & x25519 & ML-KEM-768 & X-Wing & Classic McEliece \\
        \hline 
        Public Key & 32 & 1 184 & 1 216 & 524 160 \\
        Ciphertext & 32 & 1 088 & 1 120 & 188 \\
        \hline 
    \end{tabular}
    \caption{Size in bytes of public keys and ciphertexts}
    \label{tab:public_keys_ciphertexts}
\end{table}

\paragraph{Storage cost.} In terms of storage, each party needs to store a key pair of the $\KEM$ scheme. In Table~\ref{tab:public_keys_ciphertexts}, we show the size of $\KEM$ public keys for the $\KEM$ schemes mentioned above.

Senders also need to store the headers of the innermost layer of encryption of single-use reply blocks they send in order to identify reply packets. To increase efficiency, they can store a hash of the header instead of the full header.

If duplicate detection is implemented, parties need to store all the packets received in order to discard duplicates. Efficient storage mechanisms to check set membership such as a Bloom filter can be used. We remark that, every time parties update their keys, a new execution of the protocol starts and the packets received in previous executions of the protocol can be erased.

\subsection{Comparison to Sphinx}
\label{sec:sphinxcomparison}

We compare the computation, communication and storage costs of Outfox with the costs of Sphinx~\cite{DBLP:conf/sp/DanezisG09}. For a fair comparison, we use the version of Sphinx in which single-use reply blocks are encrypted in the payload, as described in~\cite{DBLP:journals/popets/SchererWS24a}. Additionally, since Sphinx uses a mechanism based on the gap Diffie-Hellman assumption, we use the version of Outfox instantiated with the x25519 $\KEM$.

\paragraph{Computation Cost.} Like in Outfox, the communication cost of Sphinx is dominated by public-key operations. Therefore, we disregard the symmetric-key encryption operations, which are similar in Outfox and in Sphinx.

A key difference between Outfox and Sphinx is the way in which the shared key of the $\KEM$ is computed. In Outfox, to compute a header, for each layer of encryption the sender generates a $\KEM$ ciphertext and a $\KEM$ shared key that are computed independently of other layers of encryption. However, in Sphinx, the sender picks up only one secret for the outermost layer of encryption, and all the ciphertexts and shared keys are derived from that secret. As a consequence, when processing a packet, the $\KEM$ ciphertext needs to be recomputed, rather than decrypted as in Outfox.

As result, in Sphinx, computing a packet of $l+1$ layers requires $2(l+1)$ exponentiations, as in Outfox. However, processing a packet requires $2$ exponentiations in Sphinx, whereas in Outfox it requires $1$ exponentiation.

\paragraph{Communication Cost.} In contrast to Outfox, Sphinx packets hide the length of the path. To achieve that, all packets must have the same bit length, and thus Sphinx adds padding to each layer so that the bit length is constant. Consequently, while in Outfox the bit length of packets decreases after a layer of encryption is processed, in Sphinx it does not change.

The bit length of a Sphinx packet for a maximum of $l+1$ layers of encryption and that encrypts a single-use reply block in the payload is $2p + (4(l+1)+4)k + |\flrmmessagerequest| = 8k + |\flrmmessagerequest| + 2p + 4lk$, whereas the one of Outfox is $10k + |\flrmmessagerequest| + 2(l+1)p + 4lk$. As can be seen, the main difference is that, in Outfox, $l+1$ $\KEM$ ciphertexts are transmitted for each header, whereas in Sphinx, as explained above, only one $\KEM$ ciphertext is transmitted for each header.

We stress that the bit length of Outfox packets decreases when removing layers of encryption. The bit length of the packet at layer $l-i$ for $i \allowbreak \in \allowbreak [0,l]$ is $10k + |\flrmmessagerequest| + 2(i+1)p + 2(i+l)k$. For layer $l$, the bit length is $12k + |\flrmmessagerequest| + 2p$. This efficiency gain is derived from the fact that Outfox does not hide the length of the path.






%% file: 9Conclusion.tex
\section{Conclusion}
\label{sec:conclusion}

We have proposed the packet format Outfox. Outfox is suitable for mixnets in which all paths are of the same length and where the positions of nodes in a path are known. Outfox optimizes the cost of packet processing by nodes and is quantum-safe when instantiated with a quantum-safe KEM scheme. We have analyzed its security and its efficiency. To this end, we have proposed an ideal functionality for a layered replyable mixnet and a construction that uses Outfox to realize it. 

%% file: main.bbl

%% file: WPES1introduction.tex
\section{Introduction}
\label{wpes:sec:previouswork}


Anonymous communication requires packet formats that prevent adversaries from tracing messages as they pass through a network. A key property of such formats is \emph{bitwise unlinkability}, which ensures that a message entering a node cannot be correlated with any message leaving it based on bit patterns. The \emph{Sphinx} packet format~\cite{DBLP:conf/sp/DanezisG09} offers strong anonymity guarantees through bitwise unlinkability and has become the de facto standard for anonymous multi-hop messaging in systems such as mixnets~\cite{DBLP:journals/cacm/Chaum81, DBLP:conf/uss/PiotrowskaHEMD17, nym2024whitepaper} and Bitcoin’s Lightning Network~\cite{kiayias2020composable}.

Sphinx was designed in response to failures in earlier anonymous packet formats, such as those based on simple RSA encryption~\cite{DBLP:journals/cacm/Chaum81}, which were broken by active tagging attacks~\cite{pfitzmann1990break}. Subsequent designs like Minx addressed some issues but relied on informal arguments, leaving them vulnerable to RSA bit-oracle attacks~\cite{shimshock2008breaking}. Sphinx improved on these with a rigorous cryptographic design and wide applicability across mixnet architectures, including cascade, layered, and free-routing topologies~\cite{moller2003provably, bohme2004pet, danezis2003mixminion}. As a result, it has been adopted by most modern mixnets~\cite{DBLP:conf/uss/PiotrowskaHEMD17, chen2015hornet, kate2010using, nym2024whitepaper}. Sphinx also supports \emph{anonymous replies}, a key feature which was omitted in the original Chaumian design. 

Over time, Sphinx has inspired several extensions, including adaptations for multicast systems~\cite{hugenroth2021rollercoaster, schadt2024polysphinx}. However, some of these modifications have weakened its privacy properties. For example, proposals to incorporate authenticated encryption for the payload~\cite{beato2016improving} can cause packets to be dropped on tampering, rather than alerting the recipient as originally intended. Similarly, the EROR variant of Sphinx~\cite{klooss2024eror} revealed a subtle tagging attack if the first hop (gateway) and last hop (service provider) collude.

Despite its strengths, Sphinx remains inefficient in key respects. Each hop performs a full public-key operation, and the header size remains constant throughout the route, a design choice originally intended to support variable-length, free-routing paths. However, both theoretical and empirical studies have shown that \emph{layered topologies with fixed-length routes} offer stronger anonymity guarantees~\cite{bohme2004pet, diaz2010impact, piotrowska2021studying}. These insights open the door to simpler and more efficient packet formats.

While Sphinx remains secure under standard computational assumptions, the emergence of quantum computing threatens its long-term viability. A sufficiently powerful quantum computer running Shor's algorithm could break the public key cryptography Sphinx relies on. An alternative is the use of \emph{Key Encapsulation Mechanisms (KEMs)}, which offer post-quantum security and have been recommended by ongoing NIST standardization efforts~\cite{bos2018crystals, connolly-cfrg-xwing-kem-02, bernstein2006curve25519}. Although Sphinx variants using KEMs have been proposed~\cite{cryptoeprint:2023/1960}, they fall short of providing formal security guarantees or evaluating performance implications.

Motivated by these shortcomings, we propose Outfox, a redesign of the Sphinx packet format, a post-quantum secure evolution of the Sphinx packet format. Outfox replaces traditional key exchange with Key Encapsulation Mechanisms (KEMs)\cite{galbraith2012mathematics} and is optimized for mixnets with constant-length routes, such as Loopix\cite{DBLP:conf/uss/PiotrowskaHEMD17}. Its design builds on two key insights: (1) stratified topologies enhance anonymity by enforcing fixed route lengths~\cite{piotrowska2021studying}, and (2) KEMs enable robust security against quantum-capable adversaries. In this work, we provide a complete specification of Outfox and formal security proofs within the Universal Composability (UC) framework~\cite{DBLP:conf/focs/Canetti01}.

%% file: WPES1introduction2OurContribution.tex
\noindent \textbf{Our Contributions:} Our work makes the following contributions:
\begin{itemize}[leftmargin=10pt, topsep=2pt, itemsep=1pt]
    \item We present Outfox, a simplified variant of the well-established Sphinx packet format, tailored specifically for layered mixnet topologies.
    \item We show how Outfox can be made post-quantum secure by replacing traditional key exchange with quantum-resistant Key Encapsulation Mechanisms (KEMs), a critical property in light of recent developments in quantum computing and NIST's ongoing standardisation efforts\footnote{https://nvlpubs.nist.gov/nistpubs/ir/2024/NIST.IR.8547.ipd.pdf}.
    \item We provide a formal specification of Outfox. We also define the ideal functionality and a full security proof within the Universal Composability (UC) framework.
    \item We evaluate the performance improvements of Outfox over Sphinx, both in computational cost and communication overhead.
    \item We analyse Outfox’s efficiency across different KEM choices, including X25519, ML-KEM, Classic McEliece, and Xwing-KEM, highlighting the tradeoffs between post-quantum security and performance.
\end{itemize}

%% file: WPES1introduction3Outline.tex

%% file: WPES2systemmodel.tex
\section{Layered Mixnet with Replies in the client model}
\label{wpes:sec:lmc}

We describe layered mixnets with replies in the client model ($\LRM$). 
In~\S\ref{wpes:sec:parties}, we describe the parties in an $\LRM$. 
In~\S\ref{wpes:sec:phases}, we describe the phases of an $\LRM$ protocol. 
In~\S\ref{wpes:sec:securityproperties}, we describe the security properties that an $\LRM$ protocol should provide.

%% file: WPES2systemmodel1Parties.tex
\subsection{Parties}
\label{wpes:sec:parties}

The parties involved in an $\LRM$ are users, gateways and nodes.
\begin{description}[leftmargin=0cm,labelindent=0cm]

    \item[Users.] A user $\flrmUser$ is a party that uses the mixnet to send messages to other users and to receive messages from other users. Therefore, a user acts both as a sender $\flrmSender$ and a receiver $\flrmReceiver$. A packet produced by a sender is referred to as a \emph{request}, whereas a packet produced by a receiver is referred to as a \emph{reply}. 

    \begin{description}[leftmargin=0.5cm,labelindent=0.5cm]

        \item[Senders.] A sender $\flrmSender$ performs two tasks. On the one hand, a sender $\flrmSender$ computes a request to communicate a message $\flrmmessagerequest$ to a receiver $\flrmReceiver$ and sends the request to an entry gateway $\flrmGatewayentry$. On the other hand, a sender $\flrmSender$ receives a reply from an entry gateway $\flrmGatewayentryreply$ and processes it in order to retrieve a message $\flrmmessageresponse$.  When a sender $\flrmSender$ computes a request, the sender $\flrmSender$ decides whether the receiver $\flrmReceiver$ can reply to it or not.  If the sender $\flrmSender$ is corrupt, it can choose that the reply be sent to another user $\flrmSenderreply$, although an honest $\flrmSenderreply$ is unable to process the reply.

        \item[Receivers.] A receiver $\flrmReceiver$ performs two tasks. On the one hand, a receiver $\flrmReceiver$ receives a request from an exit gateway $\flrmGatewayexit$ and processes it to retrieve a message $\flrmmessagerequest$. On the other hand, a receiver $\flrmReceiver$ computes a reply to communicate a message $\flrmmessageresponse$ to a sender $\flrmSender$ and sends the reply to an exit gateway $\flrmGatewayexitreply$. We remark that a receiver $\flrmReceiver$ can compute a reply only if the sender $\flrmSender$ enabled that possibility when computing the request associated with that reply.

    \end{description}

     \item[Gateways.] A gateway $\flrmGateway$ is a party that provides senders and receivers with access to the mixnet. Every gateway acts as both an entry gateway and an exit gateway. 

    \begin{description}[leftmargin=0.5cm,labelindent=0.5cm]

        \item[Entry gateway.] An entry gateway performs two tasks. On the one hand, an entry gateway $\flrmGatewayentry$ receives a request from a sender $\flrmSender$ and relays it to a first-layer node $\flrmNodeArequest$. On the other hand, an entry gateway $\flrmGatewayentryreply$ receives a reply from a third-layer node $\flrmNodeCresponse$, processes it, and sends it to a sender $\flrmSender$. We remark that entry gateways do not process requests, i.e.\ they simply relay them from a sender to a first-layer node.

        \item[Exit gateway.] An exit gateway performs two tasks. On the one hand, an exit gateway $\flrmGatewayexit$ receives a request from a third-layer node $\flrmNodeCrequest$, processes it, and sends it to a receiver $\flrmReceiver$. On the other hand, an exit gateway $\flrmGatewayexitreply$ receives a reply from a receiver $\flrmReceiver$ and relays it to a first-layer node $\flrmNodeAresponse$. We remark that exit gateways do not process replies, i.e.\ they simply relay them from a receiver to a first-layer node.    
    
    \end{description}


    \item[Nodes.] A node $\flrmNode$ is a party that receives, processes and sends request packets and reply packets. Our mixnet consists of three layers, and each node is assigned to one of the layers. (Our model can easily be generalized to any number of layers.) We denote nodes assigned to the first, second and third layers by  $\flrmNodeArequest$, $\flrmNodeBrequest$, and $\flrmNodeCrequest$, when they are processing a request, and by $\flrmNodeAresponse$, $\flrmNodeBresponse$, and $\flrmNodeCresponse$, when they are processing a reply, but we stress that every node processes both request and reply packets.
    \begin{description}[leftmargin=0.5cm,labelindent=0.5cm]

        \item[First-layer nodes.] First-layer nodes perform two tasks. On the one hand, a first-layer node $\flrmNodeArequest$ receives a request from an entry gateway $\flrmGatewayentry$, processes it and sends it to a second-layer node $\flrmNodeBrequest$. On the other hand, a first-layer node $\flrmNodeAresponse$ receives a reply from an exit gateway $\flrmGatewayexitreply$, processes it and sends it to a second-layer node $\flrmNodeBresponse$.

        \item[Second-layer nodes.] Second-layer nodes perform two tasks. On the one hand, a second-layer node $\flrmNodeBrequest$ receives a request from a first-layer node $\flrmNodeArequest$, processes it and sends it to a third-layer node $\flrmNodeCrequest$. On the other hand, a second-layer node $\flrmNodeBresponse$ receives a reply from a first-layer node $\flrmNodeAresponse$, processes it and sends it to a third-layer node $\flrmNodeCresponse$.

        \item[Third-layer nodes.] Third-layer nodes perform two tasks. On the one hand, a third-layer node $\flrmNodeCrequest$ receives a request from a second-layer node $\flrmNodeBrequest$, processes it and sends it to an exit gateway $\flrmGatewayexit$. On the other hand, a third-layer node $\flrmNodeCresponse$ receives a reply from a second-layer node $\flrmNodeBresponse$, processes it and sends it to an entry gateway $\flrmGatewayentryreply$. 

    \end{description}

 \end{description}
In Table~\ref{wpes:tab:parties}, we describe the notation for the parties involved in an $\LRM$.

\begin{table}
    \caption{Parties involved in an  $\LRM$}
    \label{wpes:tab:parties}
    \centering
    \begin{tabular}{|l|l|} \hline
        \multicolumn{2}{|c|}{\textbf{Parties}} \\ \hline
        $\flrmUser$ & User \\ \hline 
        $\flrmSender$ & Sender \\ \hline
        $\flrmReceiver$ & Receiver \\ \hline
        $\flrmSenderreply$ & Receiver of a reply ($\flrmSender = \flrmSenderreply$ if $\flrmSender$ is honest) \\ \hline
        $\flrmGateway$ & Gateway \\ \hline
        $\flrmGatewayentry$ & Entry Gateway processing a request \\ \hline
        $\flrmGatewayentryreply$ & Entry Gateway processing a reply \\ \hline
        $\flrmGatewayexit$ & Exit Gateway processing a request \\ \hline
        $\flrmGatewayexitreply$ & Exit Gateway processing a reply \\ \hline
        $\flrmNode$ & Node \\ \hline
        $\flrmNodeArequest$ & First-layer node processing a request \\ \hline
        $\flrmNodeAresponse$ & First-layer node processing a reply \\ \hline
        $\flrmNodeBrequest$ & Second-layer node processing a request \\ \hline
        $\flrmNodeBresponse$ & Second-layer node processing a reply \\ \hline
        $\flrmNodeCrequest$ & Third-layer node processing a request \\ \hline
        $\flrmNodeCresponse$ & Third-layer node processing a reply \\ \hline
        $\flrmParty$ & A party \\ \hline
    \end{tabular}
\end{table}

%% file: WPES2systemmodel2Phases.tex
\subsection{Phases}
\label{wpes:sec:phases}

An $\LRM$ protocol consists of the following phases: setup, registration, request, reply and forward. 

\begin{description}[leftmargin=0.5cm,labelindent=0cm]
    
    \item[Setup.] The setup phase is run by every node and gateway to generate their keys.

    \item[Registration.] The registration phase is run by every user to generate her keys and to retrieve the keys of nodes and gateways.

    \item[Request.] In the request phase, a sender $\flrmSender$ receives as input a request message $\flrmmessagerequest$ and a packet route  $[\flrmSender, \allowbreak \flrmGatewayentry, \allowbreak \flrmNodeArequest, \allowbreak \flrmNodeBrequest, \allowbreak \flrmNodeCrequest, \allowbreak \flrmGatewayexit, \allowbreak \flrmReceiver, \langle \flrmGatewayexitreply, \allowbreak \flrmNodeAresponse, \allowbreak \flrmNodeBresponse, \allowbreak \flrmNodeCresponse, \allowbreak  \flrmGatewayentryreply, \allowbreak \flrmSenderreply \rangle]$. When replies are not enabled, the route $\langle \flrmGatewayexitreply, \allowbreak \flrmNodeAresponse, \allowbreak \flrmNodeBresponse, \allowbreak \flrmNodeCresponse, \allowbreak  \flrmGatewayentryreply, \allowbreak \flrmSenderreply \rangle$ for the reply packet is set to $\bot$. When replies are enabled, if the sender is honest, $\flrmSenderreply$ must be equal to $\flrmSender$.
    
    First, the sender $\flrmSender$ checks if replies are enabled and, in that case, creates a single-use reply block $\surb$, an identifier $\surbidentifier$ and some secrets $\surbsecrets$. $\surbidentifier$ and $\surbsecrets$ allow  $\flrmSender$ to identify a reply packet and to process it respectively. Then $\flrmSender$ creates a request packet $\packet$. When replies are enabled, the request packet encrypts a single-use reply block $\surb$ in addition to the request message $\flrmmessagerequest$. Finally, $\flrmSender$ sends $\packet$ and the identifier $\flrmNodeArequest$ to the entry gateway $\flrmGatewayentry$. $\flrmGatewayentry$ stores $\packet$ and $\flrmNodeArequest$. $\flrmGatewayentry$ does not process $\packet$, i.e., it does not remove a layer of encryption from $\packet$.
    
    \item[Reply.] In the reply phase, a receiver $\flrmReceiver$ receives as input a reply message  $\flrmmessageresponse$, a single-use reply block $\surb$, and the identifiers of an exit gateway $\flrmGatewayexitreply$ and a first-layer node $\flrmNodeAresponse$. The single-use reply block $\surb$ and the identifiers $\flrmGatewayexitreply$ and $\flrmNodeAresponse$ were obtained by $\flrmReceiver$ after processing a request packet from a sender. $\flrmReceiver$ creates a reply packet $\packet$ and sends $\packet$ along with the identifier $\flrmNodeAresponse$ to an exit gateway $\flrmGatewayexitreply$. $\flrmGatewayexitreply$ stores $\packet$ and $\flrmNodeAresponse$. $\flrmGatewayexitreply$ does not process $\packet$.

    \item[Forward.] In the forward phase, a node or a gateway $\flrmParty$ uses as input a packet $\packet$ and the identifier of the next party $\flrmParty'$ in the route. Both $\packet$ and the identifier $\flrmParty'$ were stored by $\flrmParty$ after receiving a packet from the previous party in the route and processing it. $\flrmParty$ sends $\packet$ to $\flrmParty'$. Then $\flrmParty'$ processes $\packet$. The way in which $\packet$ is processed depends on whether $\flrmParty'$ is either a user, or a node or gateway.

    If $\flrmParty'$ is a user, $\flrmParty'$ first checks if $\packet$ is a reply packet by using all the identifiers $\surbidentifier$ it stores. If there is a match, then $\flrmParty'$ uses $\surbsecrets$ to process the reply packet and obtain the reply message $\flrmmessageresponse$. Otherwise $\flrmParty'$ processes $\packet$ as a request packet to obtain the request message $\flrmmessagerequest$ and, if the sender enables a reply, a single-use reply block $\surb$ and the identifiers $\flrmGatewayexitreply$ and $\flrmNodeAresponse$ for the route of the reply packet.

    If $\flrmParty'$ is a node or a gateway, $\flrmParty'$ processes $\packet$ to obtain a packet $\packet'$ and the identifier of the next party in the route. $\flrmParty'$ stores both values. $\packet'$ is obtained by removing one layer of encryption from $\packet$.
   
\end{description}

%% file: WPES2systemmodel3SecurityProperties.tex
\subsection{Security Properties}
\label{wpes:sec:securityproperties}

In this section, we describe informally the security properties that a packet format for an $\LRM$ should provide. In~\ref{sec:functionalityDefinition}, we define formally an ideal functionality for $\LRM$ that guarantees those properties.

We consider an adversary that controls the communication channels between parties, and that therefore can drop, delay and inject messages. Additionally, the adversary is allowed to corrupt any subset of users, nodes and gateways.

We remark that our security definition and our security analysis focuses on the properties that a secure packet format should provide. This means that we focus on the information that the packets themselves may leak to the adversary, and not on other sources of information leakage (e.g.\ the times when packets are sent and received), which the adversary may exploit through traffic analysis.

As a consequence, our protocol in~\S\ref{wpes:sec:protocol} shows how packets are computed, communicated and processed in the mixnet, but omits the protections that a secure layered mixnet~\cite{DBLP:conf/uss/PiotrowskaHEMD17} should provide against traffic analysis. In particular, the nodes in our protocol do not perform any mixing of packets, and they do not provide protection against reply attacks (i.e., they do not check whether packets are being resent). Additionally, cover traffic is not used. We remark that all those protections can and should be added to construct a secure layered mixnet that uses Outfox.

\paragraph{Confidentiality.} This property guarantees that the adversary does not obtain any information about a layer of encryption that (1) should be processed by an honest party, (2) is not received by the adversary, and (3) is not obtained by a corrupt party after processing a layer of encryption associated to that party. This property only holds for request packets computed by honest senders, and by reply packets computed on input a single-use reply block previously generated by an honest sender. For other packets, the adversary learns all the layers of encryption when computing them.  

Consider e.g.\ a request packet that does not enable replies with route  $[\flrmSender, \allowbreak \flrmGatewayentry, \allowbreak \flrmNodeArequest, \allowbreak \flrmNodeBrequest, \allowbreak \flrmNodeCrequest, \allowbreak \flrmGatewayexit, \allowbreak \flrmReceiver]$. Let $\flrmSender$, $\flrmNodeBrequest$ and $\flrmNodeCrequest$ be the only honest parties in the route. In this case, confidentiality ensures that the adversary does not learn any information about the layer of encryption associated with $\flrmNodeCrequest$, because that layer of encryption is associated to the honest party $\flrmNodeCrequest$ and the adversary never receives nor obtains it through processing another layer (because $\flrmNodeBrequest$ is also honest).

Confidentiality on its own does not guarantee that the adversary cannot trace a packet through the mixnet. Consider e.g.\ a request packet with route  $[\flrmSender, \allowbreak \flrmGatewayentry, \allowbreak \flrmNodeArequest, \allowbreak \flrmNodeBrequest, \allowbreak \flrmNodeCrequest, \allowbreak \flrmGatewayexit, \allowbreak \flrmReceiver]$ in which only $\flrmSender$ and $\flrmNodeCrequest$ are honest. In this case, the adversary can learn all the layers of encryption of the packet. To guarantee that the adversary cannot trace the packet through the mixnet (and thereby learn the identity of the honest sender $\flrmSender$ that communicates with the receiver $\flrmReceiver$), we require the property of bitwise unlinkability.

\paragraph{Bitwise unlinkability.} This property ensures that, once a packet $\packet$ is processed by an honest party, the output packet $\packet'$ cannot be linked to $\packet$. Consider e.g.\ a setting where an honest sender computes a packet $\packet$ that does not enable replies on input a message $\flrmmessagerequest$ and a packet route  $[\flrmSender, \allowbreak \flrmGatewayentry, \allowbreak \flrmNodeArequest, \allowbreak \flrmNodeBrequest, \allowbreak \flrmNodeCrequest, \allowbreak \flrmGatewayexit, \allowbreak \flrmReceiver]$. Consider that the entry gateway $\flrmGatewayentry$ is corrupt, the first-layer node $\flrmNodeArequest$ is honest, and the second-layer node $\flrmNodeBrequest$ is corrupt. Bitwise unlinkability guarantees that, when $\flrmNodeBrequest$ receives a packet $\packet'$, it cannot tell whether $\packet'$ is the result of the processing of $\packet$ by $\flrmNodeArequest$, or a new packet computed for the second layer. 

In general, bitwise unlinkability ensures that, when a packet $\packet$ is processed (i.e.\ a layer of encryption is removed) by an honest party, the resulting packet $\packet'$ cannot be distinguished from a newly computed packet. We remark that this property only holds for request packets computed by honest senders, and by reply packets computed on input a single-use reply block previously generated by an honest sender. If the sender is corrupt, then the sender knows all the layers of encryption in a packet and thus bitwise unlinkability does not hold.

We remark that, if nodes do not perform mixing, the adversary may still be able to link $\packet$ with $\packet'$ by checking e.g.\ the times $\packet$ and $\packet'$ are received and sent by the honest party that processes $\packet$. However, when mixing is performed, bitwise unlinkability guarantees that the adversary cannot link $\packet$ with $\packet'$.

\paragraph{Integrity.} This property guarantees that, if the adversary modifies a packet, the modification will be detected by honest parties. Therefore, integrity guarantees that the adversary cannot modify the packet route or the message encrypted by a packet. In the case of request packets that enable a reply, integrity also guarantees that the adversary cannot modify the single-use reply block encrypted in the packet.

In Outfox, packets consists of a header and a payload. The header encrypts the packet route, while the payload encrypts the message and, for request packets that enable replies, the single-use reply block. We require a different level of integrity for the header and for the payload. 

For the header, integrity guarantees that any modification is detected by the next honest party that processes a packet. For example, consider a packet with packet route $[\flrmSender, \allowbreak \flrmGatewayentry, \allowbreak \flrmNodeArequest, \allowbreak \flrmNodeBrequest, \allowbreak \flrmNodeCrequest, \allowbreak \flrmGatewayexit, \allowbreak \flrmReceiver]$ in which only $\flrmSender$, $\flrmNodeCrequest$ and $\flrmReceiver$ are honest. Integrity guarantees that, if any of the corrupt parties $(\flrmGatewayentry, \allowbreak \flrmNodeArequest, \allowbreak \flrmNodeBrequest)$ modifies the header, then the honest third-layer node $\flrmNodeCrequest$ detects the modification.  If the corrupt gateway $\flrmGatewayexit$ modifies the header, then the modification is detected by the honest receiver $\flrmReceiver$.

For the payload, integrity guarantees that any modification is detected by an honest receiver that receives a request packet, or by the sender that receives a reply packet. Therefore, other honest parties in the route of the packet do not detect the modification.

\paragraph{Request-reply indistinguishability.} Request packets should be indistinguishable from reply packets. That applies both to an external adversary that controls the communication channels, and to corrupt nodes and gateways that process those packets. Request-reply indistinguishability implies that the processing of request and reply packets is the same.

In the case of gateways, when a gateway $\flrmGateway$ receives a packet from a user $\flrmUser$, the gateway $\flrmGateway$ should not be able to distinguish whether the user acts as a sender $\flrmSender$ that sends a request (and thus the gateway acts as an entry gateway $\flrmGatewayentry$), or whether the user acts as a receiver $\flrmReceiver$ that sends a reply (and thus the gateway acts as an exit gateway $\flrmGatewayexitreply$). We remark that, in both cases, the gateway simply relays the packet to a  first-layer node without processing it. 
    
Similarly, when a gateway $\flrmGateway$ receives a packet from a third-layer node, the gateway  $\flrmGateway$ should not be able to distinguish whether it is a request from a third-layer node $\flrmNodeCrequest$ (and thus the gateway acts as an exit gateway $\flrmGatewayexit$) or a reply from a third-layer node $\flrmNodeCresponse$ (and thus the gateway acts as an entry gateway $\flrmGatewayentryreply$).  We remark that, in both cases, the gateway processes the packet and sends it to a user.

In practice, request-reply indistinguishability for gateways holds in settings where users send both requests and replies. If a user is e.g.\ a web host, then the gateway knows that any packet sent to that user is a request and any packet received from that user is a reply.

%% file: WPES3construction.tex
\section{Construction}
\label{wpes:sec:construction}

We describe the Outfox packet format and a protocol $\mathrm{\Pi}_{\LRM}$ that uses Outfox to provide a secure layered mixnet in the client model. In~\S\ref{wpes:sec:buildingblocks}, we describe the building blocks used by Outfox and by the protocol $\mathrm{\Pi}_{\LRM}$. In~\S\ref{wpes:sec:outfoxpacketformat}, we describe the Outfox packet format. In~\S\ref{wpes:sec:protocol}, we describe the protocol $\mathrm{\Pi}_{\LRM}$.

%% file: WPES3construction1buildingblocks.tex
\subsection{Building Blocks}
\label{wpes:sec:buildingblocks}

The Outfox packet format uses as building blocks a key encapsulation mechanism $\KEM$, a key derivation function $\KDF$, an authenticated encryption with associated data $\AEAD$ scheme, and a block cipher $\blockcipher$. The protocol $\mathrm{\Pi}_{\LRM}$ uses, in addition to Outfox, the functionality $\Functionality_{\Freg}$ for registration, the functionality $\Functionality_{\Fpreg}$ for registration with privacy-preserving key retrieval, and the functionality $\Functionality_{\SMT}$ for secure message transmission.

 \paragraph{\textbf{Key encapsulation mechanism.}} A key encapsulation mechanism $\KEM$~\cite{cryptoeprint:2024/039} is a triple of algorithms $(\KEMKeyGen, \allowbreak \KEMEnc, \allowbreak \KEMDec)$ with public keyspace $\KEMpublickeyspace$, secret keyspace $\KEMsecretkeyspace$, ciphertext space $\KEMciphertextspace$, and shared keyspace $\KEMsharedkeyspace$. The triple of algorithms is defined as follows:
\begin{description}
    
    \item[$\KEMKeyGen(1^\securityparameter)$.] Randomized algorithm that, given the security parameter $1^\securityparameter$, outputs a secret key $\KEMsecretkey \allowbreak \in \KEMsecretkeyspace$, and a public key $\KEMpublickey \allowbreak \in \KEMpublickeyspace$.

    \item[$\KEMEnc(\KEMpublickey)$.] Randomized algorithm that, given a public key $\KEMpublickey \allowbreak \in \KEMpublickeyspace$, outputs a shared key $\KEMsharedkey \allowbreak \in \allowbreak \KEMsecretkeyspace$, and a ciphertext $\KEMciphertext \allowbreak \in \allowbreak \KEMciphertextspace$.

    \item[$\KEMDec(\KEMciphertext, \KEMsecretkey)$.] Deterministic algorithm that, given a secret key $\KEMsecretkey \allowbreak \in \allowbreak \KEMsecretkeyspace$  and a ciphertext $\KEMciphertext \allowbreak \in \allowbreak \KEMciphertextspace$, outputs the shared key $\KEMsharedkey \in \KEMsharedkeyspace$. In case of rejection, this algorithm outputs $\bot$.
    
\end{description}
A $\KEM$ should be IND-CCA secure (see~\S\ref{sec:keyencapsulation}).

 \paragraph{\textbf{Key derivation function.}} A key derivation function $\KDF$~\cite{cryptoeprint:2010/264} accepts as input four arguments: a value $\sigma$ sampled from a source of keying material (described below), a length value $\ell$, and two additional arguments, a salt value $r$ defined over a set of possible salt values and a context variable $ctx$, both of which are optional, i.e., can be set to the null string or to a constant. The $\KDF$ output is a string of $\ell$ bits.

A source of keying material $\Sigma$ is a two-valued probability distribution $(\sigma, \allowbreak \alpha)$ generated by an efficient probabilistic algorithm. We will refer to both the probability distribution as well as the generating algorithm by $\Sigma$. The output $\sigma$ represents the (secret) source key material to be input to a $\KDF$, while $\alpha$ represents some auxiliary knowledge about $\sigma$ (or its distribution). We require a $\KDF$ to be secure on inputs $\sigma$ even when the knowledge value $\alpha$ is given to the adversary.

In Outfox, the source of keying material $\Sigma$ is the key encapsulation mechanism $\KEM$ defined above. The input $\sigma$ of the $\KDF$ is the shared key $\KEMsharedkey$ output by $\KEMEnc$, while $\alpha$ consists of the values $\KEMpublickey$ and $\KEMciphertext$ output by $\KEMKeyGen$ and $\KEMEnc$ respectively. The secret key $\KEMsecretkey$ output by $\KEMKeyGen$ should not be given to the adversary, because otherwise the adversary is able to compute $\sigma$ on input $\alpha$. However, in $\mathrm{\Pi}_{\LRM}$ the adversary computes ciphertexts on input the public key of honest users that need be decrypted by honest users. Therefore, our reduction to the security of the $\KDF$ with respect to a source of keying material $\KEM$ needs a decryption oracle. To address this issue, we adapt the game for secure $\KDF$ in~\cite{cryptoeprint:2010/264} so that the adversary can access a decryption oracle. This requires a $\KEM$ that is IND-CCA secure. The definition of secure $\KDF$ with respect to a source of keying material IND-CCA $\KEM$ is given in~\S\ref{sec:keyderivationfunction}.

\paragraph{\textbf{Authenticated encryption with Associated Data.}} An authenticated encryption  with associated data $(\AEAD)$ scheme
is a tuple  $(\AEADKeyspace, \allowbreak \AEADE, \allowbreak \AEADD)$, where $\AEADKeyspace$ is the key space. The syntax of the encryption and decryption algorithms is the following.
\begin{description}

\item[$\AEADE(k,m)$.] On input a key $k \allowbreak \in \allowbreak \AEADKeyspace$ and a message $m$, output a ciphertext $\beta$ and associated data $\gamma$.

\item[$\AEADD(k, \beta, \gamma)$.] On input a key $k \allowbreak \in \allowbreak \AEADKeyspace$, a ciphertext $\beta$ and  associated data $\gamma$, output a message $m$ or $\bot$.

\end{description}
For simplicity, we have omitted the nonce. In~\S\ref{sec:authenticatedencryption}, we recall a formal definition of $\AEAD$, with the definitions of IND\$-CPA security, unforgeability and committing $\AEAD$~\cite{DBLP:conf/ccs/Rogaway02, cryptoeprint:2022/1260}.

\paragraph{\textbf{Symmetric-key encryption.}} A symmetric-key encryption $(\blockcipher)$ scheme is a tuple  $(\AEADKeyspace, \allowbreak \blockcipherenc, \allowbreak \blockcipherdec)$. Associated to $(\AEADKeyspace, \allowbreak \blockcipherenc, \allowbreak \blockcipherdec)$ is a message space $\blockciphermessagespace$ that consists of strings of bit length $\ell$. The ciphertext space is equal to the message space. The key space $\blockcipherkeyspace$ is a finite nonempty set of strings. The algorithms $(\blockcipherenc, \allowbreak \blockcipherdec)$ are defined as follows.
\begin{description}

\item[$\blockcipherenc(k,m)$.] On input a key $k \allowbreak \in \allowbreak \blockcipherkeyspace$ and a message $m \allowbreak \in \allowbreak \blockciphermessagespace$, output a ciphertext $c$.

\item[$\blockcipherdec(k,c)$.] On input a key $k \allowbreak \in \allowbreak \blockcipherkeyspace$ and a ciphertext $c$, output a message $m$.

\end{description}
We define correctness and IND\$-CCA security in~\S\ref{sec:symmetrickey}.

\paragraph{\textbf{Functionality $\Functionality_{\Freg}$.}} The ideal functionality $\Functionality_{\Freg}$ for registration~\cite{DBLP:conf/focs/Canetti01} interacts with authorized parties $\FregT$ that register a message $\fregvalue$ and with any parties $\Party$ that retrieve the registered message. $\Functionality_{\Freg}$ consists of two interfaces $\fregregister$ and $\fregretrieve$. The interface $\fregregister$ is used by $\FregT$ to register a message $\fregvalue$ with $\Functionality_{\Freg}$. A party $\Party$ uses $\fregretrieve$ on input a party identifier $\FregT$ to retrieve the message $\fregvalue$ associated with $\FregT$ from $\Functionality_{\Freg}$. We depict $\Functionality_{\Freg}$ in~\S\ref{sec:funcREG}.

\paragraph{\textbf{Functionality $\Functionality_{\Fpreg}$.}} The ideal functionality $\Functionality_{\Fpreg}$ for registration with privacy-preserving key retrieval is similar to $\Functionality_{\Freg}$, but there are two differences. In the registration interface, $\Functionality_{\Fpreg}$ allows registrations from any party. In the retrieval interface, $\Functionality_{\Fpreg}$ does not leak the identity of the party whose registered message is being retrieved to the adversary. We depict $\Functionality_{\Fpreg}$ in~\S\ref{sec:funcPREG}.

\paragraph{\textbf{Functionality $\Functionality_{\SMT}$.}} The ideal functionality $\Functionality_{\SMT}$ for secure message transmission~\cite{DBLP:conf/focs/Canetti01} interacts with a sender $\Sender$ and a receiver $\Receiver$, and consists of one interface $\fsmtsend$. $\Sender$ uses the $\fsmtsend$ interface to send a message $\SMTmessage$ to $\Receiver$. We depict $\Functionality_{\SMT}$ in~\S\ref{sec:IdealFunctionalitySMT}.

%% file: WPES3construction2Outfox.tex
\subsection{Outfox Packet Format}
\label{wpes:sec:outfoxpacketformat}

We describe the Outfox packet format. In~\S\ref{wpes:sec:definitionalgorithms}, we define the algorithms of Outfox. In~\S\ref{wpes:sec:constructionalgorithms}, we describe our implementation of those algorithms.  

\subsubsection{Definition of Outfox Algorithms}
\label{wpes:sec:definitionalgorithms}
\label{sec:definitionalgorithms}

We define the algorithms of the Outfox packet format and we describe how those algorithms are used in $\mathrm{\Pi}_{\LRM}$.

\begin{description}

    \item[$\lrmPacketCreate(\route, \flrmmessagerequest, \receiverinfo, \surb)$.] On input a route $\route$, a message $\flrmmessagerequest$, the receiver information $\receiverinfo$, and optionally a single-use reply block $\surb$, output a packet $\packet$. This algorithm is used in the request interface by a sender to compute a request packet. The route $\route$ consists of a tuple $\langle(\node{i}, \allowbreak \KEMpublickey_{\node{i}}, \allowbreak \routing{i})\rangle_{i\in[0,k-1]}$, where $k+1$ denotes the number of layers of encryption, $\node{i}$ is the identifier of the party that processes the packet at layer $i$, $\KEMpublickey_{\node{i}}$ is the public key of $\node{i}$, and $\routing{i}$ is the routing information that enables $\node{i}$ to forward the packet to the next party that should process it. The receiver information $\receiverinfo$ consists of a tuple $(\flrmReceiver, \allowbreak pk_{\flrmReceiver}, \allowbreak \routing{\flrmReceiver})$, where $\flrmReceiver$ is the identifier of the receiver, $pk_{\flrmReceiver}$ is the public key of the receiver, and $\routing{\flrmReceiver}$ is either $\bot$, if $\surb \allowbreak = \allowbreak \bot$, or contains routing information that enables the receiver to send a reply packet, if $\surb \allowbreak \neq \allowbreak \bot$.

    \item[$\lrmPacketProcess(\KEMsecretkey, \packet, a)$.] On input a packet $\packet$, a secret key $\KEMsecretkey$,  and a bit $a$ that indicates whether the last layer is being processed, when the processing of the packet header fails, output $\top$, and when the processing of the payload fails, output $\bot$. Otherwise, if $a \allowbreak = \allowbreak 1$, output a message $\flrmmessagerequest$ and optionally a single-use reply block $\surb$ and the routing information $\routing{\flrmReceiver}$. If $a \allowbreak \neq \allowbreak 1$, output a packet $\packet'$ and the routing information $\routing{i}$.  This algorithm is used in the forward interface by a node or a gateway to process a packet. Moreover, it is used in the forward interface by a receiver to retrieve the message encrypted in a request packet, and optionally also a single-use reply block $\surb$ and the routing information $\routing{\flrmReceiver}$.
    
    \item[$\lrmSurbCreate(\routeresponse, \senderinfo)$.] On input a route $\routeresponse$ and sender information $\senderinfo$, output a single-use reply block $\surb$, a single-use reply block identifier $\surbidentifier$ and single-use reply block secret information $\surbsecrets$. This algorithm is used in the request interface by a sender in order to compute a single-use reply block when the sender computes a request packet and wishes to enable replies to that packet. The route $\routeresponse$ consists of a tuple $\langle(\node{i}', \allowbreak \KEMpublickey_{\node{i}'}, \allowbreak \routing{i}')\rangle_{i\in[0,k-1]}$, where $\node{i}'$ is the identifier of the party that processes the reply packet at layer $i$, $\KEMpublickey_{\node{i}'}$ is the public key of $\node{i}'$, and $\routing{i}'$ is the routing information that enables $\node{i}'$ to forward the reply packet to the next party that should process it. The sender information $\senderinfo$ consists of a tuple $(\flrmSender, \allowbreak \KEMpublickey_{\flrmSender}, \allowbreak \allowbreak \bot)$, where $\flrmSender$ is the sender identifier, $\KEMpublickey_{\flrmSender}$ is the public key of the sender, and $\bot$ is a symbol that indicates that the packet should not be forwarded.

    \item[$\lrmSurbUse(\surb, \flrmmessageresponse)$.] On input a single-use reply block $\surb$ and a message $\flrmmessageresponse$, output a packet $\packet$. This algorithm is used in the reply interface by a receiver in order to compute a reply packet.

    \item[$\lrmSurbCheck(\packet, \surbidentifier)$.] On input a packet $\packet$ and a single-use reply block identifier $\surbidentifier$, output $1$ if $\packet$  was created from a single-use reply block identified by $\surbidentifier$, else output $0$. This algorithm is used in the forward interface by a user in order to check whether a packet is a reply packet and, if that is the case, in order to match a reply packet with the request packet that enabled that reply.

    \item[$\lrmSurbRecover(\packet, \surbsecrets)$.] On input a packet $\packet$ and single-use reply block secret information $\surbsecrets$, output $\bot$ if processing fails or if  $\packet$ was not created from a single-use reply block associated with the secrets $\surbsecrets$. Otherwise output a message $\flrmmessageresponse$. This algorithm is used in the forward interface by a sender in order to retrieve the message encrypted by a reply packet.

\end{description}

\subsubsection{Construction of Outfox Algorithms}
\label{wpes:sec:constructionalgorithms}
\label{sec:constructionalgorithms}

Outfox is based on layered encryption. To compute a packet, a layer of encryption is added for each party in the route. A packet $\packet$ consists of a header $h_i$ and a payload $\payload{i}$. To compute the encryption layer that is later processed by the party $\node{i}$, the encryption algorithm $\KEMEnc$ of the key encapsulation mechanism is executed on input the public key $\KEMpublickey_{\node{i}}$ to generate a shared key $\KEMsharedkey_i$ and a ciphertext $\KEMciphertext_i$. Then a key derivation function $\KDF$ is executed on input $\KEMsharedkey_i$ to create two keys: $s_i^{h}$, which is used to compute the header, and $s_i^{p}$, which is used to compute the payload. Next we proceed as follows:
\begin{description}

    \item[Header computation.] To compute the header $h_i$, the encryption algorithm $\AEADE$ of an authenticated encryption scheme is run on input the secret key $s_i^{h}$ and  $\routing{i} \ || \ h_{i+1}$, where $\routing{i}$ is the routing information and $h_{i+1}$ is the header of the packet to be processed by $\node{i+1}$. The algorithm $\AEADE$ outputs a ciphertext  $\beta_{i}$ and an authentication tag $\gamma_{i}$. The header $h_i$ consists of $\KEMciphertext_i$,  $\beta_{i}$ and $\gamma_{i}$.

    \item[Payload computation.] To compute the payload $\payload{i}$, a block cipher $\blockcipherenc$ is executed on input the secret key $s_i^{p}$ and the payload $\payload{i+1}$. 
    
\end{description}

To process a packet with a header $h_i = (\KEMciphertext_i, \allowbreak \beta_{i}, \allowbreak \gamma_{i})$ and a payload $\payload{i}$, the party $\node{i}$ proceeds as follows. First, $\node{i}$ runs the decryption algorithm $\KEMDec$ of the key encapsulation mechanism on input the secret key of $\node{i}$ and the ciphertext $\KEMciphertext_i$ in order to obtain the shared key $\KEMsharedkey_i$. Then $\node{i}$ computes the keys $s_i^{h}$ and $s_i^{p}$ by running $\KDF$ on input $\KEMsharedkey_i$. Next $\node{i}$ proceeds as follows:
\begin{description}

    \item[Header processing.] To process the header, $\node{i}$ runs the algorithm $\AEADD$ on input the secret key $s_{i}^h$, the ciphertext $\beta_{i}$, and the authentication tag $\gamma_{i}$. If decryption fails, the processing of the packet fails. Otherwise $\node{i}$ retrieves the routing information $\routing{i}$ and the header $h_{i+1}$.

    \item[Payload processing.] To process the payload, $\node{i}$ runs the algorithm $\blockcipherdec$ of the block cipher on input the secret key $s_i^{p}$ and the payload $\payload{i}$ in order to obtain the payload $\payload{i+1}$.

\end{description}

Request packets and reply packets are computed and processed similarly. The only difference is the following. To compute a request packet, the algorithm  $\lrmPacketCreate$ follows the procedure described above. To compute a reply packet, the algorithm $\lrmSurbUse$ uses a single-use reply block $\surb$, which consists of a header $h_{0}$ and a secret key $s_k^{p}$. The header $h_{0}$ is computed by the algorithm $\lrmSurbCreate$ and follows the same procedure described above. To compute the payload $\payload{0}$, the algorithm $\lrmSurbUse$ encrypts the reply message $\flrmmessageresponse$ by running $\blockcipherenc$ on input $s_k^{p}$. Therefore, while the payload of a request packet has been encrypted $k$ times, the payload of a reply packet is encrypted just once. The reply packet is processed in the same way as a request packet by nodes and gateways. When the reply packet is received by the sender, the sender first runs $\lrmSurbCheck$ to associate the reply packet to its corresponding request packet, and then runs $\lrmSurbRecover$ on input $\surbsecrets = (s_0^{p}, \allowbreak \ldots, \allowbreak s_k^{p})$ to reverse all the operations performed on the payload. The secrets $\surbsecrets$ are stored by the sender when running $\lrmSurbCreate$.

\paragraph{\textbf{Construction.}} We describe our construction for the algorithms defined in~\S\ref{wpes:sec:definitionalgorithms}. The algorithms are parameterized by the message space $\flrmMessageSpace$.

\begin{description}[leftmargin=2mm]
    
    \item[$\lrmPacketCreate(\route, \flrmmessagerequest, \receiverinfo, \surb)$.] Execute the following steps:
    
    \begin{itemize}[leftmargin=*]

        \item Abort if $\flrmmessagerequest \notin \flrmMessageSpace$.
    
       \item Parse the route $\route$ as $\langle(\node{i}, \KEMpublickey_{\node{i}}, \routing{i})\rangle_{i\in[0,k-1]}$.
       
       \item Parse the information of the receiver $\receiverinfo = (\flrmReceiver, pk_{\flrmReceiver}, \routing{\flrmReceiver}) = (\node{k}, \allowbreak \KEMpublickey_{\node{k}}, \allowbreak \routing{k})$. If $\surb = \bot$, set $\routing{\flrmReceiver} = \allowbreak \routing{k} \gets \bot$. Append $(\node{k}, \allowbreak \KEMpublickey_{\node{k}}, \allowbreak \routing{k})$ to $\route$.
       
       \item For each $i \in \{0, \ldots, k\}$, compute $(\KEMsharedkey_i, \allowbreak \KEMciphertext_i) \allowbreak \gets \allowbreak \KEMEnc(\KEMpublickey_{\node{i}})$.
       
       \item  For each $i \in \{0, \ldots, k\}$, compute $(s_i^{h}, \allowbreak s_i^{p}) \allowbreak \gets \allowbreak \KDF(\KEMsharedkey_i, \ell_i, ctx_i)$, where $\ell_i$ is the key length required for this layer (see~\S\ref{wpes:sec:efficiencyanalysisOutfox}) and $ctx_i$ contains $\KEMciphertext_i$, $\KEMpublickey_{\node{i}}$ and the session identifier of the protocol.
       
       \item Compute $(\beta_{k}, \allowbreak \gamma_{k}) \gets \AEADE(s_{k}^h,\routing{k})$.
       
       \item Set  as $h_{k} \gets (\KEMciphertext_k \ || \ \beta_{k} \ || \ \gamma_{k})$ the most inner-layer of the packet header.
       
       \item For $i \in \{0, \ldots, k-1\}$, run $(\beta_{i}, \gamma_{i}) \gets \AEADE(s_i^h,\routing{i} \ || \ h_{i+1})$ and set $h_{i} \gets (\KEMciphertext_i \ || \  \beta_{i} \ || \ \gamma_{i})$.

       \item If $\surb \allowbreak = \allowbreak \bot$, set $\flrmmessagerequest \allowbreak \gets \allowbreak 0^{\securityparameter}||0^{s}||\flrmmessagerequest$, else set $\flrmmessagerequest \allowbreak \gets \allowbreak 0^{\securityparameter}||\surb||\flrmmessagerequest$. $\securityparameter$ is the security parameter and $s$ is the bit length of a single-use reply block. 
       
       \item Compute $\payload{k} \gets \blockcipherenc(s_{k}^p, \flrmmessagerequest)$.
       
       \item For $i \in \{0, \ldots, k-1\}$, compute $\payload{i} \allowbreak \gets \allowbreak \blockcipherenc(s_{i}^p, \payload{i+1})$.

       \item Output a packet $\packet \allowbreak \gets \allowbreak (h_{0}, \allowbreak \payload{0})$.
       
    \end{itemize}

    \item[$\lrmPacketProcess(\KEMsecretkey, \packet, a)$.]  Execute the following steps:
    
    \begin{itemize}[leftmargin=*]

        \item Parse the packet $\packet$ as $(h_i, \allowbreak \payload{i})$.

        \item Parse the header $h_{i}$ as $(\KEMciphertext_{i}, \beta_{i}, \gamma_{i})$.

        \item Compute the shared key $\KEMsharedkey_i \allowbreak \gets \allowbreak \KEMDec(\KEMciphertext_{i}, \KEMsecretkey)$.

        \item Compute $(s_i^{h}, \allowbreak s_i^{p}) \allowbreak \gets \allowbreak \KDF(\KEMsharedkey_i, \ell_i, ctx_i)$, where $\ell_i$ is the key length required for this layer (see~\S\ref{wpes:sec:efficiencyanalysisOutfox}) and $ctx_i$ contains $\KEMciphertext_i$, $\KEMpublickey$ (which can be obtained from $\KEMsecretkey$) and the session identifier of the protocol. 
    
        \item Compute $d_h \allowbreak \gets \allowbreak \AEADD(s_{i}^h, \beta_{i}, \gamma_{i})$. If $d_h = \bot$, output $\top$.

        \item If $a \neq 1$, parse $d_h$ as $(\routing{i} \ || \ h_{i+1})$. If $a = 1$, parse $d_h$ as $\routing{k}$. Output $\top$ if parsing fails.

        \item Compute $\payload{i+1} \gets \blockcipherdec(s_{i}^p, \payload{i})$. 
        
        \item If $a \allowbreak = \allowbreak 1$, do the following:
        \begin{itemize}

            \item If $\routing{k} = \bot$, parse $\payload{i+1}$ as $0^{\securityparameter}||0^{s}||\flrmmessagerequest$. If parsing does not work, output $\bot$, else output $\routing{k} \gets \bot$, $\surb \gets \bot$, and $\flrmmessagerequest$.

            \item If $\routing{k} \neq \bot$, parse $\payload{i+1}$ as $0^{\securityparameter}||\surb||\flrmmessagerequest$. If parsing does not work, output $\bot$, else output $\routing{k}$, $\surb$ and $\flrmmessagerequest$.

        \end{itemize}
        \item If $a \allowbreak \neq \allowbreak 1$, output the packet $\packet'  \allowbreak \gets \allowbreak (h_{i+1}, \allowbreak \payload{i+1})$ and the routing information $\routing{i}$.

    \end{itemize}

    \item[$\lrmSurbCreate(\routeresponse, \senderinfo)$.] Execute the following steps:

    \begin{itemize}[leftmargin=*]

        \item Parse $\routeresponse$ as $\langle(\node{i}', \KEMpublickey_{\node{i}'}, \routing{i}')\rangle_{i\in[0,k-1]}$.

        \item Append the information of the sender $\senderinfo = (\flrmSender, \KEMpublickey_{\flrmSender}, \allowbreak \bot) = \allowbreak (\node{k}, \allowbreak \KEMpublickey_{\node{k}}, \allowbreak \routing{k})$ to $\routeresponse$. 

        \item For each $i \in \{0, \ldots, k\}$, compute $(\KEMsharedkey_i, \allowbreak \KEMciphertext_i) \allowbreak \gets \allowbreak \KEMEnc(\KEMpublickey_{\node{i}})$.

        \item  For each $i \in \{0, \ldots, k\}$, compute $(s_i^{h}, \allowbreak s_i^{p}) \allowbreak \gets \allowbreak \KDF(\KEMsharedkey_i, \ell_i, ctx_i)$, where $\ell_i$ is the key length required for this layer (see~\S\ref{wpes:sec:efficiencyanalysisOutfox}) and $ctx_i$ contains $\KEMciphertext_i$, $\KEMpublickey_{\node{i}}$ and the session identifier of the protocol. 
               
        \item Compute $(\beta_{k}, \allowbreak \gamma_{k}) \gets \AEADE(s_{k}^h,\routing{k})$.
       
        \item Set  as $h_{k} \gets (\KEMciphertext_k \ || \ \beta_{k} \ || \ \gamma_{k})$ the most inner-layer of the packet header.

       
        \item For $i \in \{0, \ldots, k-1\}$, run $(\beta_{i}, \gamma_{i}) \gets \AEADE(s_i^h,\routing{i} \ || \ h_{i+1})$ and set $h_{i} \gets (\KEMciphertext_i \ || \  \beta{i} \ || \ \gamma_{i})$.

        \item Output a single-use reply block $\surb \allowbreak \gets \allowbreak (h_{0}, s_k^{p})$, a single-use reply block identifier  $\surbidentifier \allowbreak \gets \allowbreak h_k$, and single-use reply block secret information $\surbsecrets \allowbreak \gets \allowbreak (s_0^{p}, \allowbreak \ldots, \allowbreak s_k^{p})$.

    \end{itemize}

    \item[$\lrmSurbUse(\surb, \flrmmessageresponse)$.] Execute the following steps:

    \begin{itemize}[leftmargin=*]

        \item Abort if $\flrmmessageresponse \notin \flrmMessageSpace$.

        \item Parse $\surb$ as $(h_{0}, s_k^{p})$.

        \item Set $\flrmmessageresponse \allowbreak \gets \allowbreak 0^{k+s}||\flrmmessageresponse$  and compute $\payload{0} \gets \blockcipherenc(s_{k}^p, \flrmmessageresponse)$. The bit length of $\payload{0}$ should be the same as in a request packet.

        \item Output a packet $\packet \allowbreak \gets \allowbreak (h_{0}, \payload{0})$.

    \end{itemize}
    
    \item[$\lrmSurbCheck(\packet, \surbidentifier)$.] Execute the following steps:

     \begin{itemize}[leftmargin=*]

        \item Parse $\packet$ as $(h_{k}, \payload{k})$ and $\surbidentifier$ as $h'_k$.


        \item If $h_{k} \allowbreak = \allowbreak h'_{k}$, output $1$, else output $0$.

     \end{itemize}
    
     \item[$\lrmSurbRecover(\packet, \surbsecrets)$.] Execute the following steps:

     \begin{itemize}[leftmargin=*]

        \item Parse $\surbsecrets$ as $(s_0^{p}, \allowbreak \ldots, \allowbreak s_k^{p})$ and $\packet$ as $(h_{k},\payload{k})$.

        \item For $i = k-1$ to $0$, compute $\payload{i} \gets \blockcipherenc(s_{i}^p, \payload{i+1})$.
 
        \item Compute $\flrmmessageresponse \gets \blockcipherdec(s_{k}^p, \payload{0})$.

        \item Parse $\flrmmessageresponse$ as $0^{k+s}||\flrmmessagerequest$. If parsing does not work, output $\bot$, else set $\flrmmessageresponse \gets \flrmmessagerequest$  and output $\flrmmessageresponse$. 

     \end{itemize}

\end{description}

%% file: WPES3construction3Protocol.tex
\subsection{Protocol}
\label{wpes:sec:protocol}

We describe the protocol $\mathrm{\Pi}_{\LRM}$. In~\S\ref{wpes:sec:construction1setup}, we describe and discuss the setup interface.
In~\S\ref{wpes:sec:construction2registration}, we describe  the registration interface.
In~\S\ref{wpes:sec:construction3request}, we describe  the request interface.
In~\S\ref{wpes:sec:construction4reply}, we describe the reply interface.
In~\S\ref{wpes:sec:construction5forward}, we describe  the forward interface.
In~\S\ref{sec:descriptionconstruction}, we give a formal description of $\mathrm{\Pi}_{\LRM}$ that realizes the ideal functionality $\Functionality_{\LRM}$ in~\S\ref{sec:functionalityDefinition}.

\subsubsection{Setup Phase}
\label{wpes:sec:construction1setup}

The setup phase works as follows:
    \begin{itemize}

        \item A node or a gateway $\flrmParty$ runs $(\KEMsecretkey, \allowbreak \KEMpublickey) \allowbreak \gets \allowbreak \KEMKeyGen(1^\securityparameter)$, where $1^\securityparameter$ is the security parameter.




        \item $\flrmParty$ uses the $\fregregister$ interface  to register the public key $\KEMpublickey$ with $\Functionality_{\Freg}$. The use of $\Functionality_{\Freg}$ guarantees that all honest users retrieve the same public keys for each of the nodes and gateways.

        \item $\flrmParty$ stores $(\KEMsecretkey, \allowbreak \KEMpublickey)$.

    \end{itemize}

\subsubsection{Registration Phase}
\label{wpes:sec:construction2registration}

The registration phase works as follows:
    \begin{itemize}

        \item For every node or gateway $\flrmParty$, a user  $\flrmUser$ does the following:
        \begin{itemize}

            \item $\flrmUser$ uses the $\fregretrieve$ interface of $\Functionality_{\Freg}$ to retrieve the public key $\KEMpublickey$ of $\flrmParty$.

            \item If $\KEMpublickey \allowbreak = \allowbreak \bot$, $\flrmUser$ outputs an abortion message, else $\flrmUser$ stores $(\allowbreak \flrmParty, \allowbreak \KEMpublickey)$.
        
        \end{itemize}

        \item $\flrmUser$ runs $(\KEMsecretkey, \allowbreak \KEMpublickey) \allowbreak \gets \allowbreak \KEMKeyGen(1^\securityparameter)$.

        \item $\flrmUser$ uses the $\fregregister$ interface to register the public key $\KEMpublickey$ with $\Functionality_{\Fpreg}$.  $\Functionality_{\Fpreg}$, like $\Functionality_{\Freg}$, guarantees that all honest users retrieve the same public key $\KEMpublickey$ for $\flrmUser$. However, $\Functionality_{\Fpreg}$, in contrast to $\Functionality_{\Freg}$, allows users to retrieve $\KEMpublickey$ without leaking $\KEMpublickey$ or $\flrmUser$ to the adversary, which is needed to avoid the leakage of the intended receiver of a packet in the request interface.

        \item $\flrmUser$ stores $(\allowbreak \KEMsecretkey, \allowbreak \KEMpublickey)$.

    \end{itemize}

\paragraph{Remark.} It could be possible to use $\Functionality_{\Fpreg}$ to also register the public keys of nodes and gateways, so that users could retrieve those public keys in the request interface when they need them to compute a packet, instead of retrieving all of them in the register interface. We chose to use a different functionality $\Functionality_{\Freg}$ because it can be realized by more efficient protocols than $\Functionality_{\Fpreg}$. In a typical mixnet, there are less than 1000 nodes and gateways, so retrieving all their public keys is practical. In contrast, there are potentially millions of users, so using $\Functionality_{\Freg}$ to register user public keys is not practical because a user would need to retrieve all their keys in the registration phase.

\subsubsection{Request Phase}
\label{wpes:sec:construction3request}

The request phase works as follows.

    \begin{itemize}

        \item A sender $\flrmSender$ receives as input a message $\flrmmessagerequest$ and a packet route  $[\flrmSender, \allowbreak \flrmGatewayentry, \allowbreak \flrmNodeArequest, \allowbreak \flrmNodeBrequest, \allowbreak \flrmNodeCrequest, \allowbreak \flrmGatewayexit, \allowbreak \flrmReceiver, \langle \flrmGatewayexitreply, \allowbreak \flrmNodeAresponse, \allowbreak \flrmNodeBresponse, \allowbreak \flrmNodeCresponse, \allowbreak  \flrmGatewayentryreply, \allowbreak \flrmSenderreply \rangle]$.

        \item $\flrmSender$ finds the stored tuples $(\flrmGatewayentry, \allowbreak \KEMpublickey_{\flrmGatewayentry})$, $(\flrmGatewayexit, \allowbreak \KEMpublickey_{\flrmGatewayexit})$,  $(\flrmNodeArequest, \allowbreak \KEMpublickey_{\flrmNodeArequest})$, $(\flrmNodeBrequest, \allowbreak \KEMpublickey_{\flrmNodeBrequest})$ and $(\flrmNodeCrequest, \allowbreak \KEMpublickey_{\flrmNodeCrequest})$.

        \item If $\langle  \allowbreak \flrmGatewayentryreply, \allowbreak \flrmGatewayexitreply, \allowbreak \flrmNodeAresponse, \allowbreak \flrmNodeBresponse, \allowbreak \flrmNodeCresponse, \allowbreak \flrmSenderreply \rangle \neq \bot$, $\flrmSender$ retrieves the stored tuples $(\flrmGatewayentryreply, \allowbreak \KEMpublickey_{\flrmGatewayentryreply})$, $(\flrmGatewayexitreply, \allowbreak \KEMpublickey_{\flrmGatewayexitreply})$,  $(\flrmNodeAresponse, \allowbreak \KEMpublickey_{\flrmNodeAresponse})$, $(\flrmNodeBresponse, \allowbreak \KEMpublickey_{\flrmNodeBresponse})$ and finally $(\flrmNodeCresponse, \allowbreak \KEMpublickey_{\flrmNodeCresponse})$.

        \item If $(\flrmReceiver', \allowbreak \KEMpublickey)$ such that $\flrmReceiver' = \flrmReceiver$ is not stored, $\flrmSender$ does the following: 
        \begin{itemize}
        
             \item $\flrmSender$ uses the $\fpregretrieve$ interface of $\Functionality_{\Fpreg}$ to retrieve the public key $\KEMpublickey$ of $\flrmReceiver$.

            \item If $\KEMpublickey \allowbreak = \allowbreak \bot$, $\flrmSender$ outputs an abortion message, else $\flrmSender$ stores $(\flrmReceiver, \allowbreak \KEMpublickey)$.

        \end{itemize}
        
        \item If $\langle  \allowbreak \flrmGatewayentryreply, \allowbreak \flrmGatewayexitreply, \allowbreak \flrmNodeAresponse, \allowbreak \flrmNodeBresponse, \allowbreak \flrmNodeCresponse, \allowbreak \flrmSenderreply \rangle \neq \bot$, $\flrmSender$ does the following:

        \begin{itemize}


            \item $\flrmSender$ sets the routing information as follows:
            \begin{itemize}

            
                \item Set $\routinginformation_{\flrmNodeAresponse} \allowbreak \gets \allowbreak \flrmNodeBresponse$.

                \item Set $\routinginformation_{\flrmNodeBresponse} \allowbreak \gets \allowbreak \flrmNodeCresponse$.

                \item Set $\routinginformation_{\flrmNodeCresponse} \allowbreak \gets \allowbreak \flrmGatewayentryreply$.

                \item Set $\routinginformation_{\flrmGatewayentryreply} \allowbreak \gets \allowbreak \flrmSender$. (We recall that $\flrmSender \allowbreak = \allowbreak \flrmSenderreply$.)
                
            \end{itemize}

            \item Set the route $\routeresponse \allowbreak \gets \allowbreak (\langle \flrmNodeAresponse, \allowbreak \KEMpublickey_{\flrmNodeAresponse}, \allowbreak \routinginformation_{\flrmNodeAresponse} \rangle, \allowbreak \langle \flrmNodeBresponse, \KEMpublickey_{\flrmNodeBresponse}, \allowbreak \routinginformation_{\flrmNodeBresponse} \rangle, \allowbreak \langle \flrmNodeCresponse, \allowbreak \KEMpublickey_{\flrmNodeCresponse}, \allowbreak \routinginformation_{\flrmNodeCresponse} \rangle, \allowbreak \langle \flrmGatewayentryreply, \allowbreak \KEMpublickey_{\flrmGatewayentryreply}, \allowbreak \routinginformation_{\flrmGatewayentryreply} \rangle)$.  We recall that, in a reply, the exit gateway $\flrmGatewayexitreply$ does not process the packet, so it is not included in $\routeresponse$.

            \item Set the sender information $\senderinfo \allowbreak \gets \allowbreak \langle \flrmSender, \allowbreak \KEMpublickey, \allowbreak \bot \rangle$, where $\KEMpublickey$ is stored in the tuple $(\KEMsecretkey, \allowbreak \KEMpublickey)$ create during the registration phase.

            \item Compute $(\surb, \allowbreak \surbidentifier, \allowbreak \surbsecrets) \allowbreak \gets \allowbreak \lrmSurbCreate(\routeresponse, \allowbreak \senderinfo)$.

            \item Store $(\surbidentifier, \allowbreak \surbsecrets, \allowbreak \flrmGatewayentryreply)$.


        \end{itemize}

        \item $\flrmSender$ sets the following routing information:

        \begin{itemize}

            \item Set $\routinginformation_{\flrmNodeArequest} \allowbreak \gets \allowbreak \flrmNodeBrequest$.

            \item Set $\routinginformation_{\flrmNodeBrequest} \allowbreak \gets \allowbreak \flrmNodeCrequest$.

            \item Set $\routinginformation_{\flrmNodeCrequest} \allowbreak \gets \allowbreak \flrmGatewayexit$.

            \item Set $\routinginformation_{\flrmGatewayexit} \allowbreak \gets \allowbreak \flrmReceiver$.

        \end{itemize}

        \item $\flrmSender$ sets the route $\route \allowbreak \gets \allowbreak (\langle \flrmNodeArequest, \allowbreak \KEMpublickey_{\flrmNodeArequest}, \allowbreak \routinginformation_{\flrmNodeArequest} \rangle, \allowbreak \langle \flrmNodeBrequest, \KEMpublickey_{\flrmNodeBrequest}, \allowbreak \routinginformation_{\flrmNodeBrequest} \rangle, \allowbreak \langle \flrmNodeCrequest, \allowbreak \KEMpublickey_{\flrmNodeCrequest}, \allowbreak \routinginformation_{\flrmNodeCrequest} \rangle, \allowbreak \langle \flrmGatewayexit, \allowbreak \KEMpublickey_{\flrmGatewayexit}, \allowbreak \routinginformation_{\flrmGatewayexit} \rangle)$. We recall that, in a request, the entry gateway $\flrmGatewayentry$ does not process the packet, so it is not included in $\route$.

        \item If $\langle  \allowbreak \flrmGatewayentryreply, \allowbreak \flrmGatewayexitreply, \allowbreak \flrmNodeAresponse, \allowbreak \flrmNodeBresponse, \allowbreak \flrmNodeCresponse, \allowbreak \flrmSenderreply \rangle \neq \bot$, $\flrmSender$ sets the receiver information $\receiverinfo \allowbreak \gets \allowbreak \langle \flrmReceiver, \allowbreak \KEMpublickey, \allowbreak (\flrmGatewayexitreply, \allowbreak  \flrmNodeAresponse) \rangle$, else $\flrmSender$ sets  $\receiverinfo \allowbreak \gets \allowbreak \langle \flrmReceiver, \allowbreak \KEMpublickey, \allowbreak \bot \rangle$ and $\surb \allowbreak \gets \allowbreak \bot$, where $\KEMpublickey$ is in the tuple  $(\flrmReceiver, \allowbreak \KEMpublickey)$. When a reply is enabled, $\flrmNodeAresponse$ is included because exit gateways do not process reply packets, so the receiver needs to tell the exit gateway the identifier of the first-layer node to which the reply packet should be forwarded.

        \item $\flrmSender$ runs $\packet \allowbreak \gets \allowbreak \lrmPacketCreate(\route, \flrmmessagerequest, \receiverinfo, \surb)$.

        \item $\flrmSender$ uses the $\fsmtsend$ of $\Functionality_{\SMT}$ to send the message $\langle \packet, \allowbreak \flrmNodeArequest \rangle$ to the entry gateway $\flrmGatewayentry$. $\flrmNodeArequest$ is included so that $\flrmGatewayentry$ knows the first-layer node to which the packet should be forwarded. Thanks to the use of $\Functionality_{\SMT}$, an adversary that controls the communication channel cannot learn or modify $\langle \packet, \allowbreak \flrmNodeArequest \rangle$.

        \item $\flrmGatewayentry$ creates a fresh random packet identifier $\tid$ and stores $(\tid, \allowbreak \packet, \allowbreak \flrmNodeArequest)$.

    \end{itemize}

\subsubsection{Reply Phase}
\label{wpes:sec:construction4reply}

The reply phase works as follows.
    \begin{itemize}

        \item A receiver $\flrmReceiver$ receives as input a packet identifier $\tid$ and a message $\flrmmessageresponse$.

        \item $\flrmReceiver$ retrieves the tuple $(\tid', \allowbreak \surb, \allowbreak \flrmGatewayexitreply, \allowbreak \flrmNodeAresponse)$ such that $\tid' = \tid$. This tuple is stored in the forward phase when $\flrmReceiver$ receives a request packet that enables a reply.

        \item $\flrmReceiver$ runs $\packet \allowbreak \gets \allowbreak \lrmSurbUse(\surb, \allowbreak \flrmmessageresponse)$.
        
        \item $\flrmReceiver$ uses the $\fsmtsend$ interface of $\Functionality_{\SMT}$ to send the message $\langle \packet, \allowbreak \flrmNodeAresponse \rangle$ to the exit gateway $\flrmGatewayexitreply$.
        
        \item $\flrmGatewayexitreply$ creates a fresh random packet identifier $\tid$ and stores $(\tid, \allowbreak \packet, \allowbreak \flrmNodeAresponse)$. 

    \end{itemize}

\subsubsection{Forward Phase}
\label{wpes:sec:construction5forward}

The forward phase works as follows.
    \begin{itemize}

        \item A node or gateway $\flrmParty$ receives as input a packet identifier $\tid$.

        \item $\flrmParty$ retrieves the tuple $(\tid', \allowbreak \packet, \allowbreak \flrmParty')$ such that $\tid' \allowbreak = \allowbreak \tid$. This tuple was stored when $\flrmParty$ received a packet in the request, reply or forward phases.

        \item $\flrmParty$ uses the $\fsmtsend$ interface of $\Functionality_{\SMT}$ to send the packet $\packet$ to $\flrmParty'$.

        \item If $\flrmParty'$ is a user, $\flrmParty'$ does the following:

        \begin{itemize}

            \item Set $b \gets 0$.

            \item For all the stored tuples  $(\surbidentifier, \allowbreak \surbsecrets, \allowbreak \flrmGatewayentryreply)$, do the following:

            \begin{itemize}

                \item Run $b \allowbreak \gets \allowbreak \lrmSurbCheck(\packet, \allowbreak \surbidentifier)$.

                \item If $b = 1$, do the following:

                \begin{itemize}

                    \item Abort if $\flrmParty \allowbreak \neq \allowbreak \flrmGatewayentryreply$. $\flrmParty'$ learns $\flrmParty$ from $\Functionality_{\SMT}$.

                    \item Run $\flrmmessageresponse \allowbreak \gets \allowbreak \lrmSurbRecover(\packet, \surbsecrets)$. 

                    \item Output $\flrmmessageresponse$. 

                \end{itemize}

            \end{itemize}

            \item If $b = 0$, do the following:

            \begin{itemize} 
            
                \item Run $(\routing{k}, \surb, \flrmmessagerequest)  \allowbreak \gets \allowbreak \lrmPacketProcess(\KEMsecretkey, \allowbreak \packet, \allowbreak 1)$.

                \item If $(\routing{k}, \surb, \flrmmessagerequest) \allowbreak = \allowbreak \top$, abort. The output $\top$ means that the processing of the header was not successful, which could happen e.g.\ because the packet was not intended to be processed by $\flrmParty'$.

                \item Else, if $(\routing{k}, \surb, \flrmmessagerequest) \allowbreak = \allowbreak \bot$, output $\bot$. This output means that the processing of the payload was not successful. 

                \item Else, if $\routing{k} = \surb = \bot$, output $\flrmmessagerequest$.
                
                \item Else, do the following:
                \begin{itemize}
                    
                    \item Create a fresh random packet identifier $\tid$.

                    \item Parse $\routing{k}$ as $(\flrmGatewayexitreply, \allowbreak \flrmNodeAresponse)$.

                    \item Store $(\tid, \allowbreak \surb, \allowbreak \flrmGatewayexitreply, \allowbreak \flrmNodeAresponse)$.

                    \item Output $\flrmmessagerequest$.
                    
                \end{itemize}
                
            \end{itemize}
            
        \end{itemize}

        \item If $\flrmParty'$ is a node or a gateway, $\flrmParty'$ does the following:
        
        \begin{itemize}

            \item Compute $(\packet', \allowbreak \routinginformation) \allowbreak \gets \allowbreak \lrmPacketProcess(\KEMsecretkey, \allowbreak \packet, \allowbreak 0)$.

            \item If $(\packet', \allowbreak \routinginformation) \allowbreak = \allowbreak \top$, abort.

            \item Else, create a fresh local packet identifier $\tid$ and store $(\tid, \allowbreak \packet', \allowbreak \routinginformation)$.

            \end{itemize}

        \end{itemize}

%% file: WPES4securityanalysis.tex
\section{Security Analysis}
\label{wpes:sec:securityanalysis}

In this section, we explain informally why our protocol in~\S\ref{wpes:sec:protocol} satisfies the security properties described in~\S\ref{wpes:sec:securityproperties}. In our security analysis in~\S\ref{sec:securityanalysis}, we prove that the formal description of our protocol in~\S\ref{sec:descriptionconstruction} realizes our ideal functionality in~\S\ref{sec:functionalityDefinition}.

\paragraph{Confidentiality.} Confidentiality relies on the use of a $\KDF$ that is secure with respect to a source of keying material IND-CCA $\KEM$, on the IND\$-CPA property of the $\AEAD$ scheme and on the IND\$-CCA property of the $\blockcipher$ scheme. In our security analysis, we use those properties to show that the adversary cannot distinguish a layer of encryption
associated to an honest party from a random layer of encryption, for layers of encryption of request packets and single-use reply blocks computed by honest senders.

First, we use the fact that the $\KDF$ is secure with respect to a source of keying material IND-CCA $\KEM$ to show that the $\KEM$ ciphertext $\KEMciphertext_i$ included in the header of a layer of encryption does not leak any information about the keys $(s_i^{h}, \allowbreak s_i^{p})$ used to encrypt the $\AEAD$ ciphertext $\beta_{i}$ and the payload $\payload{i}$. Thanks to that, afterwards we can use the IND\$-CPA property of the $\AEAD$ scheme to prove that the adversary cannot distinguish the ciphertext $\beta_{i}$ from a random string. Similarly, we use the IND\$-CCA property of the $\blockcipher$ scheme to prove that the adversary cannot distinguish the payload $\payload{i}$ from a random string. In the case of the payload, IND\$-CCA security is needed because payload integrity cannot be checked by honest nodes and gateways, so the adversary can send modified payloads to them. 

\paragraph{Bitwise unlinkability.} Bitwise unlinkability relies on the same properties as confidentiality, i.e.\ the adversary is not able to link a packet $\packet$ sent to an honest party $\flrmParty$ to the packet $\packet'$ computed by $\flrmParty$ after processing $\packet$ because the adversary cannot distinguish a layer of encryption associated to $\flrmParty$ from a random layer that does not leak any information about $\packet'$. Additionally, it relies on the fact that the layer of encryption of $\packet'$ is computed independently of the layers of encryption above it, i.e., $\packet'$ does not contain any information about the layer in $\packet$ removed by $\flrmParty$.

\paragraph{Integrity.} Integrity of the header of a layer of encryption relies on the unforgeability and the committing properties of the $\AEAD$ scheme. In our security analysis, we use those properties to show that the adversary cannot modify the header of a layer of encryption associated to an honest party, for layers of encryption of request packets and single-use reply blocks computed by honest senders.

First, we use the unforgeability property of the $\AEAD$ scheme to show that, if the adversary modifies the $\AEAD$ ciphertext $\beta_{i}$ in the header $(\KEMciphertext_i \ || \  \beta_{i} \ || \ \gamma_{i})$, then the honest party that processes the header detects the modification with all but negligible probability. After that, we use the committing property of the $\AEAD$ scheme to show that the adversary cannot modify the $\KEM$ ciphertext $\KEMciphertext_i$ and/or the associated data $\gamma_{i}$ is such a way that decryption of $\beta_{i}$ does not fail. We remark that modification of $\KEMciphertext_i$ leads to the computation of a key $\hat{s}_i^{h}$ different from the key $s_i^{h}$ one used to encrypt $\beta_{i}$. The committing property guarantees that decryption with a different key fails.

As explained in~\S\ref{wpes:sec:securityproperties}, integrity of the payload only requires that, after removing the last layer of encryption of a request or reply packet, the user can detect whether the encrypted message has been modified. (In the case of request packet that enables a reply, the user can also detect that the encrypted single-use reply block has not been modified.) We use zero padding of enough length to allow the user to detect that the payload has been modified after removing the last layer of encryption. Other methods to protect the integrity of the encrypted message could be used instead.

\paragraph{Request-reply indistinguishability.} This property holds, first of all, thanks to the fact that the processing of a packet by gateways and nodes is the same for both request and reply packets. Second, corrupt gateways and nodes are not able to distinguish request packets computed by honest senders from reply packets computed by honest receivers on input single-use reply blocks computed by honest senders thanks to the IND\$-CCA property of the $\blockcipher$ scheme. 

The header of request and reply packets is computed identically. The payload, however, is computed differently. A sender that computes a request packet encrypts the message once for each layer of encryption. In contrast, a receiver that computes a reply packet only encrypts the message once. Nevertheless, thanks to the IND\$-CCA property of the $\blockcipher$ scheme, we show that, in both cases, the adversary cannot distinguish the payload from a random string.

%% file: WPES5efficiencyanalysis.tex
\section{Efficiency Analysis}
\label{wpes:sec:efficiencyanalysis}

In~\S\ref{wpes:sec:efficiencyanalysisOutfox}, we analyze the computation, communication, and storage cost of Outfox. We also show performance measurements when Outfox is instantiated with several $\KEM$. In~\S\ref{wpes:sec:comparisonsphinx}, we compare the efficiency of Outfox with that of Sphinx.

%% file: WPES5efficiencyanalysis1Outfox.tex
\subsection{Efficiency Analysis of Outfox}
\label{wpes:sec:efficiencyanalysisOutfox}

\paragraph{Computation cost.} For an algorithm $B$, let $|B|$ denote the average running time in seconds of that algorithm for a given security parameter $1^k$. Let $l+1$ denote the number of layers of encryption of a packet or a single-use reply block. In Table~\ref{tab:computationcostOutfox}, we show the computation cost of each of the algorithms of Outfox.

\begin{table}
    \centering
    \resizebox{\columnwidth}{!}{
    \begin{tabular}{|l|l|} \hline 
        Algorithm    & Running time \\ \hline
        $\lrmPacketCreate$ & $(l+1) \cdot (|\KEMEnc| + |\KDF| + |\AEADE| + |\blockcipherenc|)$  \\ \hline
        $\lrmPacketProcess$ & $|\KEMDec| + |\KDF| + |\AEADD| + |\blockcipherdec|$ \\ \hline
        $\lrmSurbCreate$ & $(l+1) \cdot (|\KEMEnc| + |\KDF| + |\AEADE|)$ \\ \hline
        $\lrmSurbUse$ & $|\blockcipherenc|$ \\ \hline
        $\lrmSurbCheck$ & Negligible \\ \hline
        $\lrmSurbRecover$ & $l \cdot |\blockcipherenc| + |\blockcipherdec|$  \\ \hline
    \end{tabular}
    }
    \caption{Computation cost of Outfox}
    \label{tab:computationcostOutfox}
\end{table}

The computation cost of Outfox is dominated by the KEM scheme used. To compute a packet $\packet$ or a single-use reply block $\surb$ of $l+1$ layers, algorithm $\KEMEnc$ is run $l+1$ times. To process a packet, algorithm $\KEMDec$ is run once. The computation costs of the key derivation function $\KDF$, of the symmetric-key encryption scheme $\blockcipher$,  or of the $\AEAD$ scheme are small compared to the $\KEM$, which requires public-key encryption operations. 

In Table~\ref{tab:kem_performance} and in Table~\ref{tab:kem_performance2}, we show performance measurements for the implementation of algorithms $\KEMKeyGen$, $\KEMEnc$ and $\KEMDec$ for the following $\KEM$ schemes. 
\begin{itemize}

    \item X25519~\cite{langley2016elliptic,bernstein2006curve25519}, a $\KEM$  based on elliptic curve Diffie-Hellman. It is not post-quantum secure. The algorithm $\KEMEnc$ requires 2 exponentiations, whereas $\KEMDec$ requires 1 exponentiation. 

    \item ML-KEM-768~\cite{nistmlkem}, a lattice-based post-quantum $\KEM$.

    \item Classic McEliece~\cite{josefsson-mceliece-02}, a code-based post-quantum $\KEM$ with parameter set mceliece460896f.

    \item X-Wing $\KEM$ Draft 02~\cite{connolly-cfrg-xwing-kem-02}, a hybrid post-quantum $\KEM$ that combines the $\KEM$ schemes based on X25519 and ML-KEM-768.

\end{itemize}
\begin{table*}[t]
    \begin{minipage}[t]{0.48\textwidth}
        \centering
        \resizebox{\columnwidth}{!}{
        \begin{tabular}{|l|c|c|c|c|}
            \hline
            & x25519 & ML-KEM-768 & X-Wing & Classic McEliece \\
            \hline
            $\KEMKeyGen$ & \SI{34.428}{\micro\second}  & \SI{13.732}{\micro\second}  & \SI{48.438}{\micro\second}  & \SI{264.65}{\milli\second}  \\
            $\KEMEnc$ & \SI{69.587}{\micro\second}  & \SI{14.485}{\micro\second}  & \SI{84.337}{\micro\second}  & \SI{50.280}{\milli\second}  \\
            $\KEMDec$ & \SI{34.331}{\micro\second}  & \SI{15.896}{\micro\second}  & \SI{50.218}{\micro\second}  & \SI{30.881}{\milli\second}  \\
            \hline
        \end{tabular}
        }
        \caption{Performance comparison of $\KEM$ schemes on server}
        \label{tab:kem_performance}
    \end{minipage}
    \hfill
    \begin{minipage}[t]{0.48\textwidth}
        \centering
        \resizebox{\columnwidth}{!}{
        \begin{tabular}{|l|c|c|c|c|}
            \hline
            & x25519 & ML-KEM-768 & X-Wing & Classic McEliece \\
            \hline
            $\KEMKeyGen$ & \SI{29.080}{\micro\second} & \SI{10.985}{\micro\second} & \SI{40.647}{\micro\second} & \SI{114.77}{\milli\second} \\
            $\KEMEnc$ & \SI{58.189}{\micro\second} & \SI{11.406}{\micro\second} & \SI{71.844}{\micro\second} & \SI{54.287}{\milli\second} \\
            $\KEMDec$ & \SI{30.054}{\micro\second} & \SI{12.670}{\micro\second} & \SI{42.628}{\micro\second} & \SI{24.067}{\milli\second} \\
            \hline
        \end{tabular}
        }
        \caption{Performance comparison of $\KEM$ schemes on laptop}
        \label{tab:kem_performance2}
    \end{minipage}
\end{table*}


In Table~\ref{tab:kem_performance}, the performance measurements were obtained by using a server running Ubuntu Linux on an AMD Ryzen 5 3600 6-Core processor (3.6 GHz). In Table~\ref{tab:kem_performance2}, the measurements were obtained by using a Lenovo
T14s laptop running Arch Linux on an AMD Ryzen 7 PRO 5850U (1.9 GHz). The high-performance and formally verified libcrux library was used for testing Curve 25519 and ML-KEM~\cite{libcrux}.\footnote{\url{https://github.com/cryspen/libcrux}} Classic McEliece was tested using the standard C library.\footnote{\url{@https://github.com/sz3/libmcleece}}

\paragraph{Communication cost.} Each packet consists of a header and a payload. The payload length is equal to the length of a ciphertext of the symmetric-key encryption scheme. The header consists of three elements: a ciphertext and associated data for the $\AEAD$ scheme, and a ciphertext of the $\KEM$ scheme. 

The bit length of a header is given by the bit lengths of the $\KEM$ ciphertext, of the $\AEAD$ ciphertext and of the associated data. Let $p$ denote the bit length of the $\KEM$ ciphertext. For the security parameter $1^k$, the bit length of the associated data of the $\AEAD$ scheme is $k$. (This allows a fair comparison with Sphinx in~\S\ref{wpes:sec:comparisonsphinx}.) The bit length of the $\AEADE$ ciphertext depends on the layer in which the ciphertext is located. It can be computed as follows:
\begin{itemize}
    
    \item For layer $l$, the ciphertext encrypts the party identifiers of the exit gateway $\flrmGatewayexitreply$ and of the first-layer node $\flrmNodeAresponse$. In general, we consider that the bit length of a party identifier is $k$, and thus the bit length of this ciphertext is $2k$. However, to allow a fair comparison with Sphinx, we use $3k$ as size.\footnote{Sphinx operates in the service model and considers that destination addresses are of size $2k$, whereas other party identifiers are of size $k$.}

    \item For layer $l-1$, the ciphertext encrypts the identifier of the receiver $\flrmReceiver$, of size $k$, and the header of the next layer, which consists of a $\KEM$ ciphertext of size $p$, an $\AEAD$ ciphertext of size $3k$, and associated data of size $k$. The bit length is thus $k + p + 3k + k = p + 5k$.

    \item In general, for the layer $l-i$ (for $i \allowbreak \in \allowbreak [0,l]$), the bit length is $3k + ip + 2ik$.

    \item Therefore, for layer $0$, the bit length is $3k + lp + 2lk$.
    
\end{itemize}
Consequently, the bit length of the header for layer $l-i$ is $4k + (i+1)p + 2ik$, which is obtained by adding the bit length $p$ of the $\KEM$ ciphertext and the bit length $k$ of the associated data to the bit length $3k + ip + 2ik$ of the $\AEAD$ ciphertext.

The bit length of a payload ciphertext $\payloa$ is given by the bit length of the values that need to be encrypted in the innermost layer of encryption. For the security parameter $1^k$, those elements are a padding of $k$ zeroes, a message $\flrmmessagerequest$ of bit length $|\flrmmessagerequest|$, and a single-use reply block $\surb$, which consists of a header of bit length $4k + (l+1)p + 2lk$ and a key of length $k$. Consequently, the bit length of a payload is $k + |\flrmmessagerequest| + 4k + (l+1)p + 2lk + k = 6k + |\flrmmessagerequest| + (l+1)p + 2lk$.

\begin{table}[t]
    \centering
    \resizebox{\columnwidth}{!}{
    \begin{tabular}{|l|l|} \hline 
        Value    & Bit length \\ \hline
        $\KEM$ ciphertext & p  \\ \hline
        Associated data & k \\ \hline
        $\AEAD$ ciphertext at layer $l-i$ & $3k + ip + 2ik$ \\ \hline
        Header at layer $l-i$ & $4k + (i+1)p + 2ik$ \\ \hline
        Header at layer $0$ & $4k + (l+1)p + 2lk$ \\ \hline
        Single-use reply block &  $5k + (l+1)p + 2lk$  \\ \hline
        Payload & $6k + |\flrmmessagerequest| + (l+1)p + 2lk$ \\ \hline
        Packet at layer $l-i$ & $10k + |\flrmmessagerequest| + 2(i+1)p + 2(i+l)k$ \\ \hline
        Packet at layer $0$ & $10k + |\flrmmessagerequest| + 2(l+1)p + 4lk$ \\ \hline
    \end{tabular}
    }
    \caption{Communication cost of Outfox. The security parameter is $1^k$ and the number of layers is $l+1$.}
    \label{tab:communicationcostOutfox}
\end{table}

\begin{table}[t]
    \centering
    \resizebox{\columnwidth}{!}{
    \begin{tabular}{|l|c|c|c|c|}
        \hline
        & x25519 & ML-KEM-768 & X-Wing & Classic McEliece \\
        \hline
        Public Key & 32 & 1 184 & 1 216 & 524 160 \\
        Ciphertext & 32 & 1 088 & 1 120 & 188 \\
        \hline
    \end{tabular}
    }
    \caption{Size in bytes of public keys and ciphertexts}
    \label{tab:public_keys_ciphertexts}
\end{table}

In Table~\ref{tab:communicationcostOutfox}, we summarize the communication cost of Outfox. In Table~\ref{tab:public_keys_ciphertexts}, we show the bit length $p$ of $\KEM$ ciphertexts for the $\KEM$ schemes mentioned above.

\paragraph{Storage cost.} In terms of storage, each party needs to store a key pair of the $\KEM$ scheme. In Table~\ref{tab:public_keys_ciphertexts}, we show the size of $\KEM$ public keys for the $\KEM$ schemes mentioned above.

Senders also need to store the headers of the innermost layer of encryption of single-use reply blocks they send in order to identify reply packets. To increase efficiency, they can store a hash of the header instead of the full header.

If duplicate detection is implemented, parties need to store all the packets received in order to discard duplicates. Efficient storage mechanisms to check set membership, such as a Bloom filter, can be used. We remark that, every time parties update their keys, a new execution of the protocol starts, and the packets received in previous executions of the protocol can be erased. 

%% file: WPES5efficiencyanalysis2Sphinxcomparison.tex
\subsection{Comparison to Sphinx}
\label{wpes:sec:comparisonsphinx}

We compare the computation, communication and storage costs of Outfox with the costs of Sphinx~\cite{DBLP:conf/sp/DanezisG09}. For a fair comparison, we use the version of Sphinx in which single-use reply blocks are encrypted in the payload, as described in~\cite{DBLP:journals/popets/SchererWS24a}. Additionally, since Sphinx uses a mechanism based on the gap Diffie-Hellman assumption, we use the version of Outfox instantiated with the x25519 $\KEM$.

\paragraph{Computation Cost.} Like in Outfox, the communication cost of Sphinx is dominated by public-key operations. Therefore, we disregard the symmetric-key encryption operations, which are similar in Outfox and in Sphinx.

A key difference between Outfox and Sphinx is the way in which the shared key of the $\KEM$ is computed. In Outfox, to compute a header, for each layer of encryption the sender generates a $\KEM$ ciphertext and a $\KEM$ shared key that are computed independently of other layers of encryption. However, in Sphinx, the sender picks up only one secret for the outermost layer of encryption, and all the ciphertexts and shared keys are derived from that secret. As a consequence, when processing a packet, the $\KEM$ ciphertext needs to be recomputed, rather than decrypted as in Outfox.

As result, in Sphinx, computing a packet of $l+1$ layers requires $2(l+1)$ exponentiations, as in Outfox. However, processing a packet requires $2$ exponentiations in Sphinx, whereas in Outfox it requires $1$ exponentiation.

\paragraph{Communication Cost.} In contrast to Outfox, Sphinx packets hide the length of the path. To achieve that, all packets must have the same bit length, and thus Sphinx adds padding to each layer so that the bit length is constant. Consequently, while in Outfox the bit length of packets decreases after a layer of encryption is processed, in Sphinx it does not change.

The bit length of a Sphinx packet for a maximum of $l+1$ layers of encryption and that encrypts a single-use reply block in the payload is $2p + (4(l+1)+4)k + |\flrmmessagerequest| = 8k + |\flrmmessagerequest| + 2p + 4lk$, whereas the one of Outfox is $10k + |\flrmmessagerequest| + 2(l+1)p + 4lk$. As can be seen, the main difference is that, in Outfox, $l+1$ $\KEM$ ciphertexts are transmitted for each header, whereas in Sphinx, as explained above, only one $\KEM$ ciphertext is transmitted for each header.

We stress that the bit length of Outfox packets decreases when removing layers of encryption. The bit length of the packet at layer $l-i$ for $i \allowbreak \in \allowbreak [0,l]$ is $10k + |\flrmmessagerequest| + 2(i+1)p + 2(i+l)k$. For layer $l$, the bit length is $12k + |\flrmmessagerequest| + 2p$. This efficiency gain is derived from the fact that Outfox does not hide the length of the path.

\begin{table}[t]
    \centering
    \small
    \begin{tabularx}{\columnwidth}{|l|X|X|X|X|}
        \hline
        & Sphinx x25519 & Outfox x25519 & Outfox ML-KEM-768 & Outfox X-Wing \\
        \hline
        PacketCreate        & 183.24 µs & 249.74 µs & 1.13 ms & 1.344 ms \\
        PacketProcess: node                  & 55.32 µs & 31.00 µs & 243.92 µs & 482.17 µs \\
        PacketProcess: user   & 132.55 µs & 30.85 µs & 225.30 µs & 466.99 µs \\
        \hline
    \end{tabularx}
    \caption{Sphinx vs. Outfox: End-to-End Comparison}
    \label{tab:endtoend}
\end{table}

\paragraph{End-to-End Performance.} We compared Sphinx vs. Outfox in terms of end-to-end packet performance in Table~\ref{tab:endtoend} with a mix network path of 3 hops (as has been shown to be optimal in terms of anonymity in simulations ~\cite{piotrowska2021studying}) and a final hop with a decryption of the payload at a desination server (A total of 4 hops). In order to control the network bandwidth overhead, the mix net was run locally, and in order to control for payload, the payload size was kept constant at 1000 bytes. For Sphinx, the \emph{dalek} library was used.\footnote{\url{https://github.com/dalek-cryptography/x25519-dalek}} For Outfox, the \emph{libcrux} library tested earlier was used for all KEMs tested~\cite{libcrux}. An Apple M4 Pro laptop was used for testing with 24gb RAM, 4.5 GHz (14 cores).  Sphinx and Outfox both implemented using Rust.\footnote{We do not share a link to the precise code library as it would de-anonymize the authors, but the link will be added.} 

Using X25519, Outfox is about twice as fast as Sphinx for decryption per layer, as is to be expected. The construction of Sphinx packets is somewhat faster than that of Outfox using X25519, however, the final step of packing processing is much faster in Outfox than Sphinx. Furthermore, moving Outfox to a post-quantum setting using ML-KEM provides a magnitude performance loss compared to X25519 (but still usable for many applications). Then, moving to the X-Wing hybrid scheme doubles the time required for the Outfox operations. In terms of end-to-end usage of mix networks, the packet processing time per layer dominates all other factors.  Thus, in general Outfox with KEMs is more optimal for a constant-length mix network than Sphinx using standard key exchange. \footnote{ Note that high-speed layer decoding is important, as a very slow decoding could possibly be detected in terms of per-node mixnet delay times and so be used to de-anonymize the network.} Moving to a post-quantum setting still results in considerable performance loss, but given the high general performance, the loss may be acceptable in order to achieve post-quantum security.

%% file: WPES6Conclusion.tex
\section{Conclusion}
\label{wpes:sec:conclusion}
We have presented \emph{Outfox}, a simplified and post-quantum capable variant of the Sphinx packet format, tailored for multi-hop anonymous network with fixed-length routes. Outfox addresses key concerns  of Sphinx by reducing cryptographic overhead, enabling per-hop header reduction, and removing one of the two key exchanges required for packet processing. These design choices lead to measurable gains in efficiency while preserving strong anonymity.

By integrating modern Key Encapsulation Mechanisms (KEMs), Outfox offers a path toward quantum-resistant anonymous communication. We formally specify the Outfox format and prove its security within the Universal Composability framework. Although the security proof of Outfox may be considered quite detailed, it is particularly important given how a lack of formal rigor in proofs led to concrete de-anonymization attacks in prior mixnet packet formats~\cite{shimshock2008breaking} and have revealed subtle attacks on onion-routing and mixnets in the past~\cite{DBLP:conf/sp/KuhnBS20, klooss2024eror}. Furthermore, we evaluate the performance of Outfox across a range of classical and post-quantum KEMs. Our results show that Outfox achieves both practical efficiency and long-term cryptographic resilience, making it a strong candidate for next-generation mixnet deployments.

A standardized post-quantum unlinkable packet format may have uses outside of mixnets. There has also been deployment of Sphinx outside of mixnets: Bitcon's Lightning Network uses the Sphinx packet format (with variable hops, thus opening the Lightning network to de-anonymization attacks~\cite{rohrer2020counting}), file storage network IPFS~\cite{daniel2024poster}, and there is currently an investigation of using Sphinx for anonymizing consensus protocols.\footnote{\url{https://github.com/vacp2p/mix}} There is also generic usage of Sphinx for the protection of data in the cloud~\cite{song2024cloudnet}. There may be interest in Outfox for usage in other communication systems, including hybrid onion-mixnet routing~\cite{xia2020hybrid} and onion routing with anonymous replies~\cite{10.1007/978-3-030-92075-3_20}, as well as newer anonymous communication systems and scaling techniques for mixnets~\cite{das2024divide}.  

For future work, although we provide a formal security proof in the UC framework and the KEMs used use the formally-verified cryptographic library \emph{libcrux}~\cite{libcrux}, further work could formally verify the entire Sphinx implementation. We could also extend the KEM framework and UC proofs to deal with a variable number of hops as in the original Sphinx.  Although the Outfox scheme is clearly needed for mixnets now in terms of moving mixnets to a post-quantum setting and increasing their performance, we also believe Outfox may have utility outside of mixnets as a generic privacy-enhanced data format due to its ability to deploy post-quantum cryptography with bitwise unlinkability. 

%% file: main.bbl
\begin{thebibliography}{42}


\ifx \showCODEN    \undefined \def \showCODEN     #1{\unskip}     \fi
\ifx \showDOI      \undefined \def \showDOI       #1{#1}\fi
\ifx \showISBNx    \undefined \def \showISBNx     #1{\unskip}     \fi
\ifx \showISBNxiii \undefined \def \showISBNxiii  #1{\unskip}     \fi
\ifx \showISSN     \undefined \def \showISSN      #1{\unskip}     \fi
\ifx \showLCCN     \undefined \def \showLCCN      #1{\unskip}     \fi
\ifx \shownote     \undefined \def \shownote      #1{#1}          \fi
\ifx \showarticletitle \undefined \def \showarticletitle #1{#1}   \fi
\ifx \showURL      \undefined \def \showURL       {\relax}        \fi
\providecommand\bibfield[2]{#2}
\providecommand\bibinfo[2]{#2}
\providecommand\natexlab[1]{#1}
\providecommand\showeprint[2][]{arXiv:#2}

\bibitem[\protect\citeauthoryear{Anderson and Biham}{Anderson and
  Biham}{1996}]%
        {fse-1996-2968}
\bibfield{author}{\bibinfo{person}{Ross~J. Anderson} {and} \bibinfo{person}{Eli
  Biham}.} \bibinfo{year}{1996}\natexlab{}.
\newblock \showarticletitle{Two Practical and Provably Secure Block Ciphers:
  BEARS and LION}. In \bibinfo{booktitle}{\emph{Fast Software Encryption, Third
  International Workshop, Cambridge, UK, February 21-23, 1996, Proceedings}}
  \emph{(\bibinfo{series}{Lecture Notes in Computer Science},
  Vol.~\bibinfo{volume}{1039})}. \bibinfo{publisher}{Springer},
  \bibinfo{pages}{113--120}.
\newblock
\urldef\tempurl%
\url{https://doi.org/10.1007/3-540-60865-6_48}
\showDOI{\tempurl}


\bibitem[\protect\citeauthoryear{Barbosa, Connolly, Duarte, Kaiser, Schwabe,
  Varner, and Westerbaan}{Barbosa et~al\mbox{.}}{2024}]%
        {cryptoeprint:2024/039}
\bibfield{author}{\bibinfo{person}{Manuel Barbosa}, \bibinfo{person}{Deirdre
  Connolly}, \bibinfo{person}{João~Diogo Duarte}, \bibinfo{person}{Aaron
  Kaiser}, \bibinfo{person}{Peter Schwabe}, \bibinfo{person}{Karolin Varner},
  {and} \bibinfo{person}{Bas Westerbaan}.} \bibinfo{year}{2024}\natexlab{}.
\newblock \bibinfo{title}{X-Wing: The Hybrid {KEM} You’ve Been Looking For}.
\newblock \bibinfo{howpublished}{Cryptology {ePrint} Archive, Paper 2024/039}.
\newblock
\urldef\tempurl%
\url{https://doi.org/10.62056/a3qj89n4e}
\showDOI{\tempurl}


\bibitem[\protect\citeauthoryear{Beato, Halunen, and Mennink}{Beato
  et~al\mbox{.}}{2016}]%
        {beato2016improving}
\bibfield{author}{\bibinfo{person}{Filipe Beato}, \bibinfo{person}{Kimmo
  Halunen}, {and} \bibinfo{person}{Bart Mennink}.}
  \bibinfo{year}{2016}\natexlab{}.
\newblock \showarticletitle{Improving the sphinx mix network}. In
  \bibinfo{booktitle}{\emph{International Conference on Cryptology and Network
  Security}}. Springer, \bibinfo{pages}{681--691}.
\newblock


\bibitem[\protect\citeauthoryear{Bernstein}{Bernstein}{2006}]%
        {bernstein2006curve25519}
\bibfield{author}{\bibinfo{person}{Daniel~J Bernstein}.}
  \bibinfo{year}{2006}\natexlab{}.
\newblock \showarticletitle{Curve25519: new Diffie-Hellman speed records}. In
  \bibinfo{booktitle}{\emph{Public Key Cryptography-PKC 2006: 9th International
  Conference on Theory and Practice in Public-Key Cryptography, New York, NY,
  USA, April 24-26, 2006. Proceedings 9}}. Springer, \bibinfo{pages}{207--228}.
\newblock


\bibitem[\protect\citeauthoryear{Bhargavan, Hansen, Kiefer, Schneider-Bensch,
  and Spitters}{Bhargavan et~al\mbox{.}}{2025}]%
        {libcrux}
\bibfield{author}{\bibinfo{person}{Karthikeyan Bhargavan},
  \bibinfo{person}{Lasse~Letager Hansen}, \bibinfo{person}{Franziskus Kiefer},
  \bibinfo{person}{Jonas Schneider-Bensch}, {and} \bibinfo{person}{Bas
  Spitters}.} \bibinfo{year}{2025}\natexlab{}.
\newblock \showarticletitle{Formal Security and Functional Verification of
  Cryptographic Protocol Implementations in Rust}.
\newblock \bibinfo{journal}{\emph{Cryptology ePrint Archive}}
  (\bibinfo{year}{2025}).
\newblock


\bibitem[\protect\citeauthoryear{B{\"o}hme, Danezis, Diaz, K{\"o}psell, and
  Pfitzmann}{B{\"o}hme et~al\mbox{.}}{2004}]%
        {bohme2004pet}
\bibfield{author}{\bibinfo{person}{Rainer B{\"o}hme}, \bibinfo{person}{George
  Danezis}, \bibinfo{person}{Claudia Diaz}, \bibinfo{person}{Stefan
  K{\"o}psell}, {and} \bibinfo{person}{Andreas Pfitzmann}.}
  \bibinfo{year}{2004}\natexlab{}.
\newblock \showarticletitle{On the PET workshop panel “Mix cascades versus
  peer-to-peer: Is one concept superior?”}. In
  \bibinfo{booktitle}{\emph{International Workshop on Privacy Enhancing
  Technologies}}. Springer, \bibinfo{pages}{243--255}.
\newblock


\bibitem[\protect\citeauthoryear{Bos, Ducas, Kiltz, Lepoint, Lyubashevsky,
  Schanck, Schwabe, Seiler, and Stehl{\'e}}{Bos et~al\mbox{.}}{2018}]%
        {bos2018crystals}
\bibfield{author}{\bibinfo{person}{Joppe Bos}, \bibinfo{person}{L{\'e}o Ducas},
  \bibinfo{person}{Eike Kiltz}, \bibinfo{person}{Tancr{\`e}de Lepoint},
  \bibinfo{person}{Vadim Lyubashevsky}, \bibinfo{person}{John~M Schanck},
  \bibinfo{person}{Peter Schwabe}, \bibinfo{person}{Gregor Seiler}, {and}
  \bibinfo{person}{Damien Stehl{\'e}}.} \bibinfo{year}{2018}\natexlab{}.
\newblock \showarticletitle{CRYSTALS-Kyber: a CCA-secure module-lattice-based
  KEM}. In \bibinfo{booktitle}{\emph{2018 IEEE European Symposium on Security
  and Privacy (EuroS\&P)}}. IEEE, \bibinfo{pages}{353--367}.
\newblock


\bibitem[\protect\citeauthoryear{Camenisch, Dubovitskaya, and Rial}{Camenisch
  et~al\mbox{.}}{2016}]%
        {DBLP:conf/crypto/CamenischDR16}
\bibfield{author}{\bibinfo{person}{Jan Camenisch}, \bibinfo{person}{Maria
  Dubovitskaya}, {and} \bibinfo{person}{Alfredo Rial}.}
  \bibinfo{year}{2016}\natexlab{}.
\newblock \showarticletitle{{UC} Commitments for Modular Protocol Design and
  Applications to Revocation and Attribute Tokens}. In
  \bibinfo{booktitle}{\emph{Advances in Cryptology - {CRYPTO} 2016 - 36th
  Annual International Cryptology Conference, Santa Barbara, CA, USA, August
  14-18, 2016, Proceedings, Part {III}}}. \bibinfo{pages}{208--239}.
\newblock
\urldef\tempurl%
\url{https://doi.org/10.1007/978-3-662-53015-3\_8}
\showDOI{\tempurl}


\bibitem[\protect\citeauthoryear{Canetti}{Canetti}{2001}]%
        {DBLP:conf/focs/Canetti01}
\bibfield{author}{\bibinfo{person}{Ran Canetti}.}
  \bibinfo{year}{2001}\natexlab{}.
\newblock \showarticletitle{Universally Composable Security: {A} New Paradigm
  for Cryptographic Protocols}. In \bibinfo{booktitle}{\emph{42nd Annual
  Symposium on Foundations of Computer Science, {FOCS} 2001, 14-17 October
  2001, Las Vegas, Nevada, {USA}}}. \bibinfo{publisher}{{IEEE} Computer
  Society}, \bibinfo{pages}{136--145}.
\newblock
\urldef\tempurl%
\url{https://doi.org/10.1109/SFCS.2001.959888}
\showDOI{\tempurl}


\bibitem[\protect\citeauthoryear{Chan and Rogaway}{Chan and Rogaway}{2022}]%
        {cryptoeprint:2022/1260}
\bibfield{author}{\bibinfo{person}{John Chan} {and} \bibinfo{person}{Phillip
  Rogaway}.} \bibinfo{year}{2022}\natexlab{}.
\newblock \bibinfo{title}{On Committing Authenticated Encryption}.
\newblock \bibinfo{howpublished}{Cryptology {ePrint} Archive, Paper 2022/1260}.
\newblock
\urldef\tempurl%
\url{https://eprint.iacr.org/2022/1260}
\showURL{%
\tempurl}


\bibitem[\protect\citeauthoryear{Chaum}{Chaum}{1981}]%
        {DBLP:journals/cacm/Chaum81}
\bibfield{author}{\bibinfo{person}{David Chaum}.}
  \bibinfo{year}{1981}\natexlab{}.
\newblock \showarticletitle{Untraceable Electronic Mail, Return Addresses, and
  Digital Pseudonyms}.
\newblock \bibinfo{journal}{\emph{Commun. {ACM}}} \bibinfo{volume}{24},
  \bibinfo{number}{2} (\bibinfo{year}{1981}), \bibinfo{pages}{84--88}.
\newblock
\urldef\tempurl%
\url{https://doi.org/10.1145/358549.358563}
\showDOI{\tempurl}


\bibitem[\protect\citeauthoryear{Chen, Asoni, Barrera, Danezis, and
  Perrig}{Chen et~al\mbox{.}}{2015}]%
        {chen2015hornet}
\bibfield{author}{\bibinfo{person}{Chen Chen}, \bibinfo{person}{Daniele~E
  Asoni}, \bibinfo{person}{David Barrera}, \bibinfo{person}{George Danezis},
  {and} \bibinfo{person}{Adrain Perrig}.} \bibinfo{year}{2015}\natexlab{}.
\newblock \showarticletitle{HORNET: High-speed onion routing at the network
  layer}. In \bibinfo{booktitle}{\emph{Proceedings of the 22nd ACM SIGSAC
  Conference on Computer and Communications Security}}.
  \bibinfo{pages}{1441--1454}.
\newblock


\bibitem[\protect\citeauthoryear{Connolly, Schwabe, and Westerbaan}{Connolly
  et~al\mbox{.}}{2024}]%
        {connolly-cfrg-xwing-kem-02}
\bibfield{author}{\bibinfo{person}{Deirdre Connolly}, \bibinfo{person}{Peter
  Schwabe}, {and} \bibinfo{person}{Bas Westerbaan}.}
  \bibinfo{year}{2024}\natexlab{}.
\newblock \bibinfo{booktitle}{\emph{{X-Wing: general-purpose hybrid
  post-quantum KEM}}}.
\newblock \bibinfo{type}{Internet-Draft} draft-connolly-cfrg-xwing-kem-02.
  \bibinfo{institution}{Internet Engineering Task Force}.
\newblock
\urldef\tempurl%
\url{https://datatracker.ietf.org/doc/draft-connolly-cfrg-xwing-kem/02/}
\showURL{%
\tempurl}
\newblock
\shownote{Work in Progress.}


\bibitem[\protect\citeauthoryear{Danezis, Dingledine, and Mathewson}{Danezis
  et~al\mbox{.}}{2003}]%
        {danezis2003mixminion}
\bibfield{author}{\bibinfo{person}{George Danezis}, \bibinfo{person}{Roger
  Dingledine}, {and} \bibinfo{person}{Nick Mathewson}.}
  \bibinfo{year}{2003}\natexlab{}.
\newblock \showarticletitle{Mixminion: Design of a type III anonymous remailer
  protocol}. In \bibinfo{booktitle}{\emph{2003 Symposium on Security and
  Privacy, 2003.}} IEEE, \bibinfo{pages}{2--15}.
\newblock


\bibitem[\protect\citeauthoryear{Danezis and Goldberg}{Danezis and
  Goldberg}{2009}]%
        {DBLP:conf/sp/DanezisG09}
\bibfield{author}{\bibinfo{person}{George Danezis} {and} \bibinfo{person}{Ian
  Goldberg}.} \bibinfo{year}{2009}\natexlab{}.
\newblock \showarticletitle{Sphinx: {A} Compact and Provably Secure Mix
  Format}. In \bibinfo{booktitle}{\emph{30th {IEEE} Symposium on Security and
  Privacy {(SP} 2009), 17-20 May 2009, Oakland, California, {USA}}}.
  \bibinfo{publisher}{{IEEE} Computer Society}, \bibinfo{pages}{269--282}.
\newblock
\urldef\tempurl%
\url{https://doi.org/10.1109/SP.2009.15}
\showDOI{\tempurl}


\bibitem[\protect\citeauthoryear{Daniel and Tschorsch}{Daniel and
  Tschorsch}{2024}]%
        {daniel2024poster}
\bibfield{author}{\bibinfo{person}{Erik Daniel} {and} \bibinfo{person}{Florian
  Tschorsch}.} \bibinfo{year}{2024}\natexlab{}.
\newblock \showarticletitle{Poster: On Integrating Sphinx in IPFS}. In
  \bibinfo{booktitle}{\emph{Proceedings of the 2024 ACM on Internet Measurement
  Conference}}. \bibinfo{pages}{753--754}.
\newblock


\bibitem[\protect\citeauthoryear{Das, Meiser, Mohammadi, and Kate}{Das
  et~al\mbox{.}}{2024}]%
        {das2024divide}
\bibfield{author}{\bibinfo{person}{Debajyoti Das}, \bibinfo{person}{Sebastian
  Meiser}, \bibinfo{person}{Esfandiar Mohammadi}, {and} \bibinfo{person}{Aniket
  Kate}.} \bibinfo{year}{2024}\natexlab{}.
\newblock \showarticletitle{Divide and funnel: a scaling technique for
  mix-networks}. In \bibinfo{booktitle}{\emph{2024 IEEE 37th Computer Security
  Foundations Symposium (CSF)}}. IEEE, \bibinfo{pages}{49--64}.
\newblock


\bibitem[\protect\citeauthoryear{Diaz, Halpin, and Kiayias}{Diaz
  et~al\mbox{.}}{2024}]%
        {nym2024whitepaper}
\bibfield{author}{\bibinfo{person}{Claudia Diaz}, \bibinfo{person}{Harry
  Halpin}, {and} \bibinfo{person}{Aggelos Kiayias}.}
  \bibinfo{year}{2024}\natexlab{}.
\newblock \bibinfo{title}{The Nym Network: The Next Generation of Privacy
  Infrastructure}.
\newblock \bibinfo{howpublished}{\url{https://nym.com/nym-whitepaper.pdf}}.
\newblock
\newblock
\shownote{Whitepaper.}


\bibitem[\protect\citeauthoryear{Diaz, Murdoch, and Troncoso}{Diaz
  et~al\mbox{.}}{2010}]%
        {diaz2010impact}
\bibfield{author}{\bibinfo{person}{Claudia Diaz}, \bibinfo{person}{Steven~J
  Murdoch}, {and} \bibinfo{person}{Carmela Troncoso}.}
  \bibinfo{year}{2010}\natexlab{}.
\newblock \showarticletitle{Impact of network topology on anonymity and
  overhead in low-latency anonymity networks}. In
  \bibinfo{booktitle}{\emph{International Symposium on Privacy Enhancing
  Technologies Symposium}}. Springer, \bibinfo{pages}{184--201}.
\newblock


\bibitem[\protect\citeauthoryear{Galbraith}{Galbraith}{2012}]%
        {galbraith2012mathematics}
\bibfield{author}{\bibinfo{person}{Steven~D Galbraith}.}
  \bibinfo{year}{2012}\natexlab{}.
\newblock \bibinfo{booktitle}{\emph{Mathematics of public key cryptography}}.
\newblock \bibinfo{publisher}{Cambridge University Press}.
\newblock


\bibitem[\protect\citeauthoryear{Hugenroth, Kleppmann, and Beresford}{Hugenroth
  et~al\mbox{.}}{2021}]%
        {hugenroth2021rollercoaster}
\bibfield{author}{\bibinfo{person}{Daniel Hugenroth}, \bibinfo{person}{Martin
  Kleppmann}, {and} \bibinfo{person}{Alastair~R Beresford}.}
  \bibinfo{year}{2021}\natexlab{}.
\newblock \showarticletitle{Rollercoaster: An Efficient $\{$Group-Multicast$\}$
  Scheme for Mix Networks}. In \bibinfo{booktitle}{\emph{30th USENIX Security
  Symposium (USENIX Security 21)}}. \bibinfo{pages}{3433--3450}.
\newblock


\bibitem[\protect\citeauthoryear{Josefsson}{Josefsson}{2025}]%
        {josefsson-mceliece-02}
\bibfield{author}{\bibinfo{person}{Simon Josefsson}.}
  \bibinfo{year}{2025}\natexlab{}.
\newblock \bibinfo{booktitle}{\emph{{Classic McEliece}}}.
\newblock \bibinfo{type}{Internet-Draft} draft-josefsson-mceliece-02.
  \bibinfo{institution}{Internet Engineering Task Force}.
\newblock
\urldef\tempurl%
\url{https://datatracker.ietf.org/doc/draft-josefsson-mceliece/02/}
\showURL{%
\tempurl}
\newblock
\shownote{Work in Progress.}


\bibitem[\protect\citeauthoryear{Kate and Goldberg}{Kate and Goldberg}{2010}]%
        {kate2010using}
\bibfield{author}{\bibinfo{person}{Aniket Kate} {and} \bibinfo{person}{Ian
  Goldberg}.} \bibinfo{year}{2010}\natexlab{}.
\newblock \showarticletitle{Using sphinx to improve onion routing circuit
  construction}. In \bibinfo{booktitle}{\emph{International conference on
  financial cryptography and data security}}. Springer,
  \bibinfo{pages}{359--366}.
\newblock


\bibitem[\protect\citeauthoryear{Kiayias and Litos}{Kiayias and Litos}{2020}]%
        {kiayias2020composable}
\bibfield{author}{\bibinfo{person}{Aggelos Kiayias} {and}
  \bibinfo{person}{Orfeas Stefanos~Thyfronitis Litos}.}
  \bibinfo{year}{2020}\natexlab{}.
\newblock \showarticletitle{A composable security treatment of the lightning
  network}. In \bibinfo{booktitle}{\emph{2020 IEEE 33rd Computer Security
  Foundations Symposium (CSF)}}. IEEE, \bibinfo{pages}{334--349}.
\newblock


\bibitem[\protect\citeauthoryear{Kloo{\ss}, Rupp, Schadt, Strufe, and
  Weis}{Kloo{\ss} et~al\mbox{.}}{2024}]%
        {klooss2024eror}
\bibfield{author}{\bibinfo{person}{Michael Kloo{\ss}}, \bibinfo{person}{Andy
  Rupp}, \bibinfo{person}{Daniel Schadt}, \bibinfo{person}{Thorsten Strufe},
  {and} \bibinfo{person}{Christiane Weis}.} \bibinfo{year}{2024}\natexlab{}.
\newblock \showarticletitle{EROR: Efficient Repliable Onion Routing with Strong
  Provable Privacy}.
\newblock \bibinfo{journal}{\emph{Cryptology ePrint Archive}}
  (\bibinfo{year}{2024}).
\newblock


\bibitem[\protect\citeauthoryear{Krawczyk}{Krawczyk}{2010}]%
        {cryptoeprint:2010/264}
\bibfield{author}{\bibinfo{person}{Hugo Krawczyk}.}
  \bibinfo{year}{2010}\natexlab{}.
\newblock \bibinfo{title}{Cryptographic Extraction and Key Derivation: The
  {HKDF} Scheme}.
\newblock \bibinfo{howpublished}{Cryptology {ePrint} Archive, Paper 2010/264}.
\newblock
\urldef\tempurl%
\url{https://eprint.iacr.org/2010/264}
\showURL{%
\tempurl}


\bibitem[\protect\citeauthoryear{Kuhn, Beck, and Strufe}{Kuhn
  et~al\mbox{.}}{2020}]%
        {DBLP:conf/sp/KuhnBS20}
\bibfield{author}{\bibinfo{person}{Christiane Kuhn}, \bibinfo{person}{Martin
  Beck}, {and} \bibinfo{person}{Thorsten Strufe}.}
  \bibinfo{year}{2020}\natexlab{}.
\newblock \showarticletitle{Breaking and (Partially) Fixing Provably Secure
  Onion Routing}. In \bibinfo{booktitle}{\emph{2020 {IEEE} Symposium on
  Security and Privacy, {SP} 2020, San Francisco, CA, USA, May 18-21, 2020}}.
  \bibinfo{publisher}{{IEEE}}, \bibinfo{pages}{168--185}.
\newblock
\urldef\tempurl%
\url{https://doi.org/10.1109/SP40000.2020.00039}
\showDOI{\tempurl}


\bibitem[\protect\citeauthoryear{Kuhn, Hofheinz, Rupp, and Strufe}{Kuhn
  et~al\mbox{.}}{2021}]%
        {10.1007/978-3-030-92075-3_20}
\bibfield{author}{\bibinfo{person}{Christiane Kuhn}, \bibinfo{person}{Dennis
  Hofheinz}, \bibinfo{person}{Andy Rupp}, {and} \bibinfo{person}{Thorsten
  Strufe}.} \bibinfo{year}{2021}\natexlab{}.
\newblock \showarticletitle{Onion Routing with Replies}. In
  \bibinfo{booktitle}{\emph{Advances in Cryptology -- ASIACRYPT 2021}},
  \bibfield{editor}{\bibinfo{person}{Mehdi Tibouchi} {and}
  \bibinfo{person}{Huaxiong Wang}} (Eds.). \bibinfo{publisher}{Springer
  International Publishing}, \bibinfo{address}{Cham},
  \bibinfo{pages}{573--604}.
\newblock
\showISBNx{978-3-030-92075-3}


\bibitem[\protect\citeauthoryear{Langley, Hamburg, and Turner}{Langley
  et~al\mbox{.}}{2016}]%
        {langley2016elliptic}
\bibfield{author}{\bibinfo{person}{Adam Langley}, \bibinfo{person}{Mike
  Hamburg}, {and} \bibinfo{person}{Sean Turner}.}
  \bibinfo{year}{2016}\natexlab{}.
\newblock \bibinfo{booktitle}{\emph{Elliptic curves for security}}.
\newblock \bibinfo{type}{{T}echnical {R}eport}. \bibinfo{institution}{RFC
  7748}.
\newblock


\bibitem[\protect\citeauthoryear{M{\"o}ller}{M{\"o}ller}{2003}]%
        {moller2003provably}
\bibfield{author}{\bibinfo{person}{Bodo M{\"o}ller}.}
  \bibinfo{year}{2003}\natexlab{}.
\newblock \showarticletitle{Provably secure public-key encryption for
  length-preserving chaumian mixes}. In
  \bibinfo{booktitle}{\emph{Cryptographers’ Track at the RSA Conference}}.
  Springer, \bibinfo{pages}{244--262}.
\newblock


\bibitem[\protect\citeauthoryear{Pfitzmann and Pfitzmann}{Pfitzmann and
  Pfitzmann}{1990}]%
        {pfitzmann1990break}
\bibfield{author}{\bibinfo{person}{Birgit Pfitzmann} {and}
  \bibinfo{person}{Andreas Pfitzmann}.} \bibinfo{year}{1990}\natexlab{}.
\newblock \showarticletitle{How to break the direct RSA-implementation of
  mixes}. In \bibinfo{booktitle}{\emph{Advances in Cryptology—EUROCRYPT’89:
  Workshop on the Theory and Application of Cryptographic Techniques Houthalen,
  Belgium, April 10--13, 1989 Proceedings 8}}. Springer,
  \bibinfo{pages}{373--381}.
\newblock


\bibitem[\protect\citeauthoryear{Piotrowska}{Piotrowska}{2021}]%
        {piotrowska2021studying}
\bibfield{author}{\bibinfo{person}{Ania~M Piotrowska}.}
  \bibinfo{year}{2021}\natexlab{}.
\newblock \showarticletitle{Studying the anonymity trilemma with a
  discrete-event mix network simulator}. In
  \bibinfo{booktitle}{\emph{Proceedings of the 20th Workshop on Workshop on
  Privacy in the Electronic Society}}. \bibinfo{pages}{39--44}.
\newblock


\bibitem[\protect\citeauthoryear{Piotrowska, Hayes, Elahi, Meiser, and
  Danezis}{Piotrowska et~al\mbox{.}}{2017}]%
        {DBLP:conf/uss/PiotrowskaHEMD17}
\bibfield{author}{\bibinfo{person}{Ania~M. Piotrowska}, \bibinfo{person}{Jamie
  Hayes}, \bibinfo{person}{Tariq Elahi}, \bibinfo{person}{Sebastian Meiser},
  {and} \bibinfo{person}{George Danezis}.} \bibinfo{year}{2017}\natexlab{}.
\newblock \showarticletitle{The Loopix Anonymity System}. In
  \bibinfo{booktitle}{\emph{26th {USENIX} Security Symposium, {USENIX} Security
  2017, Vancouver, BC, Canada, August 16-18, 2017}},
  \bibfield{editor}{\bibinfo{person}{Engin Kirda} {and} \bibinfo{person}{Thomas
  Ristenpart}} (Eds.). \bibinfo{publisher}{{USENIX} Association},
  \bibinfo{pages}{1199--1216}.
\newblock
\urldef\tempurl%
\url{https://www.usenix.org/conference/usenixsecurity17/technical-sessions/presentation/piotrowska}
\showURL{%
\tempurl}


\bibitem[\protect\citeauthoryear{Publication}{Publication}{2023}]%
        {nistmlkem}
\bibfield{author}{\bibinfo{person}{Federal Information Processing~Standards
  Publication}.} \bibinfo{year}{2023}\natexlab{}.
\newblock \bibinfo{booktitle}{\emph{Module-Lattice-Based Key-Encapsulation
  Mechanism Standard}}.
\newblock \bibinfo{type}{{T}echnical {R}eport}. \bibinfo{institution}{NIST}.
\newblock


\bibitem[\protect\citeauthoryear{Rogaway}{Rogaway}{2002}]%
        {DBLP:conf/ccs/Rogaway02}
\bibfield{author}{\bibinfo{person}{Phillip Rogaway}.}
  \bibinfo{year}{2002}\natexlab{}.
\newblock \showarticletitle{Authenticated-encryption with associated-data}. In
  \bibinfo{booktitle}{\emph{Proceedings of the 9th {ACM} Conference on Computer
  and Communications Security, {CCS} 2002, Washington, DC, USA, November 18-22,
  2002}}, \bibfield{editor}{\bibinfo{person}{Vijayalakshmi Atluri}} (Ed.).
  \bibinfo{publisher}{{ACM}}, \bibinfo{pages}{98--107}.
\newblock
\urldef\tempurl%
\url{https://doi.org/10.1145/586110.586125}
\showDOI{\tempurl}


\bibitem[\protect\citeauthoryear{Rohrer and Tschorsch}{Rohrer and
  Tschorsch}{2020}]%
        {rohrer2020counting}
\bibfield{author}{\bibinfo{person}{Elias Rohrer} {and} \bibinfo{person}{Florian
  Tschorsch}.} \bibinfo{year}{2020}\natexlab{}.
\newblock \showarticletitle{Counting down thunder: Timing attacks on privacy in
  payment channel networks}. In \bibinfo{booktitle}{\emph{Proceedings of the
  2nd ACM Conference on Advances in Financial Technologies}}.
  \bibinfo{pages}{214--227}.
\newblock


\bibitem[\protect\citeauthoryear{Schadt, Coijanovic, Weis, and Strufe}{Schadt
  et~al\mbox{.}}{2024}]%
        {schadt2024polysphinx}
\bibfield{author}{\bibinfo{person}{Daniel Schadt}, \bibinfo{person}{Christoph
  Coijanovic}, \bibinfo{person}{Christiane Weis}, {and}
  \bibinfo{person}{Thorsten Strufe}.} \bibinfo{year}{2024}\natexlab{}.
\newblock \showarticletitle{PolySphinx: Extending the Sphinx Mix Format With
  Better Multicast Support}. In \bibinfo{booktitle}{\emph{2024 IEEE Symposium
  on Security and Privacy (SP)}}. IEEE, \bibinfo{pages}{4386--4404}.
\newblock


\bibitem[\protect\citeauthoryear{Scherer, Weis, and Strufe}{Scherer
  et~al\mbox{.}}{2024}]%
        {DBLP:journals/popets/SchererWS24a}
\bibfield{author}{\bibinfo{person}{Philip Scherer}, \bibinfo{person}{Christiane
  Weis}, {and} \bibinfo{person}{Thorsten Strufe}.}
  \bibinfo{year}{2024}\natexlab{}.
\newblock \showarticletitle{Provable Security for the Onion Routing and Mix
  Network Packet Format Sphinx}.
\newblock \bibinfo{journal}{\emph{Proc. Priv. Enhancing Technol.}}
  \bibinfo{volume}{2024}, \bibinfo{number}{4} (\bibinfo{year}{2024}),
  \bibinfo{pages}{755--783}.
\newblock
\urldef\tempurl%
\url{https://doi.org/10.56553/POPETS-2024-0140}
\showDOI{\tempurl}


\bibitem[\protect\citeauthoryear{Shimshock, Staats, and Hopper}{Shimshock
  et~al\mbox{.}}{2008}]%
        {shimshock2008breaking}
\bibfield{author}{\bibinfo{person}{Erik Shimshock}, \bibinfo{person}{Matt
  Staats}, {and} \bibinfo{person}{Nick Hopper}.}
  \bibinfo{year}{2008}\natexlab{}.
\newblock \showarticletitle{Breaking and provably fixing minx}. In
  \bibinfo{booktitle}{\emph{Privacy Enhancing Technologies: 8th International
  Symposium, PETS 2008 Leuven, Belgium, July 23-25, 2008 Proceedings 8}}.
  Springer, \bibinfo{pages}{99--114}.
\newblock


\bibitem[\protect\citeauthoryear{Song, Yang, Peng, Liang, Liu, and Hu}{Song
  et~al\mbox{.}}{2024}]%
        {song2024cloudnet}
\bibfield{author}{\bibinfo{person}{Xiaowei Song}, \bibinfo{person}{Mingqian
  Yang}, \bibinfo{person}{Wei Peng}, \bibinfo{person}{Mengen Liang},
  \bibinfo{person}{Ling Liu}, {and} \bibinfo{person}{Ning Hu}.}
  \bibinfo{year}{2024}\natexlab{}.
\newblock \showarticletitle{CloudNet: Building a Data-Plane for Anonymous
  Communication Network Based on Cloud Service}. In
  \bibinfo{booktitle}{\emph{2024 IEEE 9th International Conference on Data
  Science in Cyberspace (DSC)}}. IEEE, \bibinfo{pages}{575--582}.
\newblock


\bibitem[\protect\citeauthoryear{Stainton}{Stainton}{2023}]%
        {cryptoeprint:2023/1960}
\bibfield{author}{\bibinfo{person}{David~Anthony Stainton}.}
  \bibinfo{year}{2023}\natexlab{}.
\newblock \bibinfo{title}{Post Quantum Sphinx}.
\newblock \bibinfo{howpublished}{Cryptology {ePrint} Archive, Paper 2023/1960}.
\newblock
\urldef\tempurl%
\url{https://eprint.iacr.org/2023/1960}
\showURL{%
\tempurl}


\bibitem[\protect\citeauthoryear{Xia, Chen, Su, Pan, and Su}{Xia
  et~al\mbox{.}}{2020}]%
        {xia2020hybrid}
\bibfield{author}{\bibinfo{person}{Yusheng Xia}, \bibinfo{person}{Rongmao
  Chen}, \bibinfo{person}{Jinshu Su}, \bibinfo{person}{Chen Pan}, {and}
  \bibinfo{person}{Han Su}.} \bibinfo{year}{2020}\natexlab{}.
\newblock \showarticletitle{Hybrid routing: towards resilient routing in
  anonymous communication networks}. In \bibinfo{booktitle}{\emph{ICC 2020-2020
  IEEE International Conference on Communications (ICC)}}. IEEE,
  \bibinfo{pages}{1--7}.
\newblock


\end{thebibliography}
